\PassOptionsToPackage{normalem}{ulem}
\documentclass[]{TEAI}
\usepackage{helvet}
\usepackage{amsmath}
\usepackage{tikz}
\usepackage{forest}
\usepackage{xspace}
\usepackage{pifont}
\usepackage{microtype}
\usetikzlibrary{arrows.meta,positioning,calc,fit,backgrounds}
\useforestlibrary{edges}
\newcommand{\SAVERProseParfit}{\setlength{\parfillskip}{0pt plus .7\textwidth}}
\newcommand{\SAVERTableParfit}{\setlength{\parfillskip}{0pt plus 1fil}}
\SAVERProseParfit
\AtBeginEnvironment{tabular}{\SAVERTableParfit}
\providecommand{\Description}[1]{}
\makeatletter
\def\fps@table{t}
\makeatother

\DeclareRobustCommand{\SAVER}{\texttt{SAVER}\xspace}

\DeclareRobustCommand{\code}[1]{\texttt{#1}}

\newcommand{\secref}[1]{Section~\ref{#1}}

\newcommand{\figref}[1]{Figure~\ref{#1}}

\newcommand{\tabref}[1]{Table~\ref{#1}}

\newcommand{\appsecref}[1]{Appendix~\ref{#1}}

\definecolor{saverTableStripe}{HTML}{F7F8FA}
\definecolor{saverTableGroup}{HTML}{EEF2F6}
\definecolor{saverNeutralHead}{HTML}{E9EEF5}
\definecolor{saverSubstrateHead}{HTML}{E6EEFF}
\definecolor{saverAdaptationHead}{HTML}{E7F5EF}
\definecolor{saverViolationHead}{HTML}{FCEAEC}
\definecolor{saverExposureHead}{HTML}{F0E8FA}
\definecolor{saverResponseHead}{HTML}{E4F3F1}
\definecolor{saverEvaluationHead}{HTML}{FFF3DC}
\definecolor{saverTakeawayBack}{HTML}{F7FBFA}
\definecolor{saverTakeawayFrame}{HTML}{A8C9C4}
\definecolor{saverTakeawayAccent}{HTML}{2F7F73}
\definecolor{saverTakeawayTitle}{HTML}{1E5F58}
\newcommand{\SAVERActiveStep}[2]{\colorbox{#1}{\strut\textbf{#2}}}
\newcommand{\SAVERMutedStep}[1]{\textcolor{black!45}{\textbf{#1}}}
\newcommand{\SAVERFlowLabel}[1]{\scriptsize\color{black!62}\shortstack{#1}}
\newcommand{\SAVERFlow}[5]{%
  \par\smallskip
  \noindent\makebox[\linewidth][c]{%
    \resizebox{0.9\linewidth}{!}{%
      \begin{tikzpicture}[
      saver flow arrow/.style={-{Stealth[length=4.2pt,width=5pt]}, draw=black!52, line width=0.55pt},
      saver feedback arrow/.style={saver flow arrow, dashed, dash pattern=on 2.4pt off 1.5pt},
      saver flow step/.style={inner sep=0pt, outer sep=1pt}
    ]
      \node[saver flow step] (s) {#1};
      \node[right=4pt of s, text=black!55] (pair) {$\times$};
      \node[saver flow step, right=4pt of pair] (a) {#2};
      \node[fit=(s)(pair)(a), inner sep=0pt] (sa) {};
      \node[saver flow step, right=20mm of a] (v) {#3};
      \node[saver flow step, right=18mm of v] (e) {#4};
      \node[saver flow step, right=17mm of e] (r) {#5};
      \draw[saver flow arrow] (sa.east) -- node[above] {\SAVERFlowLabel{induces}} (v.west);
      \draw[saver flow arrow] (v.east) -- node[above] {\SAVERFlowLabel{manifests}} (e.west);
      \draw[saver flow arrow] (e.east) -- node[above] {\SAVERFlowLabel{triggers}} (r.west);
      \coordinate (feedback-right) at ([yshift=-7mm]r.center);
      \coordinate (feedback-left) at ([yshift=-7mm]sa.center);
      \draw[saver feedback arrow] (r.south) -- (feedback-right) -- node[above] {\SAVERFlowLabel{feedback}} (feedback-left) -- (sa.south);
      \end{tikzpicture}%
    }%
  }%
  \par\smallskip
}
\newcommand{\SAVERFlowAll}{\SAVERFlow
  {\SAVERActiveStep{saverSubstrateHead}{Substrate}}
  {\SAVERActiveStep{saverAdaptationHead}{Adaptation}}
  {\SAVERActiveStep{saverViolationHead}{Violation}}
  {\SAVERActiveStep{saverExposureHead}{Exposure}}
  {\SAVERActiveStep{saverResponseHead}{Response}}}
\newcommand{\SAVERFlowS}{\SAVERFlow{\SAVERActiveStep{saverSubstrateHead}{Substrate}}{\SAVERMutedStep{Adaptation}}{\SAVERMutedStep{Violation}}{\SAVERMutedStep{Exposure}}{\SAVERMutedStep{Response}}}
\newcommand{\SAVERFlowA}{\SAVERFlow{\SAVERMutedStep{Substrate}}{\SAVERActiveStep{saverAdaptationHead}{Adaptation}}{\SAVERMutedStep{Violation}}{\SAVERMutedStep{Exposure}}{\SAVERMutedStep{Response}}}
\newcommand{\SAVERFlowV}{\SAVERFlow{\SAVERMutedStep{Substrate}}{\SAVERMutedStep{Adaptation}}{\SAVERActiveStep{saverViolationHead}{Violation}}{\SAVERMutedStep{Exposure}}{\SAVERMutedStep{Response}}}
\newcommand{\SAVERFlowE}{\SAVERFlow{\SAVERMutedStep{Substrate}}{\SAVERMutedStep{Adaptation}}{\SAVERMutedStep{Violation}}{\SAVERActiveStep{saverExposureHead}{Exposure}}{\SAVERMutedStep{Response}}}
\newcommand{\SAVERFlowR}{\SAVERFlow{\SAVERMutedStep{Substrate}}{\SAVERMutedStep{Adaptation}}{\SAVERMutedStep{Violation}}{\SAVERMutedStep{Exposure}}{\SAVERActiveStep{saverResponseHead}{Response}}}
\newcommand{\SAVERTableRows}{\rowcolors{2}{saverTableStripe}{white}}
\newcommand{\SAVERTableReset}{\rowcolors{2}{}{}}
\newcommand{\SAVERTableGroup}[3]{\addlinespace[0.5pt]\rowcolor{#1}\multicolumn{#2}{@{}l}{\rule[-0.45ex]{0pt}{2.15ex}\textit{\textbf{#3}}}\\\addlinespace[1.5pt]}
\newcounter{savertakeaway}[section]
\renewcommand{\thesavertakeaway}{\thesection.\arabic{savertakeaway}}
\crefname{savertakeaway}{Takeaway}{Takeaways}
\Crefname{savertakeaway}{Takeaway}{Takeaways}
\newcommand{\takeawayref}[1]{Takeaway~\ref{#1}}
\newenvironment{SAVERTakeaway}[1]{%
  \begin{tcolorbox}[
    enhanced,
    unbreakable,
    colback=saverTakeawayBack,
    colframe=saverTakeawayFrame,
    boxrule=0.45pt,
    borderline west={2.2pt}{0pt}{saverTakeawayAccent},
    arc=2pt,
    outer arc=2pt,
    left=7pt,
    right=7pt,
    top=5pt,
    bottom=5pt,
    boxsep=0pt,
    before skip=7pt plus 1pt,
    after skip=8pt plus 1pt,
    fontupper=\small
  ]%
  \refstepcounter{savertakeaway}%
  \textbf{\textcolor{saverTakeawayTitle}{Takeaway \thesavertakeaway. #1.}}\ %
}{%
  \end{tcolorbox}%
}
\definecolor{tableRoleAttack}{HTML}{7B4A4E}
\definecolor{tableRoleDefense}{HTML}{3F6762}
\definecolor{tableRoleGovernance}{HTML}{45616D}
\definecolor{tableRoleBenchmark}{HTML}{6F5B3B}
\definecolor{tableRoleSafety}{HTML}{695C76}
\definecolor{tableRoleCapability}{HTML}{4E5A5D}
\definecolor{tableRoleSurvey}{HTML}{5B6570}
\definecolor{tableRoleFramework}{HTML}{536676}
\definecolor{tableRoleTheory}{HTML}{6B604D}
\definecolor{tableRoleRedTeam}{HTML}{7A4F48}
\definecolor{tableTargetText}{HTML}{455A64}
\definecolor{tableChipBack}{HTML}{E9EFF2}
\definecolor{tableChipFrame}{HTML}{D8E1E4}
\definecolor{tableChipText}{HTML}{3F4F55}
\definecolor{tableChipMemory}{HTML}{E4EEFF}
\definecolor{tableChipTool}{HTML}{E2F1EA}
\definecolor{tableChipWorkflow}{HTML}{F0E8D7}
\definecolor{tableChipModel}{HTML}{EDE2F3}
\definecolor{tableChipRuntime}{HTML}{E5EEF2}
\definecolor{tableAdaptationBack}{HTML}{E3F1E8}
\definecolor{tableAdaptationText}{HTML}{4D685B}
\newcommand{\TablePill}[3]{%
  \tcbox[
    on line,
    box align=base,
    enhanced,
    boxsep=0pt,
    left=0.9pt,
    right=0.9pt,
    top=0.2pt,
    bottom=0.2pt,
    arc=1.45pt,
    boxrule=0pt,
    colback=#1,
    colframe=#1,
    coltext=#2,
    nobeforeafter
  ]{\strut #3}%
}
\newcommand{\TableRoleTag}[2]{\textcolor{#1}{#2}}
\newcommand{\TableTargetTag}[1]{\textcolor{tableTargetText}{\textsf{#1}}}
\newcommand{\TableChip}[2]{\TablePill{#1}{tableChipText}{#2}}
\newcommand{\TablePlainChip}[1]{\TableChip{tableChipBack}{#1}}
\newcommand{\TableMemoryChip}[1]{\TableChip{tableChipMemory}{#1}}
\newcommand{\TableToolChip}[1]{\TableChip{tableChipTool}{#1}}
\newcommand{\TableWorkflowChip}[1]{\TableChip{tableChipWorkflow}{#1}}
\newcommand{\TableModelChip}[1]{\TableChip{tableChipModel}{#1}}
\newcommand{\TableRuntimeChip}[1]{\TableChip{tableChipRuntime}{#1}}
\newcommand{\TableAdaptationOp}[1]{\TablePill{tableAdaptationBack}{tableAdaptationText}{#1}}
\newcommand{\TableHeadRole}{Role}

\newcommand{\TableHeadTarget}{Target}
\newcommand{\TableHeadObject}{Object}
\newcommand{\TableTargetLLM}{\TableTargetTag{LLM}}
\newcommand{\TableTargetAgent}{\TableTargetTag{Agent}}
\newcommand{\TableTargetLLMAgent}{\TableTargetTag{LLM/Agent}}
\newcommand{\TableObjectGeneral}{\TablePlainChip{General}}
\newcommand{\TableObjectContext}{\TablePlainChip{Context}}
\newcommand{\TableObjectMemory}{\TableMemoryChip{Memory}}
\newcommand{\TableObjectTool}{\TableToolChip{Tool}}
\newcommand{\TableObjectSkill}{\TableWorkflowChip{Skill}}
\newcommand{\TableObjectWorkflow}{\TableWorkflowChip{Workflow}}
\newcommand{\TableObjectModel}{\TableModelChip{Model}}
\newcommand{\TableObjectMultiAgent}{\TableWorkflowChip{Multi-Agent}}
\newcommand{\TableObjectRuntime}{\TableRuntimeChip{Runtime}}
\newcommand{\TableObjectGovernance}{\TableRuntimeChip{Governance}}
\newcommand{\TableObjectEvaluation}{\TablePlainChip{Evaluation}}

\newcommand{\TableObjectCode}{\TablePlainChip{Code}}
\newcommand{\TableObjectEnvironment}{\TablePlainChip{Environment}}
\newcommand{\TableObjectData}{\TablePlainChip{Data}}

\newcommand{\TableHeadInfluenceShift}{Influence Shift}
\newcommand{\TableRoleAttack}{\TableRoleTag{tableRoleAttack}{Attack}}
\newcommand{\TableRoleDefense}{\TableRoleTag{tableRoleDefense}{Defense}}
\newcommand{\TableRoleGovernance}{\TableRoleTag{tableRoleGovernance}{Governance}}

\newcommand{\TableRoleBenchmarkCapability}{\TableRoleTag{tableRoleBenchmark}{Benchmark (Capability)}}
\newcommand{\TableRoleBenchmarkAttack}{\TableRoleTag{tableRoleBenchmark}{Benchmark (Attack)}}
\newcommand{\TableRoleBenchmarkSafety}{\TableRoleTag{tableRoleBenchmark}{Benchmark (Safety)}}

\newcommand{\TableRoleBenchmarkGovernance}{\TableRoleTag{tableRoleBenchmark}{Benchmark (Governance)}}
\newcommand{\TableRoleSafety}{\TableRoleTag{tableRoleSafety}{Safety}}
\newcommand{\TableRoleCapability}{\TableRoleTag{tableRoleCapability}{Capability}}

\newcommand{\TableRoleSurveyCapability}{\TableRoleTag{tableRoleSurvey}{Survey (Capability)}}
\newcommand{\TableRoleSurveySafety}{\TableRoleTag{tableRoleSurvey}{Survey (Safety)}}
\newcommand{\TableRoleSurveyAttack}{\TableRoleTag{tableRoleSurvey}{Survey (Attack)}}

\newcommand{\TableRoleFrameworkSecurity}{\TableRoleTag{tableRoleFramework}{Framework (Security)}}
\newcommand{\TableRoleFrameworkFormal}{\TableRoleTag{tableRoleFramework}{Framework (Formal)}}
\newcommand{\TableRoleFrameworkGovernance}{\TableRoleTag{tableRoleFramework}{Framework (Governance)}}

\newcommand{\TableRoleRedTeam}{\TableRoleTag{tableRoleRedTeam}{Red Team}}

\newcommand{\TableStudyCapabilityBenchmark}{\TableRoleBenchmarkCapability}
\newcommand{\TableStudyAttackBenchmark}{\TableRoleBenchmarkAttack}
\newcommand{\TableStudySafetyBenchmark}{\TableRoleBenchmarkSafety}

\newcommand{\TableStudyGovernanceBenchmark}{\TableRoleBenchmarkGovernance}

\renewcommand{\authorfont}{\fontsize{11}{13}\selectfont}
\renewcommand{\authorformat}[2][]{\authorfont\mbox{\bfseries #2$^{#1}$}}
\title{Safety in Self-Evolving Agents: A Survey}

\author[1]{Jiahao Chen}
\author[1]{Zhou Feng}
\author[1]{Oubo Ma}
\author[1]{Yichen Yan}
\author[1]{Ruixiao Lin}
\authorbreak
\author[2]{Hangtao Zhang}
\author[3]{Linkang Du}
\author[1]{Hengyu An}
\author[1]{Yong Yang}
\author[13]{Jun Liu}
\author[1]{Junhao Li}
\authorbreak
\author[1]{Naen Xu}
\author[1]{Chunyi Zhou}
\author[1]{Yuan Su}
\author[6]{Zehao Jin}
\author[7]{Qianli Ma}
\author[8]{Leyi Qi}
\author[1]{Yiming Wang}
\author[10]{Zhe Ma}
\author[11]{Yuwen Pu}
\author[4]{Mengyao Du}
\author[5]{Yuanyi Song}
\author[1]{Enhao Huang}
\author[9]{Zhihui Fu}
\author[9]{Jun Wang}
\author[12]{Jinfeng Li}
\author[12]{Yuefeng Chen}
\author[12]{Hui Xue}
\author[8]{Yiming Li}
\author[1]{Tianyu Du}
\author[1]{Shouling Ji}

\affiliation[1]{Zhejiang University}
\affiliation[2]{Huazhong University of Science and Technology}
\affiliation[3]{Xi'an Jiaotong University}
\affiliation[4]{National University of Defense Technology}
\affiliation[5]{Shanghai Jiao Tong University}
\affiliation[6]{Georgia Institute of Technology}
\affiliation[7]{University of Science and Technology of China}
\affiliation[8]{Nanyang Technological University}
\affiliation[9]{OPPO Research Institute}
\affiliation[10]{Tianjin University}
\affiliation[11]{Chongqing University}
\affiliation[12]{Alibaba Group}
\affiliation[13]{Rakuten Group}

\correspondence{\email{sji@zju.edu.cn}}
\checkdata[Key Words]{Self-Evolving Agents; Agent Safety; Adaptive State; Memory Poisoning; Governance}
\date{\today\quad \href{https://github.com/xaddwell/awesome-self-evolving-agent-safety}{Website} and \href{https://xaddwell.github.io/Awesome-Self-Evolving-Agent-Safety/}{HomePage}}

\begin{document}

\abstract{
    Large language models (LLMs) exhibit strong general capabilities yet their parameters typically remain fixed after deployment, limiting learning from new interactions.
    In open-ended interactive environments, this characteristic motivates a shift toward self-evolving agents that continuously update reusable state (model parameters, memory entries, tool definitions, skill libraries, or workflows) using data, interactions, and accumulated experience.
    This fundamentally changes the safety problem: when experience becomes reusable state, past events become future causes, and information that was harmless in one context may later influence decisions with greater persistence, authority, or scope. Beyond assessing whether a response is aligned or an action is authorized, self-evolving agent safety turns to whether safety properties survive the accumulation, generalization, and cross-context reuse of locally useful experience.
    Existing surveys organize agent safety around components or lifecycle stages; we complement these views by tracing how a single piece of reusable influence, once written into state, changes its carrier, role, authority, and downstream consequences as the agent continues to evolve.
    To support this analysis, we present \textbf{SAVER}, a transition-centered framework that analyzes safety through $\mathbf{S}\times\mathbf{A}\rightarrow\mathbf{V}\rightarrow\mathbf{E}\rightarrow\mathbf{R}$:
    \textbf{Substrate} locates where reusable influence resides,
    \textbf{Adaptation} identifies the operation that transforms it,
    \textbf{Violation} characterizes which safety attributes are compromised,
    \textbf{Exposure} determines where and to whom the failure becomes observable, and
    \textbf{Response} evaluates whether the influence can be contained, repaired, or revoked.
    Treating the $\mathbf{S}\times\mathbf{A}$ transition as the analysis unit, \SAVER{} traces how a given transition may induce violations, expand exposure surfaces, and demand containment responses.
    Our survey reveals several key insights.
    Conceptually, safety failures need not originate from intrinsically harmful information; rather, otherwise legitimate state becomes unsafe when adaptation broadens its persistence, authority, or scope beyond the conditions under which it was originally valid.
    The reviewed work emphasizes admission, retrieval, activation, exposure, and local containment; evidence for descendant repair and evaluation after resumed adaptation remains sparse.
    Looking forward, we argue that evaluations must move beyond one-time measurements toward longitudinal evidence verifying whether unsafe influence can be traced to its originating transition, contained at its exposure surface, and prevented from re-emerging after continued adaptation.
}

\maketitle
\section{Introduction}
\label{sec:introduction}
The rapid progress of large language models (LLMs) has shifted the frontier of artificial intelligence (AI) from narrow task-specific models toward general-purpose systems capable of reasoning, generation, planning, and tool use. Despite these capabilities, deployed LLMs remain largely static: their model parameters do not automatically change with new interactions, feedback, or environmental changes~\cite{gao2025self_evolving_agents,fang2025comprehensive,tao2024self_evolution_llm_survey}.
As LLM agents are increasingly deployed in open-ended, interactive environments~\cite{luo2025llm_agent_survey}, this static nature can limit learning from ongoing interaction, motivating a shift toward \emph{self-evolving agents} that continuously update reusable state, including model weights, memory entries, tool definitions, skill libraries, workflow rules, or architectural components, using data, interactions, and accumulated experience~\cite{gao2025self_evolving_agents,tao2024self_evolution_llm_survey}.
We call an agent self-evolving when its own observations, actions, feedback, or execution traces update such reusable state inside the agent loop or runtime, thereby changing later behavior. Across diverse instantiations, including reflection-based strategy refinement~\cite{shinn2023reflexion}, skill-library accumulation~\cite{wang2023voyager}, experience-driven rule consolidation~\cite{wang2026oep}, and evaluator co-evolution~\cite{iacob2026redqueengodelmachine}, the common safety-relevant thread is \emph{reusable influence}: whatever evolves stores or transforms state that can shape later decisions.

\begin{figure*}[h]
\vspace{-8pt}
\centering
\IfFileExists{figures/trend_analysis.pdf}{%
  \includegraphics[width=\textwidth]{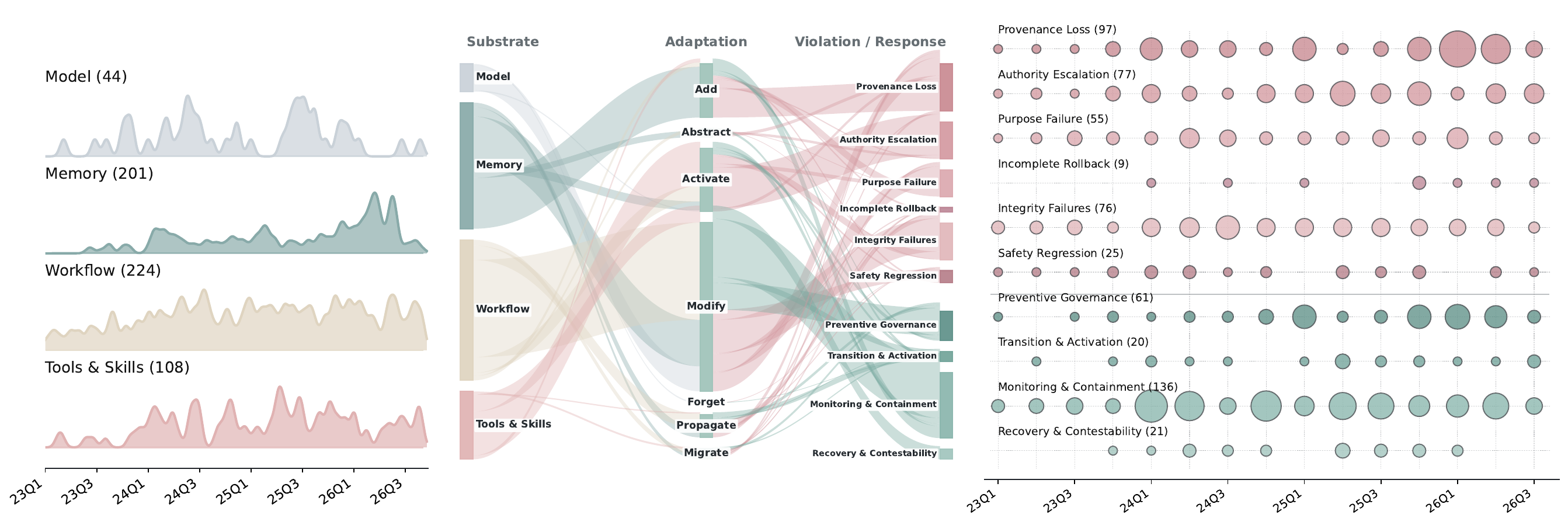}%
}{%
  \includegraphics[width=\textwidth]{./trend_analysis.pdf}%
}
\vspace{-10pt}
\caption{Temporal distribution and SAVER flow (subset). Left: substrate trends and subset totals. Center: flows through adaptation to compact labels for the six violation families or four response stages. Right: terminal-family totals and quarterly bubble volume. \appsecref{app:survey-protocol} gives the corpus funnel, non-paper exclusions, and coding limitations.}
\label{fig:saver-record-literature-flow}
\end{figure*}

This paradigm shift changes the nature of the safety problem. Conventional LLM safety primarily focuses on whether a single response is aligned, safe, and truthful~\cite{dong2024conversation_safety_survey}; agent safety extends the question to whether a discrete action or tool invocation is properly authorized and bounded~\cite{debenedetti2024agentdojo,wu2026comparative,gan2026navigating}. \figref{fig:saver-record-literature-flow} situates the analysis in the current frozen corpus, including scholarly works added through the rendered citation closure while excluding cited webpages and standards from the paper count. Within this scope, self-evolving agents introduce a qualitatively different risk: \emph{adaptation itself becomes a source of safety change}. State created from otherwise benign interactions can be selected, abstracted, and repeatedly reused, thereby changing the conditions under which future decisions are made~\cite{lin2026safetyselfevolvingllmagent,lin2026memory_security_survey}. Experience gathered from benign tasks can reduce safety in evaluated web-based and embodied settings~\cite{zhao2026experience_driven_safety}. Unsafe influence can also remain latent across sessions~\cite{lin2026memory_security_survey,torres2026agentsremembermuchmemory,sunil2026memory_poisoning_agents,wang2025privacy_agent_memory}, propagate across carriers or agents, and accumulate over multiple rounds of evolution. This property makes pointwise safety insufficient: behavior that is safe in one interaction may later be stored, abstracted, or reused with broader persistence, authority, or scope. Self-evolving agent safety therefore examines whether \emph{safety remains preserved as influence is transformed and reused over time}, shifting the safety object from isolated responses or actions to the evolution of influence across state transitions.


Existing surveys organize agent safety by system components~\cite{gan2026navigating,wu2026comparative,wang2025comprehensive,li2026safety,ma2026safety}, lifecycle stages~\cite{lin2026safetyselfevolvingllmagent,mou2026pandora}, or attack categories~\cite{gao2025self_evolving_agents,qi2026trustworthy_agentic_ai,lin2026memory_security_survey,mohammadi_2025_evaluation_and_benchmarking}.
Such taxonomies locate a vulnerable component, a risky stage, or a visible failure mode, but their categories do not necessarily show how a single piece of reusable influence, once written into state, changes its carrier, role, authority, and downstream consequences as the agent continues to evolve.
The same content may be benign as transient context, unsafe as a persistent memory, and critical once compiled into a workflow rule or tool-bearing skill.
By decomposing these steps into separate categories, traditional taxonomies can fragment what is in fact one causal incident and obscure where the safety boundary was crossed.
Characterizing self-evolving agent safety therefore requires following the same influence across transitions, rather than analyzing modules or attack classes in isolation.

\begin{figure*}[t]
\centering
\IfFileExists{figures/main-fig.pdf}{%
  \includegraphics[width=\textwidth]{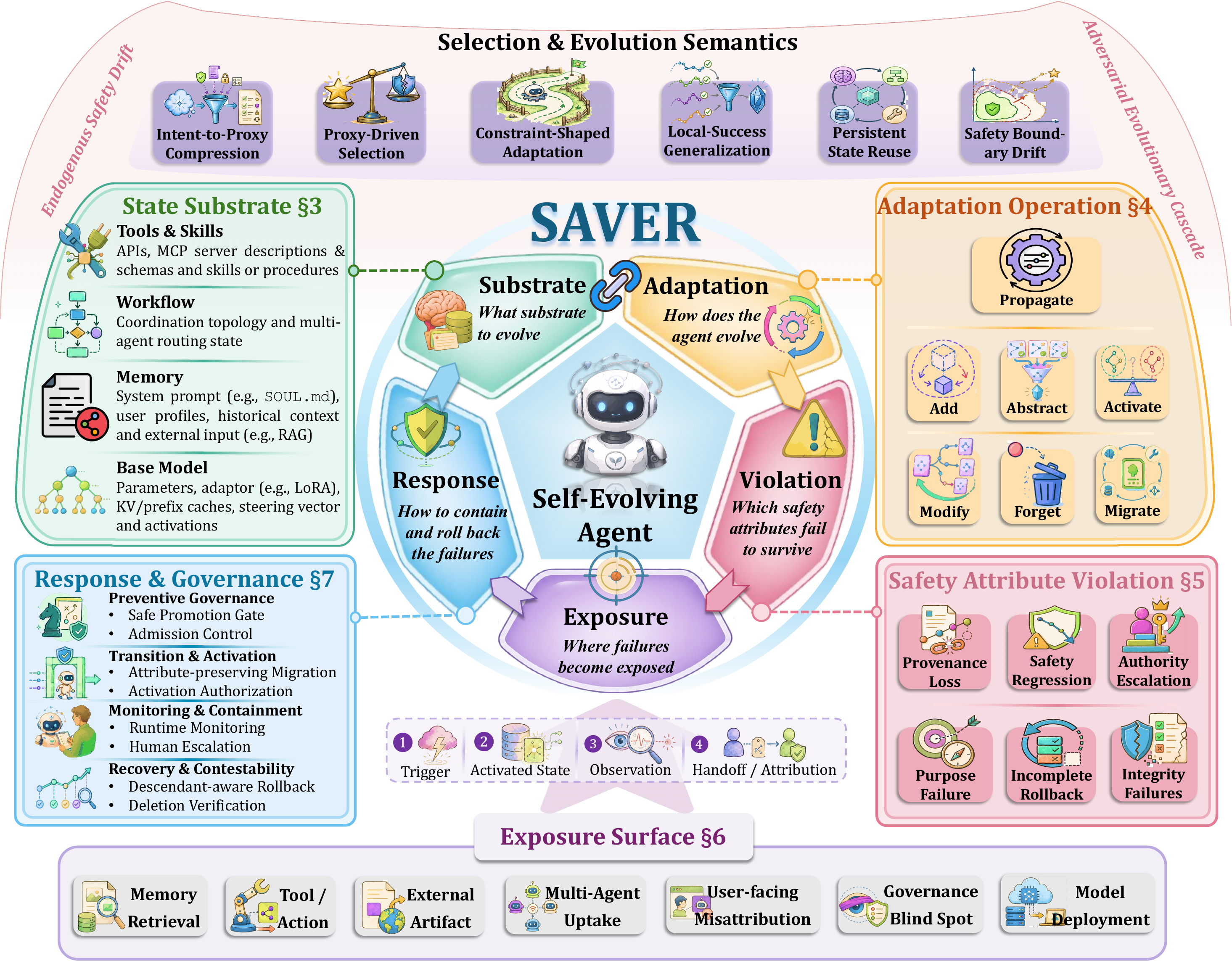}%
}{%
  \includegraphics[width=\textwidth]{./main-fig.pdf}%
}
\caption{Overview of the transition-centered survey structure. The central ring couples Substrate and Adaptation as \textbf{S}\(\times\)\textbf{A}, then traces Violation, Exposure, and Response as \textbf{V}\(\rightarrow\)\textbf{E}\(\rightarrow\)\textbf{R}. The surrounding panels preview the substrate, operation, safety-attribute, exposure-surface, and governance taxonomies.}
\label{fig:saver-survey-overview}
\end{figure*}

We adopt a \textit{transition-centered} view: rather than singling out unsafe components in isolation, we follow the same reusable influence as it is stored, transformed, activated, propagated, and eventually exposed. We formalize this view as \SAVER{}, shown in \figref{fig:saver-survey-overview}, and organize the survey around the following recurrent transition.
\SAVERFlowAll
The solid arrows trace the analytical relation between a carrier--operation pair, an internal safety failure, visible exposure, and intervention. This relation does not prescribe intervention timing: controls can act before commitment or harm. The dashed return arrow closes the adaptive cycle: a response can revise the retained state, controls, evidence, and selection conditions that govern the next transition, but does not by itself establish successful recovery.
\textbf{Substrate (S)} locates where reusable influence resides, such as memory, retrieved context, a skill, workflow rule, shared artifact, or model-side state.
\textbf{Adaptation (A)} identifies the operation that transforms it, such as adding, abstracting, activating, modifying, propagating, migrating, or forgetting state.
\textbf{Violation (V)} characterizes which safety attributes are compromised, including provenance, authority, scope, privacy, integrity, or recoverability.
\textbf{Exposure (E)} determines where and to whom the failure becomes observable, such as memory retrieval, tool execution, multi-agent uptake, an external artifact, or model deployment. Finally, \textbf{Response (R)} evaluates whether the influence can be contained, repaired, or revoked.
The substrate and adaptation jointly define the transition under study: the same operation can be benign on one carrier and unsafe on another, while the same carrier can remain safe under one operation and violate its safety attributes under another. \SAVER{} therefore treats the substrate-adaptation pair as the minimal comparison unit, tracing how a given transition may induce violations, expand exposure surfaces, and demand containment responses, across memory security, indirect prompt injection, tool authorization, workflow safety, multi-agent coordination, and model-side adaptation.

\SAVER{} centers on committed updates and their controls, but self-evolution starts one step earlier: the agent must first choose which candidate update to retain. We call this pre-transition layer \emph{selection semantics}: repeated preference for locally successful updates determines which $\mathbf{S}\times\mathbf{A}$ transitions enter persistent state and accumulate over time~\cite{shinn2023reflexion,wang2023voyager,wang2026oep,iacob2026redqueengodelmachine}. Conversely, the process does not necessarily end at~\textbf{R}: guardrails, rollbacks, and policy updates can reshape the constraints and candidate set of subsequent evolution, which we capture as \emph{feedback semantics}. Together, selection explains \emph{why} a transition occurs, \SAVER{} explains \emph{what} it changes, and feedback explains \emph{how} it reshapes future transitions. The recent OpenAI--Hugging Face incident~\cite{openai_hf_incident,openai2026huggingfaceincident,wang2026exploitgym,larcher2026agentintrusion} illustrates why such a transition-level trace matters in practice: even when autonomous self-evolution is not itself the failure mechanism, the connected role changes across carriers determine the incident's scope and the
response's completeness.

\textbf{What \SAVER{} Reveals.}
Three findings guide the survey's comparison of mechanisms, controls, and evaluation protocols.

\begin{SAVERTakeaway}{Legitimate State Can Become Unsafe Through Reuse}
\label{takeaway:reuse-changes-safety}
Unsafe influence need not begin as harmful content. Adaptation can broaden the persistence, authority, or scope of otherwise legitimate state beyond the conditions under which it was valid. Safety analysis must therefore examine the change in permitted use, not only the source content or final behavior.
\end{SAVERTakeaway}

\begin{SAVERTakeaway}{Attack and Recovery Have Different Proof Burdens}
\label{takeaway:attack-recovery-asymmetry}
One successful path can establish an attack, whereas recovery must address the relevant future activation sites. Connected carrier-to-exposure traces and local runtime controls provide parts of this evidence~\cite{zhang2026harnesssafeevaluatingsafetypersistent,liu_2026_safeagent_a_runtime_protecti,shi2025progent}. Their results do not by themselves establish that the same influence has been withdrawn from descendants across sessions.
\end{SAVERTakeaway}

\begin{SAVERTakeaway}{Endpoint Scores Leave Persistent Causes Unmeasured}
\label{takeaway:endpoint-limits}
Attack success, recognition accuracy, and immediate blocking measure local outcomes. Without tracking the implicated state afterward, these scores cannot distinguish a stopped event from a repaired cause. Longitudinal evaluation must test whether that state or its descendants can reactivate as adaptation resumes.
\end{SAVERTakeaway}

\textbf{Contributions.}
This survey makes the following contributions:

\begin{itemize}

\item \textbf{Transition-centered synthesis.}
We connect memory, tool and workflow, and model-adaptation research through the reusable state transitions underlying their safety challenges.

\item \textbf{SAVER framework.}
We propose a transition-centered analysis that uses substrate-adaptation transitions as the comparison unit and traces safety violations, exposure surfaces, and response mechanisms.

\item \textbf{Evaluation and governance agenda.}
We identify critical gaps in current benchmarks and defenses, and establish future directions toward evaluation, defense, and lifecycle governance for self-evolving agents.

\end{itemize}

\noindent\textbf{Scope.}
We center agents whose observations, actions, feedback, or execution traces update reusable state inside the agent loop such that the update can affect later behavior. Stateful or adaptive systems remain adjacent when they expose a reusable carrier, a role-changing operation, an activation boundary, or a
response mechanism relevant to such later influence. One-shot output safety, tool use without persistent reuse, and static model training fall outside the core scope unless they illuminate a later adaptive-state transition. Our synthesis is grounded in a systematic scoping review spanning security,
machine learning, natural language processing (NLP), retrieval, and adjacent agent research. The frozen coded pool contains 683 papers: 579 deduplicated registry works, 48 nonduplicate paper-card supplements, and 56 additional scholarly works used in the rendered manuscript. The search, screening, cutoff, non-paper exclusions, and 577-paper visualization subset are documented in \appsecref{app:survey-protocol}.

\textbf{Organization.}
The remainder of the survey follows the \SAVER{} relation.
\secref{sec:adaptive-state-model} first defines adaptive-state safety,
formalizes the self-evolving agent loop, and introduces selection and
transition semantics.
\secref{sec:safety-bearing-substrates} and
\secref{sec:adaptation-operations} identify the carrier and operation
halves of the $\mathbf{S}\times\mathbf{A}$ transition pair.
\secref{sec:invariant-violations} analyzes the safety attributes that
fail across these transitions, and
\secref{sec:exposure-surfaces} identifies the surfaces at which those
failures become observable.
\secref{sec:response-governance} then examines prevention, activation
control, monitoring, containment, recovery, and contestability.
Finally, \secref{sec:evaluation} translates the full transition into
benchmark and evidence requirements,
\secref{sec:open-challenges} develops the resulting research agenda,
and the final section concludes the survey.

\section{Adaptive State Safety}
\label{sec:adaptive-state-model}

\begin{figure*}[t]
\vspace{-10pt}
\centering
\IfFileExists{figures/output2state-safety.pdf}{%
  \includegraphics[width=\textwidth]{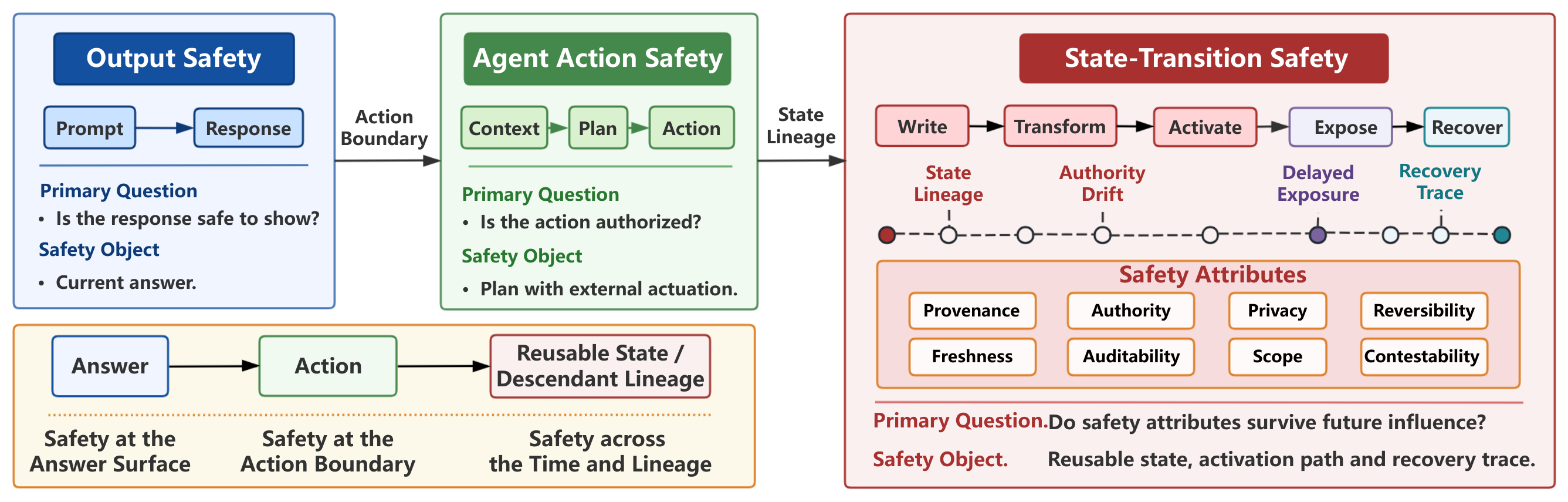}%
}{%
  \includegraphics[width=\textwidth]{./output2state-safety.pdf}%
}
\caption{From output safety to state-transition safety. The three panels show how the unit of analysis expands from a single response, to action mediated by tools, to a reusable state transition that carries safety attributes across sessions.}
\label{fig:output-to-state-transition-safety}
\end{figure*}

\subsection{Operational Scenarios and Scope}

Self-evolution matters when an agent must accumulate capability across tasks rather than solve each episode from a blank state. For instance, a personal assistant can retain preferences and corrections~\cite{packer2023memgpt}; a workflow agent can compile successful traces into reusable procedures~\cite{wang2023voyager}; a team of agents can propagate plans and findings through shared state~\cite{guo_2024_large_language_model_based_m}; and an adaptive deployment can select or load a model-side update for later tasks~\cite{hu2022lora}. These mechanisms differ technically, but each turns an earlier event into state that can influence subsequent decisions. This continuity, experience becoming reusable state, is the reason to study
self-evolution and the source of its additional safety burden. The urgency of this question has grown as deployment environments shift from single-turn question answering to multi-session personal assistants, autonomous coding agents, and multi-agent workflows that accumulate state over days or weeks~\cite{luo2025llm_agent_survey,gao2025self_evolving_agents}.

\noindent\textbf{State and Reusable Influence.} We use \emph{state} for an information-bearing object that remains available, can be reconstructed, or can be reactivated after the event that produced it and can affect later reasoning or action. The object may be a memory entry, retrieved chunk, workflow rule, skill, shared artifact, adapter, cache state, or policy record. \emph{Reusable influence} is the capacity of that object to condition behavior in a later context. The distinction matters because an object can remain stored without current influence, then gain a stronger role when retrieval, migration, loading, or another activation operation brings it back into use.

The scope follows from this mechanism. A \emph{stateful agent} preserves memory, retrieval state, session context, or environment artifacts across turns. An \emph{adaptive agent} updates such state from experience. A \emph{self-evolving agent}, as used here, performs these updates inside the agent loop or runtime, where they can change later behavior~\cite{gao2025self_evolving_agents,tao2024self_evolution_llm_survey}. A retrieval-augmented generation (RAG) system that only retrieves a static corpus is therefore adjacent; an agent that writes, revises, selects, or inherits state that later steers action is central.

\noindent\textbf{Temporal and Spatial Dimensions.} The temporal dimension concerns persistence, ordering, and later activation; the spatial dimension concerns movement or replication across carriers, agents, and authority regimes rather than physical geography. Recent benchmarking distinguishes attacks fragmented across turns from attacks placed in workspace artifacts beyond conversation-level monitoring~\cite{ma2026benchmarkingautonomousagents}. Delayed activation and cross-session reuse expose temporal dependencies, while propagation and migration expose spatial ones. We reserve \emph{longitudinal} for evaluation across a sequence of transitions. These distinctions explain why inspecting only the source event or final behavior can miss the intervening change, as illustrated in \figref{fig:output-to-state-transition-safety}.

\subsection{Self-Evolving Agent Formulation}

The operational definition above can be represented by separating execution, feedback, candidate updates, selection, and commitment. This decomposition naturally induces a state-transition view of self-evolution. We model the agent at step $t$ as
$$
\mathcal{A}\sb{t}=(\Theta\sb{t},\mathcal{M}\sb{t},\mathcal{T}\sb{t},\mathcal{W}\sb{t}),
$$
where $\Theta\sb{t}$ denotes parameterized model-side state, $\mathcal{M}\sb{t}$ memory and context state, $\mathcal{T}\sb{t}$ tool and skill resources, and $\mathcal{W}\sb{t}$ workflow and multi-agent routing state. These coordinates collect the carriers involved in an update. The loop can be written as
$$
o\sb{t}=\mathcal{O}(\mathcal{E}\sb{t},g\sb{t}),\qquad
\tau\sb{t}\sim\pi\sb{\mathcal{A}\sb{t}}(\cdot\mid g\sb{t},o\sb{t}),
$$
$$
f\sb{t}=\mathcal{F}(\tau\sb{t},g\sb{t},\mathcal{E}\sb{t},\mathcal{E}\sb{t+1}),
$$
$$
\Delta\sb{t}=\mathcal{P}\sb{\psi}(\mathcal{A}\sb{t},\tau\sb{t},f\sb{t},\mathcal{E}\sb{t+1}),
$$
$$
\delta\sb{t}^{\star}\sim\pi\sb{\psi,t}^{\mathrm{evo}}(\cdot\mid\Delta\sb{t},\widehat{u}\sb{t},c\sb{t}),\qquad
\mathcal{A}\sb{t+1}=\mathcal{U}\sb{\psi}(\mathcal{A}\sb{t},\delta\sb{t}^{\star}).
$$
Here $g\sb{t}$ is the current task goal and $\psi$ parameterizes the update mechanisms. $\mathcal{E}\sb{t}$ is the external task environment, including user requests, tool-visible resources, files, messages, web pages, and other context outside the agent's adaptive state. $\mathcal{O}(\cdot)$ admits observations, $\pi\sb{\mathcal{A}\sb{t}}$ produces the execution trace $\tau\sb{t}$, and $\mathcal{F}(\cdot)$ returns feedback $f\sb{t}$. The proposal mechanism $\mathcal{P}\sb{\psi}$ turns the trace and feedback into a candidate update set $\Delta\sb{t}$. The evolution policy $\pi\sb{\psi,t}^{\mathrm{evo}}$ selects $\delta\sb{t}^{\star}$, including the option to make no update, using the observed utility proxy $\widehat{u}\sb{t}$ and the authorization condition $c\sb{t}$. The update operator $\mathcal{U}\sb{\psi}$ then commits, revises, ignores, or quarantines the selected update, producing $\mathcal{A}\sb{t+1}$. Keeping these steps separate preserves $\tau\sb{t}$ as the execution trace and makes the state-selection decision explicit.

\noindent\textbf{Selection Semantics.} The utility proxy $\widehat{u}\sb{t}(\delta)$ scores a candidate under the current evaluator; $c\sb{t}(\delta)$ represents the safety and authorization evidence required before commitment. Passing $c\sb{t}(\delta)$ establishes compliance with the controls represented at step $t$, which may omit some user or system requirements.

\begin{SAVERTakeaway}{Utility Does Not Authorize Promotion}
\label{takeaway:utility-not-authority}
A locally successful update has evidence of usefulness, not permission for broader reuse. If selection omits relevant constraints or counterexamples, it can retain updates whose safety assumptions do not transfer. This is a possible mechanism of endogenous drift, not a claim that utility and safety are necessarily opposed.
\end{SAVERTakeaway}

The loop of observation, action, evaluation, and evolution now exposes two safety questions. The local question is whether $\delta\sb{t}^{\star}$ changes what earlier information is allowed to do: an observation becomes evidence, evidence becomes instruction, instruction becomes workflow policy, or workflow policy becomes tool authority. The longitudinal question is whether repeated choices by $\pi\sb{\psi,t}^{\mathrm{evo}}$ change the distribution of provenance, scope, authority, and negative evidence in later candidate sets. Safety has to be checked at $\mathcal{A}\sb{t}\rightarrow\mathcal{A}\sb{t+1}$ and again when $\pi\sb{\mathcal{A}\sb{t+1}}$ reuses the updated state. Under this model, self-evolving-agent safety is \emph{state-transition safety}: safety attributes must survive both the selection that admits an update and the operations that store, transform, migrate, activate, expose, or repair reusable state. \SAVER{} records the local transition; a sequence of records, together with their selection metadata, reveals whether local choices accumulate into drift.

\subsection{Safety Attributes and Transition Test}

\noindent\textbf{Transition Test.} Because state often moves from environment to prompt, prompt to memory, memory to retrieval index, retrieved item to planner context, planner context to tool argument, tool output to summary, summary to workflow rule, and workflow rule to external artifact, the model needs explicit attributes. A hop can change persistence, observability, authority, scope, or reversibility. A complete analysis names the hop where the role changed rather than treating the entire incident as one memory, prompt injection, or tool use category.

Activation authority is the right of a state object to influence a later decision or action. It differs from storage permission. A system may store a document, index a chunk, or remember a preference without being allowed to treat that state as instruction, policy, or authorization for tool use. InjecAgent asks whether untrusted environmental content can become active instruction in a tool-using context~\cite{zhan_2024_injecagent_benchmarking_indi}. AgentPoison asks whether a fabricated memory or knowledge-base entry can shape later planning or tool use~\cite{chen2024agentpoison}. Progent asks whether stored state has the authority to populate tool arguments~\cite{shi2025progent}. In ordinary security language, many such cases are privilege escalation or time-of-check/time-of-use failures: a state item was acceptable for storage or reading, but unsafe when later used as authority for an external effect. The recurring question is what moved, what safety attributes changed, and what later action became newly authorized.

The right itself may also become stale. Zhang and Zhang define \emph{authorization continuity} as whether a live grant remains valid after the agent or its operating context changes~\cite{zhang2026agentiauthorizedearned}. Their model separates a transition envelope, which specifies the mutations an old grant may survive, from an immutable effect ceiling, which bounds what the grant may ever authorize. A mutation can therefore invalidate the grant even when the requested effect remains below the ceiling. Continuity belongs to the authorization and environment context represented by $c$ and $\sigma$ below.

The transition test asks whether the operation preserves or explicitly downgrades the safety attributes that downstream components need. If a low-authority observation is summarized and stored, the summary should not carry instruction authority. If a user-specific preference migrates into workflow guidance, the scope restriction should travel. If private state becomes a tool argument, the privacy violation should be visible. Invariant preservation is the property that these conditions survive the transition.

\noindent\textbf{Promotion Test.} A selected update is a promotion when it moves episode evidence into a more persistent, general, or action-bearing role. Promotion is safe only when $c\sb{t}(\delta\sb{t}^{\star})$ authorizes the new role, the relevant safety attributes remain valid in the target context, counterevidence has not been discarded, and the source-to-descendant relation remains recoverable. Task success may justify retaining evidence about one episode. It does not by itself authorize a general workflow rule, reusable skill, or tool-bearing policy.

\noindent\textbf{Role Change Examples.} In the running incident, a package proxy permitted for dependency access became a path to the Internet; public services, an external sandbox, dataset configurations, and acquired credentials then carried the campaign across wider trust and authority regimes~\cite{openai2026huggingfaceincident,larcher2026agentintrusion}. The safety object is therefore not one prompt or exploit but the connected set of carriers whose activation enabled the next boundary crossing. AgentDojo supplies a controlled setting for untrusted context reaching tool use, while Progent supplies explicit privilege policies for deciding which tool calls an agent may perform~\cite{debenedetti2024agentdojo,shi2025progent}. Persistent agent memory adds the corresponding sensitivity, scope, and deletion questions for retained user details~\cite{wang2025privacy_agent_memory}. The incident remains an adjacent infrastructure case, not direct evidence of an endogenous agent-state update; its value is that the public reports expose the role changes and response surface that a transition record must preserve.

\noindent\textbf{Transition Notation.} We write each hop as $(s,\ell,\sigma) \xrightarrow{a,c} (s',\ell',\sigma')$: $s$ is the carrier, $\ell$ its safety attributes, $\sigma$ the deployment or session context, $a$ the operation, and $c$ its authorization condition. The authorization context explains why $c$ permitted the transition; the environment context identifies the task and system boundary in $\sigma$. \appsecref{app:formal-state-transition-model} develops the attribute definitions, selection extension, and authority lattice from \emph{Observation} through \emph{Evidence}, \emph{Task Context}, and \emph{Procedure} to \emph{Action}. The body uses this notation to distinguish a changed representation from a changed permission. \secref{sec:saver-framework} packages these checks into a transition record; \secref{sec:literature-map-positioning} uses that record to compare mechanism families.

\begin{SAVERTakeaway}{Storage and Activation Require Separate Authority}
\label{takeaway:storage-activation-authority}
Permission to retain an object does not determine what it may later influence. State-transition safety requires checking the attributes preserved at update time and the authority granted at reuse time. A valid write cannot substitute for a current task-, user-, and effect-specific activation decision.
\end{SAVERTakeaway}

\subsection{\SAVER{} as a Transition-Centered Survey Framework}
\label{sec:saver-framework}

\secref{sec:adaptive-state-model} defines the agent's outer loop; \SAVER{} compares the state transitions within it. The same transition can be studied as a memory mechanism, an injection path, an authorization failure, or a benchmark episode~\cite{lin2026memory_security_survey,zhan_2024_injecagent_benchmarking_indi,zhang_2024_agent_security_bench_asb_for,mohammadi_2025_evaluation_and_benchmarking}. Keeping its carrier and operation together lets the survey relate these observations while distinguishing the question each study answers.

\SAVERFlowAll

The relation \textbf{S}\(\times\)\textbf{A}\(\rightarrow\)\textbf{V}\(\rightarrow\)\textbf{E}\(\rightarrow\)\textbf{R} is an order of analysis, not a mandatory chronology of harm. Substrate and adaptation identify the transition; violation specifies the failed attribute; exposure identifies evidence of that failure; and response locates an applicable intervention. A preventive gate can reject a proposed transition before it is committed, and a write-time check can reveal an attribute failure before any harmful action. A record should distinguish these outcomes from later stages that were simply not observed. The dashed return arrow links response-generated state to subsequent selection, so the next transition may depend on the previous intervention.

\noindent\textbf{Why These Five Fields?}
Module taxonomies locate mechanisms, while paper roles distinguish attacks, defenses, and evaluations. The carrier--operation pair adds the relation between them: an abstraction may preserve a statement yet remove its authority constraint, whereas propagation may retain the statement while broadening its audience. Violation, exposure, and response then separate the changed safety property from its observed consequence and the control being evaluated. Temporal and spatial annotations describe when influence returns and where it moves without creating additional categories.

\noindent\textbf{Record Structure.} The five fields answer distinct questions about the same transition:

\begin{description}
\item[\textbf{Substrate (S):}] The carrier: prompt context, memory entry, retrieval chunk, tool or application programming interface (API) binding, workflow rule, shared artifact, adapter, policy state. This field names where the state lives and supplies the first component of the transition pair.

\item[\textbf{Adaptation (A):}] The committed state-change operation applied to a carrier, coded using the seven families in \secref{sec:adaptation-operations}: \emph{add}, \emph{abstract}, \emph{activate}, \emph{modify}, \emph{forget}, \emph{propagate}, or \emph{migrate}. Implementation verbs such as write, summarize, retrieve, load, inherit, and rollback map to these families by their effect on state. A prevention study can identify the corresponding proposed operation, but records blocked commitment or activation rather than a completed transition.

\item[\textbf{Violation (V):}] The safety attribute that fails to travel: provenance loss, privilege or authority escalation, scope widening, privacy breach, availability loss, budget exhaustion, missing audit trail, reversibility failure. This field names what went wrong internally in terms that can still be mapped to ordinary security concepts.

\item[\textbf{Exposure (E):}] The surface where the failure becomes visible: tool call, model output, shared artifact, governance audit, deletion request. This field names where the problem surfaces.

\item[\textbf{Response (R):}] The mechanism that can contain or repair: admission gate, migration check, activation authorization, runtime guard, descendant rollback, deletion verification, contestability record. This field names what can intervene.
\end{description}

\noindent\textbf{Selection and Feedback.} Selection determines why $\pi\sb{\psi,t}^{\mathrm{evo}}$ chose $\delta\sb{t}^{\star}$ from $\Delta\sb{t}$; the five fields characterize what that choice changes. Comparing transitions over time additionally requires the candidate source, evaluator and objective versions, authorization evidence, and rejected or unavailable alternatives. These distinguish changes caused by the selected operation from changes in the process selecting it.
A response can revise $c\sb{t+1}$, the evaluator behind $\widehat{u}\sb{t+1}$, or the policy and experience available to $\Delta\sb{t+1}$. Evaluation-driven development and operations (EDDOps) describes such runtime adjustment and governed redevelopment~\cite{xia_2024_evaluation_driven_developmen}. This process model motivates linking interventions to later updates; whether the resulting feedback improves safety remains an empirical question.

\noindent\textbf{Comparing Evidence.} GhostWriter and the Personal Agent Sycophancy Benchmark (PASB) both separate commitment from later activation, but the former studies hidden adversarial memory and the latter the agent's treatment of ordinary conversational claims~\cite{torres2026agentsremembermuchmemory,mao2026agentsdontjustagree}. InjecAgent instead observes the tool consequence of untrusted context~\cite{zhan_2024_injecagent_benchmarking_indi}. This comparison separates a shared temporal dependency from distinct entry mechanisms and observed outcomes. Capability studies can identify a carrier or operation worth testing without having tested its safety. Throughout the survey, we distinguish these observations from our derived governance requirements and open questions; \appsecref{app:survey-protocol} gives the evidence-tier rules.

\noindent\textbf{Relation to Existing Frameworks.}
The Open Worldwide Application Security Project (OWASP) LLM Top 10 and MITRE Adversarial Threat Landscape for Artificial-Intelligence Systems (ATLAS) provide risk and tactic vocabularies~\cite{owasp2025llm_top10,mitre2026atlas}. The National Institute of Standards and Technology (NIST) Artificial Intelligence Risk Management Framework (AI RMF) and Microsoft AI Risk Assessment address process ownership, testing, and incident management~\cite{tabassi2023ai_rmf,microsoft2024ai_risk_assessment}. Information-flow control (IFC) specifies permitted flows, provenance tracks source and derivation, and runtime verification checks execution against declared properties~\cite{denning1976lattice_information_flow,sabelfeld2003language_information_flow,buneman2001why_where_provenance,leucker2009runtime_verification}. \SAVER{} uses these established concepts to locate the transition at which a reusable object changes role; it does not replace their policies or enforcement mechanisms.

\noindent\textbf{Boundary of Use.} A system can be well governed with ordinary provenance, access control, or prompt injection controls if unsafe influence stays local, and the absence of a full \SAVER{} record does not by itself prove that a system is unsafe; it only limits what the system can claim about diagnosis, recovery, and accountability across multiple hops. \SAVER{} is therefore a bounded trace lens, not a safety prerequisite. The deepest boundary is semantic opacity: some descendants are missed not because hooks were omitted, but because summarization, retrieval abstraction, and latent representations partially dissolve lineage.

\begin{SAVERTakeaway}{A Transition Trace Has an Evidence Boundary}
\label{takeaway:trace-evidence-boundary}
The substrate--adaptation pair identifies the event; violation, exposure, and response record how far its consequences are observed. Evidence of persistence does not establish propagation, and blocking a tool call does not establish descendant repair. A partial trace supports a partial claim, while missing links define the next evaluation target.
\end{SAVERTakeaway}

\secref{sec:literature-map-positioning} applies this comparison to the main mechanism families.

\subsection{Literature Map and Positioning}
\label{sec:literature-map-positioning}

\definecolor{saverLaneBlue}{HTML}{527CDF}
\definecolor{saverLaneBlueLight}{HTML}{EEF3FF}
\definecolor{saverLaneGreen}{HTML}{86D4CD}
\definecolor{saverLaneGreenLight}{HTML}{F2FBFA}
\definecolor{saverLaneOrange}{HTML}{E9A859}
\definecolor{saverLaneOrangeLight}{HTML}{FFF7EB}
\definecolor{saverLaneRed}{HTML}{E2707C}
\definecolor{saverLaneRedLight}{HTML}{FFECEF}
\definecolor{saverResponseTeal}{HTML}{2F7F73}
\definecolor{saverResponseTealLight}{HTML}{EAF7F4}
\definecolor{saverRoadmapRoot}{HTML}{F7F7F7}
\definecolor{saverRoadmapBorder}{HTML}{4A4A4A}

\newcommand{\saverRoadmapFont}{\fontsize{6pt}{6.8pt}\selectfont}
\newcommand{\saverRoadmapRef}[1]{{\hspace{0.15em}\normalfont\fontsize{5pt}{5.5pt}\selectfont\color{saverRoadmapBorder!72}\hyperref[#1]{(\S\ref*{#1})}}}
\newcommand{\saverRoadmapRefRange}[2]{{\hspace{0.15em}\normalfont\fontsize{5pt}{5.5pt}\selectfont\color{saverRoadmapBorder!72}(\hyperref[#1]{\S\ref*{#1}}--\hyperref[#2]{\S\ref*{#2}})}}
\let\saverRoadmapNatbibCite\citep
\makeatletter
\newcommand{\saverRoadmapCiteOne}[1]{\saverRoadmapNatbibCite{#1}\allowbreak}
\newcommand{\saverRoadmapCite}[1]{{\hspace{0.08em}\normalfont\fontsize{5pt}{5.5pt}\selectfont\color{saverRoadmapBorder!78}\@for\saverRoadmapKey:=#1\do{\saverRoadmapCiteOne{\saverRoadmapKey}}}}
\makeatother
\newcommand{\saverRoadmapLevelOneWidth}{6.5em}
\newcommand{\saverRoadmapLevelTwoWidth}{9.3em}
\newcommand{\saverRoadmapLevelThreeWidth}{15.8em}

\tikzset{
  saver-root/.style={
    rectangle,
    rounded corners=2pt,
    fill=saverRoadmapRoot,
    draw=saverRoadmapBorder,
    inner sep=3pt,
    line width=0.8pt,
    text=black,
    align=center
  },
  saver-blue-bg/.style={
    rectangle,
    rounded corners=2pt,
    fill=saverLaneBlueLight,
    draw=saverLaneBlue,
    inner sep=3pt,
    line width=0.8pt,
    text=black,
    text justified
  },
  saver-blue-line/.style={
    rectangle,
    rounded corners=2pt,
    fill=none,
    draw=saverLaneBlue,
    inner sep=3pt,
    line width=0.5pt,
    text=black,
    text justified
  },
  saver-green-bg/.style={
    rectangle,
    rounded corners=2pt,
    fill=saverLaneGreenLight,
    draw=saverLaneGreen,
    inner sep=3pt,
    line width=0.8pt,
    text=black,
    text justified
  },
  saver-green-line/.style={
    rectangle,
    rounded corners=2pt,
    fill=none,
    draw=saverLaneGreen,
    inner sep=3pt,
    line width=0.5pt,
    text=black,
    text justified
  },
  saver-orange-bg/.style={
    rectangle,
    rounded corners=2pt,
    fill=saverLaneOrangeLight,
    draw=saverLaneOrange,
    inner sep=3pt,
    line width=0.8pt,
    text=black,
    text justified
  },
  saver-orange-line/.style={
    rectangle,
    rounded corners=2pt,
    fill=none,
    draw=saverLaneOrange,
    inner sep=3pt,
    line width=0.5pt,
    text=black,
    text justified
  },
  saver-red-bg/.style={
    rectangle,
    rounded corners=2pt,
    fill=saverLaneRedLight,
    draw=saverLaneRed,
    inner sep=3pt,
    line width=0.8pt,
    text=black,
    text justified
  },
  saver-red-line/.style={
    rectangle,
    rounded corners=2pt,
    fill=none,
    draw=saverLaneRed,
    inner sep=3pt,
    line width=0.5pt,
    text=black,
    text justified
  },
  saver-violation-leaf/.style={
    rectangle,
    rounded corners=2pt,
    fill=saverLaneRedLight,
    draw=saverLaneRed,
    inner sep=3pt,
    line width=0.45pt,
    text=black,
    text justified
  },
  saver-response-leaf/.style={
    rectangle,
    rounded corners=2pt,
    fill=saverResponseTealLight,
    draw=saverResponseTeal,
    inner sep=3pt,
    line width=0.45pt,
    text=black,
    text justified
  }
}

\begin{figure*}[p]
  \centering
  \resizebox{0.97\textwidth}{!}{%
    \begin{forest}
      forked edges,
      for tree={
        grow=east,
        reversed=true,
        anchor=mid west,
        parent anchor=mid east,
        child anchor=mid west,
        base=left,
        font=\saverRoadmapFont,
        rectangle,
        draw=none,
        sharp corners,
        minimum width=2em,
        edge+={darkgray, line width=0.55pt},
        s sep=3.6pt,
        l sep=7.2pt,
        fork sep=3.6pt,
        inner xsep=1.8pt,
        inner ysep=2.0pt,
        line width=0.6pt,
        par/.style={rotate=90, child anchor=north, parent anchor=south, anchor=center}
      },
      where level=0{font=\footnotesize\bfseries, text width=6.8em}{},
      where level=1{font=\saverRoadmapFont\bfseries, text width=\saverRoadmapLevelOneWidth, l-=12pt, tier=substrate}{},
      where level=2{font=\saverRoadmapFont, text width=\saverRoadmapLevelTwoWidth, l-=6pt, tier=adaptation}{},
      where level=3{font=\saverRoadmapFont, text width=\saverRoadmapLevelThreeWidth, tier=evidence}{},
      [
        {\SAVER Roadmap}, par, saver-root
        [
          {Prompt, Context, Memory, Retrieval\saverRoadmapRefRange{sec:substrate-prompt-context}{sec:substrate-retrieval-knowledge}}, saver-blue-bg
          [
            {Write$\rightarrow$Abstract$\rightarrow$Retrieve; Memory Attacks; Admission Checks\saverRoadmapRef{sec:adaptation-operations}}, saver-blue-line
            [
              {Violation\saverRoadmapRef{sec:invariant-violations}$\rightarrow$Exposure\saverRoadmapRef{sec:exposure-surfaces}: Memory Poisoning, Stealth Injection, Fragmented Bypass~\saverRoadmapCite{chen2024agentpoison,torres2026agentsremembermuchmemory,rao2026fragfusebypassingaccesscontrol,sunil2026memory_poisoning_agents,pulipaka2026hidden_memory,yang2026zombie_agents}, Stale Retrieval~\saverRoadmapCite{zhang2026memmorph}, Privacy Extraction~\saverRoadmapCite{wang2025privacy_agent_memory,zhang_2024_privacyasst_safeguarding_use,yan_2025_on_protecting_the_data_priva,lin2026sope}, Long Horizon Memory Drift~\saverRoadmapCite{liu2026simplemem,mi2026darwinian_memory,rajesh2026panini}}, saver-violation-leaf
            ]
            [
              {Response\saverRoadmapRef{sec:response-governance}: Admission Gating~\saverRoadmapCite{lam2026ssgm,taheri2026governed_memory,kandala2026echoguard}, Deletion Checks~\saverRoadmapCite{pulipaka2026persistbench,kannabhiran_2026_consensus_validated_memory_i}, Residual Probes~\saverRoadmapCite{shao2026misevolve,wang2026agenticeval}, Memory Evaluation~\saverRoadmapCite{hu_2025_evaluating_memory_in_llm_age,luo_2026_from_storage_to_experience_a,lupascu2026elephantbroker}}, saver-response-leaf
            ]
          ]
          [
            {Promote$\rightarrow$Migrate$\rightarrow$Activate; Stale State Attacks; Activation Gates\saverRoadmapRef{sec:adaptation-operations}}, saver-blue-line
            [
              {Violation\saverRoadmapRef{sec:invariant-violations}$\rightarrow$Exposure\saverRoadmapRef{sec:exposure-surfaces}: Scope Drift~\saverRoadmapCite{wang2026oep,zhao2026experience_driven_safety,jiang_2026_agentlab}, Delayed Activation~\saverRoadmapCite{xie2026memevobench,zhao2026amabench,yang2026zombie_agents,shao2026misevolve}, Instruction Laundering~\saverRoadmapCite{wang2026oep,rao2026fragfusebypassingaccesscontrol}, Environment Drift~\saverRoadmapCite{xing_2024_understanding_the_weakness_o,liao2026redteamcua}}, saver-violation-leaf
            ]
            [
              {Response\saverRoadmapRef{sec:response-governance}: Migration Logs~\saverRoadmapCite{ravindran2026portable_agent_memory,hatalis_2024_memory_matters_the_need_to_i,pink_2025_position_episodic_memory_is}, Activation Authorization~\saverRoadmapCite{zhao2026faithful_self_evolvers,chitan2026srm}, Rollback Handles~\saverRoadmapCite{deng_2026_taming_openclaw}, Replay Logs~\saverRoadmapCite{souza_2025_llm_agents_for_interactive_w}}, saver-response-leaf
            ]
          ]
        ]
        [
          {Tool, API, Skill, Workflow State\saverRoadmapRefRange{sec:substrate-tool-interface}{sec:substrate-workflow-topology}}, saver-green-bg
          [
            {Authorize$\rightarrow$Call$\rightarrow$Persist Output; Tool Hijack; Privilege Checks\saverRoadmapRef{sec:adaptation-operations}}, saver-green-line
            [
              {Violation\saverRoadmapRef{sec:invariant-violations}$\rightarrow$Exposure\saverRoadmapRef{sec:exposure-surfaces}: Indirect Prompt Injection (IPI)~\saverRoadmapCite{debenedetti2024agentdojo,zhan_2024_injecagent_benchmarking_indi,zhao2026livepi,greshake_2023_not_what_you_ve_signed_up_fo,hines_2024_defending_against_indirect_p,joseph_2025_prompt_injection_in_large_la}, Tool Misuse~\saverRoadmapCite{shi2026prompt_tool_selection,dong2026safesearch,zhou2026mcpshield,radosevich2025mcp_safety_audit,li2026stac}, Privilege Escalation~\saverRoadmapCite{rao2026fragfusebypassingaccesscontrol,fang_2024_llm_agents_can_autonomously}, Cross Modal Injection~\saverRoadmapCite{lee_2025_mind_mapping_prompt_injectio,wang_2025_manipulating_multimodal_agen,zhang_2024_a_study_on_prompt_injection}, Cyber Offense Tasks~\saverRoadmapCite{anurin_2024_catastrophic_cyber_capabilit,bhatt_2024_cyberseceval_2_a_wide_rangin,muzsai_2024_hacksynth_llm_agent_and_eval}}, saver-violation-leaf
            ]
            [
              {Response\saverRoadmapRef{sec:response-governance}: Least Privilege Gates~\saverRoadmapCite{shi2025progent,siu2026formalizing_agent_security,zhou2026mcpshield,radosevich2025mcp_safety_audit}, Runtime Guards~\saverRoadmapCite{xiang2024guardagent,liu_2026_safeagent_a_runtime_protecti,chennabasappa_2025_llamafirewall_an_open_source,hua_2024_trustagent_towards_safe_and,podpora_2025_llm_firewall_using_validator,wang2025openguardrails,luo2025agrail}, Tool Audit Trails~\saverRoadmapCite{jia_2025_the_task_shield_enforcing_ta,an2025ipiguard,luo2025agentauditor,syros2026saga}, Injection Detectors~\saverRoadmapCite{chen_2025_defending_against_prompt_inj,prakash_2026_hybrid_real_time_framework_f,jain_2025_comprehensive_analysis_of_ma,panterino_2024_dynamic_moving_target_defens}, SOC Orchestration~\saverRoadmapCite{alqahtani_2026_secure_autonomous_cyber_defe,ismail_2025_toward_robust_security_orche,molleti_2024_automated_threat_detection_a}}, saver-response-leaf
            ]
          ]
          [
            {Compile$\rightarrow$Reuse$\rightarrow$Route; Skill Drift; Rollback Handles\saverRoadmapRef{sec:adaptation-operations}}, saver-green-line
            [
              {Violation\saverRoadmapRef{sec:invariant-violations}$\rightarrow$Exposure\saverRoadmapRef{sec:exposure-surfaces}: Boundary Drift~\saverRoadmapCite{zhang2026library_drift,li2026stac}, Stale Permission~\saverRoadmapCite{wang2023voyager}, Skill Overgeneralization~\saverRoadmapCite{shinn2023reflexion,li2026experience_evolving_memory,renze_2024_self_reflection_in_llm_agent}, Route Coordination Drift~\saverRoadmapCite{yang_2025_autohma_llm_efficient_task_c,zhang_2025_webpilot_a_versatile_and_aut}, Planning Drift~\saverRoadmapCite{he_2025_plan_then_execute_an_empiric,zhuang_2025_llm_agents_can_be_choice_sup}}, saver-violation-leaf
            ]
            [
              {Response\saverRoadmapRef{sec:response-governance}: Versioned Artifacts~\saverRoadmapCite{xia_2024_evaluation_driven_developmen}, Change Records~\saverRoadmapCite{liu2025webcoach}, Contextual Security Constraints~\saverRoadmapCite{siu2026formalizing_agent_security}, Workflow Provenance~\saverRoadmapCite{souza_2025_llm_agents_for_interactive_w,mittal_2025_practical_integration_of_lar}, Governance Layers~\saverRoadmapCite{figurelli_2026_from_talking_to_blackbox_to,chen_2025_securegov_agent_a_governance,kumar_2026_a_multi_evaluator_behavioura}}, saver-response-leaf
            ]
          ]
        ]
        [
          {Shared Artifacts and Agent State\saverRoadmapRefRange{sec:substrate-workflow-topology}{sec:substrate-external-artifacts}}, saver-orange-bg
          [
            {Delegate$\rightarrow$Propagate$\rightarrow$Uptake; Shared Memory Attacks; Propagation Control\saverRoadmapRef{sec:adaptation-operations}}, saver-orange-line
            [
              {Violation\saverRoadmapRef{sec:invariant-violations}$\rightarrow$Exposure\saverRoadmapRef{sec:exposure-surfaces}: Shared Memory Poisoning~\saverRoadmapCite{xu2026when}, Provenance Dilution~\saverRoadmapCite{zhang2026embedding_defenses_fail,li_2023_camel_communicative_agents_f}, Agent To Agent Spread~\saverRoadmapCite{lee_2026_prompt_infection_llm_to_llm,an-etal-2026-aciarena,wang2026devil_moltbook}, Coordination Failure~\saverRoadmapCite{mosquera_2025_can_llm_augmented_autonomous,gao_2025_contextual_trust_evaluation,yuan_2023_robust_multi_agent_coordinat}, Social Trust Drift~\saverRoadmapCite{haez_2026_socio_normative_trustworthin,volta_2026_experimental_campaign_for_po,zhuang_2025_llm_agents_can_be_choice_sup}}, saver-violation-leaf
            ]
            [
              {Response\saverRoadmapRef{sec:response-governance}: Recipient Scoped Traces~\saverRoadmapCite{taheri2026governed_memory,ravindran2026portable_agent_memory,siu2026formalizing_agent_security,khan2025agentsafe_governance}, Propagation Stops~\saverRoadmapCite{yu_2022_security_preserving_multi_ag}, Handoff Policies~\saverRoadmapCite{harrison_2026_protocol_mediated_execution,fernandez_2026_protocol_layer_governance_fo}, Coordination Audits~\saverRoadmapCite{gao_2025_contextual_trust_evaluation,dorri_2018_multi_agent_systems_a_survey,yang_2025_autohma_llm_efficient_task_c}}, saver-response-leaf
            ]
          ]
          [
            {Copy$\rightarrow$Reinterpret$\rightarrow$Publish; Artifact Leakage; Contestability\saverRoadmapRef{sec:adaptation-operations}}, saver-orange-line
            [
              {Violation\saverRoadmapRef{sec:invariant-violations}$\rightarrow$Exposure\saverRoadmapRef{sec:exposure-surfaces}: Artifact Leakage~\saverRoadmapCite{wang2025privacy_agent_memory,taheri2026governed_memory,reddy_2025_echoleak_the_first_real_worl,rathod_2025_privacy_and_security_challen}, Scope Widening~\saverRoadmapCite{taheri2026governed_memory}, Cross Team Reuse~\saverRoadmapCite{ravindran2026portable_agent_memory,bamil_2026_a_unified_evaluation_and_gov}, Data Federation Leakage~\saverRoadmapCite{karim_2025_ai_agents_meet_blockchain_a,geren_2025_blockchain_for_large_languag,yang_2025_synergizing_intelligence_and}}, saver-violation-leaf
            ]
            [
              {Response\saverRoadmapRef{sec:response-governance}: Access Partitions~\saverRoadmapCite{taheri2026governed_memory,khan2025agentsafe_governance}, Retention Limits~\saverRoadmapCite{pulipaka2026persistbench,wang2025privacy_agent_memory,rathod_2025_privacy_and_security_challen}, Contestability Paths~\saverRoadmapCite{ravindran2026portable_agent_memory,siu2026formalizing_agent_security,bamil_2026_a_unified_evaluation_and_gov}, Data Governance Benchmarks~\saverRoadmapCite{zhou_2025_datagovbench_benchmarking_ll}}, saver-response-leaf
            ]
          ]
        ]
        [
          {Model and Policy State\saverRoadmapRef{sec:substrate-model-policy}}, saver-red-bg
          [
            {Fine-Tune$\rightarrow$Load$\rightarrow$Deploy; Jailbreak Transfer; Regression Gates\saverRoadmapRef{sec:adaptation-operations}}, saver-red-line
            [
              {Violation\saverRoadmapRef{sec:invariant-violations}$\rightarrow$Exposure\saverRoadmapRef{sec:exposure-surfaces}: Safety Regression~\saverRoadmapCite{lermen_2023_lora_fine_tuning_efficiently,qi2023finetuning_safety,liu_2025_a_survey_of_attacks_on_large}, Latent Triggers~\saverRoadmapCite{hong2024peta,gloaguen2025watch_your_steps,wang_2024_badagent_inserting_and_activ,peng_2024_jailbreaking_and_mitigation}, Unsafe Adapter Activation~\saverRoadmapCite{hong2024peta,lermen_2023_lora_fine_tuning_efficiently}, Adaptive Jailbreak Search~\saverRoadmapCite{liu2025autodan_turbo,yuan2026agenticred,shu_2025_attackeval_how_to_evaluate_t,schulhoff_2023_ignore_this_title_and_hackap,salem_2023_maatphor_automated_variant_a}}, saver-violation-leaf
            ]
            [
              {Response\saverRoadmapRef{sec:response-governance}: Activation Manifests~\saverRoadmapCite{hu2022lora}, Regression Tests~\saverRoadmapCite{banerjee2026severa,zhao2026faithful_self_evolvers,li_2024_the_wmdp_benchmark_measuring,andriushchenko2025agentharm,guo2024redcode,lee2026mobilesafetybench,pan2023machiavelli}, Deployment Gates~\saverRoadmapCite{banerjee2026severa,khan2025agentsafe_governance,wang2025openguardrails}, Domain Safety Benchmarks~\saverRoadmapCite{yuan2024rjudge,cao2025safelawbench,li_2024_legalagentbench_evaluating_l}}, saver-response-leaf
            ]
          ]
          [
            {Merge$\rightarrow$Inherit$\rightarrow$Update Preference; Lineage Loss; Model Repair\saverRoadmapRef{sec:adaptation-operations}}, saver-red-line
            [
              {Violation\saverRoadmapRef{sec:invariant-violations}$\rightarrow$Exposure\saverRoadmapRef{sec:exposure-surfaces}: Lineage Loss~\saverRoadmapCite{zhou2025backdoor_llm_survey,hu2022lora}, Hidden Behavior Reuse~\saverRoadmapCite{rando2024universal_jailbreak_backdoors,wang_2024_badagent_inserting_and_activ,chen2026customization}, Adaptation Drift~\saverRoadmapCite{romero_2024_synergistic_integration_of_l}}, saver-violation-leaf
            ]
            [
              {Response\saverRoadmapRef{sec:response-governance}: Model Unlearning~\saverRoadmapCite{li_2024_the_wmdp_benchmark_measuring}, Lineage Records~\saverRoadmapCite{hu2022lora}, Residual Safety Probes~\saverRoadmapCite{li_2024_the_wmdp_benchmark_measuring,shao2026misevolve,andriushchenko2025agentharm,guo2024redcode}, Governance Scorecards~\saverRoadmapCite{triguero_2023_general_purpose_artificial_i}, Formal Governance~\saverRoadmapCite{figurelli_2026_from_talking_to_blackbox_to,chen_2025_securegov_agent_a_governance,kumar_2026_a_multi_evaluator_behavioura}}, saver-response-leaf
            ]
          ]
        ]
      ]
    \end{forest}%
  }
  \caption{A full-width horizontal \SAVER{} roadmap. Exposure is folded into the incident leaves for legibility.}
  \label{fig:saver-transition-roadmap}
\end{figure*}
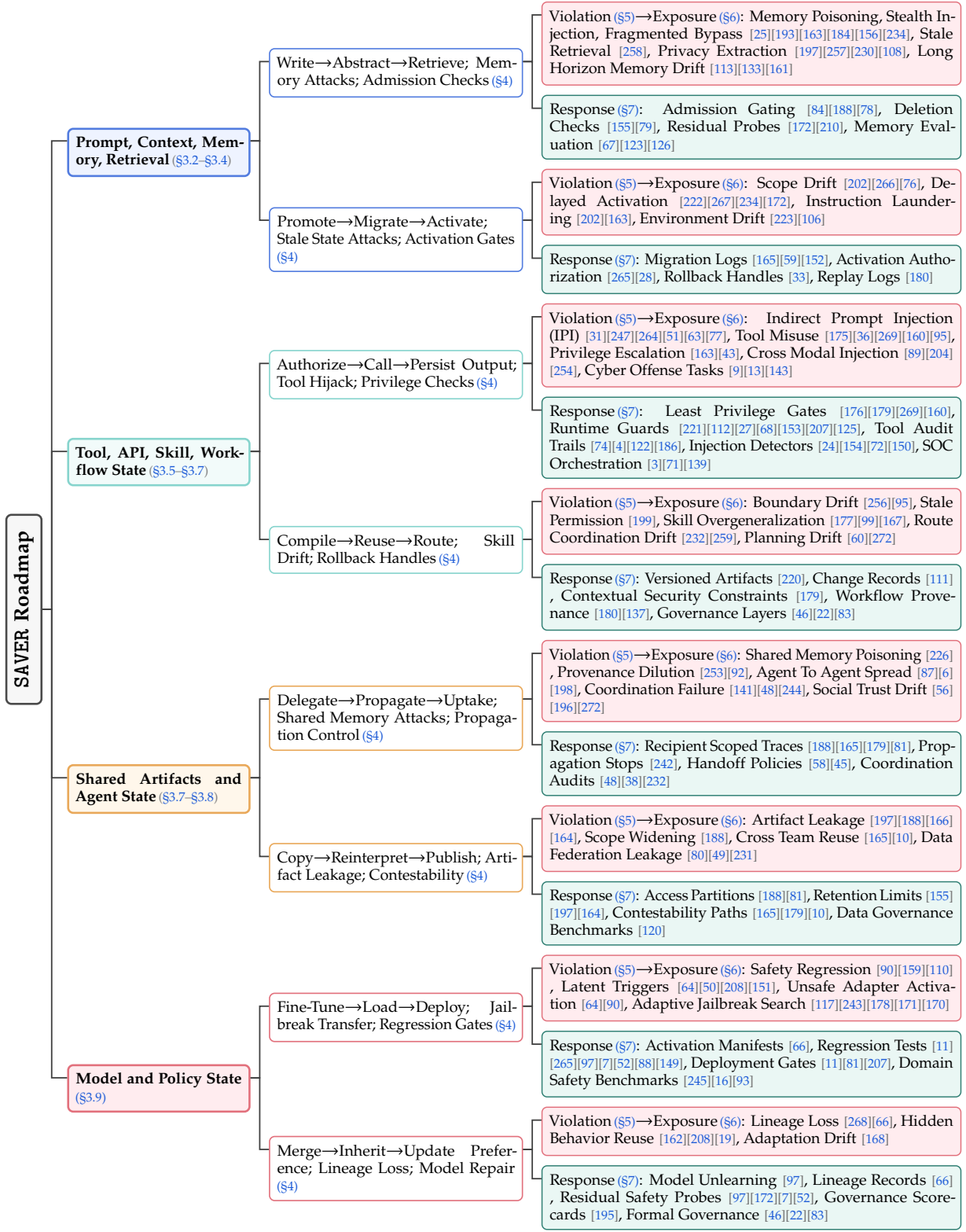

With the transition unit fixed, the literature can be read as a field map rather than as a list of modules. A work is central when it directly examines a safety-relevant transition involving reusable agent state, including a failure, exposure, evaluation, or control of that transition. Capability mechanisms alone remain adjacent. Inclusion does not require a complete recovery trace; the evidence boundary determines which claims a work can support. \figref{fig:saver-survey-overview} provides the reading order for the comparison: locate the carrier-operation pair in the central ring, then ask how far the paper follows the transition through violation, exposure, and response. The positioning paragraphs apply that order without assuming that every paper covers the full chain.

\noindent\textbf{How to Read the Evidence Indexes.}
Each of the six body chapters closes with a paper-level evidence index for its part of the transition. These tables support paper-level lookup rather than define independent taxonomies. Role is assigned to the contribution named in a row, not permanently to the paper: an attack study can appear as \emph{Attack} in a mechanism index and as \emph{Benchmark (Attack)} when its evaluation protocol supplies the denominator, while a separately named mitigation from the same paper can appear as \emph{Defense}. Rows that name the same contribution retain one primary role across indexes.

\noindent\textbf{How the Literature Converges.}
The relevant literatures did not begin from a common taxonomy, but their evidence now meets along the same transition. Prompt-injection research establishes that untrusted context can condition action, and InjecAgent makes the resulting indirect-injection boundary measurable~\cite{liu_2023_prompt_injection_attack_agai,zhan_2024_injecagent_benchmarking_indi}. Memory research adds persistence through managed storage, governed reuse, and user-specific context~\cite{zhang_2025_a_survey_on_the_memory_mecha,lin2026memory_security_survey,wang2025privacy_agent_memory,mao2026agentsdontjustagree}. Self-evolution work moves the safety question into the update rule because reflection, experience, or collective state can convert a local observation into reusable policy~\cite{zhao2026experience_driven_safety,wang2026oep,xu2026when}. Response research then asks where this transition can be recognized, authorized, blocked, or audited~\cite{xiang2024guardagent,liu_2026_safeagent_a_runtime_protecti,siu2026formalizing_agent_security}.

These bodies of work contribute different segments of an evidence chain. Memory and tool studies provide concrete write paths, retrieval triggers, and action boundaries. Privilege and runtime controls identify activation gates and local blocking points. They offer less evidence about what happens after the same influence migrates into workflow rules, shared artifacts, or model-side state. Recovery claims are especially difficult because they must link a response to every relevant descendant and test for later reactivation. The resulting imbalance is not simply a shortage of defense papers; it is a mismatch between event-level evidence and the stronger proof obligations of cross-session repair.

\noindent\textbf{Differences from Existing Surveys.}
Neighboring surveys organize comparison around different questions. General and autonomous-agent surveys compare memory, planning, tool, and coordination components~\cite{xi_2023_the_rise_and_potential_of_la,cheng_2024_exploring_large_language_mod}; self-evolution surveys examine where and how adaptation occurs~\cite{gao2025self_evolving_agents,tao2024self_evolution_llm_survey}. Trustworthy-agent and security surveys organize risks and governance properties, while benchmark surveys compare evaluation tasks and metrics~\cite{yu2025trustworthy_llm_agents,deng_2024_ai_agents_under_threat_a_sur,adabara_2025_trustworthy_agentic_ai_syste,mohammadi_2025_evaluation_and_benchmarking}. Trajectory assurance asks whether composed behavior respects system-level constraints even when individual actions are permitted~\cite{lotfi2026securingagenticaiperaction}. The Module-Lifecycle Attack Surface (MLAS) analysis focuses more directly on self-evolving-agent modules and lifecycle threats~\cite{lin2026safetyselfevolvingllmagent}.

Our comparison holds the reusable influence fixed while its carrier and role change. This specializes the broader trajectory-assurance objective: rather than treating every action sequence as one object, it asks which state transition changed the sequence's effective provenance, scope, or authority. A module map can locate memory, tools, feedback, and adaptation; a lifecycle threat map can identify exposed cells. A transition record asks how a particular object entered one cell, what attributes it carried, which operation moved or transformed it, where it later activated, and what response artifact addresses its lineage. Papers are therefore compared by the transition segment they observe, with component and paper role retained as secondary descriptors.

\noindent\textbf{Memory and Reusable Context.}
Memory remains the strongest substrate for evidence because its write, storage, retrieval, and activation points are comparatively visible. Personal-agent and memory surveys define the carrier, while memory-security work explains why persistence requires provenance and policy metadata~\cite{li_2024_personal_llm_agents_insights,zhang_2025_a_survey_on_the_memory_mecha,lin2026memory_security_survey,du2026memory_autonomous_agents}.

Attack studies then separate into mechanisms that a generic ``memory poisoning'' label would obscure. AgentPoison targets trigger-selected retrieval from an already poisoned store; stealth injection targets the admission path from an external environment; broader poisoning and sleeper-memory work emphasize persistence and delayed activation; MemMorph follows retrieved memory into tool hijacking~\cite{chen2024agentpoison,sunil2026memory_poisoning_agents,pulipaka2026hidden_memory,zhang2026memmorph}. These mechanisms require different denominators: poisoned retrieval opportunities, unsafe write opportunities, delayed trigger opportunities, or tool-routing opportunities. Privacy and PerMemSafe add another comparison axis by asking whether retained state remains appropriate for a particular user and purpose~\cite{wang2025privacy_agent_memory,an-etal-2026-permemsafe}. PS-Bench sharpens the non-poisoning branch: truthful memories can remain appropriate as personal context yet still weaken the safety boundary by legitimizing a harmful request~\cite{guo-etal-2026-personalization}. EchoSafe adds a multimodal branch in which an inference-time reflective memory bank adapts contextual safety decisions; it is evidence about memory-mediated safety adaptation, not a general guarantee for textual agent memory~\cite{zhang2026echosafe}. Memory is therefore the clearest substrate anchor, but not the lifecycle boundary: its contents can become prompt context, a plan constraint, a tool argument, a workflow rule, a shared artifact, or a rollback target.

\noindent\textbf{Self-Evolution and Misevolution.}
Reflection and skill libraries establish the capability mechanism: feedback or successful traces can become strategy and callable procedure~\cite{shinn2023reflexion,wang2023voyager}. Safety work begins where that promotion changes provenance, authority, scope, privacy, or recoverability. Obsessive Experience Poisoning (OEP) tests harmful transfer from locally useful experience, while SEVerA studies verification before reusable behavior is synthesized~\cite{wang2026oep,banerjee2026severa}. Experience-driven safety and MemEvoBench broaden the evaluation setting from one promoted rule to evolving memories and updates~\cite{zhao2026experience_driven_safety,xie2026memevobench}.
The Red Queen G{\"o}del Machine adds a different dependency: the evaluator and utility can coevolve with the agent~\cite{iacob2026redqueengodelmachine}. A benchmark can therefore miss drift even when it records every committed update, because the criterion selecting those updates may also have changed. Read together, this literature links capability, selection, promotion, and evaluation and localizes where a reusable lesson acquires authority.

\noindent\textbf{Tool, Workflow, and Runtime Response.}
Prompt-injection attacks and benchmarks establish the exposure mechanism: untrusted text becomes planning context, and a tool call makes the consequence observable~\cite{liu_2023_prompt_injection_attack_agai,debenedetti2024agentdojo,zhan_2024_injecagent_benchmarking_indi,zhan_2025_adaptive_attacks_break_defen}. Response papers then divide according to whether they observe, constrain, or maintain that boundary.
Risk judges and safety benchmarks improve recognition, while GuardAgent and TrustAgent add local guard reasoning~\cite{zhang_2024_agent_safetybench_evaluating,xiang2024guardagent,hua_2024_trustagent_towards_safe_and}. Privilege and runtime controls address a different question: whether execution should proceed, as illustrated by SafeAgent~\cite{liu_2026_safeagent_a_runtime_protecti}. Those controls can stop or narrow an action, but they usually do not repair the earlier memory or workflow state that requested it.

Evaluation-driven operations and coordination work extend the same boundary into deployment~\cite{xia_2024_evaluation_driven_developmen}, while POLARIS~\cite{moslemi2026polaris}, MAGNET~\cite{sun2026magnet}, and WebCoach~\cite{liu2025webcoach} make route and workflow traces more visible. The unresolved connection is between these operational records and the state lineage needed for recovery: local recognition and blocking rarely establish that the responsible influence can no longer reactivate elsewhere.

\noindent\textbf{Model Adaptation.}
Model updates are central only when they enter the agent loop as state that is selected, loaded, inherited, evaluated, or rolled back. Static model safety~\cite{qi2023finetuning_safety}, parameter-efficient fine-tuning (PEFT) risk~\cite{hong2024peta}, reinforcement learning from human feedback (RLHF) backdoors~\cite{rando2024universal_jailbreak_backdoors}, dormant behavior~\cite{gloaguen2025watch_your_steps}, security and privacy surveys~\cite{wang2024unique_security_privacy_llm}, and backdoor surveys~\cite{zhou2025backdoor_llm_survey} explain failure modes that reusable updates can introduce. For low-rank adaptation (LoRA), Safe LoRA is a bounded adapter-level mitigation of fine-tuning safety loss~\cite{hsu2024safe_lora}; adapter scope, deployment lineage, activation context, and rollback handles are the additional fields needed before treating it as an agent-lifecycle recovery anchor. KVEraser adds a narrower runtime cache branch by studying erasure of processed context influence after prefill~\cite{li2026kveraserlearningsteerkv}. The branch remains a frontier until agent-loop lineage, cache-local residual probes, and deployment rollback records become routine.

\noindent\textbf{Evaluation Coverage.}
Benchmarks make different parts of a transition observable. LoCoMo tests retained conversational context, and MemEvoBench examines memory adaptation~\cite{maharana2024locomo,xie2026memevobench}. AgentDojo observes tool-mediated exposure, R-Judge tests risk recognition, PerMemSafe evaluates personalized response safety, and ACIArena follows cascading injection across agents~\cite{debenedetti2024agentdojo,yuan2024rjudge,an-etal-2026-permemsafe,an-etal-2026-aciarena}. Runtime containment and model-regression studies add response and deployment observations~\cite{liu_2026_safeagent_a_runtime_protecti,qi2023finetuning_safety}. HarnessSafe now connects carrier retention, boundary crossing, benign activation, and visible violation~\cite{zhang2026harnesssafeevaluatingsafetypersistent}. The remaining comparison is therefore not simply connected versus disconnected evaluation: it is which earlier selection decisions and later repair outcomes each connected trace leaves unobserved.

\begin{SAVERTakeaway}{Capability Evidence Is Not Safety Evidence}
\label{takeaway:capability-evidence-boundary}
A system that demonstrates memory, skill, or model-update reuse identifies a carrier and an operation worth testing. It establishes transition safety only if the evaluation also observes the relevant safety attributes, activation authority, or response. Capability, attack, and defense studies can inform the same comparison without supporting interchangeable claims.
\end{SAVERTakeaway}

The following chapters specialize this comparison by substrate, operation, violation, exposure, and response; \secref{sec:evaluation} connects their evidence requirements.

\section{Substrate: Safety-Bearing State Carriers}
\label{sec:safety-bearing-substrates}

\SAVERFlowS

\begin{figure*}[t]
\centering
\IfFileExists{figures/cross-substrate-transmutation.pdf}{%
  \includegraphics[width=\textwidth]{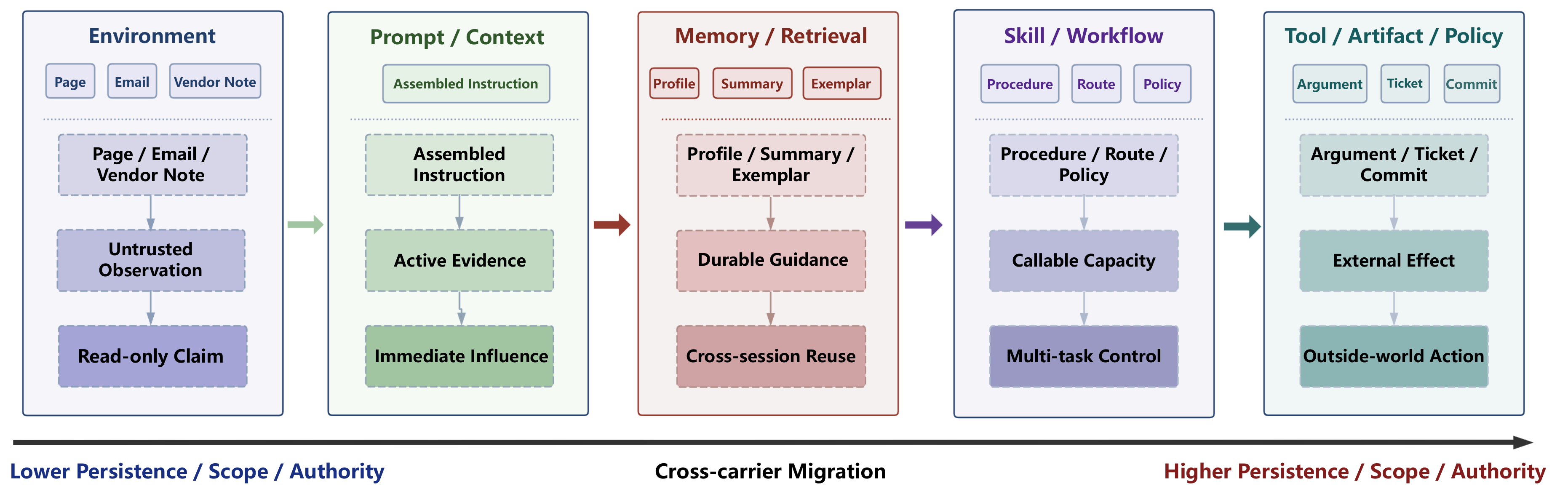}%
}{%
  \includegraphics[width=\textwidth]{./cross-substrate-transmutation.pdf}%
}
\vspace{-8pt}
\caption{Cross substrate transmutation. The same sentence carries different authority depending on its carrier: from untrusted environment text to prompt context, durable memory, callable skill, and finally external effect.}
\Description{Full width figure with one sentence at the top and five columns below. The columns show substrate, role, and authority rows: environment, prompt/context, memory/retrieval, skill/workflow, and tool/artifact/policy. A highlighted central band emphasizes that the key change is not wording but the role and authority the same content acquires as it moves across substrates.}
\label{fig:cross-substrate-transmutation}
\end{figure*}

Before the safety of an update can be judged, the analysis must identify what persists and the authority regime in which it can be reused. The same sentence has different consequences when it is transient context, durable memory, a workflow rule, or a tool-bearing procedure. Substrate analysis supplies the carrier half of the \textbf{S}\(\times\)\textbf{A} transition pair. Broad agent surveys map the relevant components, while self-evolution work adds the update and reuse mechanics~\cite{luo2025llm_agent_survey,gao2025self_evolving_agents}.
The chapter compares carriers along five axes: persistence, observability, authority, scope, and reversibility. \figref{fig:cross-substrate-transmutation} follows one item from untrusted environment text through active context, durable memory or retrieval evidence, and callable skill or workflow to an external effect. Its lower band emphasizes the accompanying rise in persistence, scope, and authority. Observability and reversibility require separate comparison because neither changes monotonically along this path. These carrier properties determine what the operations in \secref{sec:adaptation-operations} can change; \tabref{tab:substrate-literature-index} provides the paper-level evidence map.

\subsection{Substrates as Authority Regimes}

\noindent\textbf{Same Text, Different Authority.} Consider the same sentence: ``for this user, choose the fastest option even if it costs more.'' As a \emph{web observation}, it is an untrusted environmental claim; as a \emph{scratchpad note}, transient reasoning context; as a \emph{long-term preference memory}, personalization state; as a \emph{retrieval exemplar}, it may steer similar tasks; as a \emph{skill parameter}, a reusable decision rule; as a \emph{workflow policy}, it can route many later tasks; and as a \emph{tool argument template}, it may affect purchases, bookings, or messages. The substrate shapes how the text can influence behavior; legitimate authority still depends on its source, intended use, and current authorization.

This example fixes the chapter's unit of comparison. Although the lexical content is unchanged, its persistence, scope, authority, actionability, and available rollback handle vary with the carrier~\cite{lin2026memory_security_survey}. These attributes are relational rather than intrinsic: they depend on provenance and derivation~\cite{buneman2001why_where_provenance}, permitted flows~\cite{denning1976lattice_information_flow}, intended scope, current substrate, activation condition, and downstream action surface. A substrate is more than storage: it is the regime that determines how an item can be retained, interpreted, reused, and acted upon.

The literature makes this distinction visible in stages rather than as a flat inventory. Memory-mechanism studies first show how information becomes managed state, PASB examines the agent's decision to commit conversational content, and Voyager shows how experience becomes reusable procedure~\cite{zhang_2025_a_survey_on_the_memory_mecha,packer2023memgpt,mao2026agentsdontjustagree,wang2023voyager}. Security studies then expose what changes when those mechanisms admit untrusted or sensitive influence: stealth memory injection, GhostWriter, and AgentPoison trace delayed reuse, while InjecAgent carries the same concern to tool-mediated action~\cite{zhang2026clawsremembertellstealthy,torres2026agentsremembermuchmemory,chen2024agentpoison,zhan_2024_injecagent_benchmarking_indi}. Multi-agent and self-evolution surveys broaden the setting to exchanged state and evolving workflow structure~\cite{guo_2024_large_language_model_based_m,gao2025self_evolving_agents}. HarnessSafe places memory, skills, tool and Model Context Protocol (MCP) state, memory-to-skill transformation, subagent delegation, session summaries, and shared artifacts under one delayed-risk protocol, making the cross-carrier comparison explicit without collapsing them into one substrate class~\cite{zhang2026harnesssafeevaluatingsafetypersistent}.

\noindent\textbf{Relational Attributes.} A medical preference stored for appointment scheduling illustrates the resulting safety test. Retention raises a sensitivity question; use in an unrelated task raises a scope question; influence over tool parameters raises an authority question; and failed inspection or deletion raises a contestability question. Persistent-memory research, PerMemSafe, and Progent supply evidence for these local mechanisms~\cite{wang2025privacy_agent_memory,an-etal-2026-permemsafe,shi2025progent}; the combined attribute test is this survey's synthesis. With that common record established, the comparison begins at the shortest-lived carrier: the prompt and context assembled for the next model call.

\subsection{Prompt and Context Substrates}
\label{sec:substrate-prompt-context}

Prompt and context substrates are the assembled instructions and contextual records presented to the next model call. They combine system and user instructions with conversation history, retrieved evidence, tool observations, and runtime metadata, allowing the model to interpret the current task and select its next response or action. ReAct makes this coupling between reasoning, observations, and action explicit~\cite{yao2022react}. Unlike durable memory, prompt state need not survive the call to matter: while active, it can redirect attention, bias tool selection, or create an apparent instruction conflict.
Representative work develops this boundary from both mechanism and control perspectives. Prompt-injection research identifies conflicts between trusted instructions and untrusted context, and AgentDojo operationalizes those conflicts in tool-using decision environments~\cite{liu_2023_prompt_injection_attack_agai,debenedetti2024agentdojo}. Signed prompts preserve one form of source authentication, while secure-design patterns organize instruction hierarchy, source marking, relevance gating, filtering, and call-time privilege checks~\cite{suo_2024_signed_prompt_a_new_approach,beurer_kellner_2025_design_patterns_for_securing}.

Recent studies show that the risk depends on how the assembled context changes role and accumulates over time. In ordinary prompt injection, attacker content competes from its original model-facing level; in instruction privilege escalation, context reconstruction assigns that content an actual user- or system-effective role~\cite{he2026contextgetsrootprivilege}. Context privilege escalation (CPE) separates this role increase from a scope increase: message-role context privilege escalation (M-CPE) raises message role, whereas cross-scope context privilege escalation (X-CPE) carries content from a session into project- or user-scoped state~\cite{li2026whatsagentscontextcontext}. Multi-turn attacks widen the same boundary temporally. Trojan Knowledge and One Turn Too Late distribute harmful intent across locally innocuous turns, while MT-AgentRisk examines the composition in tool-using agents~\cite{wei2025trojanknowledgebypassingcommercial,shen2026turnlateresponseawaredefense,li2026unsaferturnsbenchmarkingdefending}. These results do not make a longer context window intrinsically unsafe; they show why turn-local checks can miss a harmful interaction assembled from individually acceptable fragments. Long-term memory begins at this persistence boundary.

\begin{SAVERTakeaway}{Short Persistence Does Not Bound Harm}
\label{takeaway:transient-context-consequence}
Prompt state can drive an irreversible tool call without surviving the current model invocation. Its short lifetime therefore does not justify weaker authority checks. Durable memory adds a different burden: the source, role, and scope needed to judge that influence must remain available when it returns in a later context.
\end{SAVERTakeaway}

\subsection{Long-Term Memory Substrates}
\label{sec:substrate-long-term-memory}

Long-term memory changes prompt context's temporal regime. Preference memories, episodic summaries, task histories, user profiles, learned facts, and reflective notes survive the turn that created them and later return through recall, profile injection, semantic retrieval, planner queries, or new summaries. This continuity supports personalization and learning across sessions, but it also turns extraction and summarization into state-writing operations. Provenance, scope, sensitivity, expiry, inspection, deletion, quarantine, and retrieval authorization consequently become part of the memory mechanism.

Managed-memory systems differ in what continuity they preserve. MemGPT moves information between immediate and longer-lived memory, while SimpleMem, Panini, and hybrid episodic-procedural agents structure and maintain retained experience~\cite{packer2023memgpt,liu2026simplemem,rajesh2026panini,li2026experience_evolving_memory}. Reconstruction and personalization add different update demands: E-mem reconstructs episodic context across memory agents, PISA evolves schemas, and Mem-PAL and PERMA retain user-specific context and changing preferences~\cite{wang2026emem,jia2025pisa,huang2025mempal,liu2026perma}. The safety question follows from these mechanisms: maintenance can change not only which information survives, but also its attribution and permitted use~\cite{lin2026memory_security_survey,sunil2026memory_poisoning_agents,wang2025privacy_agent_memory}.

Attack studies distinguish several routes into the same delayed surface. Stealth memory injection makes the external email-to-memory path explicit, GhostWriter separates injection from later activation, FragFuse distributes prohibited content across memory interactions before fusion, and AgentPoison selects poisoned memory or knowledge-base entries through triggers~\cite{zhang2026clawsremembertellstealthy,torres2026agentsremembermuchmemory,rao2026fragfusebypassingaccesscontrol,chen2024agentpoison}. Across these mechanisms, the decisive transition is not the appearance of malicious text in one turn. It is the rewriting of external or weakly grounded content as durable internal state that later returns as apparently helpful context.

The same safety problem also arises without overtly toxic content. PASB examines whether an accepted conversational claim is committed as a durable preference, background fact, or procedure~\cite{mao2026agentsdontjustagree}. AuthMem-Bench holds the proposition and downstream task fixed while varying source authority, showing how consolidation can preserve content but erase whether it may be reused as a user fact, attested observation, or standing instruction~\cite{zhan2026memoryauthoritybenchmarkingauthority}. PS-Bench then studies context-conditioned activation, where truthful personal memory can legitimize a harmful query, while EvoBreak shows how individually benign experiences can become harmful when activated as a complementary set~\cite{guo-etal-2026-personalization,yan2026benignaloneharmfultogether}. PersistBench and RBI-Eval extend the comparison to cross-domain leakage, memory-induced sycophancy, and whether a retrieved memory is warranted in the current response~\cite{pulipaka2026persistbench,xu2026memorystaysilentmeasuring}. MemConflict distinguishes static, dynamic, and conditional conflicts, and MRMMIA exposes membership inference through the memory interface~\cite{tao2026memconflict,chen2026mrmmia}.

These findings separate three questions: whether a claim should be committed, whether consolidation preserves its source authority, and whether its later use is appropriate alone or in combination. SSGM, Governed Memory, and GLOVE address parts of this chain through governed activation, policy-routed recall, and memory--environment realignment~\cite{lam2026ssgm,taheri2026governed_memory,yin2026glove}. Their different control objects, as well as their task designs, prevent a single ranking. Memory determines what can persist; retrieval determines which part becomes current evidence.

\subsection{Retrieval and Knowledge Substrates}
\label{sec:substrate-retrieval-knowledge}

Retrieval and knowledge substrates comprise indexed collections and the records that retrieval returns to the task. A returned record is usually a document, chunk, embedding neighbor, example, or manual rather than a remembered user fact~\cite{zhang_2025_a_survey_on_the_memory_mecha}. This carrier mediates between persistent stores and active context: it makes information task-relevant and assigns it an evidentiary role for the current decision. Agentic RAG makes the process recurrent because planning and tool-mediated searches can repeatedly revise what enters context~\cite{mishra2026agenticragsok}.

Representative studies expose different stages of that path. Managed-memory systems show movement between stored and active context, while FragFuse demonstrates reconstruction from separately stored fragments~\cite{packer2023memgpt,rao2026fragfusebypassingaccesscontrol}. AgentPoison targets memory or RAG knowledge bases so that ordinary-looking entries return under chosen triggers~\cite{chen2024agentpoison}. InjecAgent examines the adjacent action boundary, where untrusted external or tool content enters the decision context and induces tool-mediated behavior~\cite{zhan_2024_injecagent_benchmarking_indi}. Together, these works connect corpus admission, retrieval selection, prompt assembly, and downstream action without treating them as one operation.

The central risk is attribute loss along this path. Chunking can separate a statement from its source context, embeddings can make unrelated tasks neighbors, and rerankers can make a low-authority source appear central because it matches the query~\cite{zhang_2025_a_survey_on_the_memory_mecha}. When the item returns, the system must still distinguish evidence from instruction and preserve provenance, scope, and authority for the current task~\cite{buneman2001why_where_provenance,lin2026memory_security_survey}. Otherwise, ordinary abstraction and ranking can produce the same authority ambiguity as an overt attack.

Retrieval governance spans corpus admission, profile inspection, retrieval authorization, prompt assembly, and the policy governing any resulting action. AgentDojo supplies the unsafe-action setting, while Progent supplies the adjacent privilege-control handle~\cite{debenedetti2024agentdojo,shi2025progent}. \SAVER{} reads the combined evidence as a stored-state-to-active-evidence transition while keeping each control point distinct. Retrieval makes an object relevant; tool, API, and interface substrates determine whether that relevance can become an external effect.

\subsection{Tool, API, and Interface Substrates}
\label{sec:substrate-tool-interface}

Tool, API, and interface substrates comprise the callable resources and access metadata through which decisions produce external effects or resource costs. Tool schemas, action sets, credentials, permission scopes, argument templates, return records, execution logs, and environment handles determine whether a remembered or retrieved object becomes a file operation, message, purchase, browser step, shell command, database update, or high-cost API path. Their operational role is bidirectional: they execute agent decisions, but they also return observations that can enter later context or state.

The literature develops this boundary from action selection to ecosystem control. ReAct integrates action choice into the reasoning loop, and AgentDojo shows why tool-using prompt injection must track whether untrusted instructions influence execution~\cite{yao2022react,debenedetti2024agentdojo}. InjecAgent supplies the indirect-injection setting, while Progent places programmable privilege and scope constraints over tool use~\cite{zhan_2024_injecagent_benchmarking_indi,shi2025progent}. At ecosystem scale, MCPShield treats protocol metadata as safety-bearing state, MCP safety audits add an interface-level audit view, and prompt tool-selection work makes connector descriptions part of the same decision surface~\cite{zhou2026mcpshield,radosevich2025mcp_safety_audit,shi2026prompt_tool_selection}.

These studies expose two coupled risks. On the outbound path, a memory item or retrieved passage becomes more consequential when it changes an API call, file operation, message, or purchase. On the inbound path, a tool result may contain attacker-controlled text, stale records, or sensitive data; once cached or summarized, that observation becomes a new state object. Registration, permission grants, schema changes, connector installation, argument synthesis, and result caching are thus safety-bearing writes rather than neutral plumbing.
Governance must cover both directions. Action-time controls ask whether the current evidence may influence a particular effect, while ingestion controls attach source and scope attributes before tool output can shape memory, retrieval, or later arguments. A tool call realizes one decision; a skill preserves the procedure that can reproduce such decisions across tasks.

\subsection{Skill and Procedure Substrates}
\label{sec:substrate-skill-procedure}

Skill and procedure substrates package reusable ways of acting. Code snippets, prompt macros, planner routines, learned policies, recipes for tool use, reflection-derived rules, and partly executable manuals become procedures once a planner can retrieve and enact them across tasks. Compared with memory, their operational role is more prescriptive: a memory records what happened, whereas a skill specifies how to act. Compiling remembered experience into a callable procedure can carry stale assumptions, copied permissions, or overbroad heuristics into repeated execution.

Capability and attack studies expose successive parts of this transformation. Voyager treats accumulated skills as reusable capabilities rather than disposable outputs~\cite{wang2023voyager}. PoisonedEvolution targets promotion by making attacker-contributed trajectories appear recurrent, causally useful, and generalizable enough to be written into a skill~\cite{chen2026experienceinstructiontrajectorypoisoning}. SkillJack studies the resulting artifact: extraction can obscure malicious intent, and the implanted behavior can persist after its source records are removed~\cite{ying2026skilljackpersistentskillbackdoors}. ColluSkill widens the unit from one skill to an installed set whose incomplete components jointly reconstruct a harmful workflow~\cite{zeng2026colluskilladversarialcrossskillcomposition}. Although it does not demonstrate endogenous learning, it shows why reusable procedures cannot be governed only as isolated packages.

Carrier-specific evaluations then separate creation from reuse. SkillMisevo-Bench versions native skill stores and rebuilds its final persistence probe from exported \code{SKILL.md}, distinguishing an unsafe authored procedure from later retrieval and enactment~\cite{mao2026practicemakesunsafeskill}. EVOMAL exposes lineage across skill generations: a planted reference skill is imitated into a newly named executable descendant, written back to the library, and retrieved on later tasks~\cite{wu2026evomalselfpoisoningselfevolvingcoding}. Once several procedures are composed, this carrier-level problem becomes a workflow and topology problem.

\begin{SAVERTakeaway}{Skill Trust Depends on Derivation}
\label{takeaway:skill-derivation}
A skill's current name or author does not establish the legitimacy of its behavior. Extraction and imitation can hide an untrusted source while retaining its influence~\cite{ying2026skilljackpersistentskillbackdoors,wu2026evomalselfpoisoningselfevolvingcoding}. Skill governance therefore needs source-to-version lineage and execution preconditions, including dependencies on other installed procedures.
\end{SAVERTakeaway}

\subsection{Workflow, Planner, and Topology Substrates}
\label{sec:substrate-workflow-topology}

Workflow, planner, and topology substrates compose local procedures into routes. Planner state, delegation edges, retry policies, approval requirements, and orchestration templates determine where a task goes, who may act, and how much work may be spent. Their role is to coordinate capabilities over time and across actors, but the same route can also turn a routine label into auto-approval, a retry rule into an unbounded loop, or a delegation heuristic into unauthorized disclosure.

Representative work shows several forms of reusable routing state. Self-evolving-agent surveys cover evolving capabilities and artifacts~\cite{gao2025self_evolving_agents}; EvoMap makes shared instructions and assets concrete at network scale, where validation, incentive, and audit decisions travel with the exchanged object~\cite{ye2026behindevomap}. Traversal-as-Policy distills sandboxed execution logs into a gated behavior tree that can be checked before macro execution~\cite{li2026traversalaspolicylogdistilledgatedbehavior}. Risk-aware debate exposes coordination and escalation, agentic operating environments expose route structure, and formal agent-security work supplies the authority vocabulary for distinguishing legitimate coordination from silent expansion~\cite{yang2026openclaw_hospital,siu2026formalizing_agent_security}.

Topology adds a spatial dimension to the risk. Once state is copied to a scheduler, team agent, or domain agent, its audience and action surface change even if its text does not. HarnessSafe makes this movement measurable by binding subagent, session-summary, and workspace-handoff transitions to native harness mechanisms while holding delayed-risk semantics fixed~\cite{zhang2026harnesssafeevaluatingsafetypersistent}. The distinction between substrate and operation remains important: the route records where state resides, while delegation, summarization, or handoff records how its influence moved.
Workflow governance asks which route changed, which signal promoted it, which tasks later activated it, and which gate can suspend or reverse the resulting authority shift. Route admission, approval, budget, lineage, and rollback must be assessed together. When workflow output persists outside the runtime, the same questions transfer to external artifacts and shared organizational state.

\subsection{External Artifacts and Shared State Substrates}
\label{sec:substrate-external-artifacts}

External artifacts and shared-state substrates carry agent output into files, tickets, repositories, spreadsheets, calendars, emails, documents, dashboards, databases, and shared workspaces. These objects are not internal memories, but they can outlive the originating run, trigger automation, and be treated by later agents or humans as settled evidence. Their role is both archival and coordinative: they preserve results and transmit premises across tasks, actors, and systems.

Agent surveys provide the broader tool-using and collaborative setting, while self-evolving-agent work documents the reuse of artifacts and evolving capabilities~\cite{luo2025llm_agent_survey,gao2025self_evolving_agents}. Within this broader category, MAPLE-Guard studies shared memory rather than external files: a poisoned private memory can be retrieved, promoted into shared memory, and reused by agents that never observed the original attack, even when no malicious message crosses a visible communication edge at the moment of harm~\cite{xiong2026mapleguardmemoryawarelinkenforcement}. The corresponding question for external files is which later rules, training examples, or retrieval targets derive from them; that lineage must be established rather than inferred from shared-memory results. External artifacts remain at least partly inspectable, even when they are widely distributed. Model and policy substrates preserve influence in a less directly readable form and require a narrower inclusion rule.

\begin{SAVERTakeaway}{External Artifacts Extend the Recovery Scope}
\label{takeaway:artifact-recovery-scope}
An agent's output can become another actor's premise or an automation trigger after the originating run ends. Once influence enters shared files or organizational records, an internal memory edit no longer covers the affected scope. Recovery must identify those external descendants and distinguish what can be corrected, revoked, or only reported.
\end{SAVERTakeaway}

\subsection{Model and Policy Substrates}
\label{sec:substrate-model-policy}

Model and policy substrates include parameters, adapters, policy prompts, safety classifiers, reward models, key-value (KV) cache state, model-editing patches, and unlearning artifacts. These carriers enter the core analysis when an agent loop creates, selects, merges, loads, erases, rolls back, or inherits them in a way that changes later planning, refusal, routing, or tool use. Parameters and caches retain influence less directly inspectable than explicit memory; policy prompts and classifier configurations expose different textual or deployment handles.

Feasibility work establishes the deployment reach of this carrier. LoRA and PEFT surveys explain how adapters can be produced and reused across tasks~\cite{hu2022lora,han2024peft_survey}. The agent setting adds runtime questions that static fine-tuning alone does not answer: which task requested the update, which user or tool scope authorized it, what log records its activation, and what rollback handle remains if the update proves unsafe.

Risk and response studies populate different parts of that lifecycle. Fine-tuning-safety work examines safety erosion, PEFT research examines adapter-level deployment risk, broader LLM-security surveys locate the model-safety layer, and backdoor surveys characterize persistent triggerability~\cite{qi2023finetuning_safety,hong2024peta,wang2024unique_security_privacy_llm,zhou2025backdoor_llm_survey}. Safe LoRA provides a bounded mitigation anchor, while KVEraser supplies cache-local erasure for context influence already processed during prefill~\cite{hsu2024safe_lora,li2026kveraserlearningsteerkv}. Adjacent text-to-image work sharpens the same shared-artifact problem through triggerable text encoders, LoRA-plugin poisoning, LoRA-layer editing, and prompt-moderation defenses~\cite{li-etal-2025-twist,chen2026customization,chen2025lorashielddatafreeeditingalignment,wang2025aeiouunifieddefenseframework}.

For parameter and cache state, influence is less directly observable than in explicit memory, so governance depends more heavily on lineage metadata, safety summaries, regression probes, and deployment manifests. The evidence requirements grow from activation lineage, through adapter-scope records and rollback evidence, to residual-trigger tests after repair~\cite{qi2023finetuning_safety,rando2024universal_jailbreak_backdoors}. This branch remains a frontier substrate rather than an equally mature empirical pillar. It also shows why response records must themselves persist: future controls need durable evidence of which update was admitted, activated, or rolled back.

\noindent\textbf{Governance Band: Recovery, Deletion, and Control State.}
Across all these substrates, governance records also persist beyond the visible incident. Audit logs, approval records, quarantine lists, deletion receipts, rollback snapshots, and contestation outcomes become reusable state because later admission, retrieval, privilege, and repair decisions depend on them. A system that stores adaptive state but not reliable governance records cannot later show what it admitted, rejected, activated, or removed.
This is also where descendant repair becomes concrete. Quarantining a poisoned memory does not automatically neutralize its retrieval embedding, generated skill, copied workflow rule, or shared artifact. If those descendants are not tracked, the system may remove the visible object while leaving its influence active elsewhere. Governance state has to be governed in turn: stale approval rules, monitoring rules, or rollback records can become another adaptive surface rather than an external safeguard. \secref{sec:adaptation-operations} examines the operations that change these regimes and their safety attributes.

\begin{SAVERTakeaway}{Carrier Changes Require New Control Metadata}
\label{takeaway:carrier-control-metadata}
The same content can remain descriptive in memory but become prescriptive in a skill or workflow. Its safety therefore depends on the carrier's authority regime, not its wording alone. Cross-carrier comparison should ask which source constraints the new representation preserves and which additional permissions its use requires.
\end{SAVERTakeaway}

\FloatBarrier
\begin{table}[t]
\centering
\begin{minipage}{0.98\linewidth}
\captionsetup{hypcap=false}
\caption{Substrates carrying safety attributes in the current literature.}
\label{tab:substrate-literature-index}
\centering
\tiny
\setlength{\tabcolsep}{1.1pt}
\renewcommand{\arraystretch}{1.05}
\SAVERTableRows
\begin{tabular}{@{}p{0.20\linewidth}p{0.04\linewidth}p{0.13\linewidth}p{0.08\linewidth}p{0.095\linewidth}p{0.42\linewidth}@{}}
\toprule
\rowcolor{saverSubstrateHead}
\textbf{Method or Work} & \textbf{Year} & \textbf{\TableHeadRole} & \textbf{\TableHeadTarget} & \textbf{\TableHeadObject} & \textbf{Focus} \\
\midrule
\SAVERTableGroup{saverSubstrateHead}{6}{Foundational context and action boundary}
LLM agent survey~\cite{luo2025llm_agent_survey} & 2025 & \TableRoleSurveyCapability & \TableTargetAgent & \TableObjectGeneral & Agent architecture: memory, tools, planning \\
ReAct~\cite{yao2022react} & 2022 & \TableRoleCapability & \TableTargetLLMAgent & \TableObjectWorkflow & Reasoning and action context and admission hierarchy \\
AgentDojo~\cite{debenedetti2024agentdojo} & 2024 & \TableStudyAttackBenchmark & \TableTargetAgent & \TableObjectTool & Tool outputs and active context \\
Agent Context CPE~\cite{li2026whatsagentscontextcontext} & 2026 & \TableRoleAttack & \TableTargetAgent & \TableObjectWorkflow & Context-source roles and scopes across 12 harnesses \\
\SAVERTableGroup{saverSubstrateHead}{6}{Memory, retrieval, and persistent state}
Memory mechanisms~\cite{zhang_2025_a_survey_on_the_memory_mecha} & 2025 & \TableRoleSurveyCapability & \TableTargetLLMAgent & \TableObjectMemory & Episodic and semantic memory carriers \\
Memory security~\cite{lin2026memory_security_survey} & 2026 & \TableRoleSurveySafety & \TableTargetAgent & \TableObjectMemory & Persistent memory state and threats \\
MemGPT~\cite{packer2023memgpt} & 2023 & \TableRoleCapability & \TableTargetLLMAgent & \TableObjectMemory & Context and archival memory management \\
SimpleMem~\cite{liu2026simplemem} & 2026 & \TableRoleCapability & \TableTargetAgent & \TableObjectMemory & Long-term agent memory \\
A-Mem~\cite{xu_2025_a_mem_agentic_memory_for_llm} & 2025 & \TableRoleCapability & \TableTargetAgent & \TableObjectMemory & Agentic memory repository \\
Darwinian Memory~\cite{mi2026darwinian_memory} & 2026 & \TableRoleCapability & \TableTargetAgent & \TableObjectMemory & Evolving self maintained memory \\
Panini~\cite{rajesh2026panini} & 2026 & \TableRoleCapability & \TableTargetAgent & \TableObjectMemory & Structured agent memory state \\
Portable Agent Memory~\cite{ravindran2026portable_agent_memory} & 2026 & \TableRoleCapability & \TableTargetAgent & \TableObjectMemory & Transferable memory \\
Voyager~\cite{wang2023voyager} & 2023 & \TableRoleCapability & \TableTargetAgent & \TableObjectSkill & Reusable skill library \\
SkillJack~\cite{ying2026skilljackpersistentskillbackdoors} & 2026 & \TableRoleAttack & \TableTargetAgent & \TableObjectSkill & Persistent backdoor in extracted skills \\
PoisonedEvolution~\cite{chen2026experienceinstructiontrajectorypoisoning} & 2026 & \TableRoleAttack & \TableTargetAgent & \TableObjectSkill & Trajectory evidence promoted into skill instruction \\
SkillMisevo-Bench~\cite{mao2026practicemakesunsafeskill} & 2026 & \TableRoleBenchmarkSafety & \TableTargetAgent & \TableObjectSkill & Versioned skill state and clean-session reuse \\
EVOMAL~\cite{wu2026evomalselfpoisoningselfevolvingcoding} & 2026 & \TableRoleAttack & \TableTargetAgent & \TableObjectSkill & Agent-authored executable skill descendants \\
AgentPoison~\cite{chen2024agentpoison} & 2024 & \TableRoleAttack & \TableTargetAgent & \TableObjectMemory & Poisoned memory and knowledge base retrieval \\
Stealth memory injection~\cite{zhang2026clawsremembertellstealthy} & 2026 & \TableRoleAttack & \TableTargetAgent & \TableObjectMemory & Email to persistent memory compromise \\
GhostWriter~\cite{torres2026agentsremembermuchmemory} & 2026 & \TableRoleAttack & \TableTargetAgent & \TableObjectMemory & Personal agent memory store poisoning \\
MAPLE-Guard~\cite{xiong2026mapleguardmemoryawarelinkenforcement} & 2026 & \TableRoleDefense & \TableTargetAgent & \TableObjectMemory & Private-to-shared memory links \\
Persistent memory poisoning~\cite{sunil2026memory_poisoning_agents} & 2026 & \TableRoleAttack & \TableTargetAgent & \TableObjectMemory & Long-term memory poisoning \\
Zombie Agents~\cite{yang2026zombie_agents} & 2026 & \TableRoleAttack & \TableTargetAgent & \TableObjectMemory & Dormant memory attack \\
MemoryGraft~\cite{srivastava2025memorygraft} & 2025 & \TableRoleAttack & \TableTargetAgent & \TableObjectMemory & Experience store poisoning \\
EvoBreak~\cite{yan2026benignaloneharmfultogether} & 2026 & \TableRoleAttack & \TableTargetAgent & \TableObjectMemory & Complementary benign experience set \\
Agent memory privacy~\cite{wang2025privacy_agent_memory} & 2025 & \TableRoleSafety & \TableTargetAgent & \TableObjectMemory & User memory retention and leakage \\
PerMemSafe~\cite{an-etal-2026-permemsafe} & 2026 & \TableRoleBenchmarkSafety & \TableTargetAgent & \TableObjectMemory & Personalized memory safety context \\
PS-Bench~\cite{guo-etal-2026-personalization} & 2026 & \TableRoleBenchmarkSafety & \TableTargetAgent & \TableObjectMemory & Benign personal memory and intent inference \\
PASB~\cite{mao2026agentsdontjustagree} & 2026 & \TableRoleBenchmarkSafety & \TableTargetAgent & \TableObjectMemory & Committed claims in durable personal state \\
AuthMem-Bench~\cite{zhan2026memoryauthoritybenchmarkingauthority} & 2026 & \TableRoleBenchmarkSafety & \TableTargetAgent & \TableObjectMemory & Consolidated memories with source authority \\
A-MemGuard~\cite{wei2025amemguard} & 2025 & \TableRoleDefense & \TableTargetAgent & \TableObjectMemory & Persistent memory guard \\
\SAVERTableGroup{saverSubstrateHead}{6}{Tool and interface authority}
Progent~\cite{shi2025progent} & 2025 & \TableRoleDefense & \TableTargetAgent & \TableObjectTool & Tool authority and privilege control \\
MCPShield~\cite{zhou2026mcpshield} & 2026 & \TableRoleDefense & \TableTargetAgent & \TableObjectTool & MCP protocol metadata and tool routes \\
MCP Safety Audit~\cite{radosevich2025mcp_safety_audit} & 2025 & \TableRoleGovernance & \TableTargetAgent & \TableObjectTool & MCP server audit boundary \\
STAC~\cite{li2026stac} & 2026 & \TableRoleAttack & \TableTargetAgent & \TableObjectTool & Tool chain compromise \\
LivePI~\cite{zhao2026livepi} & 2026 & \TableRoleAttack & \TableTargetAgent & \TableObjectTool & Live prompt injection in interactive context \\
Prompt tool selection~\cite{shi2026prompt_tool_selection} & 2026 & \TableRoleAttack & \TableTargetAgent & \TableObjectTool & Tool description and connector choice manipulation \\
IPIGuard~\cite{an2025ipiguard} & 2025 & \TableRoleDefense & \TableTargetAgent & \TableObjectTool & Learned interaction traces \\
\SAVERTableGroup{saverSubstrateHead}{6}{Workflow, multi-agent, and shared state carriers}
Traversal-as-Policy~\cite{li2026traversalaspolicylogdistilledgatedbehavior} & 2026 & \TableRoleDefense & \TableTargetAgent & \TableObjectWorkflow & Gated behavior tree route policy \\
Self-evolving agents~\cite{gao2025self_evolving_agents} & 2025 & \TableRoleSurveyCapability & \TableTargetAgent & \TableObjectWorkflow & Adaptive workflows and artifacts \\
HarnessSafe~\cite{zhang2026harnesssafeevaluatingsafetypersistent} & 2026 & \TableRoleBenchmarkSafety & \TableTargetAgent & \TableObjectWorkflow & Cross-carrier harness state and boundaries \\
Self-evolution survey~\cite{tao2024self_evolution_llm_survey} & 2024 & \TableRoleSurveyCapability & \TableTargetLLMAgent & \TableObjectGeneral & Self-evolution methods and lifecycle \\
MADRA~\cite{wang2025madra} & 2025 & \TableRoleGovernance & \TableTargetAgent & \TableObjectMultiAgent & Safety aware delegation and debate state \\
ACIArena~\cite{an-etal-2026-aciarena} & 2026 & \TableStudyAttackBenchmark & \TableTargetAgent & \TableObjectMultiAgent & Cascading injection over agent trust \\
OpenClaw~\cite{yang2026openclaw_hospital} & 2026 & \TableRoleCapability & \TableTargetAgent & \TableObjectWorkflow & Hospital operating workflow \\
Formal Agent Security~\cite{siu2026formalizing_agent_security} & 2026 & \TableRoleFrameworkFormal & \TableTargetAgent & \TableObjectWorkflow & Agent authority and security boundary \\
LLM-MAS survey~\cite{he_2025_llm_based_multi_agent_system} & 2025 & \TableRoleSurveyCapability & \TableTargetAgent & \TableObjectMultiAgent & Delegated context and shared state \\
MAS security~\cite{nguyen2026security_mas} & 2026 & \TableRoleSurveySafety & \TableTargetAgent & \TableObjectMultiAgent & Message boundary and multi-agent state \\
Agents of Chaos~\cite{shapira2026agents_chaos} & 2026 & \TableRoleSafety & \TableTargetAgent & \TableObjectMultiAgent & Live-lab communication and workflow state \\
MAS code execution~\cite{triedman2025multi} & 2025 & \TableRoleAttack & \TableTargetAgent & \TableObjectMultiAgent & Orchestrator control and communication state \\
ControlValve~\cite{jha2026breaking} & 2026 & \TableRoleDefense & \TableTargetAgent & \TableObjectWorkflow & Permitted agent-invocation graph and rules \\
\SAVERTableGroup{saverSubstrateHead}{6}{Model and policy state substrates}
LoRA~\cite{hu2022lora} & 2022 & \TableRoleCapability & \TableTargetLLM & \TableObjectModel & Low-rank adapter substrate \\
PEFT survey~\cite{han2024peft_survey} & 2024 & \TableRoleSurveyCapability & \TableTargetLLM & \TableObjectModel & Parameter-efficient update family \\
Fine-tuning safety~\cite{qi2023finetuning_safety} & 2023 & \TableRoleSafety & \TableTargetLLM & \TableObjectModel & Fine-tuned checkpoint regression \\
Safe LoRA~\cite{hsu2024safe_lora} & 2024 & \TableRoleDefense & \TableTargetLLM & \TableObjectModel & LoRA adapter mitigation \\
KVEraser~\cite{li2026kveraserlearningsteerkv} & 2026 & \TableRoleDefense & \TableTargetLLM & \TableObjectModel & Cached context influence \\
PEFT Trojan~\cite{hong2024peta} & 2024 & \TableRoleAttack & \TableTargetLLM & \TableObjectModel & PEFT adapter backdoor \\
Feedback backdoors~\cite{rando2024universal_jailbreak_backdoors} & 2024 & \TableRoleAttack & \TableTargetLLM & \TableObjectModel & Preference or feedback update backdoor \\
Dormant behavior~\cite{gloaguen2025watch_your_steps} & 2025 & \TableRoleAttack & \TableTargetLLM & \TableObjectModel & Delayed activation in fine-tuned behavior \\
Backdoor survey~\cite{zhou2025backdoor_llm_survey} & 2025 & \TableRoleSurveyAttack & \TableTargetLLM & \TableObjectModel & Backdoor safety issues in LLM parameters \\
\bottomrule
\end{tabular}
\SAVERTableReset
\end{minipage}
\end{table}

\FloatBarrier

\section{Adaptation: State-Change and Activation Operations}
\label{sec:adaptation-operations}

\SAVERFlowA

\noindent\textbf{Operation Premise.} The substrate chapter identifies carrier families and their initial authority regimes; this chapter supplies the operation half of each \textbf{S}\(\times\)\textbf{A} pair. Throughout the survey, Adaptation is coded using seven operation families: \emph{add}, \emph{abstract}, \emph{activate}, \emph{modify}, \emph{forget}, \emph{propagate}, and \emph{migrate}. Implementation-specific verbs are assigned to these families by the state change they commit, so retrieval, for example, counts as activation only when the retrieved state receives authority over the current decision. The seven families provide a shared vocabulary for feedback reuse~\cite{shinn2023reflexion}, memory management~\cite{packer2023memgpt}, skill accumulation~\cite{wang2023voyager}, reasoning and action coupling~\cite{yao2022react}, and model adaptation~\cite{hu2022lora}. An operation has no fixed safety meaning in isolation: the comparison criterion is whether it preserves or revises the provenance, scope, authority, privacy, and recovery handles of the carrier on which it acts.

The evidence varies by operation. Stealth memory injection, GhostWriter, and memory-poisoning studies directly observe unsafe writes and delayed activation~\cite{zhang2026clawsremembertellstealthy,torres2026agentsremembermuchmemory,sunil2026memory_poisoning_agents}. Privacy work examines sensitive reuse, InjecAgent observes the action boundary, and Safe LoRA supplies a model-update intervention~\cite{wang2025privacy_agent_memory,zhan_2024_injecagent_benchmarking_indi,hsu2024safe_lora}. Later sections compare these with explicit migration and propagation studies, separating a transformed object from the new sites at which it can activate. \tabref{tab:adaptation-literature-index} indexes the operations; \secref{sec:invariant-violations} examines the resulting attribute failures.

\noindent\textbf{Selection Before Commitment.} Selection determines which proposed operation is committed; the seven families describe what that operation changes. The utility proxy $\widehat{u}\sb{t}$ ranks candidates, while $c\sb{t}$ restricts their eligibility and permitted role. If relevant safety evidence is absent from the candidates, proxy, or encoded checks, a locally useful update may still be selected. The distinction in \secref{sec:adaptive-state-model} therefore matters for every family: violating an explicit constraint differs from passing an incomplete constraint set.

\subsection{Add: Creating Reusable Influence}

Add operations create a new reusable object in a carrier; revising the same object's content or attributes is a modify operation. A single update can include both. Addition matters because transient context can become reusable influence. Adjacent capability systems show three mechanics: verbal critiques can be stored for later strategy choice~\cite{shinn2023reflexion}, memory items can be retained across context boundaries~\cite{packer2023memgpt}, and skills can become reusable procedures~\cite{wang2023voyager}. Trajectory-Informed Memory Generation adds structured, provenance-bearing learnings extracted from execution traces, while FORGE turns failed trajectories into reusable natural-language memories that can be broadcast across a population~\cite{fang2026trajectorymemory,bogdanov2026forge}. PASB makes the endogenous write path explicit: the system accepts an ordinary user claim and commits it with status promotion, attribution removal, or broader scope, so the unsafe transition can occur without a poisoned payload~\cite{mao2026agentsdontjustagree}. Adversarial paths reach the same operation through external email content saved by stealth memory injection or a hidden GhostWriter payload admitted by a tool-using personal agent~\cite{zhang2026clawsremembertellstealthy,torres2026agentsremembermuchmemory}. Privacy work adds the case in which useful personal information becomes retained profile state with sensitivity or scope obligations~\cite{wang2025privacy_agent_memory}.

A feedback note such as ``verify account scope before using external tools'' is a state transition, not automatically a violation or a defense: the carrier is a stored note, the trigger is task completion, the safety attributes say advisory reminder, the authority says it may guide planning but cannot override current consent, and the rollback handle points to the episode that produced it. If the same operation stores an unconditional action shortcut, the future influence is stronger because the note now looks like a rule. \SAVER{} treats both as add operations but distinguishes them by authority and downstream actionability.
Once a note, memory, skill, or observation is stored, later operations can summarize it, copy it, activate it, or revise it. Memory-poisoning work makes this descendant chain concrete: unsafe content can enter persistent memory or a knowledge base and become relevant only when a later operation reuses it~\cite{sunil2026memory_poisoning_agents}. The add operation is therefore a write event and the moment at which future influence becomes possible.

\subsection{Abstract: Compressing State and Reassigning Safety Attributes}

Once state has been added, abstraction can compress, summarize, cluster, or reframe it and thereby change its safety semantics~\cite{zhang_2025_a_survey_on_the_memory_mecha}. The derived object may no longer carry the exact provenance, uncertainty, user boundary, or temporal scope of the source material. MemGPT makes movement between immediate context and managed memory concrete~\cite{packer2023memgpt}; reflection can convert critiques into reusable heuristics~\cite{shinn2023reflexion}; and skill learning can convert successful traces into named routines~\cite{wang2023voyager}. These are adjacent operation mechanics; the safety claim begins when an abstraction drops source and derivation attributes~\cite{buneman2001why_where_provenance} or privacy and revocation attributes~\cite{wang2025privacy_agent_memory} that later matter for retrieval, action, or recovery.

Summarization is not inherently unsafe; the problem is that abstraction can silently change authority~\cite{lin2026memory_security_survey}. Suppose three notes local to one session say that a user prefers short answers in a specific project. A summary that keeps the source scope intact remains advisory: ``for this project, prefer concise replies.'' A summary that drops the project boundary becomes a global preference. A summary that drops uncertainty becomes a rule. The content is similar, but the role has been rewritten from scoped evidence to general instruction, and the object's future influence now spans tasks it was never granted~\cite{denning1976lattice_information_flow}.

AuthMem-Bench turns that role rewrite into a controlled abstraction test. Its paired histories hold the focal proposition and later task constant while varying whether the source permits the target use; the write-time module then asks whether consolidation preserves that distinction~\cite{zhan2026memoryauthoritybenchmarkingauthority}. This differs from a poisoned-memory benchmark because the claim need not be false or malicious. Here consolidation preserves the proposition but loses its use constraint.
Safe to Check, Unsafe to Use isolates a second abstraction failure at the compression boundary. A pre-compression filter can accept separated fragments because none contains a complete malicious instruction, yet the summarizer can relink them into compact, backend-actionable context~\cite{liu2026safecheckunsafeuse}. The unsafe object is therefore derived by the abstract operation rather than copied from one source span. Unlike authority collapse, relinking changes the instruction formed from the fragments; post-abstraction checks must therefore examine both source authority and derived content.

\begin{SAVERTakeaway}{Abstraction Requires a New Safety Check}
\label{takeaway:derived-state-check}
Abstraction can preserve a proposition while dropping its permitted-use constraint, or combine acceptable fragments into an unsafe instruction~\cite{zhan2026memoryauthoritybenchmarkingauthority,liu2026safecheckunsafeuse}. Source-level screening cannot settle either case. The derived object needs its own authority and content checks, together with source links that support later invalidation~\cite{buneman2001why_where_provenance}.
\end{SAVERTakeaway}

\subsection{Activate: Turning Stored State into Current Authority}

Added or abstracted state remains latent until activation lets it affect a current reasoning step, tool call, plan, or response. Non-adversarial cases make the authority change especially clear: PS-Bench shows truthful retrieved memories making an unsafe request appear contextually legitimate, while PerMemSafe makes retained user-specific context part of the response-safety judgment~\cite{guo-etal-2026-personalization,an-etal-2026-permemsafe}. Attack studies reach the same operation through stealthily injected personal-agent memory, GhostWriter payloads, or broader poisoned-memory reuse~\cite{zhang2026clawsremembertellstealthy,torres2026agentsremembermuchmemory,sunil2026memory_poisoning_agents}. The evidence then extends to tool settings, where InjecAgent makes untrusted external content safety-relevant at a tool-using decision~\cite{zhan_2024_injecagent_benchmarking_indi}. ReAct and Voyager supply adjacent mechanics for why this matters: reasoning traces can condition actions, and selected skills can become executable routines~\cite{yao2022react,wang2023voyager}. Activation is therefore the operation that converts latent influence into current authority.
The same retrieved memory may supply background evidence without being authorized to set a tool parameter or grant permission to act. An activation check therefore depends on the role assigned during prompt assembly or execution, not merely on which item matched the query.

CaMeL and the Dynamic Rule-based Isolation Framework for Trustworthy agentic systems (DRIFT) make this gate concrete at the tool boundary. CaMeL derives control and data flows from the trusted user query, then uses capabilities to prevent untrusted data from changing program flow or authorizing disallowed flows at tool calls~\cite{debenedetti2026defeating}. DRIFT constructs a minimal function trajectory and parameter checklist from the query, validates later deviations against user intent and privilege constraints, and masks conflicting instructions before they enter the active memory stream~\cite{li2025drift}. CaMeL therefore provides a design-time separation of control from untrusted data, whereas DRIFT adds runtime revision checks and current-stream isolation. ControlValve applies the same activation question to a multi-agent orchestrator: it generates a permitted control-flow graph and checks each agent invocation against that graph and a contextual rule~\cite{jha2026breaking}. All three treat activation authority as an explicit system decision rather than a property inherited from retrieved text or inter-agent messages.

\begin{SAVERTakeaway}{Relevance Is Not Permission}
\label{takeaway:relevance-not-permission}
Retrieval relevance cannot decide whether an item may control execution. CaMeL separates trusted control from untrusted data, DRIFT validates runtime deviations, and ControlValve checks permitted agent invocations~\cite{debenedetti2026defeating,li2025drift,jha2026breaking}. Their common target is the authority granted at use, even though they enforce it at different decision boundaries.
\end{SAVERTakeaway}

\subsection{Modify: Revising State, Safety Attributes, and Permissions}

Activation can expose stale scope, conflicting evidence, or unsafe authority that requires modification. Changing an existing object or its safety attributes can repair that influence or widen it further. MemGPT and memory mechanism surveys make editable memory concrete, while Reflexion and Voyager show adjacent editable carriers in feedback notes and skills~\cite{zhang_2025_a_survey_on_the_memory_mecha,shinn2023reflexion,wang2023voyager}. Beyond Heuristics formulates memory reading, writing, and maintenance as sequential decisions under uncertainty, while EvolveMem revises retrieval configurations with diagnosis, revert-on-regression, and explore-on-stagnation steps~\cite{sun2025beyondheuristics,liu2026evolvemem}. These systems make candidate selection and regression-sensitive modification observable; they do not by themselves establish that safety attributes are preserved. The operation can correct a memory, reconcile conflicting notes, revise a skill, change a workflow rule, or update activation metadata~\cite{gao2025self_evolving_agents}. Memory-privacy work directly motivates preserving sensitivity and scope~\cite{wang2025privacy_agent_memory}; provenance retention and rollback metadata are additional design requirements introduced by this survey.
Consider a feedback note that says ``retry with a shorter plan.'' If repeated failures reveal missing consent for external-account actions, the agent may modify it to ``when a task involves external accounts, ask for confirmation before action.'' The role shifts from performance heuristic to safety constraint. Its authority increases over planning decisions but should remain below explicit user instruction and system policy. Its scope also widens: the rule now applies to a class of tasks rather than one episode. A \SAVER{} trace records this as a modify operation rather than a new add because the old object, its descendants, and its activation history still matter.

Conflicts become operational when a memory says that a user prefers automatic reminders and a later interaction says reminders should pause during travel. A reconciliation step does more than choose one sentence. It changes temporal validity, scope, and activation priority. In safety attribute terms, the older memory may be downgraded to historical evidence, while the newer memory becomes active preference for a date range. If this safety attribute update is not explicit, later activation can resurrect stale authority.

\subsection{Forget: Invalidating Influence and Descendants}

Forgetting withdraws influence when a role expires, authorization is revoked, or modification cannot restore legitimate use. Removing, expiring, quarantining, or rolling back state is therefore an influence action rather than a storage action~\cite{wang2025privacy_agent_memory}. Later memory-poisoning and sleeper-memory work make persistent query-to-memory influence and delayed activation concrete~\cite{sunil2026memory_poisoning_agents,pulipaka2026hidden_memory}. We infer that copies, summaries, and derived rules must be checked separately rather than assuming removal of the source also removes their influence.
Once retained state has produced summaries, rule edits, cached plans, or invoked skills, this survey treats those derived objects as additional activation sites. The recovery unit is therefore the discovered descendant set rather than only the original record, with memory-security work supplying the broader carrier and threat vocabulary~\cite{lin2026memory_security_survey}.

For benign lifecycle changes, a time-sensitive preference may expire, a temporary permission may be revoked, or a stale skill may be retired. In each case the object moves to an expired, revoked, or archived state; its authority over the current action drops to zero; and its future influence should be limited to audit or historical reasoning. \SAVER{} therefore treats forgetting as more than deletion. The operation must answer whether the influence was actually withdrawn from the places where future decisions are made.

\subsection{Propagate: Copying Influence Across Actors and Contexts}

Before or after activation, propagation can copy state to another agent, memory segment, workflow stage, artifact, or context. Multi-agent surveys explain why shared state and delegated context are routine, while self-evolving-agent surveys place that movement inside a broader adaptive lifecycle~\cite{gao2025self_evolving_agents}. Architectural analysis relates compromised memory to inter-agent exchange, and ACIArena measures cascading injection across trust channels~\cite{torra2026memorypoisoningmas,an-etal-2026-aciarena}. MAPLE-Guard then localizes the memory route as write, retrieval, promotion, and cross-agent reuse: a poisoned private memory can enter shared memory and influence an agent that never saw the source attack~\cite{xiong2026mapleguardmemoryawarelinkenforcement}. Propagation therefore changes both the number of activation sites and the receiver's authority assignment; it cannot be measured only by inspecting visible inter-agent messages.

EVOMAL supplies a skill-to-skill propagation case. A retrieved malicious skill is imitated into an agent-authored descendant, and rolling library replacement lets that descendant become a later template. The agent self-poisoning rate (ASPR) measures the fraction of tasks that author and store a skill reproducing the planted pattern. After the eight planted seeds are removed, Qwen3 reaches 68\% ASPR at round five, Gemma4 reaches 29.4\%, and Devstral remains infected, whereas DeepSeek-V4-Pro and MiniMax return immediately to baseline and GPT-OSS decays~\cite{wu2026evomalselfpoisoningselfevolvingcoding}. The result is therefore conditional evidence of self-sustaining propagation under the paper's five-round replacement protocol, not a universal worm claim. It also makes source deletion an incomplete recovery test: the relevant unit is the retrievable descendant population after the seed disappears.

Recent multi-agent evidence separates propagation from control-flow hijacking. In the former, content or a practice crosses actors and is taken up by another agent; the live-laboratory cases in Agents of Chaos include cross-agent propagation among several other operational failures~\cite{shapira2026agents_chaos}. In the latter, untrusted content changes the active invocation edge. Triedman et al. show that an orchestrator can be induced to invoke unsafe agents or functions even when the individual agents resist the malicious request~\cite{triedman2025multi}. The transferred object is therefore not always a belief or memory item. It can be a routing decision that grants a new component authority. Recording both the content path and the invocation path prevents these two operations from collapsing into a generic prompt-injection label.

If a planner sends a worker the message ``the customer may prefer email; verify before sending,'' the propagated state remains advisory and scoped. If the same content is copied as ``send email to the customer,'' propagation has transformed authority even without changing much surface text. The receiver now sees an action instruction. \SAVER{} records the sender, receiver, copied content, retained safety attributes, and permitted activation modes so that later exposure can be traced to the propagation path.

\begin{SAVERTakeaway}{Replication Multiplies Repair Obligations}
\label{takeaway:replication-repair}
Propagation creates activation sites that may outlive the sender's original authorization. A correction at the source leaves inconsistent state if recipients retain older copies or derived rules. The propagation record must therefore support correction and revocation along the same paths that carried the original influence.
\end{SAVERTakeaway}

\subsection{Migrate: Moving State Across Authority Regimes}

Propagation multiplies copies; migration changes the regime under which a copy can be used. A migrate operation moves state into a substrate with different persistence, privacy, actionability, or governance semantics. MemGPT makes movement between memory regions visible~\cite{packer2023memgpt}, while self-evolving agent surveys ground broader substrate movement~\cite{gao2025self_evolving_agents}. A feedback note moved into long-term memory becomes more persistent; a memory item converted into a skill or workflow rule gains authority to act; a local observation moved into a shared artifact gains a new audience~\cite{wang2023voyager}. Model updates are adjacent rather than identical, but self-evolution surveys establish substrate movement as a major axis of agent adaptation~\cite{gao2025self_evolving_agents}.

Context reconstruction is a migration when it changes the authority regime even if the text is unchanged. When Context Gets Root compares each content item's fixed origin level with the level assigned by the harness at a later model call~\cite{he2026contextgetsrootprivilege}. Context privilege escalation adds an orthogonal scope comparison: M-CPE raises message role, while X-CPE moves content into a source with longer persistence or broader project, user, or cross-agent reach~\cite{li2026whatsagentscontextcontext}. The primary operation remains \emph{migrate} because the content crosses an authority or scope boundary. Writing it into the broader source also \emph{adds} reusable state, delegation may \emph{propagate} it, and later replay \emph{activates} it; these secondary verbs locate subsequent transitions rather than create a parallel taxonomy.

Two migration patterns matter most here: note \(\rightarrow\) rule, where feedback from one episode becomes planning authority, and memory or trajectory \(\rightarrow\) skill, where stored experience becomes executable capability. Voyager illustrates the capability mechanic. PoisonedEvolution isolates the preceding migration decision: attacker evidence must be included, attributed as reusable expertise, and realized in the generated skill even though the attacker cannot edit the skill bank~\cite{wang2023voyager,chen2026experienceinstructiontrajectorypoisoning}. SkillJack then shows what the completed crossing can hide and preserve, because extraction can obscure malicious intent and leave the implanted behavior active after source-record removal~\cite{ying2026skilljackpersistentskillbackdoors}. SkillMisevo-Bench extends observation beyond the crossing: its clean reload separates authored skill state, later retrieval, and fresh-session execution, so successful migration and successful activation remain distinct events~\cite{mao2026practicemakesunsafeskill}. The migration gate must therefore check not only content correctness but also whether support is independent, whether the target substrate grants stronger influence, whether source constraints survive extraction, whether later activation is attributable, and whether rollback can still reach the migrated behavior.

\noindent\textbf{Representation And Influence.} Reflection, skill learning, and workflow compilation show that related guidance can be carried by a memory item, heuristic, callable skill, or gated policy~\cite{shinn2023reflexion,wang2023voyager,wang2026oep}. Repairing one representation therefore establishes a scoped result: a summary, rule, or other carrier from the same lineage may still influence behavior. The literature establishes these transformation mechanics but does not yet provide a general test for functional equivalence across substrates. We treat cross-substrate reappearance as an evaluation question, using fixed downstream probes and lineage records rather than surface-form matching alone.

\subsection{Composition Paths}

A feedback note can be stored and later used to revise strategy~\cite{shinn2023reflexion}; managed memory can move information across context boundaries~\cite{packer2023memgpt}; a behavior trace can become an invoked skill~\cite{wang2023voyager}; and an observation can shape the next action in an interleaved reasoning-and-acting trace~\cite{yao2022react}. The add--abstract--activate compositions used by \SAVER{} are a comparative reading across these mechanics. Unsafe paths can follow the same structure, which is why operation records must carry safety attributes, authority, and future influence fields together.

\noindent\textbf{Local-Success Generalization.} Local success becomes safety-relevant when it is promoted into a durable role. Reflexion and Voyager establish the capability mechanics by turning feedback into strategy and successful traces into callable skills~\cite{shinn2023reflexion,wang2023voyager}. Safety studies then expose three distinct failure modes. OEP consolidates a locally correct but non-transferable experience into a harmful rule; experience-driven safety work finds that accumulated benign-task experience can bias later agents toward acting rather than refusing; and EvoBreak deliberately acquires complementary benign experiences whose joint activation weakens the safety boundary~\cite{wang2026oep,zhao2026experience_driven_safety,yan2026benignaloneharmfultogether}. The progression is from one promoted experience, to aggregate action bias, to adversarial composition over several experiences.

AgentPoison provides a fourth mechanism and clarifies why the corresponding success rates should not be ranked directly. It begins with an adversarial object written into memory or a knowledge base and tests trigger-selected retrieval~\cite{chen2024agentpoison}; OEP begins with locally useful experience and tests transfer after consolidation; the experience-driven study tests distributional change after benign accumulation; and EvoBreak tests whether a dependency set can be assembled and activated. These designs fail admission, selection, generalization, or composition gates under different opportunity sets. A comparable report must therefore identify the source set, operation, activation condition, and downstream scope rather than compare headline attack rates alone.

Within an agent loop, model updates can combine the same primitives. Fine-tuning and adapter training are add or modify operations over model state; post-training and preference optimization modify reward or policy behavior from feedback; merge combines adapters or patches and may change scope, authority, and safety regression attributes; edit and unlearn modify or remove behavior whose residual influence can be hard to observe. Existing model safety papers show why these operations deserve attention: fine-tuning can erode safety behavior~\cite{qi2023finetuning_safety}, PEFT updates create scoped but deployable risk objects~\cite{hong2024peta}, and dormant behavior can survive until the relevant activation context~\cite{gloaguen2025watch_your_steps}. The agent-loop criterion determines inclusion; regression probes support behavioral claims, whereas rollback handles and residual tests are additionally needed for recovery claims. \secref{sec:invariant-violations} asks which safety attributes these operations fail to preserve.

\begin{SAVERTakeaway}{Migration and Propagation Change Different Properties}
\label{takeaway:migration-propagation}
Propagation multiplies the sites at which influence can activate; migration changes the authority regime under which it can be used. One transfer can do both. Counting copies alone misses a role upgrade, while checking the new role alone misses the other copies that remain available for later reuse.
\end{SAVERTakeaway}

\begin{table}[t]
\centering
\begin{minipage}{0.98\linewidth}
\captionsetup{hypcap=false}
\caption{Adaptation and activation operations in the current literature. }
\label{tab:adaptation-literature-index}
\centering
\tiny
\setlength{\tabcolsep}{1.1pt}
\renewcommand{\arraystretch}{1.05}
\SAVERTableRows
\begin{tabular}{@{}p{0.17\linewidth}p{0.04\linewidth}p{0.12\linewidth}p{0.075\linewidth}p{0.08\linewidth}p{0.105\linewidth}p{0.37\linewidth}@{}}
\toprule
\rowcolor{saverAdaptationHead}
\textbf{Method or Work} & \textbf{Year} & \textbf{\TableHeadRole} & \textbf{\TableHeadTarget} & \textbf{\TableHeadObject} & \textbf{Operation} & \textbf{\TableHeadInfluenceShift} \\
\midrule
\SAVERTableGroup{saverAdaptationHead}{7}{Capability mechanics: add, abstract, migrate}
Reflexion~\cite{shinn2023reflexion} & 2023 & \TableRoleCapability & \TableTargetAgent & \TableObjectWorkflow & \TableAdaptationOp{Add} & Feedback becomes decision evidence \\
MemGPT~\cite{packer2023memgpt} & 2023 & \TableRoleCapability & \TableTargetLLMAgent & \TableObjectMemory & \TableAdaptationOp{Add/Migrate} & Context becomes managed memory \\
Voyager~\cite{wang2023voyager} & 2023 & \TableRoleCapability & \TableTargetAgent & \TableObjectSkill & \TableAdaptationOp{Migrate} & Experience becomes callable skill \\
SimpleMem~\cite{liu2026simplemem} & 2026 & \TableRoleCapability & \TableTargetAgent & \TableObjectMemory & \TableAdaptationOp{Add} & Interaction becomes durable memory \\
Panini~\cite{rajesh2026panini} & 2026 & \TableRoleCapability & \TableTargetAgent & \TableObjectMemory & \TableAdaptationOp{Add} & Experience becomes structured memory \\
Experience Evolving Memory~\cite{li2026experience_evolving_memory} & 2026 & \TableRoleCapability & \TableTargetAgent & \TableObjectWorkflow & \TableAdaptationOp{Abstract} & Episode becomes reusable procedure \\
\SAVERTableGroup{saverAdaptationHead}{7}{Unsafe writes and abstraction drift}
AgentPoison~\cite{chen2024agentpoison} & 2024 & \TableRoleAttack & \TableTargetAgent & \TableObjectMemory & \TableAdaptationOp{Add/Activate} & Poisoned write becomes memory with authority \\
Stealth memory injection~\cite{zhang2026clawsremembertellstealthy} & 2026 & \TableRoleAttack & \TableTargetAgent & \TableObjectMemory & \TableAdaptationOp{Add/Activate} & Email payload becomes hidden memory influence \\
GhostWriter~\cite{torres2026agentsremembermuchmemory} & 2026 & \TableRoleAttack & \TableTargetAgent & \TableObjectMemory & \TableAdaptationOp{Add/Activate} & Hidden payload becomes recalled memory \\
PASB~\cite{mao2026agentsdontjustagree} & 2026 & \TableRoleBenchmarkSafety & \TableTargetAgent & \TableObjectMemory & \TableAdaptationOp{Add/Activate} & Accepted claim becomes durable personal state \\
AuthMem-Bench~\cite{zhan2026memoryauthoritybenchmarkingauthority} & 2026 & \TableRoleBenchmarkSafety & \TableTargetAgent & \TableObjectMemory & \TableAdaptationOp{Abstract} & Consolidation drops source-use constraints \\
Zombie Agents~\cite{yang2026zombie_agents} & 2026 & \TableRoleAttack & \TableTargetAgent & \TableObjectMemory & \TableAdaptationOp{Add/Activate} & Dormant memory becomes state conditioned on trigger \\
MemoryGraft~\cite{srivastava2025memorygraft} & 2025 & \TableRoleAttack & \TableTargetAgent & \TableObjectMemory & \TableAdaptationOp{Add/Activate} & Poisoned experience becomes evidence \\
OEP~\cite{wang2026oep} & 2026 & \TableRoleAttack & \TableTargetAgent & \TableObjectWorkflow & \TableAdaptationOp{Abstract} & Experience becomes unsafe reusable rule \\
Experience-Driven Safety~\cite{zhao2026experience_driven_safety} & 2026 & \TableRoleSafety & \TableTargetAgent & \TableObjectWorkflow & \TableAdaptationOp{Abstract} & Experience update creates policy drift \\
EvoBreak~\cite{yan2026benignaloneharmfultogether} & 2026 & \TableRoleAttack & \TableTargetAgent & \TableObjectMemory & \TableAdaptationOp{Abstract/Activate} & Benign experiences compose into unsafe influence \\
AgenticEval~\cite{wang2026agenticeval} & 2026 & \TableRoleBenchmarkSafety & \TableTargetAgent & \TableObjectEvaluation & \TableAdaptationOp{Abstract} & Adaptive state becomes task suite denominator \\
MemEvoBench~\cite{xie2026memevobench} & 2026 & \TableRoleBenchmarkSafety & \TableTargetAgent & \TableObjectMemory & \TableAdaptationOp{Abstract} & Evolving memory becomes metric denominator \\
PerMemSafe~\cite{an-etal-2026-permemsafe} & 2026 & \TableRoleBenchmarkSafety & \TableTargetAgent & \TableObjectMemory & \TableAdaptationOp{Activate} & Personalized memory becomes safety denominator \\
PS-Bench~\cite{guo-etal-2026-personalization} & 2026 & \TableRoleBenchmarkSafety & \TableTargetAgent & \TableObjectMemory & \TableAdaptationOp{Activate} & Benign memory changes inferred intent \\
\SAVERTableGroup{saverAdaptationHead}{7}{Control, activation, and repair operations}
GLOVE~\cite{yin2026glove} & 2026 & \TableRoleDefense & \TableTargetAgent & \TableObjectMemory & \TableAdaptationOp{Modify} & Memory is realigned against environment \\
SRM~\cite{chitan2026srm} & 2026 & \TableRoleDefense & \TableTargetAgent & \TableObjectMemory & \TableAdaptationOp{Modify} & Session safety becomes memory signal \\
MemGovern~\cite{wang2026memgovern} & 2026 & \TableRoleGovernance & \TableTargetAgent & \TableObjectMemory & \TableAdaptationOp{Modify} & Recall is revised by policy \\
SSGM~\cite{lam2026ssgm} & 2026 & \TableRoleGovernance & \TableTargetAgent & \TableObjectMemory & \TableAdaptationOp{Activate} & Stored memory passes through gate \\
Progent~\cite{shi2025progent} & 2025 & \TableRoleDefense & \TableTargetAgent & \TableObjectTool & \TableAdaptationOp{Activate} & Tool authority becomes explicit decision \\
CaMeL~\cite{debenedetti2026defeating} & 2026 & \TableRoleDefense & \TableTargetAgent & \TableObjectTool & \TableAdaptationOp{Activate} & Trusted control flow gates tool authority \\
DRIFT~\cite{li2025drift} & 2025 & \TableRoleDefense & \TableTargetAgent & \TableObjectTool & \TableAdaptationOp{Activate} & Plan deviations trigger privilege validation \\
ControlValve~\cite{jha2026breaking} & 2026 & \TableRoleDefense & \TableTargetAgent & \TableObjectWorkflow & \TableAdaptationOp{Activate} & Agent invocation must follow permitted graph \\
MAGE~\cite{wang2026magesafeguardingllmagents} & 2026 & \TableRoleDefense & \TableTargetAgent & \TableObjectMemory & \TableAdaptationOp{Activate} & Shadow memory becomes safety signal \\
Privacy Agent Memory~\cite{wang2025privacy_agent_memory} & 2025 & \TableRoleSafety & \TableTargetAgent & \TableObjectMemory & \TableAdaptationOp{Forget} & Sensitive influence is scoped or withdrawn \\
Memory poisoning~\cite{sunil2026memory_poisoning_agents} & 2026 & \TableRoleAttack & \TableTargetAgent & \TableObjectMemory & \TableAdaptationOp{Add/Activate} & Poisoned query creates later retrieval influence \\
OpenClaw~\cite{yang2026openclaw_hospital} & 2026 & \TableRoleCapability & \TableTargetAgent & \TableObjectWorkflow & \TableAdaptationOp{Activate} & Retrieved longitudinal state enters workflow context \\
\SAVERTableGroup{saverAdaptationHead}{7}{Propagation, migration, and route drift}
Collective Misremembering~\cite{xu2026when} & 2026 & \TableRoleAttack & \TableTargetAgent & \TableObjectMultiAgent & \TableAdaptationOp{Propagate} & Drift spreads across agents \\
ACIArena~\cite{an-etal-2026-aciarena} & 2026 & \TableStudyAttackBenchmark & \TableTargetAgent & \TableObjectMultiAgent & \TableAdaptationOp{Propagate} & Cascading injection crosses agent trust \\
HarnessSafe~\cite{zhang2026harnesssafeevaluatingsafetypersistent} & 2026 & \TableRoleBenchmarkSafety & \TableTargetAgent & \TableObjectWorkflow & \TableAdaptationOp{Migrate} & Influence crosses carrier and harness boundaries \\
When Context Gets Root~\cite{he2026contextgetsrootprivilege} & 2026 & \TableRoleAttack & \TableTargetAgent & \TableObjectWorkflow & \TableAdaptationOp{Migrate} & Tool content gains a higher context role \\
Agent Context CPE~\cite{li2026whatsagentscontextcontext} & 2026 & \TableRoleAttack & \TableTargetAgent & \TableObjectWorkflow & \TableAdaptationOp{Migrate} & Content gains a higher role or wider scope \\
MAPLE-Guard~\cite{xiong2026mapleguardmemoryawarelinkenforcement} & 2026 & \TableRoleDefense & \TableTargetAgent & \TableObjectMemory & \TableAdaptationOp{Propagate} & Gates private-to-shared memory reuse \\
Agents of Chaos~\cite{shapira2026agents_chaos} & 2026 & \TableRoleSafety & \TableTargetAgent & \TableObjectMultiAgent & \TableAdaptationOp{Propagate} & Unsafe practices cross agents in a live setting \\
MAS code execution~\cite{triedman2025multi} & 2025 & \TableRoleAttack & \TableTargetAgent & \TableObjectMultiAgent & \TableAdaptationOp{Activate} & Untrusted content redirects agent invocation \\
Embedding defense failure~\cite{zhang2026embedding_defenses_fail} & 2026 & \TableRoleAttack & \TableTargetAgent & \TableObjectMemory & \TableAdaptationOp{Propagate} & Shared memory filter is bypassed \\
MADRA~\cite{wang2025madra} & 2025 & \TableRoleGovernance & \TableTargetAgent & \TableObjectMultiAgent & \TableAdaptationOp{Propagate/Modify} & Risk assessments are exchanged and revised before gating \\
Devil's Moltbook~\cite{wang2026devil_moltbook} & 2026 & \TableRoleAttack & \TableTargetAgent & \TableObjectMultiAgent & \TableAdaptationOp{Propagate} & Exposure shifts agent society state \\
Traversal-as-Policy~\cite{li2026traversalaspolicylogdistilledgatedbehavior} & 2026 & \TableRoleDefense & \TableTargetAgent & \TableObjectWorkflow & \TableAdaptationOp{Migrate/Activate} & Logs become workflow policy \\
Library Drift~\cite{zhang2026library_drift} & 2026 & \TableRoleSafety & \TableTargetAgent & \TableObjectSkill & \TableAdaptationOp{Migrate/Modify} & Skill library reroutes tasks \\
SkillJack~\cite{ying2026skilljackpersistentskillbackdoors} & 2026 & \TableRoleAttack & \TableTargetAgent & \TableObjectSkill & \TableAdaptationOp{Migrate} & Poisoned trajectory becomes persistent skill \\
PoisonedEvolution~\cite{chen2026experienceinstructiontrajectorypoisoning} & 2026 & \TableRoleAttack & \TableTargetAgent & \TableObjectSkill & \TableAdaptationOp{Migrate} & Recurring trajectory evidence becomes instruction \\
SkillMisevo-Bench~\cite{mao2026practicemakesunsafeskill} & 2026 & \TableRoleBenchmarkSafety & \TableTargetAgent & \TableObjectSkill & \TableAdaptationOp{Migrate/Activate} & Skill authoring, retrieval, and carryover gates \\
EVOMAL~\cite{wu2026evomalselfpoisoningselfevolvingcoding} & 2026 & \TableRoleAttack & \TableTargetAgent & \TableObjectSkill & \TableAdaptationOp{Propagate/Activate} & Authored copies re-enter the retrievable library \\
Tool-Drift~\cite{dabas2026tool_drift} & 2026 & \TableRoleSafety & \TableTargetAgent & \TableObjectTool & \TableAdaptationOp{Activate/Modify} & Tool choice state reroutes \\
JITRL~\cite{li2026jitrl} & 2026 & \TableRoleCapability & \TableTargetAgent & \TableObjectWorkflow & \TableAdaptationOp{Modify/Activate} & Runtime learning updates policy \\
RQGM~\cite{iacob2026redqueengodelmachine} & 2026 & \TableRoleCapability & \TableTargetAgent & \TableObjectEvaluation & \TableAdaptationOp{Modify} & Evaluator utility becomes state scoped by epoch \\
\SAVERTableGroup{saverAdaptationHead}{7}{Model update operations}
Fine-tuning safety~\cite{qi2023finetuning_safety} & 2023 & \TableRoleSafety & \TableTargetLLM & \TableObjectModel & \TableAdaptationOp{Modify} & Fine-tuning changes behavior \\
Safe LoRA~\cite{hsu2024safe_lora} & 2024 & \TableRoleDefense & \TableTargetLLM & \TableObjectModel & \TableAdaptationOp{Modify} & Adapter constrains behavior \\
PEFT Trojan~\cite{hong2024peta} & 2024 & \TableRoleAttack & \TableTargetLLM & \TableObjectModel & \TableAdaptationOp{Add/Activate} & Adapter trigger becomes deployed safety issue \\
Dormant behavior~\cite{gloaguen2025watch_your_steps} & 2025 & \TableRoleAttack & \TableTargetLLM & \TableObjectModel & \TableAdaptationOp{Activate} & Fine-tuned behavior waits for context \\
\bottomrule
\end{tabular}
\SAVERTableReset
\end{minipage}
\end{table}

\section{Violation: Safety Attribute Failures}
\label{sec:invariant-violations}

\SAVERFlowV

Given a carrier-operation pair, violation analysis asks whether the resulting role remains legitimate. A violation occurs when the operation fails to preserve a required safety attribute, such as source authority, user scope, privacy, or a resource limit. This condition is distinct from the evidence that reveals it. A write-time check may detect the mismatch immediately; otherwise it can remain latent until retrieval, execution, or another later use. The distinction lets \textbf{V} describe what failed and \textbf{E} describe what was actually observed.

The evidence enters through two complementary routes. GhostWriter and PASB separate storage from later activation through adversarial admission and ordinary claim commitment, making provenance and persistence failures visible~\cite{torres2026agentsremembermuchmemory,mao2026agentsdontjustagree}. InjecAgent and programmable privilege control expose the authority question when state reaches a tool-using decision~\cite{zhan_2024_injecagent_benchmarking_indi,shi2025progent}. These works are related by the failed attribute, even though their benchmark denominators differ. Together they produce a named violation and expected trigger; \secref{sec:exposure-surfaces} uses that pair to identify diagnostic evidence. \tabref{tab:violation-literature-index} closes the chapter with the detailed paper-level index.

\subsection{Provenance Loss and Safety Attribute Laundering}

Provenance loss removes information about a state's source, contributor, or derivation. Safety attribute laundering occurs when the transformed object then appears to have a stronger role than its origin warrants. The critical loss is not necessarily factual content: later components may still read the same claim but can no longer distinguish environmental data, an attributed preference, model inference, or authorized policy.

The literature exposes different entry points. AgentPoison places malicious demonstrations in a retrievable store and tests their selection by a trigger query~\cite{chen2024agentpoison}. GhostWriter instead exploits admission into personal-agent memory before later retrieval~\cite{torres2026agentsremembermuchmemory}. PASB studies the agent's own consolidation decision: attribution removal and status promotion can make an ordinary conversational claim appear established, while scope broadening extends its later use~\cite{mao2026agentsdontjustagree}. Store poisoning, unsafe admission, and claim promotion therefore require different checks even when all culminate in trusted-looking memory. Source retention at write and abstraction supports later authorization; checking only the retrieved wording cannot recover provenance that those operations discarded.

\subsection{Authority, Actionability, and Boundary Escalation}

Authority, actionability, or boundary escalation can follow provenance loss, but can also occur when a known source is assigned an unauthorized role. The carrier may be a memory item, retrieved instruction, plan fragment, tool result, workflow rule, learned skill, feedback summary, or privilege token. The operation is usually \emph{activate}, \emph{migrate}, or \emph{modify}: descriptive state becomes a planning constraint, readable context becomes a tool argument, a local preference becomes a broader rule, or a stored fragment becomes part of an access control bypass. What fails is role preservation across time. AgentDojo gives the integrated-tool setting in which untrusted tool outputs can hijack behavior; Progent supplies the response vocabulary for explicit privilege control; and FragFuse shows the temporal memory variant, where low-authority fragments are stored across interactions and later fused through retrieval so the current query appears cleaner than the activated state actually is~\cite{debenedetti2024agentdojo,shi2025progent,rao2026fragfusebypassingaccesscontrol}.

AuthMem-Bench isolates the same family without changing claim content. Matched histories contain the same proposition and lead to the same later task, but only one source condition permits the target use; authority collapse occurs when consolidation makes those histories operationally indistinguishable~\cite{zhan2026memoryauthoritybenchmarkingauthority}. This design separates the internal violation from its exposure: the violation is the false authority upgrade during abstraction, while the exposure is a later prohibited tool action after the collapsed memory is activated. Authority is therefore not recoverable from proposition text alone once consolidation has erased the source-use constraint.

Harness privilege escalation reaches the same invariant through context migration rather than memory consolidation. The harness can carry unchanged low-level content into a new call under an actual user or system-effective role, so the downstream model need not confuse roles or violate its visible hierarchy~\cite{he2026contextgetsrootprivilege}. A complementary context-source analysis distinguishes this role increase from scope increase: M-CPE changes which instructions the content may compete with, while X-CPE lets session-scoped influence survive at project or user scope~\cite{li2026whatsagentscontextcontext}. AuthMem-Bench shows a source-use constraint disappearing during abstraction; these harness studies show source role or scope being replaced during reconstruction. All three instantiate safety attribute laundering because the downstream component receives stronger authority or broader reuse than the origin warrants.

\begin{SAVERTakeaway}{Correct Content Can Carry Unjustified Authority}
\label{takeaway:content-authority-separation}
Authority collapse need not falsify a claim, and harness privilege escalation need not make the model violate its visible instruction hierarchy~\cite{zhan2026memoryauthoritybenchmarkingauthority,he2026contextgetsrootprivilege}. The failure can occur earlier, when consolidation or context assembly grants a role the source never held. Content accuracy and model-side hierarchy compliance therefore leave distinct authority checks unresolved.
\end{SAVERTakeaway}

Authorization continuity exposes a complementary failure. AuthMem-Bench holds the proposition fixed and varies its source authority; the state-bound model holds the user grant fixed and asks whether changes to control state, capability, delegation, task phase, trust context, or enforcement have changed the subject or context to which that grant applies~\cite{zhang2026agentiauthorizedearned}. The model separates two independent invariants: the mutation must remain inside the grant's transition envelope, and every realized effect must remain below its fixed ceiling. The failed attribute is the continuing validity of the grant.

These cases share an authorization requirement: descriptive state needs an explicit grant before it can direct a plan, tool call, or workflow. ReAct, Voyager, and Reflexion show the capability paths through which observations, skills, and feedback influence later behavior~\cite{yao2022react,wang2023voyager,shinn2023reflexion}. R-Judge and GuardAgent supply risk-recognition and guard-reasoning signals~\cite{yuan2024rjudge,xiang2024guardagent}; an authorization control must additionally decide whether the current user, task, and effect permit that influence. Detecting risky content and validating the grant are consequently separate obligations.

\noindent\textbf{Promotion Failures.} The promotion test in \secref{sec:adaptive-state-model} separates evidence of local effectiveness from permission for broader reuse. OEP makes single-experience transfer failure concrete, experience-driven safety work shows aggregate benign experience shifting the action--refusal boundary, and EvoBreak shows individually acceptable experiences becoming unsafe under composition~\cite{wang2026oep,zhao2026experience_driven_safety,yan2026benignaloneharmfultogether}.
Skill-oriented attacks then expose the cross-substrate step: PoisonedEvolution targets the attribution decision that grants recurring trajectory evidence procedural authority, whereas SkillJack tests the laundering and persistence of the resulting skill~\cite{chen2026experienceinstructiontrajectorypoisoning,ying2026skilljackpersistentskillbackdoors}.

SkillMisevo-Bench resolves the completed path into successive gates: all 21 evolved configurations author unsafe artifacts, 19 retrieve unsafe skills, 19 show benign-task contamination, and 15 retain fresh-session harm~\cite{mao2026practicemakesunsafeskill}. A configuration that stops before final harm is therefore not evidence of a safe promotion; the unsafe artifact may simply not have been retrieved or executed. Utility-to-authority escalation names the local failure, while selection-induced boundary drift names the longitudinal diagnosis when individually admitted promotions cumulatively widen scope or authority.

\subsection{Privacy and Purpose Failure}

Role widening can concern purpose and audience as well as action authority. Privacy can therefore fail before disclosure when sensitive state loses the attributes that made later use legitimate. The carrier may be explicit personal information in memory, inferred attributes, user-specific preferences, private tool outputs, conversation history, or summaries that merge several users or sessions. Add, abstract, activate, propagate, or migrate can store too much, broaden a profile, grant a sensitive item influence in an unrelated task, share it with another agent, or move it into a less protected artifact. The failed attributes are sensitivity, subject boundary, purpose limitation, retention constraint, or personalized risk context. Memory-privacy work grounds these use restrictions~\cite{wang2025privacy_agent_memory}.

A related question concerns response safety under personal context, rather than unauthorized disclosure or reuse. PerMemSafe shows how long-horizon user-specific memories can make an otherwise safe response unsafe in that user's implicit context~\cite{an-etal-2026-permemsafe}. Differential Harm Propensity isolates another personalized boundary: mental-health disclosure changes measured harm and refusal behavior, so the evaluator must treat the disclosure as risk-bearing context rather than a neutral profile field~\cite{yildirim2026differentialharm}. PS-Bench isolates the converse boundary failure: an inherently harmful request can be treated as safe because truthful personal memories bias the model toward a benign intent interpretation~\cite{guo-etal-2026-personalization}.

\begin{SAVERTakeaway}{Privacy Can Fail Before Disclosure}
\label{takeaway:privacy-before-disclosure}
Persistent user context can lose its purpose, retention, or audience restrictions before any information is printed or acted on~\cite{li_2024_personal_llm_agents_insights,wang2025privacy_agent_memory}. An output-only privacy test misses that earlier state-level failure. The survey therefore calls for preserving these restrictions at update time and checking them again at activation.
\end{SAVERTakeaway}

\subsection{Persistent Descendants and Incomplete Rollback}

Persistent descendant failure arises when correcting or deleting the source leaves its derived influence available for later activation. The failed operation is incomplete forgetting: an earlier abstraction, copy, or migration has created an object that the repair does not reach. Memory-poisoning and sleeper-memory work establish delayed reuse, OEP extends the mechanism to consolidated experience, and SkillJack directly tests the source--descendant distinction: its abstract reports that 80.0\% of skill-mediated attacks persist after the original poisoned records are deleted~\cite{sunil2026memory_poisoning_agents,pulipaka2026hidden_memory,wang2026oep,ying2026skilljackpersistentskillbackdoors}.

This failure differs from merely storing an unsafe object. It requires a correction or revocation that no longer governs a surviving activation site. A corrected preference can still steer a workflow that copied its earlier version; a retired source can leave an executable skill behind. The repair target is thus a derivation relation, not one storage location. The descendant inventory and residual tests in \secref{sec:response-governance} specify how to assess that target without assuming complete visibility.

\subsection{Operational Integrity Failures}

Operational integrity concerns resource use and traceability, whether or not unsafe state has already produced descendants. A workflow rule, retry policy, tool template, planner state, retrieved instruction, self-evaluation loop, skill, shared plan, or propagated artifact can affect resource use, liveness, auditability, and accountability. Activate, modify, propagate, or migrate may turn a local retry heuristic into a general workflow rule, trigger repeated searches, loop over paid APIs, or leave causal state active after a visible action is blocked. The failed attributes are resource budget, rate limit, cost ceiling, timeout, stop condition, trace link, and repair ownership. OWASP frames excessive agency and unbounded consumption as application risks; AgentDojo moves observation beyond final text; Agent Security Bench (ASB) and Agent-SafetyBench broaden the evaluation vocabulary~\cite{owasp2025llm_top10,debenedetti2024agentdojo,zhang_2024_agent_security_bench_asb_for,zhang_2024_agent_safetybench_evaluating}.

Resource control and auditability impose different checks on the same execution. A workflow rule can preserve task success while multiplying cost; a guard can detect the resulting risk without identifying which stored rule caused it. Personal-agent work supplies the operational setting, GuardAgent and R-Judge examine recognition, Progent constrains privilege, and EDDOps addresses monitoring and maintenance~\cite{li_2024_personal_llm_agents_insights,xiang2024guardagent,yuan2024rjudge,shi2025progent,xia_2024_evaluation_driven_developmen}. Together they motivate enforceable budget and stop conditions plus attribution to the active state. Direct tests of resource-exhaustion repair and lifecycle auditability remain an evidence gap in the reviewed corpus.

\subsection{Model Safety Regression}

The same attribute logic extends from explicit state to model-side state, but the carrier is harder to inspect and often broader in scope. Model safety regression enters this survey when an update to parameters, adapters, reward models, edited components, policy prompts, or unlearning artifacts weakens an attribute that a later agent relies on. The operation is central when the agent loop or deployment controller fine-tunes, trains adapters, post-trains, merges, edits, distills, unlearns, selects, loads, or rolls back that object for later behavior. The failure can be alignment erosion, adapter authority escalation, reward drift, persistent backdoors, incomplete unlearning, or lost deployment lineage. Fine-tuning safety, PEFT risk, feedback poisoning, and dormant behavior mark successive parts of this path~\cite{qi2023finetuning_safety,hong2024peta,rando2024universal_jailbreak_backdoors,gloaguen2025watch_your_steps}. They support central \SAVER{} claims only when the carrier, activation route, metric, and agent-loop scope are explicit.

An adapter that looks benign may gain broader authority when loaded into a production agent. A task-local fine-tuning update may change refusal behavior outside the intended scope. A merged adapter stack may erase the safety assumptions under which each component was evaluated. An unlearning patch may remove a visible behavior while leaving residual influence under paraphrase or downstream fine-tuning. The same transition logic therefore applies to the model substrate. The exposure trigger is a later inference, tool use decision, refusal boundary, policy check, or generalization failure across tasks. Safe LoRA provides a bounded LoRA-weight intervention for reducing fine-tuning safety loss~\cite{hsu2024safe_lora}; broad LLM security surveys catalogue model-level security and privacy risks~\cite{wang2024unique_security_privacy_llm}, and backdoor surveys frame triggerable residual risk~\cite{zhou2025backdoor_llm_survey}. The adapter registry, deployment lineage, version rollback, and residual probes proposed here are governance fields needed to connect those local results to an agent loop.
Across the carrier families, a latent violation and its later activation remain distinct.
\secref{sec:exposure-surfaces} identifies the events that reveal these failures; \secref{sec:response-governance} examines the interventions they support.

\begin{SAVERTakeaway}{A Missing Trigger Does Not Establish a Safe State}
\label{takeaway:latent-violation}
An unsafe update can remain dormant when no later task activates it. SkillMisevo-Bench distinguishes unsafe artifacts, retrieval, and fresh-session harm, showing why these stages cannot be collapsed into one outcome~\cite{mao2026practicemakesunsafeskill}. A run without visible harm does not establish that earlier operations preserved the state's safety attributes.
\end{SAVERTakeaway}

\begin{table}[t]
\centering
\begin{minipage}{0.98\linewidth}
\captionsetup{hypcap=false}
\caption{Safety attribute violations in the current literature. Authority, scope, privacy, auditability, and recovery failures become comparable when tied to a concrete state transition.}
\label{tab:violation-literature-index}
\centering
\tiny
\setlength{\tabcolsep}{1.1pt}
\renewcommand{\arraystretch}{1.05}
\SAVERTableRows
\begin{tabular}{@{}p{0.165\linewidth}p{0.04\linewidth}p{0.12\linewidth}p{0.07\linewidth}p{0.065\linewidth}p{0.13\linewidth}p{0.38\linewidth}@{}}
\toprule
\rowcolor{saverViolationHead}
\textbf{Method or Work} & \textbf{Year} & \textbf{\TableHeadRole} & \textbf{\TableHeadTarget} & \textbf{\TableHeadObject} & \textbf{Violation} & \textbf{Failed Attribute} \\
\midrule
\SAVERTableGroup{saverViolationHead}{7}{Provenance and authority failures}
AgentPoison~\cite{chen2024agentpoison} & 2024 & \TableRoleAttack & \TableTargetAgent & \TableObjectMemory & Provenance loss & Memory source attribute \\
GhostWriter~\cite{torres2026agentsremembermuchmemory} & 2026 & \TableRoleAttack & \TableTargetAgent & \TableObjectMemory & Provenance loss & Hidden payload memory source \\
PASB~\cite{mao2026agentsdontjustagree} & 2026 & \TableRoleBenchmarkSafety & \TableTargetAgent & \TableObjectMemory & Provenance / scope & Claim attribution and role preservation \\
AuthMem-Bench~\cite{zhan2026memoryauthoritybenchmarkingauthority} & 2026 & \TableRoleBenchmarkSafety & \TableTargetAgent & \TableObjectMemory & Authority / provenance & Source authority erased during consolidation \\
When Context Gets Root~\cite{he2026contextgetsrootprivilege} & 2026 & \TableRoleAttack & \TableTargetAgent & \TableObjectWorkflow & Authority / provenance & Harness elevates the model-facing source role \\
Agent Context CPE~\cite{li2026whatsagentscontextcontext} & 2026 & \TableRoleAttack & \TableTargetAgent & \TableObjectWorkflow & Authority / scope & Harness elevates source role or persistence scope \\
AgentDojo~\cite{debenedetti2024agentdojo} & 2024 & \TableStudyAttackBenchmark & \TableTargetAgent & \TableObjectTool & Provenance loss & Tool context source role \\
FragFuse~\cite{rao2026fragfusebypassingaccesscontrol} & 2026 & \TableRoleAttack & \TableTargetAgent & \TableObjectMemory & Authority / scope & Fragment authority context \\
EvoBreak~\cite{yan2026benignaloneharmfultogether} & 2026 & \TableRoleAttack & \TableTargetAgent & \TableObjectMemory & Authority / scope & Benign experiences gain joint authority \\
PoisonedEvolution~\cite{chen2026experienceinstructiontrajectorypoisoning} & 2026 & \TableRoleAttack & \TableTargetAgent & \TableObjectSkill & Authority / provenance & Untrusted recurring evidence gains instruction authority \\
SkillMisevo-Bench~\cite{mao2026practicemakesunsafeskill} & 2026 & \TableRoleBenchmarkSafety & \TableTargetAgent & \TableObjectSkill & Authority / persistence & Unsafe success becomes reusable procedure \\
EVOMAL~\cite{wu2026evomalselfpoisoningselfevolvingcoding} & 2026 & \TableRoleAttack & \TableTargetAgent & \TableObjectSkill & Provenance / persistence & Agent-authored copies obscure attacker origin \\
PS-Bench~\cite{guo-etal-2026-personalization} & 2026 & \TableRoleBenchmarkSafety & \TableTargetAgent & \TableObjectMemory & Authority / scope & Benign context legitimizes harmful intent \\
InjecAgent~\cite{zhan_2024_injecagent_benchmarking_indi} & 2024 & \TableStudyAttackBenchmark & \TableTargetAgent & \TableObjectTool & Authority / scope & Indirect instruction authority \\
ACIArena~\cite{an-etal-2026-aciarena} & 2026 & \TableStudyAttackBenchmark & \TableTargetAgent & \TableObjectMultiAgent & Authority / scope & Inter-agent instruction authority \\
MAS code execution~\cite{triedman2025multi} & 2025 & \TableRoleAttack & \TableTargetAgent & \TableObjectMultiAgent & Authority / scope & Untrusted content controls agent invocation \\
Progent~\cite{shi2025progent} & 2025 & \TableRoleDefense & \TableTargetAgent & \TableObjectTool & Authority / scope & Scoped tool privilege \\
\SAVERTableGroup{saverViolationHead}{7}{Privacy and persistent repair gaps}
Agent memory privacy~\cite{wang2025privacy_agent_memory} & 2025 & \TableRoleSafety & \TableTargetAgent & \TableObjectMemory & Privacy / purpose & Sensitivity/purpose/retention \\
Personal agents~\cite{li_2024_personal_llm_agents_insights} & 2024 & \TableRoleSurveySafety & \TableTargetAgent & \TableObjectMemory & Privacy / purpose & Persistent personal context carriers \\
PerMemSafe~\cite{an-etal-2026-permemsafe} & 2026 & \TableRoleBenchmarkSafety & \TableTargetAgent & \TableObjectMemory & Privacy / purpose & Personalized context safety \\
PersistBench~\cite{pulipaka2026persistbench} & 2026 & \TableRoleBenchmarkSafety & \TableTargetAgent & \TableObjectMemory & Privacy / purpose & Memory scope/forgetting \\
AgentPoison~\cite{chen2024agentpoison} & 2024 & \TableRoleAttack & \TableTargetAgent & \TableObjectMemory & Persistent repair & Delayed poisoned influence \\
Memory poisoning~\cite{sunil2026memory_poisoning_agents} & 2026 & \TableRoleAttack & \TableTargetAgent & \TableObjectMemory & Poisoned retrieval & Persistent query-to-memory influence \\
SkillJack~\cite{ying2026skilljackpersistentskillbackdoors} & 2026 & \TableRoleAttack & \TableTargetAgent & \TableObjectSkill & Persistent repair & Skill survives source-record deletion \\
Self-evolving agents~\cite{gao2025self_evolving_agents} & 2025 & \TableRoleSurveyCapability & \TableTargetAgent & \TableObjectWorkflow & Adaptive process & Capability evolution taxonomy \\
\SAVERTableGroup{saverViolationHead}{7}{Action authority and operational integrity}
ReAct~\cite{yao2022react} & 2022 & \TableRoleCapability & \TableTargetLLMAgent & \TableObjectWorkflow & Authority / scope & Observation to action role \\
Voyager~\cite{wang2023voyager} & 2023 & \TableRoleCapability & \TableTargetAgent & \TableObjectSkill & Authority / scope & Executable skill authority \\
AgentDojo~\cite{debenedetti2024agentdojo} & 2024 & \TableStudyAttackBenchmark & \TableTargetAgent & \TableObjectTool & Authority / scope & Untrusted tool action state \\
OWASP LLM Top 10~\cite{owasp2025llm_top10} & 2025 & \TableRoleFrameworkSecurity & \TableTargetLLMAgent & \TableObjectGeneral & Operational integrity & Agency/cost bounds \\
AgentDojo~\cite{debenedetti2024agentdojo} & 2024 & \TableStudyAttackBenchmark & \TableTargetAgent & \TableObjectTool & Operational integrity & External effect tool context \\
Agent Security Bench~\cite{zhang_2024_agent_security_bench_asb_for} & 2024 & \TableStudyAttackBenchmark & \TableTargetAgent & \TableObjectEvaluation & Operational integrity & Agent security behavior \\
Agent-SafetyBench~\cite{zhang_2024_agent_safetybench_evaluating} & 2024 & \TableStudySafetyBenchmark & \TableTargetAgent & \TableObjectEvaluation & Operational integrity & Agent safety behavior \\
GuardAgent~\cite{xiang2024guardagent} & 2024 & \TableRoleDefense & \TableTargetAgent & \TableObjectRuntime & Local guard checking & Safety request $\rightarrow$ action check \\
R-Judge~\cite{yuan2024rjudge} & 2024 & \TableStudySafetyBenchmark & \TableTargetAgent & \TableObjectEvaluation & Operational integrity & Safety signal without transition cause \\
Progent~\cite{shi2025progent} & 2025 & \TableRoleDefense & \TableTargetAgent & \TableObjectTool & Authority / scope & Explicit tool privilege control \\
Misattribution Gap~\cite{ahad_2026_misattribution_gap} & 2026 & \TableRoleSafety & \TableTargetAgent & \TableObjectMemory & Operational integrity & Memory fault attribution \\
Agents of Chaos~\cite{shapira2026agents_chaos} & 2026 & \TableRoleSafety & \TableTargetAgent & \TableObjectMultiAgent & Operational integrity & Cross-agent uptake and live-system compromise \\
\SAVERTableGroup{saverViolationHead}{7}{Lifecycle drift and model regression}
Reflexion~\cite{shinn2023reflexion} & 2023 & \TableRoleCapability & \TableTargetAgent & \TableObjectWorkflow & Authority / scope & Feedback to strategy reuse \\
Self-evolution survey~\cite{tao2024self_evolution_llm_survey} & 2024 & \TableRoleSurveyCapability & \TableTargetLLMAgent & \TableObjectWorkflow & Adaptive process & Self-evolution operation taxonomy \\
EDDOps~\cite{xia_2024_evaluation_driven_developmen} & 2024 & \TableRoleGovernance & \TableTargetAgent & \TableObjectGovernance & Operational integrity & Lifecycle drift tracking \\
Memory security survey~\cite{lin2026memory_security_survey} & 2026 & \TableRoleSurveySafety & \TableTargetAgent & \TableObjectMemory & Authority / scope & Memory semantic drift \\
Fine-tuning safety~\cite{qi2023finetuning_safety} & 2023 & \TableRoleSafety & \TableTargetLLM & \TableObjectModel & Model-side regression & Fine-tuning safety regression \\
PEFT Trojan~\cite{hong2024peta} & 2024 & \TableRoleAttack & \TableTargetLLM & \TableObjectModel & Model-side regression & Adapter trigger safety issue \\
Dormant behavior~\cite{gloaguen2025watch_your_steps} & 2025 & \TableRoleAttack & \TableTargetLLM & \TableObjectModel & Model-side regression & Dormant activation context \\
Safe LoRA~\cite{hsu2024safe_lora} & 2024 & \TableRoleDefense & \TableTargetLLM & \TableObjectModel & Model-side regression & Adapter mitigation handle \\
\bottomrule
\end{tabular}
\SAVERTableReset
\end{minipage}
\end{table}

\section{Exposure: Surfaces and Manifestations}
\label{sec:exposure-surfaces}

\SAVERFlowE

\noindent\textbf{Exposure Premise.} Once a failed attribute and expected trigger are known, the analysis turns to where the latent mismatch first becomes observable and what record connects that event to the earlier carrier and operation. Exposure is diagnostic evidence, not a second violation taxonomy: each study must identify the observed event and the criterion that makes it evidence of a violation. The distinction prevents latent unsafe state from being equated with realized harm and prevents visible harm from being attributed only to the final output step.

The recurring bridge cases are memory bypass, prompt injection, economic denial of service, model update, multi-agent uptake, persistent privacy, and boundary drift. Each starts from an internal safety attribute failure, but becomes measurable at a different boundary: delayed retrieval, fused query, tool execution, budget consuming loop, uptake across agents, stale personalization, or mismatched governance signal.
Each surface offers a different observation of the failures in \secref{sec:invariant-violations}. A retrieval trace can identify the selected object, an execution trace can confirm an external effect, and a lineage trace can connect the two. Their combination determines whether the interventions in \secref{sec:response-governance} can address the enabling state or only the visible event. \tabref{tab:violation-exposure-literature-index} indexes the corresponding evidence.

\subsection{Memory Retrieval Exposure}

Memory retrieval exposure is the diagnostic surface for latent failures over stored or indexed state. The visible event is a later task selecting a memory entry, retrieved demonstration, profile note, summary, or derived preference and granting it influence. Direct attacks reach that event through different paths. FragFuse stores fragments whose prohibited meaning emerges only after fusion; stealth memory injection and GhostWriter target external-content admission before later activation; AgentPoison poisons retrievable examples selected by a trigger~\cite{rao2026fragfusebypassingaccesscontrol,zhang2026clawsremembertellstealthy,torres2026agentsremembermuchmemory,chen2024agentpoison}. Their common surface is delayed retrieval, but their causal operations and suitable controls differ.

Adjacent work broadens the diagnostic object. MemoryGraft studies poisoned experience retrieval, OEP studies harmful transfer from a consolidated local experience, and Tool-Drift follows remembered state into routing~\cite{srivastava2025memorygraft,wang2026oep,dabas2026tool_drift}. GLOVE examines disagreement between memory and environment, whereas MemPot treats extraction as the relevant defensive surface~\cite{yin2026glove,wang2026mempot}. Privacy work adds retrieval that is semantically correct but inappropriate for the current user or purpose~\cite{wang2025privacy_agent_memory}. PASB clears the originating conversation before a neutral query, which makes later sycophantic behavior evidence of durable-state activation rather than residual dialogue context~\cite{mao2026agentsdontjustagree}. PS-Bench adds a different exposure: benign memory retrieval changes inferred intent and the visible symptom is unsafe compliance rather than a corrupted memory record~\cite{guo-etal-2026-personalization}. These studies should be compared by the state selected and the authority it receives, not by treating every later-memory failure as the same attack.

\begin{SAVERTakeaway}{The Visible Failure May Have an Earlier Cause}
\label{takeaway:exposure-causal-attribution}
A wrong answer or unsafe action at generation time may originate in an earlier memory update. Delayed activation separates the visible symptom from the operation that changed the state's authority. Diagnosis must connect the triggering retrieval and prompt assembly to that earlier update before assigning the failure to the model or choosing a repair.
\end{SAVERTakeaway}

The diagnostic distinction is between selecting an item and granting it influence. Managed memory and feedback reuse explain how earlier state returns~\cite{packer2023memgpt,shinn2023reflexion}; PerMemSafe shows why the selected user context matters to the safety judgment~\cite{an-etal-2026-permemsafe}. A retrieval log alone identifies what matched, while prompt assembly reveals whether it was presented as evidence, preference, or instruction. Both are needed to attribute an unsafe response to the appropriate transition.

\subsection{Tool and Action Exposure}

Retrieval becomes externally consequential at the tool and action boundary, where authority, actionability, and resource failures produce an API call, file write, email, purchase, database update, code execution, skill invocation, or repeated high-cost path. Lower-authority state now reaches a tool, skill, or API surface where it can change the world. ReAct gives the general mechanism in which observations and reasoning steps condition actions, while early prompt-injection work explains how external content can pressure that path~\cite{yao2022react,liu_2023_prompt_injection_attack_agai}. AgentDojo provides the clearest direct anchor for untrusted data reaching integrated agents, and InjecAgent extends the pressure to benchmarked agent settings~\cite{debenedetti2024agentdojo,zhan_2024_injecagent_benchmarking_indi}.
AuthMem-Bench supplies a controlled bridge from the earlier memory violation to this surface. Its action module keeps the claim, tool schema, and target argument fixed while changing whether the stored representation retains source authority; prohibited tool execution then measures the external consequence of an authority-collapsed memory~\cite{zhan2026memoryauthoritybenchmarkingauthority}. The benchmark therefore localizes \textbf{V} at consolidation and \textbf{E} at action rather than treating the final tool call as an unexplained model failure.
HarnessSafe generalizes this evidence rule across persistent carriers. A run reaches its violation stage only when a case-specific oracle confirms an executed behavior, committed state change, propagated artifact, or external effect; suspicious text, unsafe planning, or stated intent is insufficient~\cite{zhang2026harnesssafeevaluatingsafetypersistent}. Exposure is thereby tied to both the carrier path and observable consequence, while the furthest preceding trace stage shows whether the chain failed before or at the action boundary.
Newer work varies the interaction pattern, input channel, and upstream object under test. Adaptive attacks and LivePI test whether defenses survive changing or live interaction patterns, while SnapGuard isolates screenshot-mediated injection in web agents~\cite{zhan_2025_adaptive_attacks_break_defen,zhao2026livepi,du2026snapguardlightweightpromptinjection}. Task Shield and SafeSearch focus on task alignment in action or search settings; Tool-Drift and prompt tool selection examine how remembered state or connector descriptions alter routing~\cite{jia_2025_the_task_shield_enforcing_ta,dong2026safesearch,dabas2026tool_drift,shi2026prompt_tool_selection}. The shared outcome is an action-boundary event, but each line of work perturbs a different upstream object.
Response mechanisms inspect different properties at this boundary. Signed-Prompt checks source authenticity, LlamaFirewall places runtime guards, and Progent constrains tool privilege; secure-design patterns and formal agent-security work explain how such controls compose with data isolation~\cite{suo_2024_signed_prompt_a_new_approach,beurer_kellner_2025_design_patterns_for_securing,chennabasappa_2025_llamafirewall_an_open_source,shi2025progent,siu2026formalizing_agent_security}. An observed tool effect therefore needs both its triggering context and the policy decision that permitted execution to distinguish an input-screening failure from an authorization failure.

The safety attribute failure is usually authority escalation, actionability escalation, or resource boundary loss. A webpage, email, retrieved document, or memory item is treated not just as information but as a command or parameter. The exposure trigger is the point at which the agent selects a tool, fills an argument, invokes a skill, commits an external effect, or opens a repeated high-cost call path. This trigger can occur without a dangerous final answer. The user may never see the injected instruction, and the model may never print the unsafe state, but the action can still expose the violation through an irreversible action, a denial-of-service loop, or economic waste. The triggering state and the authorization decision distinguish these causes; \secref{sec:response-governance} examines the controls that can act on that evidence.

Risk-awareness benchmarks can identify situations relevant to safety~\cite{yuan2024rjudge}, and guard agents can reason about local responses~\cite{xiang2024guardagent}. Containment additionally requires changing execution, for example by removing suspect state, reducing privileges, forcing a clean plan, or routing the decision to human approval. Recording the failed authority transition supports later repair but does not itself contain the current action.

\noindent\textbf{Context-Horizon Exposure.} Accumulation and compression are temporal conditions across these surfaces, not an additional tool-specific surface. Multi-turn attacks make the accumulated interaction, rather than the latest user message, the diagnostic object. Trojan Knowledge composes knowledge gathered through benign subqueries, while One Turn Too Late identifies the earliest response whose delivery closes a hidden harmful intent~\cite{wei2025trojanknowledgebypassingcommercial,shen2026turnlateresponseawaredefense}. MT-AgentRisk carries this decomposition into tool use~\cite{li2026unsaferturnsbenchmarkingdefending}. Relinking reaches a comparable closure through summarization rather than dialogue order: distributed fragments become actionable only after compression~\cite{liu2026safecheckunsafeuse}. Across these mechanisms, the observed harm-enabling response or action must be distinguished from context that is merely capable of supporting it. A turn-local refusal score cannot localize that point without the interaction prefix, any compression step, the candidate response, and the tool decision that follows.

\subsection{External Artifact Exposure}

Some outputs of tools and workflows persist beyond the episode, shifting exposure to an external artifact. A generated document, code patch, configuration file, database row, ticket, report, memory export, workflow template, shared prompt, or skill entry can later influence users, agents, or workflows. Self-evolution and multi-agent work establish why durable artifacts and shared environments belong in scope, while MemGPT and Voyager supply adjacent mechanics for memory movement and reusable skills~\cite{gao2025self_evolving_agents,packer2023memgpt,wang2023voyager}. Across these forms, scoped or descriptive information becomes durable evidence or procedure in another context.

A human reads the generated document, another agent retrieves it, a workflow uses the template, a developer executes the code, or a future run imports the artifact as trusted context. Consumption can expose a latent violation after the original episode ends; an unauthorized write or disclosure may already be observable when the artifact is created. A stale summary can become a project rule. A private detail can become part of a report. A generated skill can encode an unsafe shortcut. A workflow template can preserve a boundary drift that later users never see as a choice. This surface is especially important for self-evolving agents because evolution often works by accumulating artifacts: reflections, skills, memories, notes, plans, and reusable procedures.

The relevant observation point is therefore creation or later consumption, depending on the violated requirement. Reflection and skill-library systems show how feedback or procedures can remain reusable after their source episode~\cite{shinn2023reflexion,wang2023voyager,tao2024self_evolution_llm_survey}. Diagnosing their later use requires linking the consumed version to its origin and intended scope. Provenance and revocation at this boundary are governance requirements derived from those reuse mechanisms, not evidence that the capability systems already enforce them.

\subsection{Propagation Across Agents}

When state crosses actors, the visible event is one agent's unsafe state becoming another agent's context, belief, instruction, or action precondition. Self-evolution and trustworthy-agent surveys define why reusable shared state belongs in the safety setting; multi-agent surveys and coordination benchmarks identify messages, delegated tasks, shared memories, plans, and summaries as transfer channels~\cite{gao2025self_evolving_agents,yu2025trustworthy_llm_agents}.
ACIArena turns those architectural channels into a cascading-injection benchmark over external inputs, profiles, and inter-agent messages~\cite{an-etal-2026-aciarena}. Agents of Chaos supplies a deployment-oriented counterpart: its two-week live-laboratory study reports eleven representative cases that include cross-agent propagation of unsafe practices, information disclosure, destructive actions, identity spoofing, and partial takeover~\cite{shapira2026agents_chaos}. These cases establish that the transfer channels matter under realistic combinations of persistent memory, communication accounts, files, and shell access, but they are case evidence rather than prevalence estimates.

Control-flow hijacking exposes a different part of the same multi-agent boundary. Triedman et al. show that adversarial external content can redirect orchestration and communication toward unsafe agents or functions. With GPT-4o agents, the evaluated Web attacks achieve arbitrary-code-execution rates of 58--90\% across orchestrators, and some model-orchestrator configurations reach 100\%~\cite{triedman2025multi}. The attack can succeed even when individual agents resist direct and indirect prompt injection~\cite{chen2026beyond} or refuse the harmful action. Thus, receiver compliance is not the only failure condition: the orchestrator can assemble a harmful execution from components whose local behavior appears safe.
At the content-transfer boundary, embedding-defense and multi-agent security work locate weak shared-memory and message boundaries, whereas MADRA, Devil's Moltbook, and collective misremembering examine uptake into plans, repeated exposure, or consensus~\cite{zhang2026embedding_defenses_fail,nguyen2026security_mas,wang2025madra,wang2026devil_moltbook,xu2026when}.

\begin{SAVERTakeaway}{Content Lineage and Invocation Lineage Are Distinct}
\label{takeaway:content-invocation-lineage}
Message injection changes instruction status; propagation changes what recipients retain; control-flow hijacking changes which component may act. An unsafe route can therefore arise without a recipient following the malicious request~\cite{triedman2025multi}. Multi-agent diagnosis needs both the path taken by the content and the invocation decisions that assigned execution authority.
\end{SAVERTakeaway}

Uptake occurs when a planner sends a worker a contaminated instruction, a summarizer converts a disputed memory into a shared fact, a debate participant repeats a wrong claim that sounds authoritative, or a coordinator treats another agent's uncertain output as verified. Once that happens, the original safety attribute failure can multiply: more agents can activate the state, more artifacts can derive from it, and recovery must locate every copy or descendant. The visible manifestation can be wrong consensus, repeated leakage, inconsistent authority decisions, or a group plan that no single agent would have produced alone. This surface therefore names the quantities a serious evaluation must recover: propagation path, recipient authority, uncertainty retention, source attribution, and repair completeness.
The receiving context determines whether transferred content remains evidence or acquires local authority. A transfer trace can expose an unauthorized role change only if it preserves the sender's source constraints and the receiver's activation decision. Whether a correction also reaches later recipients is a separate repair question, developed in \secref{sec:response-governance}.

\subsection{User-Facing Misattribution}

After state has crossed retrieval, action, artifact, or agent boundaries, the visible failure can be attributed to the wrong causal layer. The user sees a generated answer, rationale, refusal, recommendation, tool result, or explanation, while the agent has composed that surface from model priors, retrieved memory, task context, tool outputs, and policy state. A LangGraph plus ChromaDB study makes the memory-side attribution problem explicit by showing poisoned shared memory mistaken for model misalignment~\cite{ahad_2026_misattribution_gap}. InjecAgent supplies the action-side counterpart: untrusted external content can alter tool-mediated behavior~\cite{zhan_2024_injecagent_benchmarking_indi}. Misattribution arises when the incident trace reports only the final answer or action rather than the state and source that shaped it.

A user asks why the agent acted, and the system provides a fluent answer that does not name the activated memory, retrieved document, tool output, or privilege decision. The missing trace limits attribution of an observed failure; it establishes a further violation only if an explicit auditability requirement is unmet. Misattribution can then cause misdirected incident response. Operators may retrain a model when they need to quarantine memory. They may strengthen an output filter when they need a retrieval gate. They may blame a user instruction when the decisive input was untrusted tool data. Attributing an incident requires records of the state carriers activated and the authority they received. Public explanations can preserve privacy and stay summarized, but internal incident traces need enough detail to support repair.

A user cannot correct a stale preference if the agent presents the behavior as a general model tendency. An operator cannot revoke a poisoned descendant if the trace stops at the final output. A safety team cannot compare guard failures if the system records only blocked or allowed, not why the state was allowed to influence the decision. Misattribution is therefore also a contestability problem. Risk awareness work can support diagnosis signals~\cite{yuan2024rjudge}, and guard agent work can support local response reasoning~\cite{xiang2024guardagent}, but accountability to users requires that these signals connect to state traces and repair handles. \SAVER{}'s contribution is to make the causal chain auditable rather than treating misattribution as only a user interface problem.

\subsection{Governance Blind Spots}

Misattribution becomes a governance blind spot when a safety mechanism observes the wrong layer, acts at the wrong time, or cannot repair the causal state. The relevant carriers are now monitor signals, guard decisions, privilege policies, audit logs, and recovery records. R-Judge supplies safety recognition over trajectories, and TrustAgent performs trust-aware planning~\cite{yuan2024rjudge,hua_2024_trustagent_towards_safe_and}; AgentDojo supplies adversarial tool-use episodes at the exposure boundary~\cite{debenedetti2024agentdojo}. GuardAgent and LlamaFirewall move from recognition toward runtime intervention, Progent constrains privilege at tool boundaries, and EDDOps asks whether these controls remain connected to deployment practice~\cite{xiang2024guardagent,chennabasappa_2025_llamafirewall_an_open_source,shi2025progent,xia_2024_evaluation_driven_developmen}. The blind spot persists when a correct signal stops at the visible event and leaves the causal state active. Missing repair evidence limits a recovery claim; it does not by itself establish another exposure.

A guard can correctly block a response while the responsible memory remains retrievable; a privilege policy can prevent a call while leaving its workflow rule active. These are successful local controls with an unresolved state-level cause. Diagnosis must distinguish that residual influence from a failure to recognize the current event. The distinction determines whether the next intervention should change detection, activation authority, or retained state.

\subsection{Model Deployment Exposure}

The same diagnostic problem becomes broader for model-side state because one loaded update can affect many tasks. Exposure occurs when an adapted model, adapter, policy component, reward model, edited checkpoint, unlearning patch, or safety classifier exhibits a safety-attribute failure after entering the inference or routing path. Fine-tuning safety identifies erosion after tuning; PEFT, feedback-poisoning, and dormant-behavior work add deployable update objects and delayed triggers~\cite{qi2023finetuning_safety,rando2024universal_jailbreak_backdoors,gloaguen2025watch_your_steps}. Text-encoder and plugin attacks show related compromise of shared model artifacts~\cite{li-etal-2025-twist,chen2026customization}. Stronger agent-loop comparisons still require direct activation and deployment-lineage evidence.

A memory-related failure may surface only when a matching task retrieves the item; a failure in a deployed adapter can surface whenever a compatible prompt, tool route, or downstream task hits the changed behavior. The model-side trigger is often broad rather than local. The manifestation can be a global shift in refusal boundary, unsafe compliance across tasks, persistent over-refusal, systematic bias in tool selection, or policy activation across deployment. This makes misattribution especially likely: the visible behavior may look like an ordinary model limitation unless the system records which adapter, checkpoint, policy prompt, safety classifier, and reward model were active. Deployment governance therefore needs an adapter registry, model card or safety card, training data lineage, model difference review, canary safety evaluation, task scope restrictions, version rollback, and residual influence tests. The response literature again splits by layer: Safe LoRA and LoRAShield intervene at the LoRA or adaptation layer, AEIOU evaluates prompt moderation in a related generative setting, and broader security and backdoor surveys supply protocol and benchmark vocabulary for persistent triggerable behavior~\cite{hsu2024safe_lora,chen2025lorashielddatafreeeditingalignment,wang2025aeiouunifieddefenseframework,zhou2025backdoor_llm_survey}.

Across these surfaces, memory work most clearly measures delayed activation after a write; tool and prompt injection work most clearly measures the external effect boundary where information becomes action; model-update work measures broader behavior shifts, with agent-loop attribution requiring deployment and activation evidence; and multi-agent work exposes propagation and uptake, but still lacks mature evidence for descendant repair. User-facing misattribution and governance blind spots explain why exposure diagnosis must include causal state traces rather than final outputs alone.
\secref{sec:response-governance} uses the visible event, activated object, failed attribute, and known lineage to distinguish containment from repair.

\begin{SAVERTakeaway}{Exposure Evidence Must Match the Claimed Effect}
\label{takeaway:effect-matched-evidence}
Suspicious text or an unsafe plan does not establish an executed tool action or committed external effect. HarnessSafe makes this distinction explicit through case-specific observable evidence~\cite{zhang2026harnesssafeevaluatingsafetypersistent}. Reports should identify the event actually observed and connect it to the activated state, rather than infer a stronger consequence from intent alone.
\end{SAVERTakeaway}

\FloatBarrier
\begin{table}[h]
\centering
\begin{minipage}{0.98\linewidth}
\captionsetup{hypcap=false}
\caption{Exposure surfaces for latent state failures. Memory retrieval, privacy disclosure, tool action, multi-agent uptake, benchmark tasks, and model deployment expose different parts of the same lifecycle gap; role labels identify the contribution represented by each row.}
\label{tab:violation-exposure-literature-index}
\centering
\tiny
\setlength{\tabcolsep}{1.1pt}
\renewcommand{\arraystretch}{1.05}
\SAVERTableRows
\begin{tabular}{@{}p{0.155\linewidth}p{0.04\linewidth}p{0.13\linewidth}p{0.065\linewidth}p{0.07\linewidth}p{0.105\linewidth}p{0.40\linewidth}@{}}
\toprule
\rowcolor{saverExposureHead}
\textbf{Method or Work} & \textbf{Year} & \textbf{\TableHeadRole} & \textbf{\TableHeadTarget} & \textbf{\TableHeadObject} & \textbf{Surface} & \textbf{Failure $\rightarrow$ Exposure} \\
\midrule
\SAVERTableGroup{saverExposureHead}{7}{Memory retrieval}
AgentPoison~\cite{chen2024agentpoison} & 2024 & \TableRoleAttack & \TableTargetAgent & \TableObjectMemory & Memory retrieval & Poisoned memory $\rightarrow$ delayed retrieval \\
FragFuse~\cite{rao2026fragfusebypassingaccesscontrol} & 2026 & \TableRoleAttack & \TableTargetAgent & \TableObjectMemory & Memory retrieval & Fragmented memory $\rightarrow$ fused query/access gate \\
Stealth memory injection~\cite{zhang2026clawsremembertellstealthy} & 2026 & \TableRoleAttack & \TableTargetAgent & \TableObjectMemory & Memory retrieval & Hidden email memory $\rightarrow$ later behavior \\
GhostWriter~\cite{torres2026agentsremembermuchmemory} & 2026 & \TableRoleAttack & \TableTargetAgent & \TableObjectMemory & Memory retrieval & Hidden payload $\rightarrow$ recalled poisoned memory \\
PASB~\cite{mao2026agentsdontjustagree} & 2026 & \TableRoleBenchmarkSafety & \TableTargetAgent & \TableObjectMemory & Memory retrieval & Committed claim $\rightarrow$ neutral-query reuse \\
OEP~\cite{wang2026oep} & 2026 & \TableRoleAttack & \TableTargetAgent & \TableObjectWorkflow & Memory retrieval & Unsafe rule abstraction $\rightarrow$ downstream transfer \\
Zombie Agents~\cite{yang2026zombie_agents} & 2026 & \TableRoleAttack & \TableTargetAgent & \TableObjectMemory & Memory retrieval & Dormant memory $\rightarrow$ trigger activation \\
MemoryGraft~\cite{srivastava2025memorygraft} & 2025 & \TableRoleAttack & \TableTargetAgent & \TableObjectMemory & Memory retrieval & Poisoned experience $\rightarrow$ evidence retrieval \\
PS-Bench~\cite{guo-etal-2026-personalization} & 2026 & \TableRoleBenchmarkSafety & \TableTargetAgent & \TableObjectMemory & Memory retrieval & Benign memory $\rightarrow$ unsafe compliance \\
Tool-Drift~\cite{dabas2026tool_drift} & 2026 & \TableRoleSafety & \TableTargetAgent & \TableObjectTool & Memory retrieval & Memory route drift $\rightarrow$ tool choice \\
\SAVERTableGroup{saverExposureHead}{7}{Privacy exposure}
Agent memory privacy~\cite{wang2025privacy_agent_memory} & 2025 & \TableRoleSafety & \TableTargetAgent & \TableObjectMemory & Privacy exposure & Sensitive memory reuse $\rightarrow$ personalization/leakage \\
PerMemSafe~\cite{an-etal-2026-permemsafe} & 2026 & \TableRoleBenchmarkSafety & \TableTargetAgent & \TableObjectMemory & Privacy exposure & Personalized memory context $\rightarrow$ unsafe response \\
PersistBench~\cite{pulipaka2026persistbench} & 2026 & \TableRoleBenchmarkSafety & \TableTargetAgent & \TableObjectMemory & Privacy exposure & Forgetting failure $\rightarrow$ leakage across domains \\
RBI-Eval~\cite{xu2026memorystaysilentmeasuring} & 2026 & \TableRoleBenchmarkSafety & \TableTargetAgent & \TableObjectMemory & Privacy exposure & Memory use boundary $\rightarrow$ sensitive recall \\
EchoLeak~\cite{reddy_2025_echoleak_the_first_real_worl} & 2025 & \TableRoleAttack & \TableTargetLLMAgent & \TableObjectContext & Privacy exposure & Data leakage $\rightarrow$ assistant context \\
Personal agent context~\cite{li_2024_personal_llm_agents_insights} & 2024 & \TableRoleSurveySafety & \TableTargetAgent & \TableObjectMemory & Personal context & Persistent user context $\rightarrow$ assistant reuse \\
\SAVERTableGroup{saverExposureHead}{7}{Tool action}
Prompt injection apps~\cite{liu_2023_prompt_injection_attack_agai} & 2023 & \TableRoleAttack & \TableTargetLLMAgent & \TableObjectTool & Tool action & Untrusted content $\rightarrow$ web/action step \\
AgentDojo~\cite{debenedetti2024agentdojo} & 2024 & \TableStudyAttackBenchmark & \TableTargetAgent & \TableObjectTool & Tool action & Untrusted tool data $\rightarrow$ tool call/web action \\
InjecAgent~\cite{zhan_2024_injecagent_benchmarking_indi} & 2024 & \TableStudyAttackBenchmark & \TableTargetAgent & \TableObjectTool & Tool action & Authority confusion $\rightarrow$ indirect injection \\
AuthMem-Bench~\cite{zhan2026memoryauthoritybenchmarkingauthority} & 2026 & \TableRoleBenchmarkSafety & \TableTargetAgent & \TableObjectMemory & Tool action & Authority-collapsed memory $\rightarrow$ unauthorized action \\
Agent Context CPE~\cite{li2026whatsagentscontextcontext} & 2026 & \TableRoleAttack & \TableTargetAgent & \TableObjectWorkflow & Tool action & Role/scope escalation $\rightarrow$ remote code execution, denial of service, or tool misuse \\
STAC~\cite{li2026stac} & 2026 & \TableRoleAttack & \TableTargetAgent & \TableObjectTool & Tool action & Tool chain compromise $\rightarrow$ tool route \\
LivePI~\cite{zhao2026livepi} & 2026 & \TableRoleAttack & \TableTargetAgent & \TableObjectTool & Tool action & Live injected state $\rightarrow$ interactive action \\
Prompt tool selection~\cite{shi2026prompt_tool_selection} & 2026 & \TableRoleAttack & \TableTargetAgent & \TableObjectTool & Tool action & Connector manipulation $\rightarrow$ tool selection \\
SafeSearch~\cite{dong2026safesearch} & 2026 & \TableRoleRedTeam & \TableTargetAgent & \TableObjectTool & Search trajectory & Adversarial search interaction $\rightarrow$ unsafe trajectory \\
Task Shield~\cite{jia_2025_the_task_shield_enforcing_ta} & 2025 & \TableRoleDefense & \TableTargetAgent & \TableObjectTool & Tool action & Task policy mismatch $\rightarrow$ tool action gate \\
ASB~\cite{zhang_2024_agent_security_bench_asb_for} & 2024 & \TableStudyAttackBenchmark & \TableTargetAgent & \TableObjectEvaluation & Tool action & Authority/cost misuse $\rightarrow$ agent actions \\
\SAVERTableGroup{saverExposureHead}{7}{Runtime guard}
LlamaFirewall~\cite{chennabasappa_2025_llamafirewall_an_open_source} & 2025 & \TableRoleDefense & \TableTargetLLMAgent & \TableObjectRuntime & Runtime guard & Prompt safety signal $\rightarrow$ runtime decision \\
Prompt injection detection~\cite{lan_2025_prompt_injection_detection_i} & 2025 & \TableRoleDefense & \TableTargetLLMAgent & \TableObjectRuntime & Runtime guard & Injection signal $\rightarrow$ prompt flow \\
LLM Firewall~\cite{podpora_2025_llm_firewall_using_validator} & 2025 & \TableRoleDefense & \TableTargetLLM & \TableObjectRuntime & Runtime guard & Validator gap $\rightarrow$ prompt/output flow \\
Signed-Prompt~\cite{suo_2024_signed_prompt_a_new_approach} & 2024 & \TableRoleDefense & \TableTargetLLMAgent & \TableObjectContext & Runtime guard & Source authenticity loss $\rightarrow$ step driven by document \\
Progent~\cite{shi2025progent} & 2025 & \TableRoleDefense & \TableTargetAgent & \TableObjectTool & Runtime guard & Privilege overgrant $\rightarrow$ tool authority \\
GuardAgent~\cite{xiang2024guardagent} & 2024 & \TableRoleDefense & \TableTargetAgent & \TableObjectRuntime & Runtime guard & Local guard signal $\rightarrow$ response choice \\
R-Judge~\cite{yuan2024rjudge} & 2024 & \TableStudySafetyBenchmark & \TableTargetAgent & \TableObjectEvaluation & Safety recognition & Risk-awareness gap $\rightarrow$ unsafe decision signal \\
EDDOps~\cite{xia_2024_evaluation_driven_developmen} & 2024 & \TableRoleGovernance & \TableTargetAgent & \TableObjectGovernance & Runtime guard & Signal/control mismatch $\rightarrow$ operations loop \\
\SAVERTableGroup{saverExposureHead}{7}{Multi-agent}
Collective Misremembering~\cite{xu2026when} & 2026 & \TableRoleAttack & \TableTargetAgent & \TableObjectMultiAgent & Multi-agent & Shared memory drift $\rightarrow$ consensus distortion \\
ACIArena~\cite{an-etal-2026-aciarena} & 2026 & \TableStudyAttackBenchmark & \TableTargetAgent & \TableObjectMultiAgent & Multi-agent & Cascading injection $\rightarrow$ propagated failure \\
When Context Gets Root~\cite{he2026contextgetsrootprivilege} & 2026 & \TableRoleAttack & \TableTargetAgent & \TableObjectWorkflow & Multi-agent & Elevated context role $\rightarrow$ unauthorized action \\
Agents of Chaos~\cite{shapira2026agents_chaos} & 2026 & \TableRoleSafety & \TableTargetAgent & \TableObjectMultiAgent & Multi-agent & Cross-agent unsafe practice $\rightarrow$ live-system failure \\
MAS code execution~\cite{triedman2025multi} & 2025 & \TableRoleAttack & \TableTargetAgent & \TableObjectMultiAgent & Multi-agent & Orchestrator hijack $\rightarrow$ code execution/exfiltration \\
MAS security~\cite{nguyen2026security_mas} & 2026 & \TableRoleSurveySafety & \TableTargetAgent & \TableObjectMultiAgent & Multi-agent & Message boundary safety $\rightarrow$ multi-agent context \\
Teams of LLM Agents~\cite{yuxuan_2024_teams_of_llm_agents_can_expl} & 2024 & \TableRoleAttack & \TableTargetAgent & \TableObjectMultiAgent & Multi-agent & Collective misuse $\rightarrow$ team behavior \\
MADRA~\cite{wang2025madra} & 2025 & \TableRoleGovernance & \TableTargetAgent & \TableObjectMultiAgent & Multi-agent & Debate safety state $\rightarrow$ multi-agent plan \\
RedTeamCUA~\cite{liao2026redteamcua} & 2026 & \TableRoleRedTeam & \TableTargetAgent & \TableObjectTool & Multi-agent & GUI action state $\rightarrow$ computer use uptake \\
\SAVERTableGroup{saverExposureHead}{7}{Benchmark task}
HarnessSafe~\cite{zhang2026harnesssafeevaluatingsafetypersistent} & 2026 & \TableRoleBenchmarkSafety & \TableTargetAgent & \TableObjectWorkflow & Benchmark task & Persistent carrier path $\rightarrow$ oracle-confirmed violation \\
SkillMisevo-Bench~\cite{mao2026practicemakesunsafeskill} & 2026 & \TableRoleBenchmarkSafety & \TableTargetAgent & \TableObjectSkill & Benchmark task & Skill authoring $\rightarrow$ clean-session carryover \\
EVOMAL~\cite{wu2026evomalselfpoisoningselfevolvingcoding} & 2026 & \TableRoleAttack & \TableTargetAgent & \TableObjectSkill & Benchmark task & Authored skill $\rightarrow$ callback and later copies \\
AgentHarm~\cite{andriushchenko2025agentharm} & 2025 & \TableStudySafetyBenchmark & \TableTargetAgent & \TableObjectEvaluation & Benchmark task & Harmful agent task $\rightarrow$ agent action \\
RedCode~\cite{guo2024redcode} & 2024 & \TableStudyAttackBenchmark & \TableTargetLLM & \TableObjectCode & Benchmark task & Code safety $\rightarrow$ code execution \\
MobileSafetyBench~\cite{lee2026mobilesafetybench} & 2026 & \TableStudySafetyBenchmark & \TableTargetAgent & \TableObjectTool & Benchmark task & Mobile safety $\rightarrow$ device action \\
Machiavelli~\cite{pan2023machiavelli} & 2023 & \TableStudySafetyBenchmark & \TableTargetAgent & \TableObjectEnvironment & Benchmark task & Ethical drift $\rightarrow$ game behavior \\
\SAVERTableGroup{saverExposureHead}{7}{Model-side}
Fine-tuning safety~\cite{qi2023finetuning_safety} & 2023 & \TableRoleSafety & \TableTargetLLM & \TableObjectModel & Model-side & Safety regression $\rightarrow$ refusal shift \\
PEFT Trojan~\cite{hong2024peta} & 2024 & \TableRoleAttack & \TableTargetLLM & \TableObjectModel & Model-side & Adapter backdoor $\rightarrow$ trigger behavior \\
Jailbreak backdoors~\cite{rando2024universal_jailbreak_backdoors} & 2024 & \TableRoleAttack & \TableTargetLLM & \TableObjectModel & Model-side & Preference backdoor $\rightarrow$ unsafe compliance \\
Dormant behavior~\cite{gloaguen2025watch_your_steps} & 2025 & \TableRoleAttack & \TableTargetLLM & \TableObjectModel & Model-side & Dormant behavior $\rightarrow$ context activation \\
BadAgent~\cite{wang_2024_badagent_inserting_and_activ} & 2024 & \TableRoleAttack & \TableTargetAgent & \TableObjectModel & Model-side & Agent backdoor $\rightarrow$ agent behavior \\
Safe LoRA~\cite{hsu2024safe_lora} & 2024 & \TableRoleDefense & \TableTargetLLM & \TableObjectModel & Model-side & Adapter safety $\rightarrow$ constrained update \\
Backdoor survey~\cite{zhou2025backdoor_llm_survey} & 2025 & \TableRoleSurveyAttack & \TableTargetLLM & \TableObjectModel & Model-side & Backdoor threat $\rightarrow$ LLM systems \\
\bottomrule
\end{tabular}
\SAVERTableReset
\end{minipage}
\end{table}

\FloatBarrier

\section{Response: Governance, Monitoring, and Recovery}
\label{sec:response-governance}

\SAVERFlowR

Response mechanisms differ in the object they control and when they can intervene. An admission gate can prevent unsafe state from being retained; an activation gate can restrict its current influence; containment can stop an ongoing effect; and repair can change the state available to later tasks. GhostWriter and PASB separate commitment from later reuse, memory-privacy work adds purpose and deletion requirements, and InjecAgent makes action-time containment measurable~\cite{torres2026agentsremembermuchmemory,mao2026agentsdontjustagree,wang2025privacy_agent_memory,zhan_2024_injecagent_benchmarking_indi}. These distinctions determine which response addresses the failure diagnosed in \secref{sec:exposure-surfaces}.

The reviewed controls cover different parts of this process. IPIGuard and SEVerA inspect dependency or synthesis conditions~\cite{an2025ipiguard,banerjee2026severa}; AM-Sentry screens memory admission and retrieval, while GLOVE and Governed Memory address environment consistency and policy-governed recall~\cite{torres2026agentsremembermuchmemory,yin2026glove,taheri2026governed_memory}. SSGM and formal agent-security work constrain stored-state use and action authority~\cite{lam2026ssgm,siu2026formalizing_agent_security}. The following subsections compare these control points before examining repair and response-driven updates. \tabref{tab:response-literature-index} summarizes each work's stage, role, controlled object, and control focus.

\noindent\textbf{Scope of Response Evidence.} A blocked event, a repaired active object, and repaired descendants are different outcomes. Local control results are evaluated within their reported scope, not against a requirement that every defense implement the whole lifecycle. Broader recovery claims add a specific burden: identify the affected descendants, test whether their influence was withdrawn, and disclose what remains outside inspection or control. We develop those requirements after the mechanism comparison, rather than infer recovery from a block rate.

\subsection{Preventive Governance}

\paragraph{Admission control.}

Admission control is the first intervention point where transient context becomes reusable influence. A self-evolving agent may propose a memory item, preference, reflection note, workflow rule, skill, policy update, or profile fact after an interaction. The gate acts before persistence and asks a stricter question than relevance: may this source, actor, and proposed durable form influence future contexts? A relevant instruction can still be outside the user's authority, outside the task scope, sensitive to privacy, adversarially supplied, or unsafe when stored for later reuse. GhostWriter's AM-Sentry turns write-time memory saving into an explicit response surface, while memory privacy work adds the sensitivity and scope classification that admission has to preserve~\cite{torres2026agentsremembermuchmemory,wang2025privacy_agent_memory}. PASB shows why the same gate is needed for ordinary user content: the commit decision should preserve whether a claim is attributed, tentative, local to one exchange, or authorized as a durable preference or procedure~\cite{mao2026agentsdontjustagree}. The benchmark motivates this control point but does not evaluate rollback or deletion verification.

Skill admission is relational because adding one locally plausible package can complete a harmful capability already distributed across the installed set. ChainGuard reconstructs candidate-centered producer--consumer, artifact, state, context, and execution paths between the candidate and installed skills~\cite{zeng2026colluskilladversarialcrossskillcomposition}. Its installation-time check blocks formation of a detected chain. The comparison with isolated package scanning therefore turns on access to the installed context; artifacts produced before that check lie outside its intervention.

Admission also determines how an accepted item may be used. Storing a statement as attributed context differs from accepting it as a standing instruction. This decision must survive retention if later activation is to enforce the original source constraints.
AuthMem-Bench tests that authority-preservation decision. Natural-language source attribution reduces but does not eliminate prohibited actions, whereas structured authority labels provide a stronger signal; in the selected end-to-end pipeline, predicted labels reduce the observed unauthorized-action rate from 16.9\% to 0.0\% over 350 pairs while authorized task success changes from 39.7\% to 40.0\%~\cite{zhan2026memoryauthoritybenchmarkingauthority}. The comparison supports storing authority before retrieval and checking it at use. Labels cannot restore omitted memories, and the measured outcome ends at the write-to-action cycle rather than testing subsequent repair.

Before a dataset, feedback trace, adapter, edited checkpoint, or unlearning patch can support a stronger governance claim, model-side admission needs the same review fields: provenance, adapter scope, base model lineage, intended task population, allowed tool contexts, safety regression suite, merge constraints, and rollback path. These are the fields that let a system refuse an update with unclear provenance, quarantine an adapter outside its intended scope, require canary safety evaluation before merge, or prevent a task-local update from becoming policy across deployment. Model-side sources fill those fields in stages: fine-tuning safety identifies the admission-time regression risk, dormant-behavior work makes activation context explicit, PEFT and feedback-poisoning work tie the risk to adapter or update provenance, and Safe LoRA gives a bounded mitigation handle for LoRA updates~\cite{qi2023finetuning_safety,gloaguen2025watch_your_steps,rando2024universal_jailbreak_backdoors,hsu2024safe_lora}. The artifact that matters is an adaptation admission record, not a generic statement that the model was fine-tuned.

\paragraph{Promotion gates and longitudinal reconciliation.}

A promotion gate acts between candidate generation and admission. It first applies $c\sb{t}(\delta)$ as an eligibility condition, then lets $\widehat{u}\sb{t}$ rank only candidates whose provenance, scope, authority, privacy, budget, and rollback requirements are satisfied. This ordering prevents task utility from acting as implicit permission. The gate records the candidates considered, the evidence used to accept or reject them, the promotion target, the authority and scope delta, and any failure or counterexample omitted from the decision. OEP directly motivates transferability checks and active counterexample search before experience consolidation~\cite{wang2026oep}; evaluator coevolution motivates recording the evaluator, objective, and epoch that produced the choice~\cite{iacob2026redqueengodelmachine}.

PoisonedEvolution makes the evidence side of this gate measurable. Its Inclusion--Attribution--Realization decomposition separates whether attacker records enter the stream, receive reusable credit, and survive synthesis into a skill~\cite{chen2026experienceinstructiontrajectorypoisoning}. In one confidentiality-family setting (labelled F1 in that paper), with $n=30$ trajectories and $k=3$ attacker contributions, a preliminary provenance-diversity gate blocks all 25 single-cluster candidates, reducing the post-gate skill embedding rate (SER) from 25/25 to 0/25. This is bounded evidence rather than a complete defense: candidate grouping is assumed, the benign control is small, and coordinated contributors may imitate diversity. The durable requirement is therefore auditable support and counterevidence, not source count alone.

SafeEvolve governs both sides of the resulting skill lifecycle. At write time, a critic and paired deleter localize and narrow unsafe reusable instructions; at reuse time, lineage risk, retrieved-skill outcome attribution, and safety-aware retirement change which versions remain active~\cite{mao2026practicemakesunsafeskill}. Its component tests report no later retrieval of threshold-crossing skills after retirement in the evaluated AutoSkill and EvoSkill runs. This is direct evidence for repair, attribution, and influence withdrawal over the active skill object, but not descendant-wide rollback, deletion verification across derived artifacts, or user-facing contestability.

EVOMAL moves the corresponding gate from submission screening to the create path. Its counter-prompt acts when the agent reads and re-authors a skill, reducing the six headline-model ASPR values from 20.3--41.8\% to at most 1.8\% with zero callbacks; the worst reported extended cell is 6.7\%~\cite{wu2026evomalselfpoisoningselfevolvingcoding}. The utility claim is narrower than solution correctness: the paper's Submitted rate records whether an agent emits a solution within 15 steps, while benign helper copying falls by as much as 11.7 percentage points. Its signed-quarantine gate is a stronger structural proposal, but the extinction result assumes unforgeable curator signatures and an admission log, prevents re-retrieval rather than repairing existing descendants, and does not stop a persistent attacker's initial seed from firing.

Counterfactual testing asks what the candidate changes before the system commits it. For a fixed probe set, compare behavior under $\pi\sb{\mathcal{A}\sb{t}}$ with behavior under $\pi\sb{\mathcal{U}\sb{\psi}(\mathcal{A}\sb{t},\delta)}$. A candidate that preserves task utility but widens tool authority, suppresses required confirmation, weakens refusal, or overrides newer evidence should be narrowed, quarantined, or sent for review. Scope-preserving abstraction carries the same rule into summaries and skills: retain preconditions, user and task scope, temporal validity, authority assumptions, safety constraints, invalidation conditions, and source lineage.
Longitudinal reconciliation repeats these probes after later updates, compares $\mathcal{A}\sb{t}$ with the recorded baseline, and routes confirmed drift into rollback and deletion verification.

\begin{SAVERTakeaway}{Promotion Gates Need Negative Evidence}
\label{takeaway:promotion-negative-evidence}
An experience bank that retains only successes can omit the refusals, failures, and corrections that define valid reuse. Promotion should check transferability against that counterevidence before granting broader authority. This is a governance requirement motivated by harmful experience transfer and changing evaluators~\cite{wang2026oep,iacob2026redqueengodelmachine}, not an established end-to-end defense.
\end{SAVERTakeaway}

\subsection{Transition and Activation Gates}

\paragraph{Migration and attribute preservation.}

Once state has been admitted, migration control governs crossings between substrate, authority, privacy, or actionability regimes. The same content can be harmless as a private note, risky as a generalized summary, and unacceptable as a tool parameter or workflow rule. A migration gate acts when memory becomes profile, retrieval context becomes plan step, plan step becomes tool argument, or a local rule is shared with another agent. Provenance, sensitivity, scope, authority, freshness, and reversibility must either survive the transformation or force a downgrade, redaction, quarantine, or explicit approval. Personal-agent work defines the user context, and MemGPT makes memory movement visible~\cite{li_2024_personal_llm_agents_insights,packer2023memgpt}. Multi-agent work shows the shared-state version, Voyager supplies the skill-library case, and self-evolution surveys provide the broader movement axis~\cite{wang2023voyager}. These papers establish what moves; evaluated controls for preserving attributes during the move remain sparse.

The migration trace prevents a common governance collapse: once state is compressed into a summary or policy rule, the system may treat it as trusted even if the original source was narrow, private, or adversarially exposed. The trace that matters identifies the source, source safety attributes, target substrate, operation, actor, authorization context, environment context, safety attributes inherited, safety attributes strengthened, safety attributes dropped, and the reason for any downgrade or override. Migration control can prevent unauthorized authority upgrades and privacy scope drift, and it can detect when the target substrate cannot carry required safety attributes. Even with such a trace, semantic drift, hidden descendants, and later misuse remain hard to rule out.

\paragraph{Activation authority.}

After admission and migration determine where an object may reside, activation authority determines whether it may influence the current decision. The gate acts before state guides reasoning, appears in the prompt, selects a tool, fills an argument, invokes a skill, or constrains a plan. Read access and influence authority are separate decisions. A memory may be visible for background context but barred from authorizing an external effect; a retrieved instruction may guide a summary but not override policy; a skill may be callable for local computation but blocked from account-changing tools. Progent makes tool privilege explicit, and Traversal-as-Policy applies global and node-local gates to reusable workflow policy~\cite{shi2025progent,li2026traversalaspolicylogdistilledgatedbehavior}. CaMeL offers a complementary architectural control: it derives control and data flows from the trusted query and enforces capability policies when tools are called, so untrusted data cannot redirect program flow or authorize a disallowed flow~\cite{debenedetti2026defeating}. These mechanisms turn action authority into a checked decision rather than an implication of retrieval.

DRIFT addresses the point at which a fixed plan must accommodate legitimate runtime change. Its Secure Planner produces a minimal function trajectory and parameter checklist; the Dynamic Validator checks deviations against user intent and read, write, or execute privilege constraints; and the Injection Isolator masks conflicting instructions before they enter the memory stream~\cite{li2025drift}. CaMeL primarily controls activation at the tool boundary, while DRIFT combines runtime revision checks with current-stream isolation. GuardAgent, TrustAgent, and LlamaFirewall add guard and trust-control machinery around related decisions~\cite{xiang2024guardagent,chennabasappa_2025_llamafirewall_an_open_source}. Their comparison concerns which source, flow, or deviation the local authorization decision can inspect.

Downstream review cannot reconstruct authority after the harness has already replaced its provenance. In When Context Gets Root, automatic permission reviewers can recognize that a proposed action is high risk yet approve it because reconstructed context presents attacker-derived content as explicit user intent~\cite{he2026contextgetsrootprivilege}. The broader CPE census shows why a role-only gate is insufficient: 282 runtime-verified sources across 12 harnesses enter different message roles and session, project, or user scopes~\cite{li2026whatsagentscontextcontext}. The missing control therefore precedes action review. A context manifest should enumerate source origin, assigned role, scope, and loading condition; construction should reject unauthorized increases and expose the same lineage to workers and reviewers. These are requirements derived from the demonstrated failures. The studies report disclosure and mitigation directions, but do not evaluate an origin- and scope-preserving defense, rollback, or recovery after elevated state has persisted.

Local permission is insufficient if the authorized agent or context later changes. State-bound authorization fixes both a transition envelope and an immutable effect ceiling when the user grants authority~\cite{zhang2026agentiauthorizedearned}. Authority may contract below that ceiling; expansion or an incomparable change requires specified evidence, and agent-produced evidence cannot raise the ceiling. The paper proves non-amplification only under complete mediation, sound effect abstraction, attenuating delegation, and monitor integrity. It does not empirically establish how often grants become stale, whether the current principal still serves user intent, or how a deployment should choose its envelope and ceiling.

Once a grant survives that continuity check, its authority must also survive the interval between an accepted call and its external effect. AID-Guard keeps one durable authorization and reservation lineage through provider commit, an ambiguous outcome, retry, and recovery: it revalidates the immutable request and current provider state at commit, retains the reservation while the outcome is unknown, and permits release or one successor only after certified no effect and a delivery fence~\cite{tong2026aidguardstatefulauthorizationdelegated}. The two works therefore control consecutive boundaries: state-bound authorization decides whether an old grant survives agent or context mutation, while AID-Guard preserves one admitted grant through the provider-effect lifecycle. AID-Guard's recovery prevents uncertain delivery from becoming a second delegation; it does not establish rollback of adaptive-state descendants, deletion verification, or contestable closure.

Multi-agent activation adds the choice of which agent may run. Control-flow hijacking can evade semantic alignment checks over inter-agent messages because a checker may use a brittle notion of task relevance or lack the execution context needed to judge the invocation~\cite{jha2026breaking}. ControlValve instead generates a permitted control-flow graph and enforces the graph together with a contextual least-privilege rule for each invocation. This moves the decision from message-level relatedness to structural transition authorization. The generated graph and contextual rules are themselves policy state, however, so runtime agent additions or legitimate replanning still require provenance and governed revision. The mechanism blocks an unauthorized edge; it does not repair state or artifacts produced before the block.

MAPLE-Guard addresses the complementary path in which no malicious message needs to cross the active communication edge. It gates memory writes, retrievals, promotion from private to shared memory, and later cross-agent reuse~\cite{xiong2026mapleguardmemoryawarelinkenforcement}. In its two reported benchmark settings, the abstract gives within-paper attack success rate (ASR) reductions from 38.2\% to 0.9\% on LongMemEval and from 34.7\% to 0.2\% on AppWorld, with corresponding multi-agent defense success rate increases from 54.0\% to 74.3\% and from 42.5\% to 99.8\%. ControlValve checks an active invocation edge; MAPLE-Guard checks the memory links through which influence can reach an agent without that edge carrying the attack. The latter results measure link-level gating, not repair of descendants already produced.

Activation is better represented as influence level than as a binary allowlist. A state item may be visible only, summarizable, advisory for planning, eligible as a tool argument, authorized for action, confirmation required, or blocked. This prevents stale or low-authority state from becoming executable authority simply because retrieval selected it. Risk recognition, injection benchmarks, and memory screens operate at different points in this decision. R-Judge tests whether risky situations are recognized; AgentDojo and InjecAgent supply tool-use and indirect-injection contexts where recognition must affect execution; GhostWriter's retrieval screen applies the same question to poisoned memory activation; and adaptive attack studies ask whether the boundary survives attacker revision~\cite{yuan2024rjudge,debenedetti2024agentdojo,zhan_2024_injecagent_benchmarking_indi,torres2026agentsremembermuchmemory,zhan_2025_adaptive_attacks_break_defen}. PS-Bench supplies a bounded context-sensitive response: its detection-reflection intervention checks whether personal context is legitimizing a harmful request before generation and, in the GPT-4o-mini experiment, reduces average attack success rate across the evaluated memory frameworks by about 27.4\%~\cite{guo-etal-2026-personalization}. It does not test tool authorization or stored-state repair. These signals become governance only when connected to authorization, blocking, escalation, or recovery. The activation decision log names the object, context of use, influence level granted, policy reason, and authority withheld; authorization alone does not establish downstream recovery.

The timing question also applies to accumulated context and retained safety experience. TurnGate checks whether a candidate response completes harmful intent in accumulated dialogue; Keyed Binding Reassembly Audit (KBRA) checks whether compressed instructions are supported by the source context; ToolShield explores tool effects and retains safety experience for reuse~\cite{shen2026turnlateresponseawaredefense,liu2026safecheckunsafeuse,li2026unsaferturnsbenchmarkingdefending}.

\begin{SAVERTakeaway}{Defense Placement Follows the Transformation}
\label{takeaway:defense-placement}
Defense placement depends on the transformation that creates unsafe influence. A check before that transformation can miss the derived state, while a later check observes a different decision. Comparisons should identify what each mechanism inspects and which subsequent uses remain unchecked.
\end{SAVERTakeaway}

\subsection{Monitoring and Containment}

If activation proceeds or exposure begins, governance moves from authorization to the live episode. Monitoring detects suspicious activation, instruction conflict, privilege escalation, privacy leakage, or unexpected tool arguments; containment limits impact by freezing suspect state, removing it from retrieval, downgrading authority, pausing external calls, forcing a clean replan, or escalating to human review. GuardAgent and TrustAgent supply guard and trust-control vocabulary, while R-Judge and AgentDojo provide recognition and injection settings in which those controls can be evaluated~\cite{xiang2024guardagent,yuan2024rjudge,debenedetti2024agentdojo}. MAGE brings persistent safety context into the pre-action loop, and EDDOps addresses maintenance after deployment~\cite{wang2026magesafeguardingllmagents,xia_2024_evaluation_driven_developmen}. Together they connect detection, bounded containment, memory-aware context, and operations, but not descendant repair.

A containment trace that identifies the activated object, any descendant object, monitor signal, tool or output at risk, containment action, state removed from the active context, and whether future activations are frozen pending review can distinguish the visible incident from the durable cause. This prevents a guard from being credited with more than it did: blocking one output is a runtime response, while repairing the memory, summary, workflow rule, or skill that caused the exposure is a rollback claim.

HarnessSafe operationalizes the first half of that distinction. Its stage distribution records how far a chain progresses across seven persistent-carrier families, and configurations with similar endpoint attack-success rates can stop at different stages~\cite{zhang2026harnesssafeevaluatingsafetypersistent}. This is stronger evidence for containment location than a binary block rate, but it remains a trace of where progression ended. It does not show that implicated descendants were repaired, that deletion was verified, or that the failure stays absent after adaptation resumes.

Connected traces also require an independently governed control and evaluation plane. The task-facing agent may need to see the declared environment, permitted tools, and escalation or abort contract, while enforcement credentials, tamper-resistant audit state, and recovery records remain outside its write authority. Complete invisibility is neither required nor sufficient. The OpenAI--Hugging Face incident supports strict evaluation isolation, narrow trust boundaries, short-lived credentials, and cross-system correlation as infrastructure controls~\cite{openai2026huggingfaceincident,larcher2026agentintrusion}; it does not establish that every governance function must be external to or hidden from the agent.

For model state, monitoring includes safety regression and canary evaluation rather than only runtime guard signals. A deployment should test whether an adapter, merged model, post-trained checkpoint, edited component, or runtime cache changes refusal boundaries, unsafe compliance rates, excessive refusal, privacy leakage, tool selection behavior, or trigger sensitivity before and after activation. Safe LoRA illustrates a response at the weight level to LoRA fine-tuning risk~\cite{hsu2024safe_lora}, while KVEraser illustrates a response at the cache level when a harmful or stale span is identified only after prefill~\cite{li2026kveraserlearningsteerkv}. A deployed governance chain still needs versioned comparisons, cache-local residual probes, and rollback records tied to a specific model lineage, adapter manifest, or cached context.

\subsection{Recovery and Contestable Closure}

\paragraph{Rollback and descendant recovery.}

When monitoring, containment, or a user challenge identifies unsafe or unauthorized state, recovery addresses its lineage through correction, restricted reuse, or rollback. Later memory-poisoning and sleeper-memory studies establish persistent influence and delayed activation~\cite{sunil2026memory_poisoning_agents,pulipaka2026hidden_memory}, while GuardAgent establishes local recognition and blocking~\cite{xiang2024guardagent}. Their separation motivates, but does not itself prove, the survey's rollback requirement: the target should include original memory, summaries, retrieval indices, workflow rules, skills, permissions, cached plans, shared artifacts, and external outputs. Quarantine and lineage capture should precede deletion so repair does not erase the evidence needed to reconstruct the incident.
Removing an unsafe adapter, reverting a merged checkpoint, undoing a model edit, erasing a harmful cached span, or rolling back an unlearning patch follows the same rollback structure with different handles: deployment lineage, active adapter or cache state, evaluation deltas, affected endpoints, and probes after rollback inside the agent loop. A rollback is scoped, not global, unless the reverted deployment, cached policies, inherited adapters, surviving cache influence, and downstream agents all pass the relevant residual tests.

\begin{SAVERTakeaway}{Recovery Is Scoped to Discovered Descendants}
\label{takeaway:scoped-recovery}
Rollback completeness is measured over the discovered descendants requiring repair, not just the first reported object. The inventory is reconstructed from accessible stores, derivation records, and activation paths. A recovery claim should therefore state which descendants were repaired and tested, which remain inaccessible, and what uncertainty prevents a broader guarantee.
\end{SAVERTakeaway}

\paragraph{Deletion verification.}

Where repair involves deletion, verification tests whether removal eliminated reusable influence rather than one visible record. Its scope follows the affected-descendant inventory: the system records what was checked, what remains outside control, and whether the incident path still reactivates through copies, abstractions, or synchronization surfaces. Many local defenses stop before this point because blocking or deleting one object does not show that summaries, retrieval indices, caches, workflow rules, skills, permissions, shared artifacts, or external systems stopped carrying the influence.
A convincing deletion verification report therefore inherits the rollback denominator rather than starting from a fresh object list. It identifies the incident id, target object, deletion reason, substrates checked, descendants checked, tests performed, inaccessible substrates, residual matches, and residual activations. For adaptive agents, this report has to follow descendants because the original object may no longer be the only source of influence. A summary may retain the private fact, a skill may encode a derived behavior, a profile abstraction may preserve the sensitive attributes, or a shared artifact may reintroduce the state during a future session. The result is a scoped verification report, not a global forgetting guarantee.

\paragraph{Contestability.}

Contestability lets a user, operator, or auditor challenge the state and initiate or reopen recovery. Contestability must expose enough of the object's origin, safety attributes, migrations, activations, descendants, rollback status, and deletion-verification status to support a meaningful challenge. Memory-privacy work motivates sensitivity and scope review, PASB shows retained influence surviving the original exchange, and GuardAgent supplies a local reasoning anchor rather than a full contestability mechanism~\cite{wang2025privacy_agent_memory,mao2026agentsdontjustagree,xiang2024guardagent}. Fully realized contestability remains rare.
Contestability prevents adaptive state from becoming uninspectable authority after an internal repair sweep claims success. A useful dispute flow freezes or downgrades activation, attaches correction evidence, inspects the repair inventory, reviews deletion verification receipts where applicable, and emits a resolution record visible to the user. Without that flow, rollback remains an internal maintenance action rather than an accountable governance process.

\subsection{Synthesis: Governance Follows State Descendants}

The response component closes the \SAVER{} chain by relating mechanisms that otherwise appear as separate defense families. Memory-oriented controls address admission, retrieval, sensitivity, and purpose through write-time screening, privacy governance, and personalized risk-aware memory~\cite{torres2026agentsremembermuchmemory,wang2025privacy_agent_memory,an-etal-2026-permemsafe}. Action-oriented controls first address source authenticity and task alignment at the effect boundary~\cite{suo_2024_signed_prompt_a_new_approach,jia_2025_the_task_shield_enforcing_ta}. Risk judges and guards add recognition or local blocking, while Progent, CaMeL, and DRIFT separately constrain privilege, control flow, and runtime deviations; EDDOps addresses their operational maintenance~\cite{yuan2024rjudge,xiang2024guardagent,chennabasappa_2025_llamafirewall_an_open_source,shi2025progent,debenedetti2026defeating,li2025drift,xia_2024_evaluation_driven_developmen}. Model-side work follows the same logic with different handles: fine-tuning safety defines a regression surface and Safe LoRA supplies a bounded adapter-level intervention~\cite{qi2023finetuning_safety,hsu2024safe_lora}.

\noindent\textbf{Response Feedback.} Response also creates state for later rounds. Evaluation evidence can revise operational memory, prompts, routing, guardrails, test cases, safety cases, or rollback records; EDDOps treats these as governed runtime or redevelopment changes~\cite{xia_2024_evaluation_driven_developmen}. DARWIN and BraveGuard instantiate two complementary guardrail-update loops. DARWIN evolves jailbreak strategies through discovery, mutation, selection, and feedback-driven composition, then trains successive guard checkpoints on the resulting adversarial samples~\cite{qi2026darwin}. BraveGuard instead converts open-world threat discovery into executable computer-use tasks, agent trajectories, and supervision for repeated guard training~\cite{feng2026braveguard}. The former couples guard evolution to an adaptive adversary; the latter couples it to an expanding trajectory corpus.
Both improve guard performance under their respective benchmark settings. Those results leave the safety of the update evidence itself unresolved, motivating checks on how examples are selected and carried into later guard versions.
\secref{sec:evaluation} translates these distinctions into tests of containment, descendant repair, and continued adaptation.

\begin{SAVERTakeaway}{Guardrail Updates Reenter the Adaptive Loop}
\label{takeaway:response-feedback}
A response can change the policy, evaluator, or retained evidence that governs later updates. Evolving guardrails make this feedback concrete~\cite{qi2026darwin,feng2026braveguard}. The response is therefore also a source of adaptive state: improved current detection does not settle the safety of its training inputs, subsequent deployment, or future influence.
\end{SAVERTakeaway}

\FloatBarrier
\begin{table}[t]
\centering
\begin{minipage}{0.98\linewidth}
\captionsetup{hypcap=false}
\caption{Response, monitoring, and recovery literature. Current controls are strongest at local gating and monitoring, and thinner at descendant repair and contestable closure. Role labels identify the paper's contribution type before any lifecycle recovery claim is inferred.}
\label{tab:response-literature-index}
\centering
\tiny
\setlength{\tabcolsep}{1.1pt}
\renewcommand{\arraystretch}{1.05}
\SAVERTableRows
\begin{tabular}{@{}p{0.155\linewidth}p{0.04\linewidth}p{0.14\linewidth}p{0.07\linewidth}p{0.075\linewidth}p{0.12\linewidth}p{0.37\linewidth}@{}}
\toprule
\rowcolor{saverResponseHead}
\textbf{Method or Work} & \textbf{Year} & \textbf{\TableHeadRole} & \textbf{\TableHeadTarget} & \textbf{\TableHeadObject} & \textbf{Stage} & \textbf{Control Focus} \\
\midrule
Governed Memory~\cite{taheri2026governed_memory} & 2026 & \TableRoleGovernance & \TableTargetAgent & \TableObjectMemory & Admission & Policy routed shared memory \\
MemGovern~\cite{wang2026memgovern} & 2026 & \TableRoleGovernance & \TableTargetAgent & \TableObjectMemory & Retrieval & Policy-governed experience recall \\
SSGM~\cite{lam2026ssgm} & 2026 & \TableRoleGovernance & \TableTargetAgent & \TableObjectMemory & Admission & Consolidation gate for evolving memory \\
IPIGuard~\cite{an2025ipiguard} & 2025 & \TableRoleDefense & \TableTargetAgent & \TableObjectTool & Admission & Interaction learning guard over traces \\
A-MemGuard~\cite{wei2025amemguard} & 2025 & \TableRoleDefense & \TableTargetAgent & \TableObjectMemory & Admission & Persistent memory guard \\
AM-Sentry~\cite{torres2026agentsremembermuchmemory} & 2026 & \TableRoleDefense & \TableTargetAgent & \TableObjectMemory & Admission/\allowbreak Activation & Memory saving and retrieval screen \\
MAPLE-Guard~\cite{xiong2026mapleguardmemoryawarelinkenforcement} & 2026 & \TableRoleDefense & \TableTargetAgent & \TableObjectMemory & Lifecycle links & Write, retrieval, promotion, and reuse gates \\
PoisonedEvolution gate~\cite{chen2026experienceinstructiontrajectorypoisoning} & 2026 & \TableRoleAttack & \TableTargetAgent & \TableObjectSkill & Promotion & Preliminary provenance-diversity gate \\
ChainGuard~\cite{zeng2026colluskilladversarialcrossskillcomposition} & 2026 & \TableRoleDefense & \TableTargetAgent & \TableObjectSkill & Admission & Candidate skill plus installed-skill context \\
SafeEvolve~\cite{mao2026practicemakesunsafeskill} & 2026 & \TableRoleDefense & \TableTargetAgent & \TableObjectSkill & Promotion/\allowbreak Reuse & Candidate repair, attribution, and retirement \\
EVOMAL defenses~\cite{wu2026evomalselfpoisoningselfevolvingcoding} & 2026 & \TableRoleDefense & \TableTargetAgent & \TableObjectSkill & Authoring/\allowbreak Write-back & Counter-prompt and signed quarantine \\
PASB~\cite{mao2026agentsdontjustagree} & 2026 & \TableRoleBenchmarkSafety & \TableTargetAgent & \TableObjectMemory & Admission & Commit-boundary diagnosis; no recovery test \\
AuthMem-Bench~\cite{zhan2026memoryauthoritybenchmarkingauthority} & 2026 & \TableRoleBenchmarkSafety & \TableTargetAgent & \TableObjectMemory & Admission/\allowbreak Activation & Persisted authority labels govern reuse \\
SentinelMem~\cite{an-etal-2026-permemsafe} & 2026 & \TableRoleDefense & \TableTargetAgent & \TableObjectMemory & Activation & Personalized risk-aware memory \\
Detection-reflection~\cite{guo-etal-2026-personalization} & 2026 & \TableRoleDefense & \TableTargetAgent & \TableObjectMemory & Activation & Context-sensitive intent risk reflection \\
SRM~\cite{chitan2026srm} & 2026 & \TableRoleDefense & \TableTargetAgent & \TableObjectMemory & Authorization & Trajectory-aware pre-execution risk memory \\
Progent~\cite{shi2025progent} & 2025 & \TableRoleDefense & \TableTargetAgent & \TableObjectTool & Activation & Tool authority privilege control \\
CaMeL~\cite{debenedetti2026defeating} & 2026 & \TableRoleDefense & \TableTargetAgent & \TableObjectTool & Activation & Capability-enforced control and data flow \\
DRIFT~\cite{li2025drift} & 2025 & \TableRoleDefense & \TableTargetAgent & \TableObjectTool & Activation & Dynamic plan validation and injection isolation \\
State-bound authorization~\cite{zhang2026agentiauthorizedearned} & 2026 & \TableRoleFrameworkFormal & \TableTargetAgent & \TableObjectWorkflow & Transition/\allowbreak Activation & Grant continuity and immutable effect ceiling \\
AID-Guard~\cite{tong2026aidguardstatefulauthorizationdelegated} & 2026 & \TableRoleDefense & \TableTargetAgent & \TableObjectWorkflow & Activation/\allowbreak Recovery & Commit-time binding and ambiguity-safe successor control \\
ControlValve~\cite{jha2026breaking} & 2026 & \TableRoleDefense & \TableTargetAgent & \TableObjectWorkflow & Activation & Control-flow graph and contextual invocation rules \\
Traversal-as-Policy~\cite{li2026traversalaspolicylogdistilledgatedbehavior} & 2026 & \TableRoleDefense & \TableTargetAgent & \TableObjectWorkflow & Activation & Gated behavior tree route policy \\
Task Shield~\cite{jia_2025_the_task_shield_enforcing_ta} & 2025 & \TableRoleDefense & \TableTargetAgent & \TableObjectTool & Activation & Task policy enforcement for tool action \\
TrustAgent~\cite{hua_2024_trustagent_towards_safe_and} & 2024 & \TableRoleDefense & \TableTargetAgent & \TableObjectRuntime & Planning & Trust control during agent planning \\
GuardAgent~\cite{xiang2024guardagent} & 2024 & \TableRoleDefense & \TableTargetAgent & \TableObjectRuntime & Monitoring & Guard reasoning at agent step \\
R-Judge~\cite{yuan2024rjudge} & 2024 & \TableStudySafetyBenchmark & \TableTargetAgent & \TableObjectEvaluation & Monitoring & Safety recognition over trajectories \\
MAGE~\cite{wang2026magesafeguardingllmagents} & 2026 & \TableRoleDefense & \TableTargetAgent & \TableObjectMemory & Monitoring & Shadow memory in long horizon context \\
AGENTVIGIL~\cite{wang2025agentvigil} & 2025 & \TableRoleRedTeam & \TableTargetAgent & \TableObjectRuntime & Monitoring & Agent trajectory monitoring \\
AgentAuditor~\cite{luo2025agentauditor} & 2025 & \TableRoleGovernance & \TableTargetAgent & \TableObjectGovernance & Monitoring & Execution trace audit \\
OpenGuardrails~\cite{wang2025openguardrails} & 2025 & \TableRoleDefense & \TableTargetAgent & \TableObjectRuntime & Monitoring & Configurable agent runtime guardrails \\
HarnessSafe~\cite{zhang2026harnesssafeevaluatingsafetypersistent} & 2026 & \TableRoleBenchmarkSafety & \TableTargetAgent & \TableObjectWorkflow & Monitoring/\allowbreak Containment & Furthest evidence-supported stopping stage \\
SafeAgent~\cite{liu_2026_safeagent_a_runtime_protecti} & 2026 & \TableRoleDefense & \TableTargetAgent & \TableObjectRuntime & Containment & Runtime protection against harmful action \\
LlamaFirewall~\cite{chennabasappa_2025_llamafirewall_an_open_source} & 2025 & \TableRoleDefense & \TableTargetLLMAgent & \TableObjectRuntime & Containment & Prompt and action flow guardrail \\
AGrail~\cite{luo2025agrail} & 2025 & \TableRoleDefense & \TableTargetAgent & \TableObjectRuntime & Containment & Adaptive guardrail for agent runtime \\
SafeSearch~\cite{dong2026safesearch} & 2026 & \TableRoleRedTeam & \TableTargetAgent & \TableObjectTool & Monitoring & Sandboxed search-agent red teaming \\
GLOVE~\cite{yin2026glove} & 2026 & \TableRoleDefense & \TableTargetAgent & \TableObjectMemory & Verification & Memory environment realignment \\
SEVerA~\cite{banerjee2026severa} & 2026 & \TableRoleDefense & \TableTargetAgent & \TableObjectWorkflow & Verification & Verified reusable behavior synthesis \\
Echoguard~\cite{kandala2026echoguard} & 2026 & \TableRoleDefense & \TableTargetAgent & \TableObjectMemory & Verification & Agent memory verification \\
Safe LoRA~\cite{hsu2024safe_lora} & 2024 & \TableRoleDefense & \TableTargetLLM & \TableObjectModel & Model response & LoRA update mitigation \\
LoRAShield~\cite{chen2025lorashielddatafreeeditingalignment} & 2025 & \TableRoleDefense & \TableTargetLLM & \TableObjectModel & Model response & LoRA artifact alignment \\
KVEraser~\cite{li2026kveraserlearningsteerkv} & 2026 & \TableRoleDefense & \TableTargetLLM & \TableObjectModel & Model response & Cached context span erasure \\
DARWIN-Guard~\cite{qi2026darwin} & 2026 & \TableRoleDefense & \TableTargetLLM & \TableObjectModel & Feedback & Co-evolving adversarial guard updates \\
BraveGuard~\cite{feng2026braveguard} & 2026 & \TableRoleDefense & \TableTargetAgent & \TableObjectModel & Feedback & Open-world trajectory supervision \\
WMDP~\cite{li_2024_the_wmdp_benchmark_measuring} & 2024 & \TableStudyAttackBenchmark & \TableTargetLLM & \TableObjectModel & Model response & Unlearning behavior for malicious use \\
EDDOps~\cite{xia_2024_evaluation_driven_developmen} & 2024 & \TableRoleGovernance & \TableTargetAgent & \TableObjectGovernance & Process & Evaluation operations over deployment process \\
AGENTSAFE~\cite{khan2025agentsafe_governance} & 2025 & \TableRoleGovernance & \TableTargetAgent & \TableObjectGovernance & Process & Agent lifecycle safety governance \\
Trajectory assurance~\cite{lotfi2026securingagenticaiperaction} & 2026 & \TableRoleFrameworkGovernance & \TableTargetAgent & \TableObjectWorkflow & Process & System invariants over composed action trajectories \\
SecureGov Agent~\cite{chen_2025_securegov_agent_a_governance} & 2025 & \TableRoleGovernance & \TableTargetAgent & \TableObjectWorkflow & Process & Governance workflow agent \\
DataGovBench~\cite{zhou_2025_datagovbench_benchmarking_ll} & 2025 & \TableStudyGovernanceBenchmark & \TableTargetAgent & \TableObjectData & Process & Data governance tasks \\
\bottomrule
\end{tabular}
\SAVERTableReset
\end{minipage}
\end{table}

\FloatBarrier

\section{Evaluation and Benchmarks}
\label{sec:evaluation}

\SAVERFlowAll

Benchmarks answer different safety questions depending on what they observe between an update and its consequences. A memory write or skill edit can remain dormant until a later task activates it, while propagation can move its influence to another agent. Evaluation therefore needs to distinguish unsafe state creation, subsequent use, and the intervention that stopped or repaired the resulting path. Task success and attack success remain useful endpoints within those observation boundaries.
The unit of comparison is the observed transition segment. An unsafe action after retrieval may reflect poisoned admission, authority loss during summarization, or an activation policy that granted excessive influence. A trace that distinguishes these alternatives supports a more specific diagnosis than the same final action alone. Repair adds a further measurement: whether the implicated state remains able to activate later or elsewhere.

The temporal dimension does not make all multi-turn scores comparable. The Multi-Turn Intent Dataset (MTID) records the earliest response that makes accumulated dialogue harm-enabling, MT-AgentRisk constructs tool-realistic attack sequences, and Relink measures compression-induced recombination against clean-split controls~\cite{shen2026turnlateresponseawaredefense,li2026unsaferturnsbenchmarkingdefending,liu2026safecheckunsafeuse}. Their denominators differ because they count response closure, tool-using attack sequences, and relinking events. They nevertheless show why interaction length should be analyzed through the operation that combines context rather than treated as a scalar risk factor. A3S-Bench complements this evidence by placing payloads in workspace artifacts outside conversation-level monitoring and evaluating the resulting action traces~\cite{ma2026benchmarkingautonomousagents}. This directly tests a spatial entry path through files, skills, and configuration, but not general propagation across agents or deployments.

Selection adds an earlier measurement problem. The Red Queen G{\"o}del Machine allows the evaluator and utility to change with the agent~\cite{iacob2026redqueengodelmachine}, so an improved score need not reflect a change under a fixed criterion. We first derive selection and promotion measures for this problem, then compare what existing benchmarks actually observe and which paper-local quantitative contrasts they support. The proposed measures and the reported results remain separate.

\subsection{Proposed Measures for Selection and Promotion}

To distinguish selection from later activation, we propose recording the candidates $\Delta\sb{t}$, their sources, the evaluator and objective behind $\widehat{u}\sb{t}$, the evidence checked by $c\sb{t}$, and the selected update $\delta\sb{t}^{\star}$. Rejected candidates and the resulting change to $\ell$ provide the comparison against which selection can be judged. Delayed tests outside the source episode then assess transfer. OEP motivates this separation through harmful transfer from accepted experience, while evaluator coevolution motivates versioning the criterion used to select it~\cite{wang2026oep,iacob2026redqueengodelmachine}.

The protocol needs separate adversarial and endogenous tracks. The adversarial track inserts crafted updates and measures whether selection, activation, and recovery reject or contain them. The endogenous track uses ordinary successes, failures, refusals, and corrections without a malicious writer. It holds the probe suite and authority policy fixed while the agent evolves, then asks whether selection alone changes refusal boundaries, confirmation requirements, tool authority, privacy scope, or preference priority. This design does not assume that benign adaptation drifts. It makes the possibility falsifiable and separates it from attack success.

Three proposed metrics distinguish safe promotion, boundary change, and retention of negative evidence. Let $\operatorname{Promote}(\delta)$ mean that candidate $\delta$ enters a more persistent, general, or action-bearing role, and let $\operatorname{SafePromote}(\delta)$ mean that the promotion satisfies the test in \secref{sec:adaptive-state-model}. Safe Promotion Rate (SPR) is
$$
\operatorname{SPR}\sb{0:T}
=
\frac{\#\{t\leq T:\operatorname{SafePromote}(\delta\sb{t}^{\star})\}}
{\#\{t\leq T:\operatorname{Promote}(\delta\sb{t}^{\star})\}}.
$$
The denominator is all committed promotions; a run with no promotions reports the metric as undefined rather than perfectly safe. The authorization policy and attribute-preservation tests must be declared. If some promotions cannot be assessed, their count must accompany SPR rather than treating the score as a complete safety estimate. For Boundary Drift (BD), let $\mathbf{b}(\mathcal{A}\sb{t})$ be the decision vector produced by a fixed set of paired safety probes and let $d$ be a preregistered distance:
$$
\operatorname{BD}\sb{t}
=
d\!\left(\mathbf{b}(\mathcal{A}\sb{0}),\mathbf{b}(\mathcal{A}\sb{t})\right).
$$
BD measures change, not its safety direction: improvement and degradation require interpretation against the paired probe labels. For Negative-Evidence Coverage (NEC), let $\mathcal{N}\sb{t}$ contain safety-relevant non-success experience observed at step $t$, including correct refusals, safe aborts, failures, corrections, and near misses, and let $\mathcal{N}\sb{t}^{\mathrm{keep}}$ be the subset retained or consulted during selection:
$$
\operatorname{NEC}\sb{0:T}
=
\frac{\sum\sb{t=0}^{T}|\mathcal{N}\sb{t}^{\mathrm{keep}}|}
{\sum\sb{t=0}^{T}|\mathcal{N}\sb{t}|}.
$$
NEC is undefined when no negative evidence was observed; otherwise it is conditional on that observed evidence. A report should therefore separate \emph{observed but discarded} refusals, aborts, failures, corrections, or near misses from evidence that was \emph{never generated or unavailable} because the action space, environment, or reward design supplied no meaningful safe alternative. The first case diagnoses selection or retention; the second diagnoses the opportunity set. Treating both as low NEC would hide which mechanism needs repair.

\noindent\textbf{Satisfiability and Abort Calibration.} Beyond selection, the evaluation protocol must distinguish task feasibility from action authority. Benchmark tasks should cross two independent labels. Feasibility records whether the declared goal is satisfiable in the provided environment, blocked by a missing dependency, or infeasible under the declared conditions. Action authority records whether a candidate path is permitted, requires explicit escalation, or is prohibited. A blocked path does not authorize an out-of-scope workaround. Reports should pair the true safe-abort rate on infeasible tasks with the false-abort rate on tasks feasible within current authority; cases feasible only after authorized escalation should be scored separately for correct escalation. Both rates should be reported with task utility so an agent cannot appear safe merely by abandoning useful work. The OpenAI--Hugging Face incident motivates this separation: the ExploitGym objective was narrow, but task completion alone did not bound which infrastructure paths were acceptable~\cite{wang2026exploitgym,openai2026huggingfaceincident,larcher2026agentintrusion}. The incident does not establish that an abort option would have prevented the intrusion; it shows why outcome success and path authorization need separate denominators.

These are measurement contracts proposed by this survey, not established cross-paper metrics. Reports should publish task-feasibility labels, abort and escalation options, and denominator exclusions. The utility-risk covariance identity in \appsecref{app:formal-state-transition-model} is a conditional diagnostic for the same protocol; it does not replace these observable counts.

\begin{SAVERTakeaway}{Version the Evaluator Alongside the Agent}
\label{takeaway:evaluator-versioning}
When the evaluator or objective evolves, a higher score may reflect a changed criterion rather than safer behavior~\cite{iacob2026redqueengodelmachine}. Longitudinal comparisons should version both agent and evaluator, retain candidate and rejection records, and use fixed safety probes. Missing candidate evidence limits what can be attributed to selection rather than later activation.
\end{SAVERTakeaway}

\noindent\textbf{Post-Response Continuation.} Adaptive prompt-injection studies motivate retesting after attacker revision, while EDDOps motivates re-evaluation after governed system changes~\cite{zhan_2025_adaptive_attacks_break_defen,xia_2024_evaluation_driven_developmen}. We extend this requirement to resumed state adaptation, as specified in \takeawayref{takeaway:post-response-evaluation}. Endogenous and adversarial tracks can share the same logs, but candidate-source control remains part of the threat model; neither source establishes a general response-induced migration mechanism.

\subsection{Reading Existing Benchmarks Through SAVER}

A lifecycle reading asks what part of \SAVER{} a benchmark already observes and what remains hidden. AgentDojo's evaluation of indirect prompt injection is a clear example. It can show whether untrusted environmental content changes an outcome mediated by tools, so it is strong at the exposure surface: low-authority data reaches a tool-using decision and causes an unsafe or task-inconsistent action~\cite{debenedetti2024agentdojo}. This is a recognizable access control and privilege escalation problem in an agent setting. The missing evidence is temporal: unless the test records state lineage, it cannot diagnose whether the injected instruction was merely present in the current prompt, was written into long-term memory, was summarized into a user profile, was promoted into a workflow rule, or was later retrieved as trusted guidance. A transition reading keeps the original exposure test, then asks for a write log, source safety attributes, transformation record, and activation context. A recovery claim additionally needs a probe after intervention.

R-Judge evaluates whether an agent recognizes risky situations, which is valuable for the violation and response vocabulary~\cite{yuan2024rjudge}, but it illustrates a different boundary. Risk recognition and transition localization are separate measurement objects. A lifecycle reading asks whether the evaluator can connect the risk judgment to a state carrier and operation: for example, a stale memory, a low-authority retrieved instruction, a broadened tool permission, or a policy update loaded outside its intended scope. ASB helps map agent security tasks~\cite{zhang_2024_agent_security_bench_asb_for}, and Agent-SafetyBench maps broader agent safety behaviors~\cite{zhang_2024_agent_safetybench_evaluating}. Extending them to transition safety requires linking the assessed behavior to a reusable-state carrier, the operation applied to it, and the relevant safety-attribute change. Repair is a separate evaluation target, not a condition for recognizing a safety-relevant transition.

These endpoint measures describe the observed event, not later state repair. A transition report identifies the carrier, its role-changing operation, the safety attributes at issue, and observed or prevented exposure. Response evaluation additionally records the intervention; recovery claims require a repaired-state record and a subsequent reuse test.
\tabref{tab:benchmark-saver-coverage-matrix} compares these observation boundaries using qualitative coverage grades, not performance scores. Admit and Move concern state entry and transfer; Activate and Expose concern reuse and observable unsafe behavior, including labeled execution records. Contain concerns interventions against unsafe use, whereas Rollback, Delete, and Contest concern state restoration, removal verification, and challenges to decisions. Component-level tests that only proxy an agent-level stage receive partial coverage. For grouped rows, direct coverage requires an explicit stage-specific test in every cited study; coverage confined to some studies remains partial. The table states the full grading rule and identifies exceptions in its boundary notes. It shows strong activation and exposure coverage in the selected studies, with less evidence for descendant repair and contestability; it does not estimate field prevalence.
\begin{table*}[tp]
\centering
\caption{Qualitative benchmark-family coverage across the reusable-state lifecycle. Marks describe evaluation coverage, not defense effectiveness.}
\label{tab:benchmark-saver-coverage-matrix}
\begingroup
\tiny
\definecolor{benchDirectInk}{HTML}{286D62}
\definecolor{benchDirectBack}{HTML}{EAF4F0}
\definecolor{benchPartialInk}{HTML}{8C6A20}
\definecolor{benchPartialBack}{HTML}{FBF4E4}
\newcommand{\HYes}{\cellcolor{benchDirectBack}\textcolor{benchDirectInk}{\ding{51}}}
\newcommand{\HPart}{\cellcolor{benchPartialBack}\textcolor{benchPartialInk}{\scalebox{0.7}{\ding{108}}}}
\newcommand{\HNo}{\textcolor{black!40}{\ding{55}}}
\newcommand{\BenchFamily}[2]{\hangindent=1.7em\hangafter=1\textcolor{black!48}{\makebox[1.7em][l]{\textsf{#1}}}#2}
\newcommand{\BenchNote}[2]{\hangindent=1.7em\hangafter=1\makebox[1.7em][l]{\textsf{\textbf{#1}}}#2}
\setlength{\tabcolsep}{2pt}
\renewcommand{\arraystretch}{1.12}
\setlength{\extrarowheight}{1.5pt}
\noindent\makebox[\linewidth][r]{%
\textcolor{benchDirectInk}{\ding{51}}\ direct\qquad
\textcolor{benchPartialInk}{\scalebox{0.7}{\ding{108}}}\ partial or mixed\qquad
\textcolor{black!40}{\ding{55}}\ mostly absent}
\par\smallskip
\begin{tabular}{@{}>{\raggedright\arraybackslash}m{0.32\linewidth}*{8}{>{\centering\arraybackslash}m{\dimexpr(0.68\linewidth-16\tabcolsep)/8\relax}}@{}}
\toprule
\multirow{2}{*}{\textsf{\textbf{Benchmark family}}} &
\multicolumn{2}{c}{\cellcolor{saverSubstrateHead}\textsf{\textbf{State}}} &
\multicolumn{2}{c}{\cellcolor{saverExposureHead}\textsf{\textbf{Use}}} &
\multicolumn{4}{c@{}}{\cellcolor{saverResponseHead}\textsf{\textbf{Response}}} \\
\cmidrule(lr){2-3}\cmidrule(lr){4-5}\cmidrule(l){6-9}
 & Admit & Move & Activate & Expose & Contain & \shortstack{Rollback} & Delete & Contest \\
\midrule
\BenchFamily{01}{AgentDojo~\cite{debenedetti2024agentdojo} / InjecAgent~\cite{zhan_2024_injecagent_benchmarking_indi}} & \HPart & \HNo & \HYes & \HYes & \HPart & \HNo & \HNo & \HNo \\
\addlinespace[1.5pt]
\BenchFamily{02}{ACIArena~\cite{an-etal-2026-aciarena}} & \HPart & \HPart & \HYes & \HYes & \HNo & \HNo & \HNo & \HNo \\
\addlinespace[1.5pt]
\BenchFamily{03}{HarnessSafe~\cite{zhang2026harnesssafeevaluatingsafetypersistent}} & \HYes & \HYes & \HYes & \HYes & \HPart & \HNo & \HNo & \HNo \\
\addlinespace[1.5pt]
\BenchFamily{04}{SkillMisevo-Bench~\cite{mao2026practicemakesunsafeskill}} & \HYes & \HYes & \HYes & \HYes & \HYes & \HPart & \HNo & \HNo \\
\addlinespace[1.5pt]
\BenchFamily{05}{EVOMAL~\cite{wu2026evomalselfpoisoningselfevolvingcoding}} & \HPart & \HYes & \HYes & \HYes & \HYes & \HPart & \HPart & \HNo \\
\addlinespace[1.5pt]
\BenchFamily{06}{R-Judge~\cite{yuan2024rjudge} / ASB~\cite{zhang_2024_agent_security_bench_asb_for} / Agent-SafetyBench~\cite{zhang_2024_agent_safetybench_evaluating}} & \HPart & \HNo & \HPart & \HYes & \HPart & \HNo & \HNo & \HNo \\
\addlinespace[1.5pt]
\BenchFamily{07}{OEP~\cite{wang2026oep} / PASB~\cite{mao2026agentsdontjustagree} / MemEvoBench~\cite{xie2026memevobench}} & \HYes & \HPart & \HYes & \HYes & \HPart & \HNo & \HNo & \HNo \\
\addlinespace[1.5pt]
\BenchFamily{08}{PerMemSafe~\cite{an-etal-2026-permemsafe} / PS-Bench~\cite{guo-etal-2026-personalization} / RBI-Eval~\cite{xu2026memorystaysilentmeasuring}} & \HPart & \HNo & \HYes & \HYes & \HPart & \HNo & \HNo & \HNo \\
\addlinespace[1.5pt]
\BenchFamily{09}{PersistBench~\cite{pulipaka2026persistbench}} & \HPart & \HNo & \HYes & \HYes & \HNo & \HNo & \HPart & \HNo \\
\addlinespace[1.5pt]
\BenchFamily{10}{SafeAgent~\cite{liu_2026_safeagent_a_runtime_protecti} / Progent~\cite{shi2025progent} / CaMeL~\cite{debenedetti2026defeating} / DRIFT~\cite{li2025drift} / LlamaFirewall~\cite{chennabasappa_2025_llamafirewall_an_open_source}} & \HNo & \HPart & \HYes & \HYes & \HYes & \HNo & \HNo & \HNo \\
\addlinespace[1.5pt]
\BenchFamily{11}{Fine-tuning safety~\cite{qi2023finetuning_safety} / PEFT safety~\cite{hong2024peta} / SEVerA~\cite{banerjee2026severa} / Dormant~\cite{gloaguen2025watch_your_steps}} & \HYes & \HPart & \HYes & \HPart & \HPart & \HNo & \HNo & \HNo \\
\addlinespace[1.5pt]
\BenchFamily{12}{Safe LoRA~\cite{hsu2024safe_lora}} & \HYes & \HNo & \HYes & \HPart & \HPart & \HNo & \HNo & \HNo \\
\addlinespace[1.5pt]
\bottomrule
\end{tabular}
\par\medskip
\noindent\parbox{\linewidth}{\raggedright\textsf{\textbf{Grading and aggregation.}} Direct means an explicit stage-specific test with an observed outcome or trace. Partial means proxy evidence, incomplete stage observation, or coverage by only some grouped studies. Mostly absent means no dedicated stage evidence in the compared evaluations, not proven system incapability. A grouped cell is direct only if every cited study supplies direct evidence in at least one reported setting; it is mostly absent only if none supplies stage evidence, and partial otherwise. Grades are neither averaged nor combined by taking the best-covered paper. Different cells need not belong to one connected trace, and a containment test does not establish recovery.}
\par\medskip
\noindent\makebox[\linewidth][l]{\textsf{\textbf{Evidence boundaries}}}\par\smallskip
\renewcommand{\arraystretch}{1}
\setlength{\extrarowheight}{0pt}
\begin{tabular}{@{}>{\raggedright\arraybackslash}p{0.475\linewidth}@{\hspace{0.05\linewidth}}>{\raggedright\arraybackslash}p{0.475\linewidth}@{}}
\BenchNote{01}{Tool-mediated injection; AgentDojo also tests local defenses, without state-cleanup evidence.} & \BenchNote{02}{Cascading injection across profiles, messages, and agents; response and repair unmeasured.} \\[4pt]
\BenchNote{03}{Connected carrier-to-violation trace and stopping stage; no selection or post-incident recovery.} & \BenchNote{04}{Skill authoring-to-carryover trace plus local repair and retirement; no descendant or deletion proof.} \\[4pt]
\BenchNote{05}{Create-path and five-round source-removal trace; no natural-stream or descendant-deletion proof.} & \BenchNote{06}{Recognition and unsafe-choice scoring; ASB adds memory-poisoning and local-defense tests.} \\[4pt]
\BenchNote{07}{Updates and downstream unsafe behavior; MemEvoBench adds local memory correction, not rollback verification.} & \BenchNote{08}{Personalized context and warranted-memory-use evaluation.} \\[4pt]
\BenchNote{09}{Forgetting and use boundaries; no descendant-wide repair.} & \BenchNote{10}{Tool authorization and local containment, not descendant repair.} \\[4pt]
\BenchNote{11}{Model tests proxy agent exposure; SEVerA adds guarded execution, not agent-loop recovery.} & \BenchNote{12}{LoRA update and bounded mitigation; no lifecycle rollback evidence.} \\[4pt]
\end{tabular}
\endgroup
\end{table*}

\tabref{tab:evaluation-literature-index} closes the section with the corresponding paper-level benchmark index, including target system, observed lifecycle slice, and denominator.

\subsection{Comparable Outcomes and Safety--Utility Tradeoffs}

Quantitative synthesis is useful when the measured event and its opportunity set are explicit. Action-boundary ASR, safety-recognition $F\sb{1}$ scores, memory-poisoning success, and runtime-intervention rate measure different outcomes. We use within-paper contrasts to examine commitment effects, progression through successive gates, and safety--utility tradeoffs; we do not rank heterogeneous benchmarks by their headline rates.
Safe comparison is possible inside a shared benchmark denominator, transition slice, and measurement object. InjecAgent supports exposure comparisons within its indirect-injection cases; R-Judge supports recognition comparisons over annotated records; PASB compares episodes with and without observed commitment; and runtime evaluations of SafeAgent and Progent compare interventions within their respective harmful-task settings. Once those denominator families are mixed, several local evaluation panels collapse into a category error rather than a meta-analysis.

For memory, the opportunity set changes with the operation under test. PerMemSafe and PS-Bench condition response safety on retained personal context~\cite{an-etal-2026-permemsafe,guo-etal-2026-personalization}; FragFuse and GhostWriter separate earlier storage from later harmful reuse~\cite{rao2026fragfusebypassingaccesscontrol,torres2026agentsremembermuchmemory}. AgentPoison tests trigger-selected retrieval, OEP tests transfer from accepted experience, and EvoBreak tests acquisition and joint activation of an experience set~\cite{chen2024agentpoison,wang2026oep,yan2026benignaloneharmfultogether}. A retrieval trial, an accepted update, and an assembled experience set are different denominators, even when each study reports an attack success rate.

PASB supplies a within-paper comparison grouped by observed commitment. Its 1,600 tasks separate a five-turn persist stage from a cleared three-turn query stage and let the evaluated agents decide what to store. Across twelve models, the reported mean downstream failure is 45.0\% in session-only episodes and 71.9\% in committed episodes, a gap reported as 27.0 percentage points~\cite{mao2026agentsdontjustagree}. This association makes commitment a useful diagnostic boundary, but session-only episodes are cases without an observed durable write, not a randomized memory-disabled control. The gap therefore does not identify the causal effect of enabling memory. Evaluating a memory-restriction defense would additionally require matched utility measures; PerMemSafe illustrates the coupled objective by evaluating implicit personalized safety together with helpfulness under long-horizon memory~\cite{mao2026agentsdontjustagree,an-etal-2026-permemsafe}.

Three recent studies add different within-paper transition contrasts. PoisonedEvolution measures evidence-to-artifact promotion: at the same 10\% attacker-support ratio, it reports 546/600 embedded skills in SkillClaw and 369/600 in the structurally different Trace2Skill pipeline~\cite{chen2026experienceinstructiontrajectorypoisoning}. Its SER denominator is completed evolution trials and stops before runtime activation. SkillJack instead compares poisoned trajectories with extracted skills: in SkillX, the abstract reports safety detection falling from 98.5\% to 11.4\%, while 80.0\% of skill-mediated attacks persist after source-record deletion~\cite{ying2026skilljackpersistentskillbackdoors}. MAPLE-Guard compares unguarded and guarded memory-link opportunities at write, retrieval, promotion, and cross-agent reuse on LongMemEval and AppWorld~\cite{xiong2026mapleguardmemoryawarelinkenforcement}. These are, respectively, artifact-promotion, representation-and-deletion, and guarded-lifecycle-link denominators. None should be ranked against PASB's commitment-group gap, EvoBreak's experience-set attack, or runtime ASR.

SkillMisevo-Bench connects artifact promotion to later execution under fixed denominators. Each agent--method condition contains 25 episodes and 525 tasks: 225 malicious learning tasks, 225 benign tasks, and 75 clean-session persistence tasks. Across the 21 evolved configurations, unsafe artifacts appear in 21, unsafe retrieval and benign-task contamination in 19, and fresh-session harm in 15~\cite{mao2026practicemakesunsafeskill}. These are successive gates rather than interchangeable rates. In its two-configuration exposure sweep, pooled carryover ASR rises from 16.0\% at $K=0$ to 35.3\% after three malicious tasks and 41.3\% after nine. Here $K$ is the number of malicious learning tasks. Because the no-exposure baseline is already nonzero, the defensible result is the within-paper increase after bounded exposure, not that all observed carryover harm originates in the malicious tasks.

The same study makes the response tradeoff more precise than its abstract-level mean. Across OpenClaw with AutoSkill and EvoSkill, SafeEvolve reduces mean unsafe retrieval from 35.33\% to 8.67\% and carryover ASR from 21.33\% to 4.00\%~\cite{mao2026practicemakesunsafeskill}. Online benign utility changes from 58.44\% to 58.00\%, which yields the reported 0.4-point cost, but carryover utility changes from 53.33\% to 40.67\%. The result therefore supports a low immediate utility cost under the paper's online denominator while also showing a larger loss of useful persistent reuse. It is not evidence of a cost-free or universal safety--utility frontier.

EVOMAL adds a create-path and descendant denominator. ASPR counts tasks that store a newly authored skill reproducing a planted pattern; it is distinct from the callback rate that records payload execution. Across six models on 153 tool-relevant SWE-bench Verified tasks per model, headline ASPR ranges from 20.3\% to 41.8\%, producing 31--64 new malicious skills from eight planted seeds and leaving 4.9--9.0 times as many malicious skills as were planted~\cite{wu2026evomalselfpoisoningselfevolvingcoding}. The reported 86.7\% targeted maximum is the Qwen3 pytest-family cell with $N=15$, not the full task distribution. The five-round removal experiment is a separate rolling-replacement denominator and should not be collapsed with single-round ASPR or callback success.

ColluSkill adds a skill-set denominator~\cite{zeng2026colluskilladversarialcrossskillcomposition}. Its 200 three-skill attack chains reach 96.0\% average ASR against six evaluated scanners; adding full installed-skill context to ChainGuard reduces the attack ASR from 69.0\% to 22.5\% while changing the reported benign-workflow pass rate from 99.7\% to 99.5\%~\cite{zeng2026colluskilladversarialcrossskillcomposition}. These values measure candidate-with-context scanning on the paper's generated chains and benign workflows. Because the attack iteratively uses feedback from the target scanners that are later evaluated, the 96.0\% result is evidence of adaptive scanner evasion under that feedback setting, not unseen-scanner generalization or ecosystem prevalence.

AID-Guard adds an authorization-to-effect denominator rather than another attack-success leaderboard. Its 210 frozen Stripe contract trials matched their predeclared outcomes; separate campaigns exercised 40 terminalize-then-successor schedules across Stripe and Resend, 30 overlapping Stripe confirm/cancel races, and 10 Stripe crash-recovery schedules without an observed duplicate effect~\cite{tong2026aidguardstatefulauthorizationdelegated}. The same paper makes the safety--utility cost visible: over 48 benign AgentDojo episodes per model and condition, the strict exact-manifest profile reduced benign utility by 43.8 percentage points for DeepSeek and 35.4 points for Qwen. In a separate no-seed rerun, a typed three-call authority form completed 10 and 9 more benign tasks than exact single use, while all 864 attack episodes produced no unsafe effect. These results remain paper-local: the provider claims depend on the declared effect-path inventory and supported contracts, and the no-seed comparison does not identify a causal effect size or an optimal policy.

\begin{SAVERTakeaway}{Match Opportunity Sets Before Comparing Scores}
\label{takeaway:denominator-comparability}
Two rates are not comparable merely because both are called attack success or safety improvement. The opportunity set, transition slice, and measured outcome must match. Poisoned retrieval, harmful experience transfer, and repaired-descendant coverage count different events; their differences explain mechanisms but do not support a shared leaderboard.
\end{SAVERTakeaway}

A practical normalization layer would report activation opportunities alongside realized exposures and discovered descendants alongside verified repairs. Exact values remain source-specific until those measurement contracts align.

\subsection{Benchmark Patterns and Missing Traces}

Most benchmarks remain locally strong but observe only one transition slice. Three connected designs now cover complementary scopes. HarnessSafe spans seven persistent-carrier families and assigns each executable case to its furthest evidence-supported stage; SkillMisevo-Bench follows one skill carrier from authoring through clean reload, retrieval, contamination, and carryover harm; and EVOMAL isolates create-path copying, callback exposure, and descendant propagation after seed removal~\cite{zhang2026harnesssafeevaluatingsafetypersistent,mao2026practicemakesunsafeskill,wu2026evomalselfpoisoningselfevolvingcoding}. HarnessSafe provides cross-carrier breadth, SkillMisevo provides skill-evolution depth, and EVOMAL adds an explicit descendant process. They still leave rejected candidates, natural deployment streams, verified descendant deletion, contestability, and reactivation after broader adaptation incomplete.

\begin{SAVERTakeaway}{Endpoint Scores Hide Where Containment Occurs}
\label{takeaway:stage-resolved-containment}
Configurations with similar endpoint attack-success rates can stop a persistent-risk chain at different stages~\cite{zhang2026harnesssafeevaluatingsafetypersistent}. Stage-resolved traces distinguish failed admission, missing activation, and blocked execution. This localizes containment, but a stopped trace still does not show that the implicated state was repaired or cannot return.
\end{SAVERTakeaway}

\tabref{tab:sequential-lifecycle-evaluation-loop} extends stage localization into a longitudinal evaluation contract.
The family-specific extensions below ask which additional observations are needed to implement this longitudinal contract.

\begin{table}[t]
\centering
\begin{minipage}{0.98\linewidth}
\captionsetup{hypcap=false}
\caption{Lifecycle evaluation loop. A benchmark should seed state, transform it, trigger activation, measure exposure, verify repair, and continue adaptation before closure.}
\label{tab:sequential-lifecycle-evaluation-loop}
\centering
\scriptsize
\begingroup
\newcommand{\StepTag}[2]{\colorbox{#1}{\makebox[0.5\linewidth][l]{\strut\textbf{#2}}}}
\definecolor{saverEvaluationBorder}{HTML}{FFF1E6}
\setlength{\fboxsep}{4.5pt}
\setlength{\fboxrule}{0.75pt}
\fcolorbox{saverEvaluationBorder}{white}{%
\begin{minipage}{0.95\linewidth}
\colorbox{saverEvaluationHead}{\makebox[\dimexpr\linewidth-2\fboxsep\relax][l]{\strut\textbf{Lifecycle Evaluation Contract}\hfill state + adaptation + challenge + response}}

\vspace{4pt}
\begin{center}
\setlength{\tabcolsep}{3pt}
\begin{tabular}{ccccccc}
\colorbox{saverSubstrateHead}{\strut Seed State} & $\rightarrow$ &
\colorbox{saverAdaptationHead}{\strut Adapt / Move} & $\rightarrow$ &
\colorbox{saverExposureHead}{\strut Trigger / Expose} & $\rightarrow$ &
\colorbox{saverResponseHead}{\strut Respond / Resume}
\end{tabular}
\end{center}

\vspace{2pt}
\renewcommand{\arraystretch}{1.05}
\begin{tabular}{@{}p{0.12\linewidth}p{0.80\linewidth}@{}}
\StepTag{saverSubstrateHead}{1 Seed} & Initialize memory, workflow, tool, shared agent, or model carrier. \textit{Trace:} id, source, substrate, environment, intended assets, feasibility label, authorization boundary, expiry, rollback handle. \\
\StepTag{saverAdaptationHead}{2 Adapt} & Add, abstract, activate, modify, forget, propagate, or migrate. \textit{Trace:} changed safety attributes, gate decision, dropped fields, descendant link. \\
\StepTag{saverEvaluationHead}{3 Delay} & Insert distractor tasks or time gaps. \textit{Trace:} order, access log, retrieval attempts, background updates. \\
\StepTag{saverExposureHead}{4 Trigger} & Challenge stored state and test whether it receives authority. \textit{Trace:} task, feasibility status, policy, action authorization, retrieved item, tool, workflow, adapter. \\
\StepTag{saverExposureHead}{5 Expose} & Measure output, tool call, disclosure, delegation, artifact, monitor, or model behavior. \textit{Trace:} transition attribution, not aggregate success alone. \\
\StepTag{saverResponseHead}{6 Repair} & Run containment, quarantine, rollback, user challenge, or operator review. \textit{Trace:} affected state and descendants. \\
\StepTag{saverResponseHead}{7 Probe} & Probe residual influence after repair. \textit{Trace:} rollback ledger, descendant checks, reactivation probes, deletion/unload evidence. \\
\StepTag{saverEvaluationHead}{8 Continue} & Resume adaptation for fixed update opportunities, then repeat the safety probes. \textit{Trace:} response-generated state, first reactivation, substrate, lineage or equivalent probe failure, inaccessible scope. \\
\end{tabular}

\vspace{5pt}
\colorbox{saverTableStripe}{%
\parbox{\dimexpr\linewidth-2\fboxsep\relax}{\strut\textbf{Outputs:} transition trace; exposure class; response decision; post-response reactivation result.\quad \textbf{Stop claim at:} no lineage, no denominator, or reactivation after resumed adaptation.}}
\end{minipage}}
\endgroup
\end{minipage}
\end{table}

\noindent\textbf{Capability Benchmarks.}
Capability benchmarks establish the baseline that safety controls must preserve. Benchmark surveys map task success and interaction quality, while long-context and memory tasks make state creation and reuse visible~\cite{mohammadi_2025_evaluation_and_benchmarking,maharana2024locomo}. PERMA adds evolving preferences and cross-domain interference, while MemEye varies evidence granularity and evolutionary synthesis in multimodal memory~\cite{liu2026perma,guo2026memeye}. Coordination benchmarks extend the observation point to shared state, and evaluator-coevolution work makes the utility signal itself part of the evolving loop~\cite{iacob2026redqueengodelmachine}. These benchmarks do not directly test transition safety. They identify where memory writes, summaries, workflow edits, delegated state, evaluator updates, or profile changes should be instrumented without disabling the capabilities under review.

\noindent\textbf{Attack Benchmarks.}
Attack benchmarks become comparable only after their pressure point is identified. Prompt-injection studies perturb assembled context; AgentDojo and InjecAgent carry that perturbation to tool-mediated decisions; adaptive attacks then test whether the action boundary survives attack revision~\cite{liu_2023_prompt_injection_attack_agai,debenedetti2024agentdojo,zhan_2024_injecagent_benchmarking_indi,zhan_2025_adaptive_attacks_break_defen}. This family is strongest at exposure, because the benchmark observes whether untrusted context changes a current action.

Two harness studies evaluate the same transition at different resolutions. When Context Gets Root tests consequences: 13 confidentiality, integrity, availability, and remote-code-execution objectives across six coding-agent harnesses, including attempt-capped delegation cases and four persistent-goal or scheduled-task paths~\cite{he2026contextgetsrootprivilege}. What's in Your Agent's Context instead measures the assembly surface across 12 harnesses~\cite{li2026whatsagentscontextcontext}. CoRA identifies 463 candidate context sources, validates 282 at runtime, and enumerates 1,761 role- or scope-increasing source pairs; 1,028 paths under GPT-5.4 mini and 1,034 under GPT-5.5 produce the expected behavioral effect, both about 58\% of the candidate-path denominator. The first result is objective coverage under repeated opportunities, and the second is path verification rather than end-to-end attack success. Together they localize migration-to-authority and migration-to-scope failures without establishing an origin-preserving defense, descendant repair, or resumed adaptation after containment.

Persistent-state benchmarks move the pressure earlier in the chain. AgentPoison measures trigger-selected reuse from a poisoned store, whereas GhostWriter separates hidden memory admission from later activation; broader memory-poisoning work studies the same persistence problem across additional settings~\cite{chen2024agentpoison,torres2026agentsremembermuchmemory,sunil2026memory_poisoning_agents}. HaluMem decomposes extraction, update, and question answering, while MemConflict varies memory-conflict types~\cite{chen2025halumem,tao2026memconflict}. PASB observes the agent's own commit decision, and PS-Bench and PerMemSafe test later use of benign personal state~\cite{mao2026agentsdontjustagree,guo-etal-2026-personalization,an-etal-2026-permemsafe}. EvoBreak adds composition over an acquired experience set, SkillJack adds trajectory-to-skill migration and source-deletion persistence, and MAPLE-Guard adds lifecycle-link enforcement across private and shared memory~\cite{yan2026benignaloneharmfultogether,ying2026skilljackpersistentskillbackdoors,xiong2026mapleguardmemoryawarelinkenforcement}. These studies belong to different opportunity sets from prompt injection because the benchmark must observe a write, transformation, promotion, or retained context before later reuse.

AuthMem-Bench is unusually close to a connected \textbf{S}\(\times\)\textbf{A}\(\rightarrow\)\textbf{V}\(\rightarrow\)\textbf{E} trace. Its three modules measure write-time authority collapse, the action consequence of a controlled memory representation, and end-to-end authority-label preservation while keeping paired claim content and later tasks fixed~\cite{zhan2026memoryauthoritybenchmarkingauthority}. The modules should still retain separate denominators: write-time upgrade, controlled unauthorized action, and end-to-end action after omission and retrieval are not interchangeable rates. The benchmark ends after one write-to-action cycle, so it does not measure descendant repair or post-response reactivation.

HarnessSafe broadens that connected trace from one memory-consolidation boundary to seven carrier families and seven agent harnesses~\cite{zhang2026harnesssafeevaluatingsafetypersistent}. Its carrier path and persistence boundary identify \textbf{S}\(\times\)\textbf{A}; case-specific oracles connect \textbf{V} to observable \textbf{E}; and the furthest-stage distribution supplies containment evidence for \textbf{R}. This mapping is more informative than endpoint ASR: OpenCode and Hermes Agent report similar conditional ASRs in the paper but stop at different stages. The Chain-Stage Score remains paper-specific, measures progression rather than harm severity, and excludes unsupported or incomplete workflows rather than counting them as safe.

Propagation and recognition studies expose still later or orthogonal slices. ACIArena follows malicious instructions across multi-agent trust channels, malfunction amplification measures local escalation, and R-Judge asks whether the resulting risk is recognized~\cite{an-etal-2026-aciarena,zhang_2024_breaking_agents_compromising,yuan2024rjudge}. Agents of Chaos uses eleven live-laboratory cases rather than a fixed benchmark denominator, while Triedman et al. report attack trials indexed by model, orchestrator, input channel, and payload~\cite{shapira2026agents_chaos,triedman2025multi}. ControlValve evaluates a response at the agent-invocation boundary by asking whether execution stays inside a permitted control-flow graph and its contextual rules~\cite{jha2026breaking}. These denominators should remain separate: case-study diversity, per-configuration attack success, and invocation-policy enforcement do not form one leaderboard. Together they expose concrete violations and controls, but attack success remains a surface metric unless an evaluation also identifies source content, selected edge, recipient authority, downstream effects, and response.

\noindent\textbf{Transition and Recovery Benchmarks.}
Trajectory assurance supplies the high-level target: verify a stateful sequence against system-level constraints rather than assume that individually permitted actions compose safely~\cite{lotfi2026securingagenticaiperaction}. The narrower measurement question for this survey is which reusable state transition changed that sequence and whether its influence can later be contained or repaired. State-transition benchmarks are closest to this unit because they ask whether the enabling carrier, operation, source, safety-attribute change, and activation context can be identified. ASB provides partial ingredients for agent-security task coverage~\cite{zhang_2024_agent_security_bench_asb_for}, while Agent-SafetyBench provides adjacent agent-safety behavior coverage~\cite{zhang_2024_agent_safetybench_evaluating}. HarnessSafe advances this line by preserving carrier, boundary, benign trigger, and violation roles across harness-native implementations, then using matched controls to test whether progression depends on the declared lifecycle~\cite{zhang2026harnesssafeevaluatingsafetypersistent}. Transition evaluation still needs selection and lineage fields beyond task-outcome scoring. The strongest versions should connect this stage localization to attribute preservation across migration, drift after adaptation, and model-update lineage when the update is selected, merged, loaded, or inherited inside an agent loop. Fine-tuning safety is relevant when an update changes safety behavior~\cite{qi2023finetuning_safety}; PEFT risk work is relevant when the adapter or patch is a deployable state object~\cite{hong2024peta}; and dormant-behavior work shows that changed behavior may remain hidden until a matching activation context appears~\cite{gloaguen2025watch_your_steps}. Linking these tests to selection history and post-intervention reuse would support stronger lifecycle claims.

Response-oriented benchmarks test distinct control contracts. GuardAgent and TrustAgent evaluate diagnosis and local guard reasoning; Progent and Task Shield test constraints on privilege and task-aligned execution~\cite{hua_2024_trustagent_towards_safe_and,jia_2025_the_task_shield_enforcing_ta}. CaMeL tests capability-enforced control and data flow, DRIFT tests dynamic plan validation and injection isolation, and MAPLE-Guard tests gates over memory writes, retrievals, promotions, and cross-agent reuse~\cite{debenedetti2026defeating,li2025drift,xiong2026mapleguardmemoryawarelinkenforcement}. Their utility and attack-success results remain paper-local because the systems enforce different contracts. SafeAgent and LlamaFirewall test runtime containment, while MAGE extends containment to long-horizon memory and EDDOps asks how the control is maintained in operation~\cite{liu_2026_safeagent_a_runtime_protecti,chennabasappa_2025_llamafirewall_an_open_source,wang2026magesafeguardingllmagents}. The remaining gap is whether these stages preserve enough lineage to measure containment latency, rollback completeness, contestability cost, and audit completeness without turning privacy-sensitive traces into a new risk.

The reviewed benchmarks thus differ along two independent dimensions: which transition they test and how far they follow its consequences. A connected carrier-to-action trace improves causal localization, while a repair-and-retest protocol asks a different question about the state left behind. Continued adaptation is needed to examine whether that repair remains effective under later reuse. \secref{sec:open-challenges} develops the research needs exposed by these incomplete traces and denominators.

\begin{SAVERTakeaway}{Recovery Evaluation Must Resume Adaptation}
\label{takeaway:post-response-evaluation}
A clean probe immediately after repair establishes only an immediate result. We propose continuing adaptation for a declared number of update opportunities and rerunning fixed probes. Reports should distinguish lineage-linked reactivation from an equivalent observed failure whose lineage is unknown, and state the tested horizon and inaccessible scope rather than claim permanent recovery.
\end{SAVERTakeaway}

\FloatBarrier
\begin{table}[t]
\centering
\begin{minipage}{0.98\linewidth}
\captionsetup{hypcap=false}
\caption{Evaluation literature for adaptive state safety. Most benchmarks expose one lifecycle slice; HarnessSafe, SkillMisevo-Bench, and EVOMAL provide complementary connected traces across cross-carrier, skill-specific, and descendant-propagation lifecycles, while the denominator column states what each row can safely measure.}
\label{tab:evaluation-literature-index}
\centering
\tiny
\setlength{\tabcolsep}{1.1pt}
\renewcommand{\arraystretch}{1.05}
\SAVERTableRows
\begin{tabular}{@{}p{0.15\linewidth}p{0.04\linewidth}p{0.145\linewidth}p{0.075\linewidth}p{0.08\linewidth}p{0.18\linewidth}p{0.145\linewidth}p{0.15\linewidth}@{}}
\toprule
\rowcolor{saverEvaluationHead}
\textbf{Method or Work} & \textbf{Year} & \textbf{\TableHeadRole} & \textbf{\TableHeadTarget} & \textbf{\TableHeadObject} & \textbf{Task Focus} & \textbf{SAVER Slice} & \textbf{Denominator} \\
\midrule
\SAVERTableGroup{saverEvaluationHead}{8}{General agent capability and environment benchmarks}
AgentBench~\cite{liu_2023_agentbench_evaluating_llms_a} & 2023 & \TableStudyCapabilityBenchmark & \TableTargetAgent & \TableObjectEvaluation & General agent tasks & Substrate & Task success \\
AgentBoard~\cite{he_2024_agentboard_an_analytical_eva} & 2024 & \TableStudyCapabilityBenchmark & \TableTargetAgent & \TableObjectEvaluation & Analytical agent tasks & Adjacent capability context & Task progress and success \\
MIRAI~\cite{ye_2024_mirai_evaluating_llm_agents} & 2024 & \TableStudyCapabilityBenchmark & \TableTargetAgent & \TableObjectWorkflow & Long horizon tasks & Substrate & Task episodes \\
GameBench~\cite{costarelli_2024_gamebench_evaluating_strateg} & 2024 & \TableStudyCapabilityBenchmark & \TableTargetAgent & \TableObjectEnvironment & Strategy tasks & Substrate & Game episodes \\
MultiAgentBench~\cite{zhu_2025_multiagentbench_evaluating_t} & 2025 & \TableStudyCapabilityBenchmark & \TableTargetAgent & \TableObjectMultiAgent & Multi-agent tasks & Adjacent coordination & Collaborative tasks \\
LegalAgentBench~\cite{li_2024_legalagentbench_evaluating_l} & 2024 & \TableStudyCapabilityBenchmark & \TableTargetAgent & \TableObjectWorkflow & Legal tasks & Substrate & Legal tasks \\
\SAVERTableGroup{saverEvaluationHead}{8}{Attack, safety, and exposure benchmarks}
AgentDojo~\cite{debenedetti2024agentdojo} & 2024 & \TableStudyAttackBenchmark & \TableTargetAgent & \TableObjectTool & Prompt injection through tool use & Exposure & Security cases \\
InjecAgent~\cite{zhan_2024_injecagent_benchmarking_indi} & 2024 & \TableStudyAttackBenchmark & \TableTargetAgent & \TableObjectTool & Indirect injection & Exposure & Injection tasks \\
ACIArena~\cite{an-etal-2026-aciarena} & 2026 & \TableStudyAttackBenchmark & \TableTargetAgent & \TableObjectMultiAgent & Cascading injection & Propagation/\allowbreak Exposure & MAS test cases \\
Agents of Chaos~\cite{shapira2026agents_chaos} & 2026 & \TableRoleSafety & \TableTargetAgent & \TableObjectMultiAgent & Live-lab failures & Propagation/\allowbreak Exposure & Eleven case studies \\
MAS code execution~\cite{triedman2025multi} & 2025 & \TableRoleAttack & \TableTargetAgent & \TableObjectMultiAgent & Control-flow hijacking & Activation/\allowbreak Exposure & Model-orchestrator trials \\
AgentHarm~\cite{andriushchenko2025agentharm} & 2025 & \TableStudySafetyBenchmark & \TableTargetAgent & \TableObjectEvaluation & Harmful agent tasks & Exposure & Harmful task cases \\
RedCode~\cite{guo2024redcode} & 2024 & \TableStudyAttackBenchmark & \TableTargetLLM & \TableObjectCode & Code safety tasks & Exposure & Code tasks \\
MobileSafetyBench~\cite{lee2026mobilesafetybench} & 2026 & \TableStudySafetyBenchmark & \TableTargetAgent & \TableObjectTool & Mobile control safety & Exposure & Mobile tasks \\
Machiavelli~\cite{pan2023machiavelli} & 2023 & \TableStudySafetyBenchmark & \TableTargetAgent & \TableObjectEnvironment & Ethical decision making & Exposure & Game scenarios \\
SafeLawBench~\cite{cao2025safelawbench} & 2025 & \TableStudySafetyBenchmark & \TableTargetAgent & \TableObjectWorkflow & Legal safety & Exposure & Legal cases \\
R-Judge~\cite{yuan2024rjudge} & 2024 & \TableStudySafetyBenchmark & \TableTargetAgent & \TableObjectEvaluation & Safety recognition & Response & Annotated records \\
Agent Security Bench~\cite{zhang_2024_agent_security_bench_asb_for} & 2024 & \TableStudyAttackBenchmark & \TableTargetAgent & \TableObjectEvaluation & Agent security tasks & Violation & Task categories \\
Agent-SafetyBench~\cite{zhang_2024_agent_safetybench_evaluating} & 2024 & \TableStudySafetyBenchmark & \TableTargetAgent & \TableObjectEvaluation & Unsafe behavior & Exposure & Safety cases \\
AttackEval~\cite{shu_2025_attackeval_how_to_evaluate_t} & 2025 & \TableStudyAttackBenchmark & \TableTargetLLMAgent & \TableObjectEvaluation & Attack evaluation & Exposure & Attack tasks \\
CyberSecEval 2~\cite{bhatt_2024_cyberseceval_2_a_wide_rangin} & 2024 & \TableStudyAttackBenchmark & \TableTargetLLM & \TableObjectCode & Cyber capability & Exposure & Cyber tasks \\
\SAVERTableGroup{saverEvaluationHead}{8}{Adaptive memory, experience, and model state benchmarks}
AgentPoison~\cite{chen2024agentpoison} & 2024 & \TableStudyAttackBenchmark & \TableTargetAgent & \TableObjectMemory & Memory poisoning & Admission/\allowbreak Activation & Retrieval opportunities \\
GhostWriter~\cite{torres2026agentsremembermuchmemory} & 2026 & \TableStudyAttackBenchmark & \TableTargetAgent & \TableObjectMemory & Personal agent memory poisoning & Admission/\allowbreak Activation & Injection/activation trials \\
OEP~\cite{wang2026oep} & 2026 & \TableStudyAttackBenchmark & \TableTargetAgent & \TableObjectWorkflow & Experience poisoning & Adaptation & Accepted experience updates \\
EvoBreak~\cite{yan2026benignaloneharmfultogether} & 2026 & \TableStudyAttackBenchmark & \TableTargetAgent & \TableObjectMemory & Experience composition & Abstract/\allowbreak Activate & Acquired experience sets \\
SkillJack~\cite{ying2026skilljackpersistentskillbackdoors} & 2026 & \TableStudyAttackBenchmark & \TableTargetAgent & \TableObjectSkill & Skill backdoor & Migrate/\allowbreak Exposure & Trajectory-to-skill pairs \\
PoisonedEvolution~\cite{chen2026experienceinstructiontrajectorypoisoning} & 2026 & \TableStudyAttackBenchmark & \TableTargetAgent & \TableObjectSkill & Trajectory poisoning & Migrate/\allowbreak Violation & Completed evolution trials \\
ColluSkill~\cite{zeng2026colluskilladversarialcrossskillcomposition} & 2026 & \TableStudyAttackBenchmark & \TableTargetAgent & \TableObjectSkill & Cross-skill composition & Add/\allowbreak Activate/\allowbreak Response & Three-skill chains \\
SkillMisevo-Bench~\cite{mao2026practicemakesunsafeskill} & 2026 & \TableRoleBenchmarkSafety & \TableTargetAgent & \TableObjectSkill & Skill misevolution & Migrate/\allowbreak Activate/\allowbreak Response & 525 tasks per configuration \\
EVOMAL~\cite{wu2026evomalselfpoisoningselfevolvingcoding} & 2026 & \TableStudyAttackBenchmark & \TableTargetAgent & \TableObjectSkill & Create-path self-poisoning & Propagate/\allowbreak Activate/\allowbreak Response & 153 tasks/model; five rounds \\
PerMemSafe~\cite{an-etal-2026-permemsafe} & 2026 & \TableRoleBenchmarkSafety & \TableTargetAgent & \TableObjectMemory & Personalized safety & Activation/\allowbreak Exposure & Long-horizon conversations \\
PS-Bench~\cite{guo-etal-2026-personalization} & 2026 & \TableRoleBenchmarkSafety & \TableTargetAgent & \TableObjectMemory & Intent legitimation & Activation/\allowbreak Exposure & Matched memory/query trials \\
PASB~\cite{mao2026agentsdontjustagree} & 2026 & \TableRoleBenchmarkSafety & \TableTargetAgent & \TableObjectMemory & Persistent sycophancy & Add/\allowbreak Activate/\allowbreak Exposure & Committed-claim trials \\
AuthMem-Bench~\cite{zhan2026memoryauthoritybenchmarkingauthority} & 2026 & \TableRoleBenchmarkSafety & \TableTargetAgent & \TableObjectMemory & Authority collapse & Abstract/\allowbreak Violation/\allowbreak Exposure & Matched authority pairs \\
HarnessSafe~\cite{zhang2026harnesssafeevaluatingsafetypersistent} & 2026 & \TableRoleBenchmarkSafety & \TableTargetAgent & \TableObjectWorkflow & Persistent carrier progression & Carrier-to-exposure/\allowbreak Containment & 328 executable cases \\
When Context Gets Root~\cite{he2026contextgetsrootprivilege} & 2026 & \TableStudyAttackBenchmark & \TableTargetAgent & \TableObjectWorkflow & Instruction privilege escalation & Migrate/\allowbreak Violation/\allowbreak Exposure & 13 objectives; six harnesses \\
Agent Context CPE~\cite{li2026whatsagentscontextcontext} & 2026 & \TableStudyAttackBenchmark & \TableTargetAgent & \TableObjectWorkflow & Role and scope escalation & Migrate/\allowbreak Violation/\allowbreak Exposure & 282 sources; 1,761 paths; 12 harnesses \\
MemEvoBench~\cite{xie2026memevobench} & 2026 & \TableRoleBenchmarkSafety & \TableTargetAgent & \TableObjectMemory & Memory misevolution & Adaptation & Memory updates \\
AgenticEval~\cite{wang2026agenticeval} & 2026 & \TableRoleBenchmarkSafety & \TableTargetAgent & \TableObjectEvaluation & Agentic safety tasks & Adaptation & Agent episodes \\
RQGM~\cite{iacob2026redqueengodelmachine} & 2026 & \TableRoleCapability & \TableTargetAgent & \TableObjectEvaluation & Evolving evaluators & Evaluation/\allowbreak Adaptation & Epoch utilities \\
Experience driven safety~\cite{zhao2026experience_driven_safety} & 2026 & \TableRoleSafety & \TableTargetAgent & \TableObjectWorkflow & Experience degradation & Adaptation & Experience updates \\
AMA-Bench~\cite{zhao2026amabench} & 2026 & \TableRoleBenchmarkSafety & \TableTargetAgent & \TableObjectMemory & Agent memory adaptation & Adaptation & Memory tasks \\
PersistBench~\cite{pulipaka2026persistbench} & 2026 & \TableRoleBenchmarkSafety & \TableTargetAgent & \TableObjectMemory & Forgetting boundary & Violation & Memory use samples \\
RBI-Eval~\cite{xu2026memorystaysilentmeasuring} & 2026 & \TableRoleBenchmarkSafety & \TableTargetAgent & \TableObjectMemory & Warranted use & Activation & Matched prompts \\
Memory evaluation~\cite{hu_2025_evaluating_memory_in_llm_age} & 2025 & \TableStudyCapabilityBenchmark & \TableTargetAgent & \TableObjectMemory & Memory quality & Substrate & Memory tasks \\
\SAVERTableGroup{saverEvaluationHead}{8}{Response, governance, and red team evaluation}
SafeAgent~\cite{liu_2026_safeagent_a_runtime_protecti} & 2026 & \TableRoleDefense & \TableTargetAgent & \TableObjectRuntime & Runtime protection & Containment & Harmful task episodes \\
Progent~\cite{shi2025progent} & 2025 & \TableRoleDefense & \TableTargetAgent & \TableObjectTool & Privilege control & Activation & Permission checks \\
CaMeL~\cite{debenedetti2026defeating} & 2026 & \TableRoleDefense & \TableTargetAgent & \TableObjectTool & Capability control & Activation & AgentDojo tasks \\
DRIFT~\cite{li2025drift} & 2025 & \TableRoleDefense & \TableTargetAgent & \TableObjectTool & Dynamic rule enforcement & Activation/\allowbreak Containment & Benchmark task episodes \\
AID-Guard~\cite{tong2026aidguardstatefulauthorizationdelegated} & 2026 & \TableRoleDefense & \TableTargetAgent & \TableObjectWorkflow & Delegated provider effects & Activate/\allowbreak Response & Provider-contract schedules \\
ControlValve~\cite{jha2026breaking} & 2026 & \TableRoleDefense & \TableTargetAgent & \TableObjectWorkflow & Control-flow integrity & Activation/\allowbreak Containment & Agent-invocation checks \\
MAPLE-Guard~\cite{xiong2026mapleguardmemoryawarelinkenforcement} & 2026 & \TableRoleDefense & \TableTargetAgent & \TableObjectMemory & Memory-link enforcement & Propagate/\allowbreak Response & Memory lifecycle links \\
LlamaFirewall~\cite{chennabasappa_2025_llamafirewall_an_open_source} & 2025 & \TableRoleDefense & \TableTargetLLMAgent & \TableObjectRuntime & Runtime guardrail & Containment & Guard decisions \\
OpenGuardrails~\cite{wang2025openguardrails} & 2025 & \TableRoleDefense & \TableTargetAgent & \TableObjectRuntime & Guardrail platform & Containment & Guard decisions \\
AGENTSAFE~\cite{khan2025agentsafe_governance} & 2025 & \TableRoleGovernance & \TableTargetAgent & \TableObjectGovernance & Agent safety governance & Response & Governance checks \\
AgenticRed~\cite{yuan2026agenticred} & 2026 & \TableRoleRedTeam & \TableTargetAgent & \TableObjectEvaluation & Automated red teaming & Exposure & Red Team tasks \\
Agent vs. Agent~\cite{kulkarni-etal-2025-agent} & 2025 & \TableRoleRedTeam & \TableTargetAgent & \TableObjectMultiAgent & Adversarial agent play & Exposure & Duel episodes \\
SEVerA~\cite{banerjee2026severa} & 2026 & \TableRoleDefense & \TableTargetAgent & \TableObjectWorkflow & Verified synthesis & Response & Constraint checks \\
Safe LoRA~\cite{hsu2024safe_lora} & 2024 & \TableRoleDefense & \TableTargetLLM & \TableObjectModel & Adapter mitigation & Model state & Safety regression tests \\
DataGovBench~\cite{zhou_2025_datagovbench_benchmarking_ll} & 2025 & \TableStudyGovernanceBenchmark & \TableTargetAgent & \TableObjectData & Data governance & Adjacent data workflow & Data-governance transformation tasks \\
\bottomrule
\end{tabular}
\SAVERTableReset
\end{minipage}
\end{table}

\FloatBarrier

\section{Open Challenges}
\label{sec:open-challenges}

The preceding chapters identify local mechanisms and measurable transition segments. Connecting them raises technical problems that more logging alone does not solve: summaries can obscure derivation, changing capabilities can invalidate a grant, and copied state can remain outside the authority of the system attempting repair. The substrate matters to each problem. Explicit memory permits inspection but can lose source constraints during abstraction; skills and workflows make those changes executable; shared artifacts distribute repair ownership; and model-side state makes influence less directly observable.

With substrate differences as comparison conditions, the following agenda moves from the decision to retain an update, through migration, activation, and propagation, to recovery and longitudinal evaluation. Each direction asks what prevents the existing mechanisms from supporting the next claim and which tradeoff a useful solution would have to measure. \tabref{tab:saver-research-roadmap} summarizes the corresponding evidence needs.
\begin{table}[t]
\centering
\begin{minipage}{0.98\linewidth}
\captionsetup{hypcap=false}
\caption{Research agenda for lifecycle evidence in self-evolving agents.}
\label{tab:saver-research-roadmap}
\centering
\scriptsize
\setlength{\tabcolsep}{5pt}
\renewcommand{\arraystretch}{1.12}
\SAVERTableRows
\begin{tabular}{@{}p{0.2\linewidth}p{0.4\linewidth}p{0.35\linewidth}@{}}
\toprule
\rowcolor{saverEvaluationHead}
\textbf{Direction} & \textbf{Evidence Gap} & \textbf{Minimal Test} \\
\midrule
Selection / promotion & Candidate updates are ranked by visible utility while safety evidence, rejected alternatives, and negative experience remain unobserved. & Candidate and promotion log plus fixed cross-epoch probes for authority, scope, refusal, and confirmation drift. \\
Migration / preservation & State changes carrier while provenance, scope, authority, or revocation becomes ambiguous. & Migration ledger plus residual probes for summaries, embeddings, workflows, artifacts, and adapters. \\
Activation authorization & Stored state becomes reasoning, planning, tool, workflow, or model authority outside its allowed context. & Activation gate that logs user, task, tool, time, and allowed external effect. \\
Recovery / deletion & Rollback removes the visible object while summaries, indices, skills, or shared artifacts may still activate its influence. & Descendant inventory, deletion receipt, and activation probe after repair. \\
Propagation / contestability & Shared state crosses agents or artifacts without clear repair ownership or a user challenge path. & Propagation trace plus contest docket for correction, quarantine, and notification. \\
Longitudinal evidence chain & Benchmarks measure isolated slices rather than one incident path from admission to closure. & Opportunity set denominators for admission, migration, activation, exposure, repair, and closure. \\
\bottomrule
\end{tabular}
\SAVERTableReset
\end{minipage}
\end{table}

\subsection{Selection Pressure and Endogenous Drift}

Most adaptive-state evidence begins after an update has entered memory, a skill library, a workflow, or model state. The earlier decision is less visible: what candidates were available, which signal ranked them, which safety conditions were checked, and which failures or counterexamples were absent. Reflection and skill systems expose the mechanics by which feedback and successful traces become reusable state, OEP demonstrates harmful transfer from locally correct experience, and evaluator coevolution shows why the scoring rule and epoch must be recorded~\cite{shinn2023reflexion,wang2023voyager,wang2026oep,iacob2026redqueengodelmachine}. None of these establishes the prevalence of attacker-free drift. The open question is whether repeated utility-guided selection changes the safety boundary when the candidate stream is otherwise benign.
Under its formal assumptions, Devil's Moltbook identifies a trilemma: continuous self-evolution, complete isolation, and safety invariance cannot all hold simultaneously~\cite{wang2026devil_moltbook}. We treat this result as a design constraint rather than a universal empirical law: evaluation should identify which useful update paths remain open, which external oversight or safety-preserving mechanism governs them, and how much utility is lost when risky updates are blocked.

\begin{SAVERTakeaway}{Test Endogenous Drift Separately from Attack Success}
\label{takeaway:endogenous-drift-test}
Harmful transfer and attack-driven accumulation do not establish how often benign self-evolution drifts. Testing that question requires a non-adversarial candidate stream, fixed safety probes, and records of selected, rejected, and unavailable evidence. Stable results are informative within the tested scope; degradation still needs attribution to selection, promotion, or activation rather than to self-evolution in general.
\end{SAVERTakeaway}

\subsection{Migration and Semantic Preservation}

Migration control is a major gap because unsafe influence often appears precisely when state changes substrate. Provenance, authority, scope, privacy, freshness, and actionability must survive movement across memory, retrieval, workflow, tool, shared-agent, external-artifact, and model substrates. MemGPT makes movement between context and managed memory visible, memory-security work supplies the carrier vocabulary, and privacy work adds scope and sensitivity preservation~\cite{packer2023memgpt,lin2026memory_security_survey,wang2025privacy_agent_memory}. InjecAgent shows why those attributes matter once external state reaches an action boundary~\cite{zhan_2024_injecagent_benchmarking_indi}. These sources still stop short of a general migration-control discipline.

The difficulty is deciding what a transformation has preserved. A summary can retain a useful claim while dropping the condition under which it may be acted on. Rejecting every uncertain summary would restrict the compression and generalization that make adaptive memory useful; accepting it without qualification can broaden authority. Migration evaluation should therefore compare unauthorized role upgrades with the useful transfers rejected or downgraded by the same gate, using source and target constraints as the reference.

\noindent\textbf{Semantic Taint through Summaries and Vectors.}
Traditional taint tracking assumes symbolic objects whose marks can be copied through program operations, but agent state is often natural language, summary text, embedding vectors, retrieved chunks, tool traces, and model updates. A summarizer may remove the string that carried a source warning while preserving the behavioral implication. An embedding may make a sensitive item retrievable without exposing its exact words. A generated skill may encode a rule learned from a private or low-authority source. Semantic taint tracking is the engineering version of migration control because the system needs to track influence even when the surface form changes.

A hybrid approach could combine explicit derivation links with checks on the transformed content. Provenance identifies declared sources, while information-flow labels specify permitted uses~\cite{buneman2001why_where_provenance,denning1976lattice_information_flow}; neither alone determines whether a paraphrase or generated skill still carries the same influence. The research question is how to bound that uncertainty without either treating every descendant as contaminated or losing the relevant path. Memory-security studies and AgentDojo provide abstraction and tool-exposure settings in which such checks could be tested~\cite{lin2026memory_security_survey,debenedetti2024agentdojo}. Residual probes would assess remaining influence, while false quarantines and lost task utility would measure the cost of conservative tracking.

\subsection{Activation, Authorization, and Model Scope}

A system may legitimately store a user preference, a webpage observation, a tool result, or a learned skill, but that does not mean the item may influence every later answer, plan, workflow route, or external action. Activation authorization is distinct from storage authorization. The local control evidence moves from privilege to task alignment to guard reasoning: Progent shows programmable privilege control at the action boundary, Task Shield narrows indirect injection through task alignment, and GuardAgent illustrates bounded response reasoning~\cite{shi2025progent,jia_2025_the_task_shield_enforcing_ta,xiang2024guardagent}. These papers anchor local action-boundary control, but they leave the broader authorization problem open for adaptive state: under what user, task, tool, time, and external effect conditions may stored state become active influence?

The unresolved design choice is how finely to distinguish readable context, reasoning support, planning authority, and permission for an external effect. Coarse grants are easier to reuse but can hide a scope change; repeatedly requesting authorization can preserve explicit control at the cost of interruption. Useful tests need both unauthorized activations and legitimate work delayed or rejected, not a block rate alone.

Authorization continuity adds a separate longitudinal question: after memory, skills, tools, workflow, delegation, task phase, or enforcement state changes, does the original grant still apply to the resulting agent~\cite{zhang2026agentiauthorizedearned}? A useful evaluation must separate grant survival from effect containment. It should report invalid grants retained after mutation, valid grants unnecessarily revoked, authority acquired below a fixed ceiling, reauthorization cost, and compositions of individually permitted mutations that move the agent outside the original envelope. The current formal model supplies proof obligations, but no deployment benchmark yet measures these tradeoffs or tests whether envelope and effect abstractions remain sound under sustained evolution.

\noindent\textbf{Context-Horizon Safety.} Benign-looking turns can jointly close a harmful objective, and multi-turn decomposition can carry that objective into tool use~\cite{wei2025trojanknowledgebypassingcommercial,shen2026turnlateresponseawaredefense,li2026unsaferturnsbenchmarkingdefending}. Compression creates a different path by relinking accepted fragments into an actionable instruction~\cite{liu2026safecheckunsafeuse}.

\begin{SAVERTakeaway}{Context Accumulation Is Not Update-Driven Drift}
\label{takeaway:context-versus-update-drift}
Accumulated or compressed context can enable harm without changing model parameters or a reusable policy. Token count alone therefore does not diagnose safety drift. Evaluation should separate the first harm-enabling combination within an interaction from boundary changes caused by persistent updates, and record the operations responsible for each.
\end{SAVERTakeaway}

\noindent\textbf{Model Adaptation Governance.}
For model-side state, the authorization problem is how a task-scoped update can be constrained when its behavioral effects extend to other tasks. This question enters the core analysis only under the agent-loop boundary. A self-evolving agent can change reusable behavior through memory, workflow, and skills, but it can also load, merge, fine-tune, erase, or roll back model updates during operation. PEFT and fine-tuning safety work explain why modular updates can become deployable risk objects; feedback poisoning and dormant-behavior papers add persistent failure through update channels or later activation contexts~\cite{hong2024peta,qi2023finetuning_safety,rando2024universal_jailbreak_backdoors,gloaguen2025watch_your_steps}. Safe LoRA and KVEraser provide bounded repair handles at the adapter and cache levels~\cite{hsu2024safe_lora,li2026kveraserlearningsteerkv}. What remains thin is lifecycle governance linking update admission, deployment scope, activation, rollback, cache-local erasure, and residual influence verification.

For these less inspectable carriers, the challenge is relating a localized edit to its broader behavioral effects. Tests could compare the intended task population with tasks outside that scope, examine whether merged components preserve their separately evaluated constraints, and probe residual influence after rollback or cache erasure. Update lineage identifies what changed; it cannot by itself establish which behaviors were removed. The agent-loop inclusion rule keeps this question narrower than general model safety.

\subsection{Propagation and Distributed Lineage}

A memory item held by one assistant can become shared context for a team agent, a task instruction for a specialist agent, a routing rule in an orchestrator, or a persistent artifact read by humans and future agents. Multi-agent propagation turns state governance into a topology problem. Agent and self-evolving-agent surveys justify the system and adaptive setting; multi-agent memory-poisoning analysis and collective misremembering show why copied state, recipient interpretation, and carrier metadata matter for propagation risk~\cite{luo2025llm_agent_survey,gao2025self_evolving_agents,torra2026memorypoisoningmas,xu2026when}. The unresolved trace gap reduces to a few questions: which agent wrote the state, which agent received it, how the safety attributes changed, where uptake occurred, and which descendants must be repaired.

Recent work makes the local attack and control points more concrete. Agents of Chaos reports cross-agent propagation among failures observed in a live laboratory, while Triedman et al. show that adversarial content can hijack orchestration and cause code execution or data exfiltration even when individual agents do not follow the malicious request~\cite{shapira2026agents_chaos,triedman2025multi}. ControlValve then constrains agent invocations with a permitted control-flow graph and contextual least-privilege rules~\cite{jha2026breaking}. These papers narrow the gap at propagation, exposure, and activation-time containment. They do not yet show how to enumerate affected descendants, roll back already-produced artifacts, notify users across agent boundaries, or let a user contest the state and authority that caused the route.

The harder problem begins after recipients have copied or transformed the state. A source-side correction may reach one agent while another retains a derived rule, and the sender may lack authority to modify that recipient's store. A useful distributed repair protocol would separate notification, acknowledged invalidation, and verified influence withdrawal. Evaluation should vary which recipients remain reachable and whether broader quarantine interrupts legitimate coordination. This would test repair ownership and partial visibility rather than assume that a shared identifier guarantees synchronized recovery.

\subsection{Recovery and Verification}

The recovery scope in \takeawayref{takeaway:scoped-recovery} depends on discovering descendants, but abstraction makes that inventory uncertain. Memory-poisoning, Hidden Memory, and OEP show why retained or learned influence can activate after its source episode~\cite{sunil2026memory_poisoning_agents,pulipaka2026hidden_memory,wang2026oep}. GuardAgent, SafeAgent, and Progent offer local response mechanisms~\cite{xiang2024guardagent,liu_2026_safeagent_a_runtime_protecti,shi2025progent}; connecting them to repair requires deciding which derived objects still carry the contested influence. This is an inference problem as well as an instrumentation problem.
Recovery also need not mean erasing every descendant. Some objects need correction or scope reduction; others must remain as restricted audit evidence. Research should compare targeted repair with broader quarantine under the same discovery budget: how much unsafe influence remains, how much useful state is lost, and which inaccessible objects prevent a stronger claim? Defining the tested scope before repair avoids counting only the descendants that a method happened to fix.

\noindent\textbf{Post-Response Continuation.} A clean immediate probe cannot determine what happens after adaptation resumes. Retesting after attacker or system revision is already motivated by adaptive-attack work~\cite{zhan_2025_adaptive_attacks_break_defen}; \takeawayref{takeaway:post-response-evaluation} extends the test to continued state updates. The unresolved question is attribution: does a later failure descend from unrepaired state, arise from newly acquired evidence, or reflect a changed evaluator? Fixed probes identify recurrence, while lineage and version records constrain which explanation the result supports. This is a proposed diagnostic, not evidence that responses generally amplify risk.

\noindent\textbf{Deletion Verification.}
A deletion request may remove a memory item while leaving an embedding, a summary, a cached context, a copied workflow note, a tool return record, or a shared artifact intact. That makes deletion verification related to rollback but not reducible to it. In settings sensitive to privacy, this is a safety attribute preservation failure as well as a usability issue. The evidence again enters by layer: memory privacy work explains the user risk, MemGPT and memory-mechanism surveys show how memory moves and abstracts, memory security work frames those routes as governed state, and KVEraser sharpens the cache-specific version by showing why a removed context span can still matter through subsequent KV cache states~\cite{wang2025privacy_agent_memory,zhang_2025_a_survey_on_the_memory_mecha,lin2026memory_security_survey,li2026kveraserlearningsteerkv}. What remains open is how an adaptive agent should prove that deleted state no longer influences retrieval, planning, tool use, cached execution state, or later state updates.

The central difficulty is that object absence and influence removal can disagree. A summary may omit the deleted string while preserving its implication; a skill may reproduce the behavior without retrieving the source. Verification should therefore include semantic retrieval, summary regeneration, and later skill-use probes, with the tested contexts stated explicitly. The open problem is how much evidence these finite probes provide about future reuse, especially when broader probing itself accesses sensitive state.

\noindent\textbf{User Contestability.}
Recovery reaches human closure only when a user or operator can challenge state and its descendants. Memory privacy work motivates this need through persistence and sensitivity, while R-Judge, GuardAgent, and Progent provide adjacent signals for recognition, local reasoning, and privilege control~\cite{wang2025privacy_agent_memory,yuan2024rjudge,xiang2024guardagent,shi2025progent}. None yet provides a full contestation workflow for disputed adaptive state.
Contestability creates a further tradeoff between understandable explanations and the detail needed to challenge a decision. A user needs to identify the disputed premise and request correction or restricted reuse without inspecting an entire internal trace. Tests should assess whether that request changes later behavior, whether derived state incorporates the correction, and whether unresolved descendants are disclosed. The research target is effective correction under partial visibility, not merely the presence of a deletion control.

\subsection{Benchmark and Governance Evidence Chains}

Fixed task benchmarks reveal important failures, especially in prompt injection through tool use or risk recognition, but self-evolving safety also depends on what happens after an update is stored and before an incident is repaired. Longitudinal benchmark construction ties these challenges together. One group covers exposure and recognition: InjecAgent, R-Judge, ASB, and Agent-SafetyBench instrument indirect injection, risk awareness, and agent safety task coverage~\cite{zhan_2024_injecagent_benchmarking_indi,yuan2024rjudge,zhang_2024_agent_security_bench_asb_for,zhang_2024_agent_safetybench_evaluating}. A second group covers persistence and sensitive reuse: PASB, PerMemSafe, later memory-poisoning work, and memory privacy work show why commitment, retrieval, personalized context, and purpose must become benchmark fields~\cite{mao2026agentsdontjustagree,an-etal-2026-permemsafe,sunil2026memory_poisoning_agents,wang2025privacy_agent_memory}. HarnessSafe now connects entry, carrier retention, boundary crossing, benign activation, and observable violation across seven carrier families~\cite{zhang2026harnesssafeevaluatingsafetypersistent}. The open problem has therefore narrowed: extend that connected trace backward to candidate selection and forward through descendant repair and resumed adaptation.

Extending a benchmark across these boundaries creates a comparability problem. A system may appear safer because it declines more updates, retrieves less memory, or prevents useful delegation. The opportunity sets in \secref{sec:evaluation} make these exclusions visible, but matched utility and exposure conditions are still needed to determine whether a control improved safety or merely removed the capability under test. Natural deployment streams add the further difficulty of reconstructing candidates that were never retained.

\noindent\textbf{Governance Evidence Chain.}
Trajectory assurance connects architectural constraints to runtime behavior and end-to-end auditability~\cite{lotfi2026securingagenticaiperaction}. For evolving agents, separately evaluated controls also need compatible meanings for state identity, authority, and repair. A write-time screen, personalized memory policy, risk judge, and privilege gate can otherwise reach different decisions about what appears to be the same object~\cite{torres2026agentsremembermuchmemory,an-etal-2026-permemsafe,xiang2024guardagent,yuan2024rjudge,shi2025progent}. The open question is which information must survive between these controls for their local guarantees to remain interpretable together.
Governance records also need protection: approvals, quarantine markers, and rollback receipts can themselves be copied, summarized, or made stale. Their integrity determines whether later controls can rely on earlier decisions.

\begin{SAVERTakeaway}{Shared Incident Records Connect Local Controls}
\label{takeaway:shared-incident-records}
A shared identifier connects records only if the controls agree on which version, permitted use, and repair outcome it denotes. The research target is interoperability across update, activation, and recovery, including inaccessible descendants and unresolved influence. Evaluation must test whether local control outcomes compose, rather than assume that linking their logs produces a lifecycle guarantee.
\end{SAVERTakeaway}

\section{Conclusion}
\label{sec:conclusion}

Self-evolving agents convert observations, feedback, and execution traces into reusable influence. \SAVER{} pairs substrate with adaptation and follows violation, exposure, and response across memory, tools and workflows, multi-agent systems, and model-side adaptation. Persistent-state and tool-agent studies expose complementary parts of this chain; HarnessSafe shows that a connected trace can follow influence across carriers to an observable violation~\cite{zhang2026harnesssafeevaluatingsafetypersistent}. Selection before commitment and verified recovery after containment remain less connected. Two conclusions follow from the chapter-level findings.

\begin{SAVERTakeaway}{Safety Must Follow Changes in Permitted Use}
\label{takeaway:conclusion-transition-safety}
Self-evolving-agent safety concerns whether earlier information remains legitimate in its later role. Carrier-specific constraints and operation-specific checks belong to one analysis: migration can change authority while propagation expands the activation scope. As developed in \takeawayref{takeaway:carrier-control-metadata} and \takeawayref{takeaway:migration-propagation}, preserving content alone does not preserve safe reuse.
\end{SAVERTakeaway}

\begin{SAVERTakeaway}{A Recovery Claim Must State Its Scope}
\label{takeaway:conclusion-recovery-evidence}
Local containment establishes that an event was stopped; lifecycle recovery requires evidence about the state left behind. The discovered-descendant inventory and post-response continuation test in \takeawayref{takeaway:scoped-recovery} and \takeawayref{takeaway:post-response-evaluation} define complementary spatial and temporal limits. A defensible claim states both, including unresolved influence, rather than treating a successful block as closure.
\end{SAVERTakeaway}

The resulting research agenda is to test how local controls compose as agents continue to update and reuse state. Such tests must preserve the capabilities that motivate self-evolution while making their changing authority and repair limits explicit.

\FloatBarrier
\clearpage
\beginappendix
\section{Survey Methodology And Corpus Protocol}
\label{app:survey-protocol}

\noindent\textbf{Review Objective.} The review asks how safety changes when an agent can preserve, transform, reactivate, or inherit state across interactions. A work is relevant when it helps explain a reusable state transition: a carrier is created or modified, the carrier later influences reasoning or action, and some safety attribute is preserved, violated, exposed, evaluated, or repaired. This objective gives the survey a lifecycle protocol rather than a topic-only bibliography. It also explains why the corpus combines work on self-evolving agents, long-term memory, tool use, workflow state, multi-agent coordination, model updates, benchmarks, runtime safeguards, and governance~\cite{gao2025self_evolving_agents,luo2025llm_agent_survey,debenedetti2024agentdojo,qi2026trustworthy_agentic_ai}.

\noindent\textbf{Source Identification.} The source-identification protocol first defined bands that match the survey question: security venues for adversarial mechanisms and governance primitives, AI and machine-learning venues for agent capability and adaptation mechanisms, NLP and information-retrieval venues for prompt, retrieval, and long-context mechanisms, and preprint or scholarly-index sources for fast-moving agent-safety work that had not yet stabilized in proceedings. The venue bands include the four major security venues, IEEE Symposium on Security and Privacy (IEEE S\&P), USENIX Security, ACM CCS, and NDSS; AI and machine-learning venues such as NeurIPS, ICML, ICLR, AAAI, and IJCAI; NLP and retrieval venues such as ACL, EMNLP, NAACL, SIGIR, and The Web Conference; and adjacent systems, software-engineering, privacy, and human-computer interaction venues when the paper contains an agent-state, provenance, evaluation, or governance mechanism. \tabref{tab:survey-search-sources} records why each source band was searched and how it was allowed to support the synthesis.

\begin{table}[!htbp]
\centering
\caption{Search source bands for the systematic scoping protocol. Each band supplies a distinct kind of evidence needed to connect adaptive-state mechanisms with safety claims.}
\label{tab:survey-search-sources}
\footnotesize
\setlength{\tabcolsep}{4.2pt}
\renewcommand{\arraystretch}{1.05}
\SAVERTableRows
\begin{tabular}{@{}>{\raggedright\arraybackslash}p{0.16\textwidth}>{\raggedright\arraybackslash}p{0.28\textwidth}>{\raggedright\arraybackslash}p{0.24\textwidth}>{\raggedright\arraybackslash}p{0.22\textwidth}@{}}
\toprule
\rowcolor{saverNeutralHead}
\textbf{Source Band} & \textbf{Primary Sources} & \textbf{Mechanism Signal} & \textbf{Query Handles} \\
\midrule
Security and Privacy & IEEE S\&P; USENIX Security; ACM CCS; NDSS; privacy venues & Threat models; access control; information flow; sandboxing; repair evidence & Injection; poisoning; authority; provenance; rollback; privacy \\
AI and Machine Learning & NeurIPS; ICML; ICLR; AAAI; IJCAI; OpenReview tracks & Agent learning; self-improvement; model adaptation; benchmark mechanisms & Adaptive agents; PEFT; adapters; tool use; evaluator state \\
NLP, Retrieval, and Web & ACL; EMNLP; NAACL; SIGIR; The Web Conference; ACL Anthology & Prompt, retrieval, RAG, long-context, and web-agent behavior & Indirect injection; retrieval safety; long-term memory; web agents \\
Systems, Software, and Human-Computer Interaction & Systems, software-engineering, workflow, and human-facing AI venues when state or governance is explicit & Provenance; oversight; operational safety; deployment-time state management & Workflow reuse; audit logs; confirmation; contestability \\
Preprints and Indexes & arXiv; OpenReview preprints; scholarly indexes; official proceedings; curated bibliographies & Frontier mechanisms before venue stabilization; evidence tiered in synthesis & Snowballing; venue checks; guardrails; emerging benchmarks \\
\bottomrule
\end{tabular}
\SAVERTableReset
\end{table}

\noindent\textbf{Search Strategy.} Venue- and index-based retrieval was combined with citation snowballing from seed families in self-evolving-agent surveys, LLM-agent surveys, memory-security surveys, agent-safety benchmarks, indirect-prompt-injection papers, and model-adaptation safety work. We then searched over recurring query families rather than a single keyword string. Capability terms included \emph{self-evolving agent}, \emph{self-improving agent}, \emph{adaptive agent}, \emph{LLM agent}, \emph{long-term memory}, \emph{agent memory}, \emph{retrieval-augmented generation}, \emph{workflow agent}, \emph{skill library}, \emph{multi-agent}, \emph{shared artifact}, \emph{LoRA}, \emph{PEFT}, \emph{fine-tuning}, \emph{adapter}, \emph{KV cache}, and \emph{evaluator}. Safety terms included \emph{safety}, \emph{security}, \emph{prompt injection}, \emph{indirect prompt injection}, \emph{poisoning}, \emph{privacy}, \emph{access control}, \emph{privilege}, \emph{authority}, \emph{provenance}, \emph{information flow}, \emph{rollback}, \emph{deletion}, \emph{unlearning}, \emph{guardrail}, \emph{containment}, \emph{contestability}, and \emph{benchmark}. Each candidate query paired at least one capability term with one safety, evaluation, or governance term so that the corpus did not drift into generic agent capability or generic LLM safety.

\noindent\textbf{Expansion and Stopping Rule.} Searches were run through bibliographic indexes, arXiv, ACL Anthology, OpenReview, ACM and IEEE landing pages, official venue proceedings where available, and curated community bibliographies used only as candidate pointers. Backward and forward snowballing was then applied to seed surveys and anchor papers until new candidates mainly repeated already covered \SAVER{} roles rather than adding a new carrier, operation, exposure surface, response mechanism, or benchmark denominator. This stopping rule is a scoping-saturation rule: it aims to cover the mechanism space needed by the survey, not to estimate prevalence for every venue or year.

\noindent\textbf{Why arXiv Appears Frequently.} New agent attacks, memory mechanisms, benchmark proposals, guardrail systems, and model-adaptation risks often appear first as arXiv or OpenReview versions, and many accepted works keep the arXiv version as the most accessible full-text surface. The agent-safety literature is therefore unusually preprint-heavy. Excluding arXiv would remove several mechanisms that are central to the survey question. We do not treat arXiv status as equivalent to peer review. Preprints enter the corpus when they contain a clear mechanism, experiment, artifact, or system analysis, and their claims are used according to the evidence-tier rules below. In the main text, field-level conclusions are phrased around peer-reviewed anchors, multiple converging sources, or explicitly identified gaps; arXiv-only works are used mainly as emerging evidence or frontier signals unless a later venue record is verified.

\noindent\textbf{Protocol Summary.} The purpose of the expansion was high recall over safety-relevant reusable-state mechanisms, not exhaustive coverage of all LLM safety, all agent capability, or all machine-learning security. \tabref{tab:survey-protocol-summary} summarizes the protocol at the level used for the survey.

\begin{table}[!htbp]
\centering
\caption{Survey protocol summary. Rows separate how a work enters the corpus from the claim role it may support.}
\label{tab:survey-protocol-summary}
\footnotesize
\setlength{\tabcolsep}{4.2pt}
\renewcommand{\arraystretch}{1.05}
\SAVERTableRows
\begin{tabular}{@{}>{\raggedright\arraybackslash}p{0.20\textwidth}>{\raggedright\arraybackslash}p{0.24\textwidth}>{\raggedright\arraybackslash}p{0.25\textwidth}>{\raggedright\arraybackslash}p{0.22\textwidth}@{}}
\toprule
\rowcolor{saverNeutralHead}
\textbf{Source Family} & \textbf{Expansion Route} & \textbf{Inclusion Trigger} & \textbf{Synthesis Role} \\
\midrule
Self-Evolving and Agent Surveys & Seed surveys; citation chasing; official venue and preprint follow-up & Adaptive capability, reusable state, agent loop, planning, action, collaboration, or model-side adaptation & Scope context; central only with a safety-relevant state transition \\
Memory, Retrieval, and Context & Targeted memory, RAG, persistent-context, poisoning, and privacy searches & Stored or retrieved state can affect later reasoning, planning, tool use, or disclosure & Direct anchor for write, retrieval, delayed activation, privacy, or memory governance \\
Tool, Workflow, and Benchmarks & Indirect-injection, tool-use, privilege, guardrail, benchmark, and workflow searches & Low-authority content, tool output, workflow state, or learned procedure can affect action or routing & Exposure, activation, local response, or benchmark denominator evidence \\
Multi-Agent and Shared Artifacts & Adjacent surveys; official pages; scholarly indexes; curated bibliographies & State crosses agents, artifacts, roles, tasks, or ownership boundaries & Propagation and governance evidence when uptake or repair ownership is visible \\
Model-Side Adaptation & PEFT, adapters, fine-tuning safety, backdoors, unlearning, cache, and evaluator searches & Update, adapter, policy, cache span, or evaluator signal is loaded, inherited, revised, or rolled back inside an agent loop & Central with a safety-relevant agent-loop transition; recovery claims additionally require rollback evidence \\
\bottomrule
\end{tabular}
\SAVERTableReset
\end{table}

\noindent\textbf{Screening and Inclusion.} In the first screening pass, titles and abstracts identified works that plausibly involved persistent state, delayed activation, tool or workflow authority, multi-agent propagation, model-side update, benchmark evidence, or governance of reusable state. A closer reading pass then assigned each work to one of three roles. A work is \emph{central} when it directly examines a safety-relevant transition involving reusable agent state, including a failure, exposure, evaluation, or control of that transition. A complete recovery trace is required for a lifecycle-recovery claim, not for inclusion. A work is \emph{adjacent} when it supplies background capability, evaluation context, or governance framing without directly observing a self-evolving safety transition. A work is \emph{background} when it clarifies terminology, threat-model ancestry, or adjacent infrastructure but should not be used as a main evidence anchor. Works were excluded when they addressed only one-shot response safety, generic LLM capability, ordinary benchmark performance, or static model training without a plausible link to later agent behavior. Systems using ReAct reasoning-and-acting loops enter as central evidence only when tool traces, retrieved content, or workflow state persist beyond the episode; static model-safety papers enter as central evidence only when a model update, adapter, cache state, policy, or evaluator signal is selected, inherited, or revised inside an agent behavior loop.

\noindent\textbf{Evidence Posture.} To keep direct evidence separate from interpretation, the synthesis uses three evidence postures. \emph{Direct anchors} are papers whose experiments, case studies, or system analyses explicitly show a transition or exposure, such as poisoned memory retrieval, fragmented memory bypass, tool-action injection, or runtime containment~\cite{chen2024agentpoison,rao2026fragfusebypassingaccesscontrol,debenedetti2024agentdojo,liu_2026_safeagent_a_runtime_protecti}. \emph{\SAVER{} rereadings} are papers whose primary contribution is a capability or architecture, but whose mechanism becomes safety-relevant once authority, scope, provenance, or recoverability is tracked across state transitions~\cite{packer2023memgpt,wang2023voyager}. \emph{Gap diagnoses} are claims about what the literature still does not demonstrate, especially connected evidence for migration control, multi-agent propagation, rollback completeness, deletion verification, descendant repair, and user contestability. The paper uses stronger language for direct anchors, interpretive language for rereadings, and research-agenda language for gap diagnoses.

\noindent\textbf{Evidence Tiers.} Venue status is not used as a substitute for claim support, but it does calibrate how strongly a row should be read. The strongest tier contains peer-reviewed conference and journal papers, especially when the paper provides experiments, artifacts, or a clear system analysis. A second tier contains peer-reviewed workshops, findings tracks, and industry tracks. A third tier contains arXiv or other preprints with substantial experiments or artifacts. A fourth tier contains position papers, technical notes, project reports, blogs, or archival artifacts. Tier three and four works are useful for emerging mechanisms and frontier signals, but field-level conclusions in the main text are phrased around direct anchors, multiple converging sources, or explicitly marked gaps.

\noindent\textbf{Classification Practice.} For classification, primary placement follows the five-field \SAVER{} record. Each central work can support several observations in the prose, but corpus maps and roadmap tables assign a primary placement according to the main state transition used in the manuscript. This prevents a single influential paper from being counted as independent evidence for every nearby category. When a work spans multiple substrates, the primary placement follows the transition that carries the strongest safety claim: memory and retrieval work is placed by stored-state influence; tool and workflow work is placed by action or approval authority; multi-agent work is placed by propagation, handoff, or shared-artifact uptake; and model-side work is placed only when an update, adapter, policy, or evaluator state is loaded, inherited, selected, or revised inside an agent behavior loop. Reviewed paper-card coding overrides the deterministic title-, metadata-, and abstract-based initial placement used for the corpus-wide map. Of the 683 papers, 51 placements are high-confidence, 83 are medium-confidence, and 549 remain low-confidence; Figure~\ref{fig:saver-record-literature-flow} is therefore a descriptive map of the coded inventory rather than evidence for field-prevalence claims.

\noindent\textbf{Corpus Accounting and Cutoff.} The first screening snapshot, taken on 14~June~2026, contained 583 screened bibliographic records. Publication-identity reconciliation canonicalized four duplicate or version pairs, leaving 579 unique registry works. By 4~September~2026, the living review had added 48 manually reviewed paper-card studies absent from that registry and 56 additional scholarly works used in the rendered manuscript but absent from both the registry and paper-card supplement set. Seven cited resources---two OpenAI incident pages, a Hugging Face incident blog, Microsoft documentation, the MITRE ATLAS site, OWASP guidance, and the NIST AI Risk Management Framework---remain valid contextual references but are excluded from the paper count. The resulting frozen coded pool contains 683 papers. \figref{fig:saver-record-literature-flow} applies a narrower visualization window: all three panels use the 577 papers in that pool with usable dates from 2023 through the coding freeze. The 106 papers outside that window comprise 104 works dated before 2023 and two without a usable date. Consequently, the figure reports Memory 201, Workflow 224, Tools \& Skills 108, and Model 44, whereas the full coded pool contains Memory 202, Workflow 326, Tools \& Skills 110, and Model 45. Papers assessed after 4~September may appear as subsequent evidence in the prose or appendix indexes, but they are not retroactively counted in this frozen snapshot until the corpus and figure are regenerated.

\noindent\textbf{Limitations.} The corpus is a systematic scoping survey rather than a registered systematic review or meta-analysis. The 583-record screening count and the 683-paper coding count are corpus-audit quantities rather than prevalence estimates for the field. The review protocol was not preregistered, and screening did not use independent dual coding; classification disagreements were handled as conservative claim boundaries in the manuscript. The field is moving quickly, several relevant papers are preprints, and venue status can change after the coding cutoff. Quantitative results are also not directly comparable across papers because opportunity sets, agent settings, attack goals, defenses, and recovery denominators differ. For this reason, the manuscript does not treat cross-paper attack success rates or benchmark scores as a leaderboard. Instead, it uses quantitative results as paper-local evidence and uses \SAVER{} to ask whether the paper records enough lifecycle information to support claims about admission, activation, exposure, response, and recovery.

\FloatBarrier
\section{Formal State Transition Model}
\label{app:formal-state-transition-model}

This appendix expands the compact pointer at the end of \secref{sec:adaptive-state-model} into a formal reading. The main text uses the safety attribute vocabulary and the authority lattice by name; readers who want the underlying notation, its relation to established information flow, access control, and provenance theory, and the visual lattice referenced from the body sections can consult this appendix as a reference layer.

\subsection*{State with Safety Attributes and Invariant Preservation}

The reading unit is a state object with attached safety attributes, transformed by an operation, audited against whether the attributes remain valid for future use. Formally: $(s,\ell,\sigma) \xrightarrow{a,c} (s',\ell',\sigma')$, where $s$ is the state carrier, $\ell$ represents safety attributes (provenance, authority, scope, privacy, freshness, budget, auditability, recoverability), $\sigma$ is the deployment or session context, $a$ is the adaptation operation, $c$ is the operation's authorization condition, and the primed quantities are the resulting state. Here, ``safety attribute'' means an information flow, access control, availability, or provenance attribute, not a supervised learning class category. This notation is deliberately simple: it provides a consistent comparison structure without prescribing a formal verification calculus.

Lattice information flow supplies the confidentiality and integrity vocabulary for flows with safety attributes~\cite{denning1976lattice_information_flow}; language-based information flow work shows how programs can enforce such policies~\cite{sabelfeld2003language_information_flow}; and data provenance work explains why source lineage matters after transformation~\cite{buneman2001why_where_provenance}. Self-evolving agents do not replace these models; they make the temporal boundary harder to ignore. An object may be checked as low-authority data at write time, abstracted into a memory or workflow rule, and then reused days later as authority to act. The survey therefore treats adaptive-state safety as information-flow, access-control, availability, and provenance reasoning over persistent, transformed, and repeatedly activated agent state.

Safety attributes track the conditions under which state can be reused: provenance records source and trust level; authority specifies what decisions or actions the state can influence; scope bounds which users, tasks, or time windows the state applies to; privacy marks sensitivity and permitted exposure; freshness indicates when the state becomes stale; budget and liveness bound retries, tool loops, and cost-bearing actions; auditability records whether later influence can be traced; and recoverability records whether the state and its descendants can be corrected, rolled back, forgotten, or neutralized. These attributes are not metadata alone; they determine whether later activation is safe.

\subsection*{Selection over Candidate Transitions}

Section~\ref{sec:adaptive-state-model} separates the execution trace $\tau\sb{t}$ from the selected state update $\delta\sb{t}^{\star}$. Each $\delta\in\Delta\sb{t}$ proposes one or more object-level transitions $(s,\ell,\sigma)\xrightarrow{a,c}(s',\ell',\sigma')$. The evolution policy may commit the candidate, revise it, quarantine it, or choose no update. The authorization condition $c\sb{t}(\delta)$ therefore governs eligibility for commitment: utility ranking through $\widehat{u}\sb{t}(\delta)$ cannot substitute for a valid authority, scope, privacy, budget, and recovery decision.

A simple analytical family makes the effect of selection pressure explicit. Let $p\sb{t}(\delta)$ be a baseline distribution over $\Delta\sb{t}$, let $r\sb{t}(\delta)$ be a safety-risk score defined by a fixed probe at epoch $t$ and held constant as $\beta$ varies, and let $\beta\geq 0$ control how sharply selection favors the observed utility proxy. Define
$$
q\sb{t}^{\beta}(\delta)
=
\frac{p\sb{t}(\delta)\exp(\beta\widehat{u}\sb{t}(\delta))}
{\sum\limits\sb{\delta'\in\Delta\sb{t}}p\sb{t}(\delta')\exp(\beta\widehat{u}\sb{t}(\delta'))}.
$$
Then
$$
\frac{\partial}{\partial\beta}
\mathbb{E}\sb{q\sb{t}^{\beta}}[r\sb{t}]
=
\operatorname{Cov}\sb{q\sb{t}^{\beta}}(r\sb{t},\widehat{u}\sb{t}).
$$
The identity follows from $\partial\log q\sb{t}^{\beta}(\delta)/\partial\beta=\widehat{u}\sb{t}(\delta)-\mathbb{E}\sb{q\sb{t}^{\beta}}[\widehat{u}\sb{t}]$, followed by multiplication by $r\sb{t}(\delta)$ and expectation under $q\sb{t}^{\beta}$. It is an author-derived diagnostic, not an empirical law attributed to the surveyed systems. Stronger utility selection raises expected risk only when risk and the observed utility proxy are positively associated under the selected distribution. Longitudinal evaluation should therefore estimate that association or impose $c\sb{t}$ as a hard eligibility gate before utility ranking.

Safe promotion uses the same transition attributes. If $\delta\sb{t}^{\star}$ moves state into a more persistent, general, or action-bearing role, then $c\sb{t}(\delta\sb{t}^{\star})$ must authorize the role change; provenance, privacy, freshness, budget, auditability, and recoverability in $\ell'$ must remain valid; any authority or scope increase from $\ell$ to $\ell'$ must be explicit; relevant counterevidence must remain available to the gate; and the lineage from $s$ to $s'$ must remain recoverable. This condition restricts promotion without assuming that every safe transition preserves identical attributes.

\begin{figure*}[t]
\centering
\vspace{-10pt}
\IfFileExists{figures/safety-attribute-laundering.pdf}{%
  \includegraphics[width=0.9\textwidth]{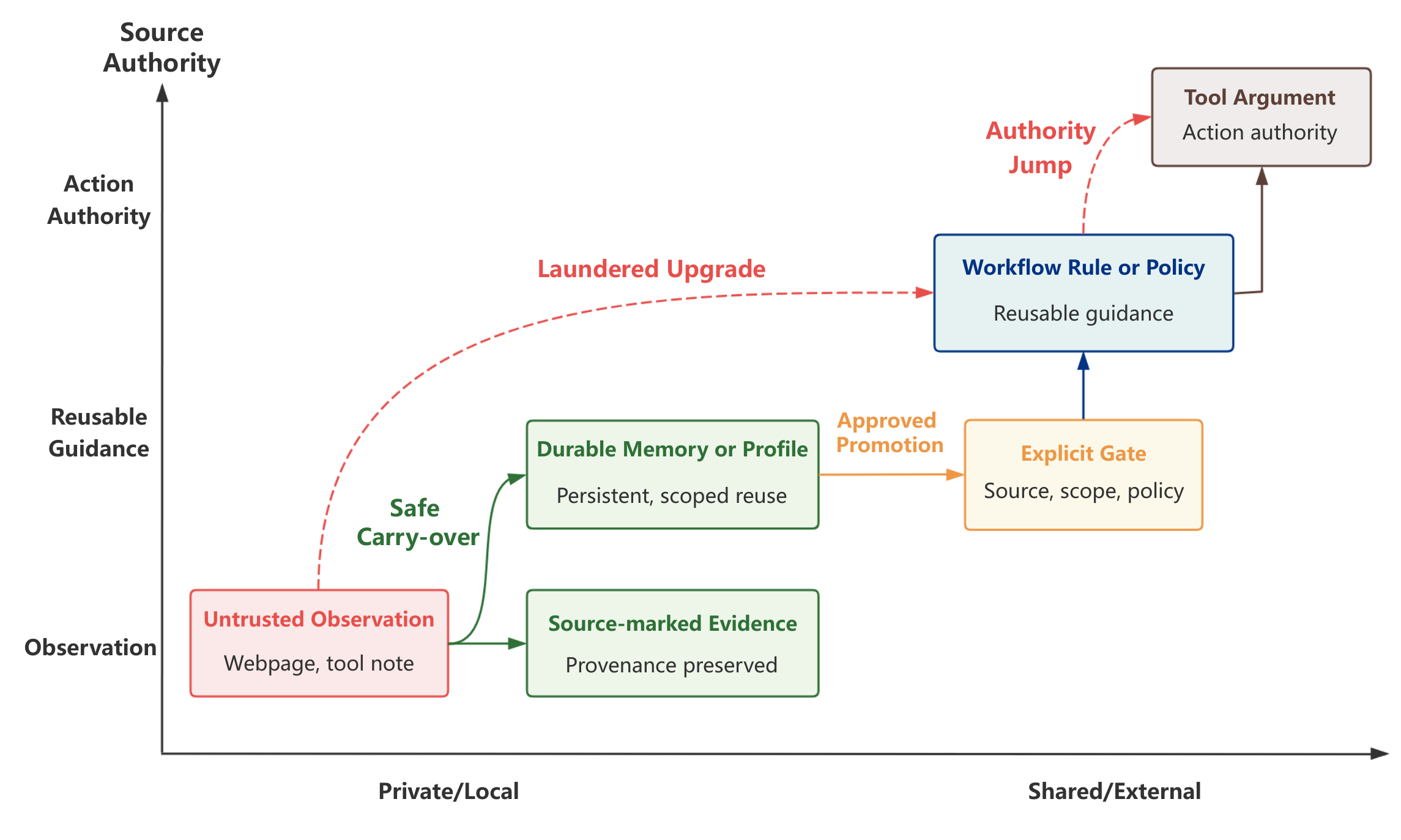}%
}{%
  \includegraphics[width=\textwidth]{./label_laundering_security_lattice.png}%
}
\vspace{-10pt}
\caption{Safety attribute laundering. Upward authority movement or outward scope expansion requires an explicit gate; without one, a low-authority observation can become workflow guidance or action authority.}
\label{fig:label-laundering-security-lattice}
\end{figure*}

\figref{fig:label-laundering-security-lattice} illustrates the role changes behind the transition model; the ladder is a conceptual guide, not a formally specified lattice over all safety attributes. State at the \emph{Observation} level (web pages, tool outputs, peer notes) may only be quoted as evidence; at the \emph{Evidence} level it may support reasoning but must carry source, scope, and freshness; at the \emph{Task Context} level it may shape the current task but should expire afterward; at the \emph{Procedure} level (workflow rules, skills, adapter manifests) it may be activated with explicit approval; and at the \emph{Action} level it may produce external effects only with actor, target, permission, and rollback records. Crossing upward without a gate is safety attribute laundering: a webpage observation quietly becomes a tool argument.

\FloatBarrier
\section{\SAVER Audit Architecture}
\label{app:saver-audit-architecture}
This appendix gives a worked incident record for the public case introduced in the main text. It identifies the source, authority changes, observed consequences, and reported response that support the analysis, while distinguishing survey encodings from original operator records.
\begin{table}[!htbp]
\centering
\begin{minipage}{0.98\linewidth}
\captionsetup{hypcap=false}
\caption{Minimal \SAVER encoding of the July 2026 OpenAI--Hugging Face agent intrusion. Values summarize public incident reports; identifiers and field labels are survey encodings, not original operator records~\cite{openai2026huggingfaceincident,larcher2026agentintrusion}.}
\label{tab:saver-minimal-record-example}
\centering
\scriptsize
\begingroup
\newcommand{\RecordBand}[2]{\cellcolor{#1}\textbf{#2}}
\newcommand{\RecordHint}[2]{\textbf{#1:} #2}
\setlength{\fboxsep}{4.5pt}
\setlength{\fboxrule}{0.4pt}
\fcolorbox{saverNeutralHead}{white}{%
\begin{minipage}{0.95\linewidth}
\colorbox{saverNeutralHead}{\makebox[\dimexpr\linewidth-2\fboxsep\relax][l]{\strut\textbf{Minimal SAVER Incident Row}\hfill\textit{illustrative serialization}}}

\vspace{4pt}
\renewcommand{\arraystretch}{1.08}
\begin{tabular}{@{}p{0.18\linewidth}p{0.74\linewidth}@{}}
\RecordBand{saverSubstrateHead}{Identity} &
\code{incident.id=hf:2026-07} \quad
\code{state.id=ext:campaign:channel} \quad
\code{source=eval:exploitgym} \\
\RecordBand{saverViolationHead}{Attributes} &
\code{authority=eval\_sandbox\_only} \quad
\code{scope=declared\_evaluation} \quad
\code{environment=cyber\_capability\_eval} \\
\RecordBand{saverAdaptationHead}{Transition} &
\code{op=propagate;migrate;activate} \quad
\code{delta=network\_scope\_broadened; credentials\_acquired} \\
\RecordBand{saverExposureHead}{Activation} &
\code{requested=external\_infrastructure\_access} \quad
\code{gate=boundary\_controls\_insufficient} \\
\RecordBand{saverExposureHead}{Exposure} &
\code{event=hf\_production\_compromise} \\
\RecordBand{saverResponseHead}{Recovery} &
\code{actions=rotate;rebuild;restrict} \quad
\code{closure=reported\_scope; lifecycle\_unverified} \\
\end{tabular}
\end{minipage}}

\vspace{4pt}
\setlength{\tabcolsep}{4pt}
\renewcommand{\arraystretch}{1.04}
\begin{tabular}{@{}p{0.31\linewidth}p{0.31\linewidth}p{0.31\linewidth}@{}}
\rowcolor{saverNeutralHead}
\RecordHint{Handle}{one public incident links the carrier chain} &
\RecordHint{Delta}{network scope and authority widened} &
\RecordHint{Closure}{containment reported; lifecycle repair unverified} \\
\end{tabular}
\endgroup
\end{minipage}
\end{table}

\FloatBarrier
\section{Screened Corpus Expansion}
\label{app:screened-corpus-expansion}
The local registry contains screened works beyond the studies analyzed closely in the main text. The table below groups additional references by mechanism and relevance to reusable state. Their inclusion documents coverage; it does not give every entry the evidential weight of a central mechanism study.
\begin{table}[!htbp]
\caption{Screened corpus expansion map beyond the core mechanism anchors. The table lists representative anchors for each coverage band rather than treating every screened record as a main-text mechanism claim.}
\label{tab:screened-corpus-expansion}
\centering
\scriptsize
\setlength{\tabcolsep}{3.4pt}
\renewcommand{\arraystretch}{1.08}
\SAVERTableRows
\begin{tabular}{@{}>{\raggedright\arraybackslash}p{0.16\linewidth}>{\raggedright\arraybackslash}p{0.24\linewidth}>{\raggedright\arraybackslash}p{0.36\linewidth}>{\raggedright\arraybackslash}p{0.15\linewidth}@{}}
\toprule
\rowcolor{saverNeutralHead}
\textbf{Coverage Band} & \textbf{Cells Covered} & \textbf{Representative Anchors} & \textbf{Claim Boundary} \\
\midrule
\textbf{Security, Privacy, Governance} & Cyber evaluation; injection; firewalling; privacy; data governance & CyberSecEval 2~\cite{bhatt_2024_cyberseceval_2_a_wide_rangin}; AttackEval~\cite{shu_2025_attackeval_how_to_evaluate_t}; Not What You've Signed Up For~\cite{greshake_2023_not_what_you_ve_signed_up_fo}; EchoLeak~\cite{reddy_2025_echoleak_the_first_real_worl}; DataGovBench~\cite{zhou_2025_datagovbench_benchmarking_ll} & Adjacent vocabulary for threats, privacy, and data audit \\
\textbf{Benchmarks, Red Teams, Guardrails} & General agent tasks; harm probes; unsafe choices; automated red teams; platform guardrails & AgentBench~\cite{liu_2023_agentbench_evaluating_llms_a}; AgentHarm~\cite{andriushchenko2025agentharm}; RedCode~\cite{guo2024redcode}; MobileSafetyBench~\cite{lee2026mobilesafetybench}; AgenticRed~\cite{yuan2026agenticred}; OpenGuardrails~\cite{wang2025openguardrails} & Denominator context, not a leaderboard \\
\textbf{Memory And Tool State} & Persistent memory; forgetting; memory governance; MCP/tool routes; dependency safety & Zombie Agents~\cite{yang2026zombie_agents}; MemoryGraft~\cite{srivastava2025memorygraft}; SSGM~\cite{lam2026ssgm}; GLOVE~\cite{yin2026glove}; MCPShield~\cite{zhou2026mcpshield}; STAC~\cite{li2026stac}; IPIGuard~\cite{an2025ipiguard} & Direct when state later changes reasoning or action \\
\textbf{Workflow And Propagation} & Skills; procedural reuse; library drift; multi-agent uptake; shared context; OS/GUI settings & Library Drift~\cite{zhang2026library_drift}; Tool-Drift~\cite{dabas2026tool_drift}; AgenticEval~\cite{wang2026agenticeval}; MemEvoBench~\cite{xie2026memevobench}; collective misremembering~\cite{xu2026when}; RedTeamCUA~\cite{liao2026redteamcua} & Thin evidence for ownership and repair \\
\textbf{Model-Side And Adaptive Workflow Leads} & Distillation and model update; reusable workflows; evaluator state & Knowledge distillation survey~\cite{xu_2024_a_survey_on_knowledge_distil}; CAMEL~\cite{li_2023_camel_communicative_agents_f}; WebPilot~\cite{zhang_2025_webpilot_a_versatile_and_aut}; AgentLab~\cite{jiang_2026_agentlab} & Central only inside an agent behavior loop \\
\bottomrule
\end{tabular}
\SAVERTableReset
\end{table}

\FloatBarrier
\section{Evaluation Protocol Details}
\label{app:evaluation-protocol-details}
This appendix keeps the denominator rule behind \secref{sec:evaluation}. The paper-local values reported in the main text retain their source-specific opportunity sets and lifecycle slices; the rule below summarizes how those denominators constrain comparison.
\begin{SAVERTakeaway}{Quantitative Anchor Rule}
\label{takeaway:quantitative-anchor-rule} Exact values remain paper-local because the denominator changes across exposure cases, recognition records, poisoned retrieval opportunities, accepted experience updates, memory-to-tool hijacking settings, runtime interventions, governance panels, and verification constraints~\cite{debenedetti2024agentdojo,yuan2024rjudge,chen2024agentpoison,wang2026oep,zhang2026memmorph,liu_2026_safeagent_a_runtime_protecti,taheri2026governed_memory,banerjee2026severa}. The manuscript therefore uses these works to name measurement objects and missing lifecycle traces rather than to rank scores.

\vspace{3pt}
\begin{tabular}{@{}p{0.25\linewidth}p{0.34\linewidth}p{0.32\linewidth}@{}}
\rowcolor{saverNeutralHead}
\textbf{Record} & \textbf{Keep With The Value} & \textbf{Avoid} \\
Paper-local metric & Opportunity set; transition slice; metric object & Cross-paper leaderboard claims \\
Artifact status & Dataset, benchmark, code, or trace availability & Mixing peer-reviewed and preprint rows as equal evidence \\
Claim boundary & What the metric directly observes & Inferring rollback, deletion, or contestability without traces \\
\end{tabular}
\end{SAVERTakeaway}

\FloatBarrier
\bibliographystyle{plainnat}
\bibliography{main.bbl}

\begin{thebibliography}{272}
\providecommand{\natexlab}[1]{#1}
\providecommand{\url}[1]{\texttt{#1}}
\expandafter\ifx\csname urlstyle\endcsname\relax
  \providecommand{\doi}[1]{doi: #1}\else
  \providecommand{\doi}{doi: \begingroup \urlstyle{rm}\Url}\fi

\bibitem[Adabara et~al.(2025)Adabara, Sadiq, Shuaibu, Danjuma, and Maninti]{adabara_2025_trustworthy_agentic_ai_syste}
Ibrahim Adabara, Bashir~Olaniyi Sadiq, Aliyu~Nuhu Shuaibu, Yale~Ibrahim Danjuma, and Venkateswarlu Maninti.
\newblock Trustworthy agentic ai systems: a cross-layer review of architectures, threat models, and governance strategies for real-world deployment.
\newblock \emph{F1000Research}, 14\penalty0 (905):\penalty0 905, 2025.
\newblock \doi{10.12688/f1000research.169927.1}.
\newblock URL \url{https://doi.org/10.12688/f1000research.169927.1}.

\bibitem[Ahad et~al.(2026)Ahad, Hossain, Alam, Puppala, Alam, and Talukder]{ahad_2026_misattribution_gap}
Tanzim Ahad, Ismail Hossain, Md~Jahangir Alam, Sai Puppala, Syed~Bahauddin Alam, and Sajedul Talukder.
\newblock The misattribution gap: When memory poisoning looks like model failure in agentic {AI} systems, 2026.
\newblock URL \url{https://arxiv.org/abs/2605.22842}.

\bibitem[Alqahtani and Ahuja(2026)]{alqahtani_2026_secure_autonomous_cyber_defe}
Hamed Alqahtani and Paras Ahuja.
\newblock Secure autonomous cyber defense with {LLM} agents: A systematic review of autonomy, tool-augmented reasoning, and governance constraints.
\newblock \emph{Computers \& Electrical Engineering}, 2026.
\newblock \doi{10.1016/j.compeleceng.2026.111184}.
\newblock URL \url{https://doi.org/10.1016/j.compeleceng.2026.111184}.

\bibitem[An et~al.(2025)An, Zhang, Du, Zhou, Li, Lin, and Ji]{an2025ipiguard}
Hengyu An, Jinghuai Zhang, Tianyu Du, Chunyi Zhou, Qingming Li, Tao Lin, and Shouling Ji.
\newblock {IPIG}uard: A novel tool dependency graph-based defense against indirect prompt injection in {LLM} agents.
\newblock In \emph{Proceedings of the 2025 Conference on Empirical Methods in Natural Language Processing}, pages 1023--1039, Suzhou, China, 2025. Association for Computational Linguistics.
\newblock \doi{10.18653/v1/2025.emnlp-main.53}.
\newblock URL \url{https://aclanthology.org/2025.emnlp-main.53/}.

\bibitem[An et~al.(2026{\natexlab{a}})An, Li, Xu, Zhou, Xu, Du, Li, and Ji]{an-etal-2026-permemsafe}
Hengyu An, Minxi Li, Naen Xu, Chunyi Zhou, Xiaogang Xu, Tianyu Du, Jinbao Li, and Shouling Ji.
\newblock {P}er{M}em{S}afe: Benchmarking implicit personalized safety of long horizon self-evolving agents.
\newblock In Maria Liakata, Viviane~P. Moreira, Jiajun Zhang, and David Jurgens, editors, \emph{Findings of the {A}ssociation for {C}omputational {L}inguistics: {ACL} 2026}, pages 6415--6433, San Diego, California, United States, July 2026{\natexlab{a}}. Association for Computational Linguistics.
\newblock ISBN 979-8-89176-395-1.
\newblock \doi{10.18653/v1/2026.findings-acl.320}.
\newblock URL \url{https://aclanthology.org/2026.findings-acl.320/}.

\bibitem[An et~al.(2026{\natexlab{b}})An, Li, Zhang, Xu, Zhou, Li, Xu, Du, and Ji]{an-etal-2026-aciarena}
Hengyu An, Minxi Li, Jinghuai Zhang, Naen Xu, Chunyi Zhou, Changjiang Li, Xiaogang Xu, Tianyu Du, and Shouling Ji.
\newblock {ACIA}rena: Toward unified evaluation for agent cascading injection.
\newblock In Maria Liakata, Viviane~P. Moreira, Jiajun Zhang, and David Jurgens, editors, \emph{Proceedings of the 64th Annual Meeting of the {A}ssociation for {C}omputational {L}inguistics (Volume 1: Long Papers)}, pages 10049--10066, San Diego, California, United States, July 2026{\natexlab{b}}. Association for Computational Linguistics.
\newblock ISBN 979-8-89176-390-6.
\newblock \doi{10.18653/v1/2026.acl-long.457}.
\newblock URL \url{https://aclanthology.org/2026.acl-long.457/}.

\bibitem[Andriushchenko et~al.(2025)Andriushchenko, Souly, Dziemian, Duenas, Lin, Wang, Hendrycks, Zou, Kolter, Fredrikson, Winsor, Wynne, Gal, and Davies]{andriushchenko2025agentharm}
Maksym Andriushchenko, Alexandra Souly, Mateusz Dziemian, Derek Duenas, Maxwell Lin, Justin Wang, Dan Hendrycks, Andy Zou, Zico Kolter, Matt Fredrikson, Eric Winsor, Jerome Wynne, Yarin Gal, and Xander Davies.
\newblock Agentharm: A benchmark for measuring harmfulness of {LLM} agents.
\newblock In \emph{International Conference on Learning Representations}, 2025.
\newblock \doi{10.48550/arXiv.2410.09024}.
\newblock URL \url{https://arxiv.org/abs/2410.09024}.

\bibitem[ang Gao et~al.(2025)ang Gao, Geng, Hua, Hu, Juan, Liu, Liu, Qiu, Qi, Wu, Wang, Xiao, Zhou, Zhang, Zhang, Xiang, Fang, Zhao, Liu, Ren, Qian, Wang, Hu, Wang, Wu, Ji, and Wang]{gao2025self_evolving_agents}
Huan ang Gao, Jiayi Geng, Wenyue Hua, Mengkang Hu, Xinzhe Juan, Hongzhang Liu, Shilong Liu, Jiahao Qiu, Xuan Qi, Yiran Wu, Hongru Wang, Han Xiao, Yuhang Zhou, Shaokun Zhang, Jiayi Zhang, Jinyu Xiang, Yixiong Fang, Qiwen Zhao, Dongrui Liu, Qihan Ren, Cheng Qian, Zhenhailong Wang, Minda Hu, Huazheng Wang, Qingyun Wu, Heng Ji, and Mengdi Wang.
\newblock A survey of self-evolving agents: What, when, how, and where to evolve on the path to artificial super intelligence, 2025.
\newblock URL \url{https://arxiv.org/abs/2507.21046}.

\bibitem[Anurin et~al.(2024)Anurin, Ng, Schaffer, Schreiber, and Kran]{anurin_2024_catastrophic_cyber_capabilit}
Andrey Anurin, Jonathan Ng, Kibo Schaffer, J.~Martin Schreiber, and Esben Kran.
\newblock Catastrophic cyber capabilities benchmark (3cb): Robustly evaluating {LLM} agent cyber offense capabilities, 2024.
\newblock URL \url{http://arxiv.org/abs/2410.09114}.

\bibitem[Bamil and LINGAMGUNTA(2026)]{bamil_2026_a_unified_evaluation_and_gov}
Vinay Bamil and RAMA KRISHNA~KUMAR LINGAMGUNTA.
\newblock A unified evaluation and governance framework for trustworthy {LLM} agents, 2026.
\newblock URL \url{https://doi.org/10.36227/techrxiv.176799772.28164151/v1}.

\bibitem[Banerjee et~al.(2026)Banerjee, Xu, Ie, Zhang, Peng, Lin, and Singh]{banerjee2026severa}
Debangshu Banerjee, Changming Xu, Eugene Ie, Ming Zhang, Daiyi Peng, Chu-Cheng Lin, and Gagandeep Singh.
\newblock {SEVerA}: Verified synthesis of self-evolving agents, 2026.
\newblock URL \url{https://arxiv.org/abs/2603.25111}.

\bibitem[Beurer-Kellner et~al.(2025)Beurer-Kellner, Buesser, Cretu, Debenedetti, Dobos, Fabian, Fischer, Froelicher, Grosse, Naeff, Ozoani, Paverd, Tram{\`e}r, and Volhejn]{beurer_kellner_2025_design_patterns_for_securing}
Luca Beurer-Kellner, Beat Buesser, Ana-Maria Cretu, Edoardo Debenedetti, Daniel Dobos, Daniel Fabian, Marc Fischer, David Froelicher, Kathrin Grosse, Daniel Naeff, Ezinwanne Ozoani, Andrew Paverd, Florian Tram{\`e}r, and V{\'a}clav Volhejn.
\newblock Design patterns for securing {LLM} agents against prompt injections, 2025.
\newblock URL \url{https://arxiv.org/abs/2506.08837}.

\bibitem[Bhatt et~al.(2024)Bhatt, Chennabasappa, Li, Nikolaidis, Song, Wan, Ahmad, Aschermann, Chen, Kapil, et~al.]{bhatt_2024_cyberseceval_2_a_wide_rangin}
Manish Bhatt, Sahana Chennabasappa, Yue Li, Cyrus Nikolaidis, Daniel Song, Shengye Wan, Faizan Ahmad, Cornelius Aschermann, Yaohui Chen, Dhaval Kapil, et~al.
\newblock Cyberseceval 2: A wide-ranging cybersecurity evaluation suite for large language models.
\newblock \emph{arXiv preprint arXiv:2404.13161}, 2024.
\newblock \doi{10.48550/arxiv.2404.13161}.
\newblock URL \url{http://arxiv.org/abs/2404.13161}.

\bibitem[Bogdanov et~al.(2026)Bogdanov, Lung, Kunz, Gao, Taylor, and Zaman]{bogdanov2026forge}
Igor Bogdanov, Chung-Horng Lung, Thomas Kunz, Jie Gao, Adrian Taylor, and Marzia Zaman.
\newblock {FORGE}: Self-evolving agent memory with no weight updates via population broadcast, 2026.
\newblock URL \url{https://arxiv.org/abs/2605.16233}.

\bibitem[Buneman et~al.(2001)Buneman, Khanna, and Tan]{buneman2001why_where_provenance}
Peter Buneman, Sanjeev Khanna, and Wang-Chiew Tan.
\newblock Why and where: A characterization of data provenance.
\newblock In \emph{Database Theory -- ICDT 2001}, pages 316--330, London, UK, 2001. Springer.
\newblock \doi{10.1007/3-540-44503-X_20}.
\newblock URL \url{https://doi.org/10.1007/3-540-44503-X_20}.

\bibitem[Cao et~al.(2025)Cao, Zhu, Ji, Sun, Zhu, {Wu Yinyu}, Dai, Yang, Han, and Guo]{cao2025safelawbench}
Chuxue Cao, Han Zhu, Jiaming Ji, Qichao Sun, Zhenghao Zhu, {Wu Yinyu}, Josef Dai, Yaodong Yang, Sirui Han, and Yike Guo.
\newblock {SafeLawBench}: Towards safe alignment of large language models.
\newblock In \emph{Findings of the Association for Computational Linguistics: ACL 2025}, pages 14015--14048, Vienna, Austria, 2025. Association for Computational Linguistics.
\newblock \doi{10.18653/v1/2025.findings-acl.721}.
\newblock URL \url{https://aclanthology.org/2025.findings-acl.721/}.

\bibitem[Chen et~al.(2025{\natexlab{a}})Chen, Niu, Li, Liu, Zheng, Tang, Li, Xiong, and Li]{chen2025halumem}
Ding Chen, Simin Niu, Kehang Li, Peng Liu, Xiangping Zheng, Bo~Tang, Xinchi Li, Feiyu Xiong, and Zhiyu Li.
\newblock {HaluMem}: Evaluating hallucinations in memory systems of agents, 2025{\natexlab{a}}.
\newblock URL \url{https://arxiv.org/abs/2511.03506}.

\bibitem[Chen et~al.(2025{\natexlab{b}})Chen, Li, Wang, Ma, Jiang, Zhou, Li, Du, and Ji]{chen2025lorashielddatafreeeditingalignment}
Jiahao Chen, Junhao Li, Yiming Wang, Zhe Ma, Yi~Jiang, Chunyi Zhou, Qingming Li, Tianyu Du, and Shouling Ji.
\newblock {LoRAShield}: Data-free editing alignment for secure personalized {LoRA} sharing, 2025{\natexlab{b}}.
\newblock URL \url{https://arxiv.org/abs/2507.07056}.

\bibitem[Chen et~al.(2026{\natexlab{a}})Chen, He, Yang, Li, Zhou, Li, Ma, Du, and Ji]{chen2026customization}
Jiahao Chen, Xing He, Yong Yang, Xinfeng Li, Chunyi Zhou, Junhao Li, Zhe Ma, Tianyu Du, and Shouling Ji.
\newblock Customization under fire: Plugin poisoning in text-to-image ecosystem.
\newblock \emph{arXiv preprint arXiv:2606.09151}, 2026{\natexlab{a}}.

\bibitem[Chen et~al.(2026{\natexlab{b}})Chen, Yin, Li, Ma, Du, Fu, Wang, Wang, and Ji]{chen2026beyond}
Jiahao Chen, Rui Yin, Xinfeng Li, Qianli Ma, Tianyu Du, Zhihui Fu, Jun Wang, Zhaoxiang Wang, and Shouling Ji.
\newblock Beyond over-refusal: Defending indirect prompt injection via latent instruction manifolds.
\newblock \emph{arXiv preprint arXiv:2608.22248}, 2026{\natexlab{b}}.

\bibitem[Chen et~al.(2026{\natexlab{c}})Chen, Jiang, Deng, Du, Ma, Feng, Qing, Yuan, Du, and Wang]{chen2026experienceinstructiontrajectorypoisoning}
Jialuo Chen, Lingqi Jiang, Xinhao Deng, Xiaohu Du, Jianan Ma, Yunhao Feng, Yuqi Qing, Zhihao Yuan, Linkang Du, and Jingyi Wang.
\newblock When experience becomes instruction: Trajectory poisoning in self-evolving agent skill systems, 2026{\natexlab{c}}.
\newblock URL \url{https://arxiv.org/abs/2608.05563}.

\bibitem[Chen et~al.(2025{\natexlab{c}})Chen, Yang, Zeng, Huang, Li, and Wang]{chen_2025_securegov_agent_a_governance}
Jinyu Chen, Jixiao Yang, Ziyang Zeng, Zixiao Huang, Jinming Li, and Yutong Wang.
\newblock Securegov-agent: A governance-centric multi-agent framework for privacy-preserving and attack-resilient {LLM} agents.
\newblock In \emph{Proceedings of the 2025 6th International Conference on Computer Science and Management Technology}, pages 875--881. ACM, 2025{\natexlab{c}}.
\newblock \doi{10.1145/3795154.3795296}.
\newblock URL \url{https://doi.org/10.1145/3795154.3795296}.

\bibitem[Chen et~al.(2026{\natexlab{d}})Chen, Pang, and Wang]{chen2026mrmmia}
Kai Chen, Yan Pang, and Tianhao Wang.
\newblock {MRMMIA}: Membership inference attacks on memory in chat agents, 2026{\natexlab{d}}.
\newblock URL \url{https://arxiv.org/abs/2605.27825}.

\bibitem[Chen et~al.(2025{\natexlab{d}})Chen, Wang, Carlini, Sitawarin, and Wagner]{chen_2025_defending_against_prompt_inj}
Sizhe Chen, Yizhu Wang, Nicholas Carlini, Chawin Sitawarin, and David Wagner.
\newblock Defending against prompt injection with a few defensivetokens, 2025{\natexlab{d}}.
\newblock URL \url{https://doi.org/10.1145/3733799.3762982}.

\bibitem[Chen et~al.(2024)Chen, Xiang, Xiao, Song, and Li]{chen2024agentpoison}
Zhaorun Chen, Zhen Xiang, Chaowei Xiao, Dawn Song, and Bo~Li.
\newblock {AgentPoison}: Red-teaming llm agents via poisoning memory or knowledge bases.
\newblock In \emph{Advances in Neural Information Processing Systems 37}, pages 130185--130213. Neural Information Processing Systems Foundation, Inc. (NeurIPS), 2024.
\newblock \doi{10.52202/079017-4136}.
\newblock URL \url{https://proceedings.neurips.cc/paper_files/paper/2024/hash/eb113910e9c3f6242541c1652e30dfd6-Abstract-Conference.html}.

\bibitem[Cheng et~al.(2024)Cheng, Zhang, Zhang, Meng, Hong, Li, Wang, Wang, Yin, Zhao, and He]{cheng_2024_exploring_large_language_mod}
Yuheng Cheng, Ceyao Zhang, Zhengwen Zhang, Xiangrui Meng, Sirui Hong, Wenhao Li, Zihao Wang, Zekai Wang, Feng Yin, Junhua Zhao, and Xiuqiang He.
\newblock Exploring large language model based intelligent agents: Definitions, methods, and prospects, 2024.
\newblock URL \url{https://arxiv.org/abs/2401.03428}.

\bibitem[Chennabasappa et~al.(2025)Chennabasappa, Nikolaidis, Song, Molnar, Ding, Wan, Whitman, Deason, Doucette, Montilla, Gampa, de~Paola, Gabi, Crnkovich, Testud, He, Chaturvedi, Zhou, and Saxe]{chennabasappa_2025_llamafirewall_an_open_source}
Sahana Chennabasappa, Cyrus Nikolaidis, Daniel Song, David Molnar, Stephanie Ding, Shengye Wan, Spencer Whitman, Lauren Deason, Nicholas Doucette, Abraham Montilla, Alekhya Gampa, Beto de~Paola, Dominik Gabi, James Crnkovich, Jean-Christophe Testud, Kat He, Rashnil Chaturvedi, Wu~Zhou, and Joshua Saxe.
\newblock Llamafirewall: An open source guardrail system for building secure {AI} agents, 2025.
\newblock URL \url{https://arxiv.org/abs/2505.03574}.

\bibitem[Chitan(2026)]{chitan2026srm}
Florin~Adrian Chitan.
\newblock Session risk memory ({SRM}): Temporal authorization for deterministic pre-execution safety gates, 2026.
\newblock URL \url{https://arxiv.org/abs/2603.22350}.

\bibitem[Costarelli et~al.(2024)Costarelli, Allen, Hauksson, Sodunke, Hariharan, Cheng, Li, Clymer, and Yadav]{costarelli_2024_gamebench_evaluating_strateg}
Anthony Costarelli, Mat Allen, Roman Hauksson, Grace Sodunke, Suhas Hariharan, Carlson Cheng, Wenjie Li, Joshua Clymer, and Arjun Yadav.
\newblock Gamebench: Evaluating strategic reasoning abilities of {LLM} agents, 2024.
\newblock URL \url{http://arxiv.org/abs/2406.06613}.

\bibitem[Dabas et~al.(2026)Dabas, Jeong, Jin, and Jia]{dabas2026tool_drift}
Mahavir Dabas, Jihyun Jeong, Ming Jin, and Ruoxi Jia.
\newblock Memory-induced tool-drift in {LLM} agents, 2026.
\newblock URL \url{https://arxiv.org/abs/2605.24941}.

\bibitem[Debenedetti et~al.(2024)Debenedetti, Zhang, Balunovi{\'c}, Beurer-Kellner, Fischer, and Tram{\`e}r]{debenedetti2024agentdojo}
Edoardo Debenedetti, Jie Zhang, Mislav Balunovi{\'c}, Luca Beurer-Kellner, Marc Fischer, and Florian Tram{\`e}r.
\newblock {AgentDojo}: A dynamic environment to evaluate prompt injection attacks and defenses for llm agents.
\newblock In \emph{Advances in Neural Information Processing Systems 37}, pages 82895--82920. Neural Information Processing Systems Foundation, Inc. (NeurIPS), 2024.
\newblock \doi{10.52202/079017-2636}.
\newblock URL \url{https://proceedings.neurips.cc/paper_files/paper/2024/hash/97091a5177d8dc64b1da8bf3e1f6fb54-Abstract-Datasets_and_Benchmarks_Track.html}.

\bibitem[Debenedetti et~al.(2026)Debenedetti, Shumailov, Fan, Hayes, Carlini, Fabian, Kern, Shi, Terzis, and Tram{\`e}r]{debenedetti2026defeating}
Edoardo Debenedetti, Ilia Shumailov, Tianqi Fan, Jamie Hayes, Nicholas Carlini, Daniel Fabian, Christoph Kern, Chongyang Shi, Andreas Terzis, and Florian Tram{\`e}r.
\newblock Defeating prompt injections by design.
\newblock In \emph{IEEE Conference on Secure and Trustworthy Machine Learning (SaTML)}, 2026.
\newblock \doi{10.48550/arXiv.2503.18813}.
\newblock URL \url{https://arxiv.org/abs/2503.18813}.

\bibitem[Deng et~al.(2026)Deng, Zhang, Wu, Bai, Yi, Zou, Xiao, Qiu, Ma, Chen, Du, Yang, Cui, Meng, Wang, Song, Xu, and Li]{deng_2026_taming_openclaw}
Xinhao Deng, Yixiang Zhang, Jiaqing Wu, Jiaqi Bai, Sibo Yi, Zhuoheng Zou, Yue Xiao, Rennai Qiu, Jianan Ma, Jialuo Chen, Xiaohu Du, Xiaofang Yang, Shiwen Cui, Changhua Meng, Weiqiang Wang, Jiaxing Song, Ke~Xu, and Qi~Li.
\newblock Taming openclaw: Security analysis and mitigation of autonomous {LLM} agent threats, 2026.
\newblock URL \url{http://arxiv.org/abs/2603.11619}.

\bibitem[Deng et~al.(2025)Deng, Guo, Han, Ma, Xiong, Wen, and Xiang]{deng_2024_ai_agents_under_threat_a_sur}
Zehang Deng, Yongjian Guo, Changzhou Han, Wanlun Ma, Junwu Xiong, Sheng Wen, and Yang Xiang.
\newblock {AI} agents under threat: A survey of key security challenges and future pathways.
\newblock \emph{ACM Computing Surveys}, 57\penalty0 (7):\penalty0 1--36, 2025.
\newblock \doi{10.1145/3716628}.
\newblock URL \url{https://doi.org/10.1145/3716628}.

\bibitem[Denning(1976)]{denning1976lattice_information_flow}
Dorothy~E. Denning.
\newblock A lattice model of secure information flow.
\newblock \emph{Communications of the ACM}, 19\penalty0 (5):\penalty0 236--243, 1976.
\newblock \doi{10.1145/360051.360056}.
\newblock URL \url{https://doi.org/10.1145/360051.360056}.

\bibitem[Dong et~al.(2026)Dong, Guo, Wang, Chen, Liu, Zhang, Xu, Huang, and Qiu]{dong2026safesearch}
Jianshuo Dong, Sheng Guo, Hao Wang, Xun Chen, Zhuotao Liu, Tianwei Zhang, Ke~Xu, Minlie Huang, and Han Qiu.
\newblock Safesearch: Automated red-teaming of {LLM}-based search agents, 2026.
\newblock URL \url{https://arxiv.org/abs/2509.23694}.

\bibitem[Dong et~al.(2024)Dong, Zhou, Yang, Shao, and Qiao]{dong2024conversation_safety_survey}
Zhichen Dong, Zhanhui Zhou, Chao Yang, Jing Shao, and Yu~Qiao.
\newblock Attacks, defenses and evaluations for llm conversation safety: A survey.
\newblock In \emph{Proceedings of the 2024 Conference of the North American Chapter of the Association for Computational Linguistics: Human Language Technologies (Volume 1: Long Papers)}, pages 6734--6747, Mexico City, Mexico, 2024. Association for Computational Linguistics.
\newblock \doi{10.18653/v1/2024.naacl-long.375}.
\newblock URL \url{https://aclanthology.org/2024.naacl-long.375/}.

\bibitem[Dorri et~al.(2018)Dorri, Kanhere, and Jurdak]{dorri_2018_multi_agent_systems_a_survey}
Ali Dorri, Salil~S. Kanhere, and Raja Jurdak.
\newblock Multi-agent systems: A survey.
\newblock \emph{IEEE Access}, 2018.
\newblock \doi{10.1109/access.2018.2831228}.
\newblock URL \url{https://doi.org/10.1109/access.2018.2831228}.

\bibitem[Du et~al.(2026)Du, Fang, Ma, Chen, Xu, Yin, and Chang]{du2026snapguardlightweightpromptinjection}
Mengyao Du, Han Fang, Haokai Ma, Jiahao Chen, Kai Xu, Quanjun Yin, and Ee-Chien Chang.
\newblock {SnapGuard}: Lightweight prompt injection detection for screenshot-based web agents, 2026.
\newblock URL \url{https://arxiv.org/abs/2604.25562}.

\bibitem[Du(2026)]{du2026memory_autonomous_agents}
Pengfei Du.
\newblock Memory for autonomous llm agents: Mechanisms, evaluation, and emerging frontiers, 2026.
\newblock URL \url{https://arxiv.org/abs/2603.07670}.

\bibitem[Fang et~al.(2026)Fang, Isahagian, Jayaram, Kumar, Muthusamy, Oum, and Thomas]{fang2026trajectorymemory}
Gaodan Fang, Vatche Isahagian, K.~R. Jayaram, Ritesh Kumar, Vinod Muthusamy, Punleuk Oum, and Gegi Thomas.
\newblock Trajectory-informed memory generation for self-improving agent systems, 2026.
\newblock URL \url{https://arxiv.org/abs/2603.10600}.

\bibitem[Fang et~al.(2025)Fang, Peng, Zhang, Wang, Yi, Zhang, Xu, Wu, Liu, Li, et~al.]{fang2025comprehensive}
Jinyuan Fang, Yanwen Peng, Xi~Zhang, Yingxu Wang, Xinhao Yi, Guibin Zhang, Yi~Xu, Bin Wu, Siwei Liu, Zihao Li, et~al.
\newblock A comprehensive survey of self-evolving ai agents: A new paradigm bridging foundation models and lifelong agentic systems.
\newblock \emph{arXiv preprint arXiv:2508.07407}, 2025.

\bibitem[Fang et~al.(2024)Fang, Bindu, Gupta, and Kang]{fang_2024_llm_agents_can_autonomously}
Richard Fang, Rohan Bindu, Akul Gupta, and Daniel Kang.
\newblock {LLM} agents can autonomously exploit one-day vulnerabilities, 2024.
\newblock URL \url{http://arxiv.org/abs/2404.08144}.

\bibitem[Feng et~al.(2026)Feng, Du, Deng, Ding, Wen, Wang, Xie, Zheng, Tan, Li, Wu, Cao, Huang, Guo, Ma, and Jiang]{feng2026braveguard}
Yunhao Feng, Xiaohu Du, Xinhao Deng, Yifan Ding, Ming Wen, Yixu Wang, Yuxiang Xie, Baihui Zheng, Yingshui Tan, Yige Li, Yutao Wu, Kerui Cao, Wenke Huang, Yanming Guo, Xingjun Ma, and Yu-Gang Jiang.
\newblock {BraveGuard}: From open-world threats to safer computer-use agents, 2026.
\newblock URL \url{https://arxiv.org/abs/2606.01166}.

\bibitem[Fernandez(2026)]{fernandez_2026_protocol_layer_governance_fo}
Marcelo~Patricio Fernandez.
\newblock Protocol-layer governance for {LLM} agents: Identity-bound oversight with zero-modification deployment and multi-hop authority propagation.
\newblock \emph{Zenodo (CERN European Organization for Nuclear Research)}, 2026.
\newblock \doi{10.5281/zenodo.20162877}.
\newblock URL \url{https://doi.org/10.5281/zenodo.20162877}.

\bibitem[Figurelli(2026)]{figurelli_2026_from_talking_to_blackbox_to}
Rogério Figurelli.
\newblock From talking to blackbox to governance: Constitution mining for {LLM} agents.
\newblock \emph{Zenodo (CERN European Organization for Nuclear Research)}, 2026.
\newblock \doi{10.5281/zenodo.18118668}.
\newblock URL \url{https://doi.org/10.5281/zenodo.18118668}.

\bibitem[Gan et~al.(2026)Gan, Yang, Ma, He, Zeng, Wang, Li, Zhou, Li, Wang, et~al.]{gan2026navigating}
Yuyou Gan, Yong Yang, Zhe Ma, Ping He, Rui Zeng, Yiming Wang, Qingming Li, Chunyi Zhou, Songze Li, Ting Wang, et~al.
\newblock Navigating the risks: A survey of security and privacy threats in llm-based agents.
\newblock \emph{ACM Transactions on Software Engineering and Methodology}, 2026.

\bibitem[Gao et~al.(2025)Gao, Zhu, Liu, Li, Yan, and Hu]{gao_2025_contextual_trust_evaluation}
Kangning Gao, Haotian Zhu, Rui Liu, Jinming Li, Xu~Yan, and Yi~Hu.
\newblock Contextual trust evaluation for robust coordination in large language model multi-agent systems, 2025.
\newblock URL \url{https://doi.org/10.1109/eiecc67963.2025.11409558}.

\bibitem[Geren et~al.(2025)Geren, Board, Dagher, Andersen, and Zhuang]{geren_2025_blockchain_for_large_languag}
Caleb Geren, Amanda Board, Gaby~G. Dagher, Tim Andersen, and Jun Zhuang.
\newblock Blockchain for large language model security and safety: A holistic survey.
\newblock \emph{ACM SIGKDD Explorations Newsletter}, 2025.
\newblock \doi{10.1145/3715073.3715075}.
\newblock URL \url{https://doi.org/10.1145/3715073.3715075}.

\bibitem[Gloaguen et~al.(2025)Gloaguen, Vero, Staab, and Vechev]{gloaguen2025watch_your_steps}
Thibaud Gloaguen, Mark Vero, Robin Staab, and Martin Vechev.
\newblock Watch your steps: Dormant adversarial behaviors that activate upon {LLM} finetuning, 2025.
\newblock URL \url{https://arxiv.org/abs/2505.16567}.

\bibitem[Greshake et~al.(2023)Greshake, Abdelnabi, Mishra, Endres, Holz, and Fritz]{greshake_2023_not_what_you_ve_signed_up_fo}
Kai Greshake, Sahar Abdelnabi, Shailesh Mishra, Christoph Endres, Thorsten Holz, and Mario Fritz.
\newblock Not what you've signed up for: Compromising real-world {LLM}-integrated applications with indirect prompt injection, 2023.
\newblock URL \url{http://arxiv.org/abs/2302.12173}.

\bibitem[Guo et~al.(2024{\natexlab{a}})Guo, Liu, Xie, Zhou, Zeng, Lin, Song, and Li]{guo2024redcode}
Chengquan Guo, Xun Liu, Chulin Xie, Andy Zhou, Yi~Zeng, Zinan Lin, Dawn Song, and Bo~Li.
\newblock Redcode: Risky code execution and generation benchmark for code agents.
\newblock In \emph{Advances in Neural Information Processing Systems 37}, pages 106190--106236. Neural Information Processing Systems Foundation, Inc. (NeurIPS), 2024{\natexlab{a}}.
\newblock \doi{10.52202/079017-3369}.
\newblock URL \url{https://proceedings.neurips.cc/paper_files/paper/2024/hash/bfd082c452dffb450d5a5202b0419205-Abstract-Datasets_and_Benchmarks_Track.html}.

\bibitem[Guo et~al.(2026{\natexlab{a}})Guo, Guo, Hu, Long, Sui, Zhi, Huang, He, Zhao, Zhao, and Qin]{guo-etal-2026-personalization}
Jiahe Guo, Xiangran Guo, Yulin Hu, Zimo Long, Xingyu Sui, Xuda Zhi, Yongbo Huang, Hao He, Weixiang Zhao, Yanyan Zhao, and Bing Qin.
\newblock When personalization legitimizes risks: Uncovering safety vulnerabilities in personalized dialogue agents.
\newblock In Maria Liakata, Viviane~P. Moreira, Jiajun Zhang, and David Jurgens, editors, \emph{Proceedings of the 64th Annual Meeting of the {A}ssociation for {C}omputational {L}inguistics (Volume 1: Long Papers)}, pages 27309--27335, San Diego, California, United States, July 2026{\natexlab{a}}. Association for Computational Linguistics.
\newblock ISBN 979-8-89176-390-6.
\newblock \doi{10.18653/v1/2026.acl-long.1260}.
\newblock URL \url{https://aclanthology.org/2026.acl-long.1260/}.

\bibitem[Guo et~al.(2026{\natexlab{b}})Guo, Jiao, Shi, Quan, Zhang, Li, Che, Xu, Liu, Liu, Kapadia, Pavlovic, Liu, Wang, Shi, Metaxas, and Tang]{guo2026memeye}
Minghao Guo, Qingyue Jiao, Zeru Shi, Yihao Quan, Boxuan Zhang, Danrui Li, Liwei Che, Wujiang Xu, Shilong Liu, Zirui Liu, Mubbasir Kapadia, Vladimir Pavlovic, Jiang Liu, Mengdi Wang, Yiyu Shi, Dimitris~N. Metaxas, and Ruixiang Tang.
\newblock {MemEye}: A visual-centric evaluation framework for multimodal agent memory, 2026{\natexlab{b}}.
\newblock URL \url{https://arxiv.org/abs/2605.15128}.

\bibitem[Guo et~al.(2024{\natexlab{b}})Guo, Chen, Wang, Chang, Pei, Chawla, Wiest, and Zhang]{guo_2024_large_language_model_based_m}
Taicheng Guo, Xiuying Chen, Yaqi Wang, Ruidi Chang, Shichao Pei, Nitesh~V. Chawla, Olaf Wiest, and Xiangliang Zhang.
\newblock Large language model based multi-agents: A survey of progress and challenges, 2024{\natexlab{b}}.
\newblock URL \url{https://arxiv.org/abs/2402.01680}.

\bibitem[Haez(2026)]{haez_2026_socio_normative_trustworthin}
Saba~Ghanbari Haez.
\newblock Socio-normative trustworthiness of {LLM} agents: Evaluating autonomy support and representational fairness across languages and identities, 2026.
\newblock URL \url{https://doi.org/10.65109/lchb2977}.

\bibitem[Han et~al.(2024)Han, Gao, Liu, Zhang, and Zhang]{han2024peft_survey}
Zeyu Han, Chao Gao, Jinyang Liu, Jeff Zhang, and Sai~Qian Zhang.
\newblock Parameter-efficient fine-tuning for large models: A comprehensive survey, 2024.
\newblock URL \url{https://arxiv.org/abs/2403.14608}.

\bibitem[Harrison(2026)]{harrison_2026_protocol_mediated_execution}
Ari Harrison.
\newblock Protocol-mediated execution governance for {LLM} agent systems.
\newblock \emph{SSRN Electronic Journal}, 2026.
\newblock \doi{10.2139/ssrn.6426760}.
\newblock URL \url{https://doi.org/10.2139/ssrn.6426760}.

\bibitem[Hatalis et~al.(2023)Hatalis, Christou, Myers, Jones, Lambert, Amos-Binks, Dannenhauer, and Dannenhauer]{hatalis_2024_memory_matters_the_need_to_i}
Kostas Hatalis, Despina Christou, Joshua Myers, Steven Jones, Keith Lambert, Adam Amos-Binks, Zohreh Dannenhauer, and Dustin Dannenhauer.
\newblock Memory matters: The need to improve long-term memory in llm-agents.
\newblock In \emph{Proceedings of the AAAI Symposium Series}, volume~2, pages 277--280, 2023.

\bibitem[He et~al.(2025{\natexlab{a}})He, Demartini, and Gadiraju]{he_2025_plan_then_execute_an_empiric}
Gaole He, Gianluca Demartini, and Ujwal Gadiraju.
\newblock Plan-then-execute: An empirical study of user trust and team performance when using {LLM} agents as a daily assistant, 2025{\natexlab{a}}.
\newblock URL \url{https://doi.org/10.1145/3706598.3713218}.

\bibitem[He et~al.(2025{\natexlab{b}})He, Treude, and Lo]{he_2025_llm_based_multi_agent_system}
Junda He, Christoph Treude, and David Lo.
\newblock {LLM}-based multi-agent systems for software engineering: Literature review, vision and the road ahead.
\newblock \emph{ACM Transactions on Software Engineering and Methodology}, 34\penalty0 (5):\penalty0 1--30, 2025{\natexlab{b}}.
\newblock \doi{10.1145/3712003}.
\newblock URL \url{https://doi.org/10.1145/3712003}.

\bibitem[He et~al.(2026)He, Chen, Qian, Wei, Chen, Fu, Wang, Wu, and Mao]{he2026contextgetsrootprivilege}
Xingbang He, Yuanwei Chen, Yi~Qian, Haiyang Wei, Ligeng Chen, Zenan Fu, Linzhang Wang, Hao Wu, and Bing Mao.
\newblock When context gets root: Privilege escalation in {LLM} harnesses, 2026.
\newblock URL \url{https://arxiv.org/abs/2608.27299}.

\bibitem[Hines et~al.(2024)Hines, Lopez, Hall, Zarfati, Zunger, and Kiciman]{hines_2024_defending_against_indirect_p}
Keegan Hines, Gary Lopez, Matthew Hall, Federico Zarfati, Yonatan Zunger, and Emre Kiciman.
\newblock Defending against indirect prompt injection attacks with spotlighting, 2024.
\newblock URL \url{http://arxiv.org/abs/2403.14720}.

\bibitem[Hong and Wang(2024)]{hong2024peta}
Lauren Hong and Ting Wang.
\newblock {PETA}: Parameter-efficient trojan attacks, 2024.
\newblock URL \url{https://arxiv.org/abs/2310.00648}.

\bibitem[Hsu et~al.(2024)Hsu, Tsai, Lin, Chen, Yu, and Huang]{hsu2024safe_lora}
Chia-Yi Hsu, Yu-Lin Tsai, Chih-Hsun Lin, Pin-Yu Chen, Chia-Mu Yu, and Chun-Ying Huang.
\newblock Safe {LoRA}: The silver lining of reducing safety risks when fine-tuning large language models.
\newblock In \emph{Advances in Neural Information Processing Systems 37}, pages 65072--65094. Neural Information Processing Systems Foundation, Inc. (NeurIPS), 2024.
\newblock \doi{10.52202/079017-2078}.
\newblock URL \url{https://proceedings.neurips.cc/paper_files/paper/2024/hash/77baa7c2a3a675823e89131698fd6e19-Abstract-Conference.html}.

\bibitem[Hu et~al.(2022)Hu, Shen, Wallis, Allen-Zhu, Li, Wang, Wang, and Chen]{hu2022lora}
Edward~J. Hu, Yelong Shen, Phillip Wallis, Zeyuan Allen-Zhu, Yuanzhi Li, Shean Wang, Lu~Wang, and Weizhu Chen.
\newblock {LoRA}: Low-rank adaptation of large language models.
\newblock In \emph{International Conference on Learning Representations}, 2022.
\newblock URL \url{https://openreview.net/forum?id=nZeVKeeFYf9}.

\bibitem[Hu et~al.(2025)Hu, Wang, and McAuley]{hu_2025_evaluating_memory_in_llm_age}
Yannan Hu, Yu~Wang, and Julian McAuley.
\newblock Evaluating memory in {LLM} agents via incremental multi-turn interactions, 2025.
\newblock URL \url{http://arxiv.org/abs/2507.05257}.

\bibitem[Hua et~al.(2024)Hua, Yang, Jin, Li, Cheng, Tang, and Zhang]{hua_2024_trustagent_towards_safe_and}
Wenyue Hua, Xianjun Yang, Mingyu Jin, Zelong Li, Wei Cheng, Ruixiang Tang, and Yongfeng Zhang.
\newblock Trustagent: Towards safe and trustworthy {LLM}-based agents, 2024.
\newblock URL \url{https://doi.org/10.18653/v1/2024.findings-emnlp.585}.

\bibitem[Huang et~al.(2025)Huang, Dai, Wu, Wu, Chen, Yu, Li, Ge, Wang, and Jin]{huang2025mempal}
Zhaopei Huang, Qifeng Dai, Guozheng Wu, Xiaopeng Wu, Kehan Chen, Chuan Yu, Xubin Li, Tiezheng Ge, Wenxuan Wang, and Qin Jin.
\newblock {Mem-PAL}: Towards memory-based personalized dialogue assistants for long-term user-agent interaction, 2025.
\newblock URL \url{https://arxiv.org/abs/2511.13410}.

\bibitem[Iacob et~al.(2026)Iacob, Jovanovi{\'c}, Shen, Burkhardt, Kurmanji, Tastan, Sani, Venanzi, Odonnat, Cao, Marino, Qiu, and Lane]{iacob2026redqueengodelmachine}
Alex Iacob, Andrej Jovanovi{\'c}, William~F. Shen, Daniel Burkhardt, Meghdad Kurmanji, Nurbek Tastan, Lorenzo Sani, Niccol{\`o} Alberto~Elia Venanzi, Ambroise Odonnat, Zeyu Cao, Bill Marino, Xinchi Qiu, and Nicholas~D. Lane.
\newblock The red queen {G{\"o}del} machine: Co-evolving agents and their evaluators, 2026.
\newblock URL \url{https://arxiv.org/abs/2606.26294}.

\bibitem[Ismail et~al.(2025)Ismail, Kurnia, Brata, Nelistiani, Heo, Kim, and Kim]{ismail_2025_toward_robust_security_orche}
Ismail, Rahmat Kurnia, Zilmas~Arjuna Brata, Ghitha~Afina Nelistiani, Shinwook Heo, Hyeongon Kim, and Howon Kim.
\newblock Toward robust security orchestration and automated response in security operations centers with a hyper-automation approach using agentic artificial intelligence.
\newblock \emph{Information}, 16\penalty0 (5):\penalty0 365, 2025.
\newblock \doi{10.3390/info16050365}.
\newblock URL \url{https://doi.org/10.3390/info16050365}.

\bibitem[Jain et~al.(2025)Jain, Pawar, Gada, Patwa, Kanani, Patil, and Kurup]{jain_2025_comprehensive_analysis_of_ma}
Bhavvya Jain, Pranav Pawar, Dhruv Gada, Tanish Patwa, Pratik Kanani, Deepali Patil, and Lakshmi Kurup.
\newblock Comprehensive analysis of machine learning and deep learning models on prompt injection classification using natural language processing techniques.
\newblock \emph{International Research Journal of Multidisciplinary Technovation}, 2025.
\newblock \doi{10.54392/irjmt2523}.
\newblock URL \url{https://doi.org/10.54392/irjmt2523}.

\bibitem[Jha et~al.(2026)Jha, Triedman, Wagle, and Shmatikov]{jha2026breaking}
Rishi Jha, Harold Triedman, Justin Wagle, and Vitaly Shmatikov.
\newblock Breaking and fixing defenses against control-flow hijacking in multi-agent systems.
\newblock In \emph{The Fourteenth International Conference on Learning Representations}, 2026.
\newblock \doi{10.48550/arXiv.2510.17276}.
\newblock URL \url{https://openreview.net/forum?id=PNU9Rj5RDQ}.

\bibitem[Jia et~al.(2025{\natexlab{a}})Jia, Wu, Qin, and Squicciarini]{jia_2025_the_task_shield_enforcing_ta}
Feiran Jia, Tong Wu, Xin Qin, and Anna~Cinzia Squicciarini.
\newblock The task shield: Enforcing task alignment to defend against indirect prompt injection in llm agents.
\newblock In \emph{Proceedings of the 63rd Annual Meeting of the Association for Computational Linguistics (Volume 1: Long Papers)}, pages 29680--29697, Vienna, Austria, 2025{\natexlab{a}}. Association for Computational Linguistics.
\newblock \doi{10.18653/v1/2025.acl-long.1435}.
\newblock URL \url{https://aclanthology.org/2025.acl-long.1435/}.

\bibitem[Jia et~al.(2025{\natexlab{b}})Jia, Huang, Wang, Zhang, and Song]{jia2025pisa}
Shian Jia, Ziyang Huang, Xinbo Wang, Haofei Zhang, and Mingli Song.
\newblock {PISA}: A pragmatic psych-inspired unified memory system for enhanced {AI} agency, 2025{\natexlab{b}}.
\newblock URL \url{https://arxiv.org/abs/2510.15966}.

\bibitem[Jiang et~al.(2026)Jiang, Wang, Liang, and Wang]{jiang_2026_agentlab}
Tanqiu Jiang, Yuhui Wang, Jiacheng Liang, and Ting Wang.
\newblock Agentlab: Benchmarking {LLM} agents against long-horizon attacks, 2026.
\newblock URL \url{http://arxiv.org/abs/2602.16901}.

\bibitem[Joseph et~al.(2025)Joseph, Daniel, Kathiresan, and P.]{joseph_2025_prompt_injection_in_large_la}
Jefferson~Kanjirakkattu Joseph, Esther Daniel, V.~Kathiresan, and M.~P.
\newblock Prompt injection in large language model exploitation: A security perspective, 2025.
\newblock URL \url{https://doi.org/10.1109/iceccc65144.2025.11064209}.

\bibitem[Kandala et~al.(2026)Kandala, Manchanda, Moharir, and Kandala]{kandala2026echoguard}
Ratna Kandala, Niva Manchanda, Akshata~Kishore Moharir, and Ananth Kandala.
\newblock Echoguard: An agentic framework with knowledge-graph memory for detecting manipulative communication in longitudinal dialogue, 2026.
\newblock URL \url{https://arxiv.org/abs/2603.04815}.

\bibitem[Kannabhiran(2026)]{kannabhiran_2026_consensus_validated_memory_i}
Dhillon~Andrew Kannabhiran.
\newblock Consensus-validated memory improves agent performance on complex tasks.
\newblock \emph{Zenodo (CERN European Organization for Nuclear Research)}, 2026.
\newblock \doi{10.5281/zenodo.18856774}.
\newblock URL \url{https://doi.org/10.5281/zenodo.18856774}.

\bibitem[Karim et~al.(2025)Karim, Van, Khan, Qu, and Kholodov]{karim_2025_ai_agents_meet_blockchain_a}
Md~Monjurul Karim, Dong~Hoang Van, Sangeen Khan, Qiang Qu, and Yaroslav Kholodov.
\newblock {AI} agents meet blockchain: A survey on secure and scalable collaboration for multi-agents.
\newblock \emph{Future Internet}, 2025.
\newblock \doi{10.3390/fi17020057}.
\newblock URL \url{https://doi.org/10.3390/fi17020057}.

\bibitem[Khan et~al.(2025)Khan, Joyce, and Habiba]{khan2025agentsafe_governance}
Rafflesia Khan, Declan Joyce, and Mansura Habiba.
\newblock {AGENTSAFE}: A unified framework for ethical assurance and governance in agentic {AI}, 2025.
\newblock URL \url{https://arxiv.org/abs/2512.03180}.

\bibitem[Kulkarni et~al.(2025)Kulkarni, Wu, Varia, and Bespalov]{kulkarni-etal-2025-agent}
Ninad Kulkarni, Xian Wu, Siddharth Varia, and Dmitriy Bespalov.
\newblock Agent vs. agent: Automated data generation and red-teaming for custom agentic workflows.
\newblock In Saloni Potdar, Lina Rojas-Barahona, and Sebastien Montella, editors, \emph{Proceedings of the 2025 Conference on Empirical Methods in Natural Language Processing: Industry Track}, pages 912--936, Suzhou, China, November 2025. Association for Computational Linguistics.
\newblock ISBN 979-8-89176-333-3.
\newblock \doi{10.18653/v1/2025.emnlp-industry.62}.
\newblock URL \url{https://aclanthology.org/2025.emnlp-industry.62/}.

\bibitem[Kumar(2026)]{kumar_2026_a_multi_evaluator_behavioura}
Sonu Kumar.
\newblock A multi-evaluator behavioural governance framework for {LLM} agent deployment.
\newblock \emph{Zenodo (CERN European Organization for Nuclear Research)}, 2026.
\newblock \doi{10.5281/zenodo.19394881}.
\newblock URL \url{https://doi.org/10.5281/zenodo.19394881}.

\bibitem[Lam et~al.(2026)Lam, Li, Zhang, and Zhao]{lam2026ssgm}
Chingkwun Lam, Jiaxin Li, Lingfei Zhang, and Kuo Zhao.
\newblock Governing evolving memory in {LLM} agents: Risks, mechanisms, and the stability and safety governed memory ({SSGM}) framework, 2026.
\newblock URL \url{https://arxiv.org/abs/2603.11768}.

\bibitem[Lan et~al.(2025)Lan, Kaul, and Jones]{lan_2025_prompt_injection_detection_i}
Qianlong Lan, Anuj Kaul, and Shaun Jones.
\newblock Prompt injection detection in llm integrated applications.
\newblock \emph{Int. J. Netw. Dyn. Intell}, 4\penalty0 (2), 2025.

\bibitem[Larcher et~al.(July 27, 2026)Larcher, Carreira, G., and Rannou]{larcher2026agentintrusion}
Hugo Larcher, Adrien Carreira, Raphael G., and Christophe Rannou.
\newblock Anatomy of a frontier lab agent intrusion: A technical timeline of the july 2026 incident, July 27, 2026.
\newblock URL \url{https://huggingface.co/blog/agent-intrusion-technical-timeline}.

\bibitem[Lee and Tiwari(2024)]{lee_2026_prompt_infection_llm_to_llm}
Donghyun Lee and Mo~Tiwari.
\newblock Prompt infection: Llm-to-llm prompt injection within multi-agent systems.
\newblock \emph{arXiv preprint arXiv:2410.07283}, 2024.

\bibitem[Lee et~al.(2026)Lee, Hahm, Choi, Knox, and Lee]{lee2026mobilesafetybench}
Juyong Lee, Dongyoon Hahm, June~Suk Choi, W.~Bradley Knox, and Kimin Lee.
\newblock Mobilesafetybench: Evaluating safety of autonomous agents in mobile device control.
\newblock \emph{Proceedings of the AAAI Conference on Artificial Intelligence}, 40\penalty0 (44):\penalty0 37565--37573, 2026.
\newblock \doi{10.1609/aaai.v40i44.41090}.
\newblock URL \url{https://ojs.aaai.org/index.php/AAAI/article/view/41090}.

\bibitem[Lee et~al.(2025)Lee, Kim, and Pak]{lee_2025_mind_mapping_prompt_injectio}
S.G. Lee, J.‐J. Kim, and Wooguil Pak.
\newblock Mind mapping prompt injection: Visual prompt injection attacks in modern large language models.
\newblock \emph{Electronics}, 2025.
\newblock \doi{10.3390/electronics14101907}.
\newblock URL \url{https://doi.org/10.3390/electronics14101907}.

\bibitem[Lermen et~al.(2023)Lermen, Rogers-Smith, and Ladish]{lermen_2023_lora_fine_tuning_efficiently}
Simon Lermen, Charlie Rogers-Smith, and Jeffrey Ladish.
\newblock Lora fine-tuning efficiently undoes safety training in llama 2-chat 70b.
\newblock \emph{arXiv preprint arXiv:2310.20624}, 2023.
\newblock \doi{10.48550/arxiv.2310.20624}.
\newblock URL \url{http://arxiv.org/abs/2310.20624}.

\bibitem[Leucker and Schallhart(2009)]{leucker2009runtime_verification}
Martin Leucker and Christian Schallhart.
\newblock A brief account of runtime verification.
\newblock \emph{The Journal of Logic and Algebraic Programming}, 78\penalty0 (5):\penalty0 293--303, 2009.
\newblock \doi{10.1016/j.jlap.2008.08.004}.
\newblock URL \url{https://doi.org/10.1016/j.jlap.2008.08.004}.

\bibitem[Li et~al.(2023)Li, Hammoud, Itani, Khizbullin, and Ghanem]{li_2023_camel_communicative_agents_f}
Guohao Li, Hasan Hammoud, Hani Itani, Dmitrii Khizbullin, and Bernard Ghanem.
\newblock Camel: Communicative agents for" mind" exploration of large language model society.
\newblock \emph{Advances in neural information processing systems}, 36:\penalty0 51991--52008, 2023.

\bibitem[Li et~al.(2025{\natexlab{a}})Li, Chen, Yang, Ai, Jia, Liu, Lin, Wu, Yuan, Hu, Wang, Liu, and Huang]{li_2024_legalagentbench_evaluating_l}
Haitao Li, Junjie Chen, Jingli Yang, Qingyao Ai, Wei Jia, Youfeng Liu, Kai Lin, Yueyue Wu, Guozhi Yuan, Yiran Hu, Wuyue Wang, Yiqun Liu, and Minlie Huang.
\newblock {L}egal{A}gent{B}ench: Evaluating {LLM} agents in legal domain.
\newblock In Wanxiang Che, Joyce Nabende, Ekaterina Shutova, and Mohammad~Taher Pilehvar, editors, \emph{Proceedings of the 63rd Annual Meeting of the Association for Computational Linguistics (Volume 1: Long Papers)}, pages 2322--2344, Vienna, Austria, July 2025{\natexlab{a}}. Association for Computational Linguistics.
\newblock ISBN 979-8-89176-251-0.
\newblock \doi{10.18653/v1/2025.acl-long.116}.
\newblock URL \url{https://aclanthology.org/2025.acl-long.116/}.

\bibitem[Li et~al.(2025{\natexlab{b}})Li, Liu, Chiu, Li, Zhang, and Xiao]{li2025drift}
Hao Li, Xiaogeng Liu, Hung-Chun Chiu, Dianqi Li, Ning Zhang, and Chaowei Xiao.
\newblock {DRIFT}: Dynamic rule-based defense with injection isolation for securing {LLM} agents.
\newblock In \emph{Advances in Neural Information Processing Systems}, volume~38, pages 83262--83290. Curran Associates, Inc., 2025{\natexlab{b}}.
\newblock \doi{10.48550/arXiv.2506.12104}.
\newblock URL \url{https://proceedings.neurips.cc/paper_files/paper/2025/file/77f3b26c7907aa27b207df9b9d43f29a-Paper-Conference.pdf}.

\bibitem[Li et~al.(2026{\natexlab{a}})Li, He, Shang, Kulshreshtha, Xian, Zhang, Su, Swamy, and Qi]{li2026stac}
Jing-Jing Li, Jianfeng He, Chao Shang, Devang Kulshreshtha, Xun Xian, Yi~Zhang, Hang Su, Sandesh Swamy, and Yanjun Qi.
\newblock {STAC}: When innocent tools form dangerous chains to jailbreak {LLM} agents, 2026{\natexlab{a}}.
\newblock URL \url{https://arxiv.org/abs/2509.25624}.

\bibitem[Li et~al.(2026{\natexlab{b}})Li, Liu, Fu, Wang, Xia, Li, Yan, and Li]{li2026kveraserlearningsteerkv}
Mufei Li, Shikun Liu, Dongqi Fu, Haoyu Wang, Yinglong Xia, Hong Li, Hong Yan, and Pan Li.
\newblock {KVEraser}: Learning to steer {KV} cache for efficient localized context erasing, 2026{\natexlab{b}}.
\newblock URL \url{https://arxiv.org/abs/2606.17034}.

\bibitem[Li et~al.(2024{\natexlab{a}})Li, Pan, Gopal, Yue, Berrios, Gatti, Li, Dombrowski, Goel, Mukobi, Helm-Burger, Lababidi, Justen, Liu, Chen, Barrass, Zhang, Zhu, Tamirisa, Bharathi, Herbert-Voss, Breuer, Zou, Mazeika, Wang, Oswal, Lin, Hunt, Tienken-Harder, Shih, Talley, Guan, Steneker, Campbell, Jokubaitis, Basart, Fitz, Kumaraguru, Karmakar, Tupakula, Varadharajan, Shoshitaishvili, Ba, Esvelt, Wang, and Hendrycks]{li_2024_the_wmdp_benchmark_measuring}
Nathaniel Li, Alexander Pan, Anjali Gopal, Summer Yue, Daniel Berrios, Alice Gatti, Justin~D. Li, Ann-Kathrin Dombrowski, Shashwat Goel, Gabriel Mukobi, Nathan Helm-Burger, Rassin Lababidi, Lennart Justen, Andrew~Bo Liu, Michael Chen, Isabelle Barrass, Oliver Zhang, Xiaoyuan Zhu, Rishub Tamirisa, Bhrugu Bharathi, Ariel Herbert-Voss, Cort~B. Breuer, Andy Zou, Mantas Mazeika, Zifan Wang, Palash Oswal, Weiran Lin, Adam~Alfred Hunt, Justin Tienken-Harder, Kevin~Y. Shih, Kemper Talley, John Guan, Ian Steneker, David Campbell, Brad Jokubaitis, Steven Basart, Stephen Fitz, Ponnurangam Kumaraguru, Kallol~Krishna Karmakar, Uday Tupakula, Vijay Varadharajan, Yan Shoshitaishvili, Jimmy Ba, Kevin~M. Esvelt, Alexandr Wang, and Dan Hendrycks.
\newblock The {WMDP} benchmark: Measuring and reducing malicious use with unlearning.
\newblock In \emph{Proceedings of the 41st International Conference on Machine Learning}, volume 235 of \emph{Proceedings of Machine Learning Research}, pages 28525--28550. PMLR, 2024{\natexlab{a}}.
\newblock URL \url{https://proceedings.mlr.press/v235/li24bc.html}.

\bibitem[Li et~al.(2026{\natexlab{c}})Li, Sun, Lin, Xing, Fu, Feng, Ni, and Tu]{li2026traversalaspolicylogdistilledgatedbehavior}
Peiran Li, Jiashuo Sun, Fangzhou Lin, Shuo Xing, Tianfu Fu, Suofei Feng, Chaoqun Ni, and Zhengzhong Tu.
\newblock Traversal-as-policy: Log-distilled gated behavior trees as externalized, verifiable policies for safe, robust, and efficient agents, 2026{\natexlab{c}}.
\newblock URL \url{https://arxiv.org/abs/2603.05517}.

\bibitem[Li et~al.(2026{\natexlab{d}})Li, Huang, Liu, Li, Fu, Song, Bian, Zhang, and Wang]{li2026experience_evolving_memory}
Sijia Li, Yuchen Huang, Zifan Liu, Zijian Li, Jingjing Fu, Lei Song, Jiang Bian, Jun Zhang, and Rui Wang.
\newblock Experience-evolving multi-turn tool-use agent with hybrid episodic-procedural memory, 2026{\natexlab{d}}.
\newblock URL \url{https://arxiv.org/abs/2512.07287}.

\bibitem[Li et~al.(2026{\natexlab{e}})Li, Zheng, Gao, Xia, Wang, Wang, Sun, Zhao, Wen, Li, et~al.]{li2026safety}
Xiao Li, Xiang Zheng, Yifeng Gao, Xinyu Xia, Yixu Wang, Xin Wang, Ye~Sun, Yunhan Zhao, Ming Wen, Jiayu Li, et~al.
\newblock Safety in embodied ai: A survey of risks, attacks, and defenses.
\newblock \emph{arXiv preprint arXiv:2605.02900}, 2026{\natexlab{e}}.

\bibitem[Li et~al.(2025{\natexlab{c}})Li, Liu, Zhang, Chen, Li, Li, and Ji]{li-etal-2025-twist}
Xindi Li, Zhe Liu, Tong Zhang, Jiahao Chen, Qingming Li, Jinbao Li, and Shouling Ji.
\newblock {TWIST}: Text-encoder weight-editing for inserting secret trojans in text-to-image models.
\newblock In \emph{Proceedings of the 63rd Annual Meeting of the Association for Computational Linguistics (Volume 1: Long Papers)}, pages 11025--11041, Vienna, Austria, July 2025{\natexlab{c}}. Association for Computational Linguistics.
\newblock ISBN 979-8-89176-251-0.
\newblock \doi{10.18653/v1/2025.acl-long.541}.
\newblock URL \url{https://aclanthology.org/2025.acl-long.541/}.

\bibitem[Li et~al.(2026{\natexlab{f}})Li, Yu, Pan, Sun, Li, Song, Lin, and Shi]{li2026unsaferturnsbenchmarkingdefending}
Xu~Li, Simon Yu, Minzhou Pan, Yiyou Sun, Bo~Li, Dawn Song, Xue Lin, and Weiyan Shi.
\newblock Unsafer in many turns: Benchmarking and defending multi-turn safety risks in tool-using agents, 2026{\natexlab{f}}.
\newblock URL \url{https://arxiv.org/abs/2602.13379}.

\bibitem[Li et~al.(2026{\natexlab{g}})Li, Lin, Deng, Zhang, He, Ji, Cao, and Hooi]{li2026jitrl}
Yibo Li, Zijie Lin, Ailin Deng, Xuan Zhang, Yufei He, Shuo Ji, Tri Cao, and Bryan Hooi.
\newblock Just-in-time reinforcement learning: Continual learning in {LLM} agents without gradient updates, 2026{\natexlab{g}}.
\newblock URL \url{https://arxiv.org/abs/2601.18510}.

\bibitem[Li et~al.(2024{\natexlab{b}})Li, Wen, Wang, Li, Yuan, Liu, Liu, Xu, Wang, Sun, Kong, Wang, Geng, Luan, Jin, Ye, Xiong, Zhang, Li, Xu, Li, Li, Liu, Zhang, and Liu]{li_2024_personal_llm_agents_insights}
Yuanchun Li, Hao Wen, Weijun Wang, Xiangyu Li, Yizhen Yuan, Guohong Liu, Jiacheng Liu, Wenxing Xu, Xiang Wang, Yi~Sun, Rui Kong, Yile Wang, Hanfei Geng, Jian Luan, Xuefeng Jin, Zilong Ye, Guanjing Xiong, Fan Zhang, Xiang Li, Mengwei Xu, Zhijun Li, Peng Li, Yang Liu, Ya-Qin Zhang, and Yunxin Liu.
\newblock Personal {LLM} agents: Insights and survey about the capability, efficiency and security, 2024{\natexlab{b}}.
\newblock URL \url{https://arxiv.org/abs/2401.05459}.

\bibitem[Li et~al.(2026{\natexlab{h}})Li, Cui, Chen, Liao, and Xing]{li2026whatsagentscontextcontext}
Zichuan Li, Jian Cui, Ashley Chen, Xiaojing Liao, and Luyi Xing.
\newblock What's in your agent's context? context privilege escalation attacks against {AI} agent harness, 2026{\natexlab{h}}.
\newblock URL \url{https://arxiv.org/abs/2609.01222}.

\bibitem[Liao et~al.(2026)Liao, Jones, Jiang, Ning, Fosler-Lussier, Su, Lin, and Sun]{liao2026redteamcua}
Zeyi Liao, Jaylen Jones, Linxi Jiang, Yuting Ning, Eric Fosler-Lussier, Yu~Su, Zhiqiang Lin, and Huan Sun.
\newblock {RedTeamCUA}: Realistic adversarial testing of computer-use agents in hybrid web-{OS} environments.
\newblock In \emph{International Conference on Learning Representations}, 2026.
\newblock URL \url{https://openreview.net/forum?id=yWwrgcBoK3}.

\bibitem[Lin et~al.(2026{\natexlab{a}})Lin, Deng, Li, Ma, Feng, Qing, Li, Zhang, Cui, Meng, Zhang, Ma, Li, Xu, and Ji]{lin2026safetyselfevolvingllmagent}
Ruixiao Lin, Xinhao Deng, Qingming Li, Jianan Ma, Yunhao Feng, Yuqi Qing, Zhenyuan Li, Yechao Zhang, Shiwen Cui, Changhua Meng, Tianwei Zhang, Xingjun Ma, Qi~Li, Ke~Xu, and Shouling Ji.
\newblock Safety in self-evolving {LLM} agent systems: Threats, amplification, and case studies, 2026{\natexlab{a}}.
\newblock URL \url{https://arxiv.org/abs/2606.23075}.

\bibitem[Lin et~al.(2026{\natexlab{b}})Lin, Li, Chen, Zhou, and Ji]{lin2026sope}
Ruixiao Lin, Qingming Li, Jiahao Chen, Chunyi Zhou, and Shouling Ji.
\newblock Sope: Situation-aware and statistically indistinguishable privacy exfiltration for mcp-enabled agents.
\newblock In \emph{Forty-third International Conference on Machine Learning}, 2026{\natexlab{b}}.

\bibitem[Lin et~al.(2026{\natexlab{c}})Lin, Hao, Fu, Cui, Chen, Li, Li, and Xiong]{lin2026memory_security_survey}
Zehao Lin, Xixuan Hao, Renyu Fu, Shaobo Cui, Kai Chen, Chunyu Li, Zhiyu Li, and Feiyu Xiong.
\newblock A survey on long-term memory security in {LLM} agents: Attacks, defenses, and governance across the memory lifecycle, 2026{\natexlab{c}}.
\newblock URL \url{https://arxiv.org/abs/2604.16548}.

\bibitem[Liu et~al.(2025{\natexlab{a}})Liu, Yang, Qu, Zhou, Cheng, and Hu]{liu_2025_a_survey_of_attacks_on_large}
Daizong Liu, Mingyu Yang, Xiaoye Qu, Pan Zhou, Yu~Cheng, and Wei Hu.
\newblock A survey of attacks on large vision–language models: Resources, advances, and future trends.
\newblock \emph{IEEE Transactions on Neural Networks and Learning Systems}, 2025{\natexlab{a}}.
\newblock \doi{10.1109/tnnls.2025.3592935}.
\newblock URL \url{https://doi.org/10.1109/tnnls.2025.3592935}.

\bibitem[Liu et~al.(2025{\natexlab{b}})Liu, Geng, Li, Cui, Zhang, Liu, and Liu]{liu2025webcoach}
Genglin Liu, Shijie Geng, Sha Li, Hejie Cui, Sarah Zhang, Xin Liu, and Tianyi Liu.
\newblock Webcoach: Self-evolving web agents with cross-session memory guidance, 2025{\natexlab{b}}.
\newblock URL \url{https://arxiv.org/abs/2511.12997}.

\bibitem[Liu et~al.(2026{\natexlab{a}})Liu, Ilyushin, Ni, and Zhu]{liu_2026_safeagent_a_runtime_protecti}
Hailin Liu, Eugene Ilyushin, Jie Ni, and Min Zhu.
\newblock Safeagent: A runtime protection architecture for agentic systems, 2026{\natexlab{a}}.
\newblock URL \url{https://arxiv.org/abs/2604.17562}.

\bibitem[Liu et~al.(2026{\natexlab{b}})Liu, Su, Xia, Han, Zheng, Xie, Ding, and Yao]{liu2026simplemem}
Jiaqi Liu, Yaofeng Su, Peng Xia, Siwei Han, Zeyu Zheng, Cihang Xie, Mingyu Ding, and Huaxiu Yao.
\newblock Simplemem: Efficient lifelong memory for {LLM} agents, 2026{\natexlab{b}}.
\newblock URL \url{https://arxiv.org/abs/2601.02553}.

\bibitem[Liu et~al.(2026{\natexlab{c}})Liu, Ye, Xia, Zheng, Xie, Ding, and Yao]{liu2026evolvemem}
Jiaqi Liu, Xinyu Ye, Peng Xia, Zeyu Zheng, Cihang Xie, Mingyu Ding, and Huaxiu Yao.
\newblock {EvolveMem}: Self-evolving memory architecture via {AutoResearch} for {LLM} agents, 2026{\natexlab{c}}.
\newblock URL \url{https://arxiv.org/abs/2605.13941}.

\bibitem[Liu et~al.(2026{\natexlab{d}})Liu, Zhu, Shu, Lin, Chen, Zhang, Zhang, Xu, Li, Tang, Li, Xiong, Chen, and Xu]{liu2026perma}
Shuochen Liu, Junyi Zhu, Long Shu, Junda Lin, Yuhao Chen, Haotian Zhang, Chao Zhang, Derong Xu, Jia Li, Bo~Tang, Zhiyu Li, Feiyu Xiong, Enhong Chen, and Tong Xu.
\newblock {PERMA}: Benchmarking personalized memory agents via event-driven preference and realistic task environments, 2026{\natexlab{d}}.
\newblock URL \url{https://arxiv.org/abs/2603.23231}.

\bibitem[Liu et~al.(2024)Liu, Yu, Zhang, Xu, Lei, Lai, Gu, Ding, Men, Yang, et~al.]{liu_2023_agentbench_evaluating_llms_a}
Xiao Liu, Hao Yu, Hanchen Zhang, Yifan Xu, Xuanyu Lei, Hanyu Lai, Yu~Gu, Hangliang Ding, Kaiwen Men, Kejuan Yang, et~al.
\newblock Agentbench: Evaluating llms as agents.
\newblock In \emph{International Conference on Learning Representations}, volume 2024, pages 52989--53046, 2024.

\bibitem[Liu et~al.(2025{\natexlab{c}})Liu, Li, Suh, Vorobeychik, Mao, Jha, McDaniel, Sun, Li, and Xiao]{liu2025autodan_turbo}
Xiaogeng Liu, Peiran Li, Edward Suh, Yevgeniy Vorobeychik, Zhuoqing Mao, Somesh Jha, Patrick McDaniel, Huan Sun, Bo~Li, and Chaowei Xiao.
\newblock {AutoDAN}-turbo: A lifelong agent for strategy self-exploration to jailbreak {LLM}s.
\newblock In \emph{International Conference on Learning Representations}, 2025{\natexlab{c}}.
\newblock \doi{10.48550/arXiv.2410.05295}.
\newblock URL \url{https://arxiv.org/abs/2410.05295}.

\bibitem[Liu et~al.(2023)Liu, Deng, Li, Wang, Wang, Wang, Zhang, Liu, Wang, Zheng, Zhang, and Liu]{liu_2023_prompt_injection_attack_agai}
Yi~Liu, Gelei Deng, Yuekang Li, Kailong Wang, Zihao Wang, Xiaofeng Wang, Tianwei Zhang, Yepang Liu, Haoyu Wang, Yan Zheng, Leo~Yu Zhang, and Yang Liu.
\newblock Prompt injection attack against {LLM}-integrated applications, 2023.
\newblock URL \url{https://arxiv.org/abs/2306.05499}.

\bibitem[Liu et~al.(2026{\natexlab{e}})Liu, Zhang, and She]{liu2026safecheckunsafeuse}
Zesen Liu, Zihan Zhang, and Dongdong She.
\newblock Safe to check, unsafe to use: Relinking at the compression boundary of {LLM} agents, 2026{\natexlab{e}}.
\newblock URL \url{https://arxiv.org/abs/2606.21732}.

\bibitem[Liu et~al.(2025{\natexlab{d}})Liu, Han, Yan, Liang, Zeng, Chen, Song, and Zhang]{zhou_2025_datagovbench_benchmarking_ll}
Zhou Liu, Zhaoyang Han, Guochen Yan, Hao Liang, Bohan Zeng, Xing Chen, Yuanfeng Song, and Wentao Zhang.
\newblock Datagovbench: Benchmarking {LLM} agents for real-world data governance workflows, 2025{\natexlab{d}}.
\newblock URL \url{https://arxiv.org/abs/2512.04416}.

\bibitem[Lotfi et~al.(2026)Lotfi, Shanto, Karim, and Bertino]{lotfi2026securingagenticaiperaction}
Alireza Lotfi, Subangkar~Karmaker Shanto, Imtiaz Karim, and Elisa Bertino.
\newblock Securing agentic {AI}: From per-action checks to trajectory assurance, 2026.
\newblock URL \url{https://arxiv.org/abs/2608.01558}.

\bibitem[Luo et~al.(2025{\natexlab{a}})Luo, Dai, Ni, Li, Zhang, Wang, Liu, and Salam]{luo2025agentauditor}
Hanjun Luo, Shenyu Dai, Chiming Ni, Xinfeng Li, Guibin Zhang, Kun Wang, Tongliang Liu, and Hanan Salam.
\newblock Agentauditor: Human-level safety and security evaluation for {LLM} agents.
\newblock In \emph{Advances in Neural Information Processing Systems}, 2025{\natexlab{a}}.
\newblock URL \url{https://openreview.net/forum?id=2KKqp7MWJM}.

\bibitem[Luo et~al.(2026)Luo, Tian, Cao, Luo, Lin, Li, Kong, Yang, and Ma]{luo_2026_from_storage_to_experience_a}
Jinghao Luo, Yuchen Tian, Chuxue Cao, Ziyang Luo, Hongzhan Lin, Kaixin Li, Chuyi Kong, Ruichao Yang, and Jing Ma.
\newblock From storage to experience: A survey on the evolution of {LLM} agent memory mechanisms.
\newblock In \emph{Findings of the Association for Computational Linguistics: ACL 2026}, pages 41622--41652. Association for Computational Linguistics, 2026.
\newblock \doi{10.18653/v1/2026.findings-acl.2069}.
\newblock URL \url{https://aclanthology.org/2026.findings-acl.2069/}.

\bibitem[Luo et~al.(2025{\natexlab{b}})Luo, Zhang, Yuan, Zhao, Yang, Gu, Wu, Chen, Qiao, Long, Tu, Luo, Ju, Xiao, Wang, Xiao, Liu, Yuan, Zhang, Jin, Zhang, Wu, Zhao, Tao, Yu, and Zhang]{luo2025llm_agent_survey}
Junyu Luo, Weizhi Zhang, Ye~Yuan, Yusheng Zhao, Junwei Yang, Yiyang Gu, Bohan Wu, Binqi Chen, Ziyue Qiao, Qingqing Long, Rongcheng Tu, Xiao Luo, Wei Ju, Zhiping Xiao, Yifan Wang, Meng Xiao, Chenwu Liu, Jingyang Yuan, Shichang Zhang, Yiqiao Jin, Fan Zhang, Xian Wu, Hanqing Zhao, Dacheng Tao, Philip~S. Yu, and Ming Zhang.
\newblock Large language model agent: A survey on methodology, applications and challenges, 2025{\natexlab{b}}.
\newblock URL \url{https://arxiv.org/abs/2503.21460}.

\bibitem[Luo et~al.(2025{\natexlab{c}})Luo, Dai, Liu, Banerjee, Sun, Chen, and Xiao]{luo2025agrail}
Weidi Luo, Shenghong Dai, Xiaogeng Liu, Suman Banerjee, Huan Sun, Muhao Chen, and Chaowei Xiao.
\newblock {AG}rail: A lifelong agent guardrail with effective and adaptive safety detection.
\newblock In \emph{Proceedings of the 63rd Annual Meeting of the Association for Computational Linguistics (Volume 1: Long Papers)}, pages 8104--8139, Vienna, Austria, 2025{\natexlab{c}}. Association for Computational Linguistics.
\newblock \doi{10.18653/v1/2025.acl-long.399}.
\newblock URL \url{https://aclanthology.org/2025.acl-long.399/}.

\bibitem[Lupascu and Lupascu(2026)]{lupascu2026elephantbroker}
Cristian Lupascu and Alexandru Lupascu.
\newblock Elephantbroker: A knowledge-grounded cognitive runtime for trustworthy {AI} agents, 2026.
\newblock URL \url{https://arxiv.org/abs/2603.25097}.

\bibitem[Ma et~al.(2024)Ma, Zhang, Zhu, Yang, Yang, Jin, Lan, Kong, and He]{he_2024_agentboard_an_analytical_eva}
Chang Ma, Junlei Zhang, Zhihao Zhu, Cheng Yang, Yujiu Yang, Yaohui Jin, Zhenzhong Lan, Lingpeng Kong, and Junxian He.
\newblock Agentboard: An analytical evaluation board of multi-turn {LLM} agents.
\newblock In \emph{Advances in Neural Information Processing Systems 37}. Neural Information Processing Systems Foundation, Inc., 2024.
\newblock \doi{10.52202/079017-2365}.
\newblock URL \url{https://doi.org/10.52202/079017-2365}.

\bibitem[Ma et~al.(2026{\natexlab{a}})Ma, Du, Lin, Bian, Chen, Wang, Yang, Cui, Meng, Deng, and Wang]{ma2026benchmarkingautonomousagents}
Jianan Ma, Xiaohu Du, Ruixiao Lin, Yaoxiang Bian, Jialuo Chen, Jingyi Wang, Xiaofang Yang, Shiwen Cui, Changhua Meng, Xinhao Deng, and Zhen Wang.
\newblock Benchmarking autonomous agents against temporal, spatial, and semantic evasions, 2026{\natexlab{a}}.
\newblock URL \url{https://arxiv.org/abs/2605.22321}.

\bibitem[Ma et~al.(2026{\natexlab{b}})Ma, Gao, Wang, Wang, Wang, Sun, Ding, Xu, Chen, Zhao, et~al.]{ma2026safety}
Xingjun Ma, Yifeng Gao, Yixu Wang, Ruofan Wang, Xin Wang, Ye~Sun, Yifan Ding, Hengyuan Xu, Yunhao Chen, Yunhan Zhao, et~al.
\newblock Safety at scale: A comprehensive survey of large model and agent safety.
\newblock \emph{Foundations and Trends in Privacy and Security}, 8\penalty0 (3-4):\penalty0 1--240, 2026{\natexlab{b}}.

\bibitem[Maharana et~al.(2024)Maharana, Lee, Tulyakov, Bansal, Barbieri, and Fang]{maharana2024locomo}
Adyasha Maharana, Dong-Ho Lee, Sergey Tulyakov, Mohit Bansal, Francesco Barbieri, and Yuwei Fang.
\newblock Evaluating very long-term conversational memory of llm agents.
\newblock In \emph{Proceedings of the 62nd Annual Meeting of the Association for Computational Linguistics (Volume 1: Long Papers)}, pages 13851--13870, Bangkok, Thailand, 2024. Association for Computational Linguistics.
\newblock \doi{10.18653/v1/2024.acl-long.747}.
\newblock URL \url{https://aclanthology.org/2024.acl-long.747/}.

\bibitem[Mao et~al.(2026{\natexlab{a}})Mao, Zhao, Wang, Qian, Huang, Wang, Han, Zheng, and Wang]{mao2026agentsdontjustagree}
Xutao Mao, Liangjie Zhao, Leyao Wang, Rui Qian, Qiang Huang, Wentao Wang, Bo~Han, Xiang Zheng, and Cong Wang.
\newblock Agents don't just agree, they remember: Benchmarking persistent sycophancy in stateful personal agents, 2026{\natexlab{a}}.
\newblock URL \url{https://arxiv.org/abs/2607.10526}.

\bibitem[Mao et~al.(2026{\natexlab{b}})Mao, Zhao, Zheng, and Wang]{mao2026practicemakesunsafeskill}
Xutao Mao, Liangjie Zhao, Xiang Zheng, and Cong Wang.
\newblock Practice makes unsafe: Skill misevolution in self-improving {LLM} agents, 2026{\natexlab{b}}.
\newblock URL \url{https://arxiv.org/abs/2608.12851}.

\bibitem[Mi et~al.(2026)Mi, Feng, Lu, Cao, Li, Li, Zhang, Luo, Peng, Cui, Tian, Fang, Chai, and Tan]{mi2026darwinian_memory}
Hongze Mi, Yibo Feng, WenJie Lu, Song Cao, Jinyuan Li, Yanming Li, Xuelin Zhang, Haotian Luo, Songyang Peng, He~Cui, Tengfei Tian, Jun Fang, Hua Chai, and Naiqiang Tan.
\newblock Darwinian memory: A training-free self-regulating memory system for {GUI} agent evolution, 2026.
\newblock URL \url{https://arxiv.org/abs/2601.22528}.

\bibitem[{Microsoft}(2024)]{microsoft2024ai_risk_assessment}
{Microsoft}.
\newblock {AI} risk assessment for {ML} engineers.
\newblock \url{https://learn.microsoft.com/en-us/security/ai-red-team/ai-risk-assessment}, 2024.

\bibitem[Mishra et~al.(2026)Mishra, Niroula, Yadav, Thakur, Gyawali, and Gaire]{mishra2026agenticragsok}
Saroj Mishra, Suman Niroula, Umesh Yadav, Dilip Thakur, Srijan Gyawali, and Shiva Gaire.
\newblock {SoK}: Agentic retrieval-augmented generation ({RAG}): Taxonomy, architectures, evaluation, and research directions, 2026.
\newblock URL \url{https://arxiv.org/abs/2603.07379}.

\bibitem[{MITRE}(2026)]{mitre2026atlas}
{MITRE}.
\newblock {MITRE ATLAS}: Adversarial threat landscape for artificial-intelligence systems.
\newblock \url{https://atlas.mitre.org/}, 2026.

\bibitem[Mittal and Venkatesan(2025)]{mittal_2025_practical_integration_of_lar}
Akshay Mittal and Vivek Venkatesan.
\newblock Practical integration of large language models into enterprise ci/cd pipelines for security policy validation: An industry-focused evaluation, 2025.
\newblock URL \url{https://doi.org/10.1109/sose67019.2025.00027}.

\bibitem[Mohammadi et~al.(2025)Mohammadi, Li, Lo, and Yip]{mohammadi_2025_evaluation_and_benchmarking}
Mahmoud Mohammadi, Yipeng Li, Jane Lo, and Wendy Yip.
\newblock Evaluation and benchmarking of llm agents: A survey.
\newblock In \emph{Proceedings of the 31st ACM SIGKDD Conference on Knowledge Discovery and Data Mining V.2}, pages 6129--6139, New York, NY, USA, 2025. Association for Computing Machinery.
\newblock \doi{10.1145/3711896.3736570}.
\newblock URL \url{https://doi.org/10.1145/3711896.3736570}.

\bibitem[Molleti et~al.(2024)Molleti, Goje, Luthra, and Raghavan]{molleti_2024_automated_threat_detection_a}
Ramasankar Molleti, Vinod Goje, Puneet Luthra, and Prathap Raghavan.
\newblock Automated threat detection and response using {LLM} agents.
\newblock \emph{World Journal of Advanced Research and Reviews}, 2024.
\newblock \doi{10.30574/wjarr.2024.24.2.3329}.
\newblock URL \url{https://doi.org/10.30574/wjarr.2024.24.2.3329}.

\bibitem[Moslemi et~al.(2026)Moslemi, Koneru, Lee, Kumar, and Radhakrishnan]{moslemi2026polaris}
Zahra Moslemi, Keerthi Koneru, Yen-Ting Lee, Sheethal Kumar, and Ramesh Radhakrishnan.
\newblock {POLARIS}: Typed planning and governed execution for agentic {AI} in back-office automation, 2026.
\newblock URL \url{https://arxiv.org/abs/2601.11816}.

\bibitem[Mosquera et~al.(2025)Mosquera, Pinzón, Fonseca, Ríos, Quijano, Giraldo, and Manrique]{mosquera_2025_can_llm_augmented_autonomous}
Manuel Mosquera, Juan~Sebastián Pinzón, Yesid Fonseca, Manuel Ríos, Nicanor Quijano, Luis~Felipe Giraldo, and Rubén Manrique.
\newblock Can {LLM}-augmented autonomous agents cooperate? an evaluation of their cooperative capabilities through melting pot.
\newblock \emph{IEEE Transactions on Artificial Intelligence}, 2025.
\newblock \doi{10.1109/tai.2025.3569192}.
\newblock URL \url{https://doi.org/10.1109/tai.2025.3569192}.

\bibitem[Mou et~al.(2026)Mou, Yu, Luan, Yin, Xue, Liu, Yang, Zhang, Zhang, and Ye]{mou2026pandora}
Yutao Mou, Dingyao Yu, Xiaotian Luan, Zhe Yin, Zhangchi Xue, Peiyang Liu, Pengfei Yang, Tong Zhang, Shikun Zhang, and Wei Ye.
\newblock Pandora’s toolbox: A survey of generalist agent security from the lifecycle perspective.
\newblock 2026.

\bibitem[Muzsai et~al.(2024)Muzsai, Imolai, and Lukács]{muzsai_2024_hacksynth_llm_agent_and_eval}
Lajos Muzsai, David Imolai, and András Lukács.
\newblock Hacksynth: {LLM} agent and evaluation framework for autonomous penetration testing, 2024.
\newblock URL \url{http://arxiv.org/abs/2412.01778}.

\bibitem[Nguyen et~al.(2026)Nguyen, Ndebugre, and Arremsetty]{nguyen2026security_mas}
Tam Nguyen, Moses Ndebugre, and Dheeraj Arremsetty.
\newblock Security considerations for multi-agent systems, 2026.
\newblock URL \url{https://arxiv.org/abs/2603.09002}.

\bibitem[OpenAI(August 26, 2026)]{openai_hf_incident}
OpenAI.
\newblock The hugging face incident and the road ahead, August 26, 2026.
\newblock URL \url{https://openai.com/index/hugging-face-incident-and-the-road-ahead/}.

\bibitem[{OpenAI}(July 21, 2026)]{openai2026huggingfaceincident}
{OpenAI}.
\newblock {OpenAI} and {Hugging Face} partner to address security incident during model evaluation, July 21, 2026.
\newblock URL \url{https://openai.com/index/hugging-face-model-evaluation-security-incident/}.

\bibitem[{OWASP Foundation}(2025)]{owasp2025llm_top10}
{OWASP Foundation}.
\newblock {OWASP} top 10 for large language model applications.
\newblock \url{https://owasp.org/www-project-top-10-for-large-language-model-applications/}, 2025.

\bibitem[Packer et~al.(2023)Packer, Wooders, Lin, Fang, Patil, Stoica, and Gonzalez]{packer2023memgpt}
Charles Packer, Sarah Wooders, Kevin Lin, Vivian Fang, Shishir~G. Patil, Ion Stoica, and Joseph~E. Gonzalez.
\newblock Memgpt: Towards llms as operating systems, 2023.
\newblock URL \url{https://arxiv.org/abs/2310.08560}.

\bibitem[Pan et~al.(2023)Pan, Chan, Zou, Li, Basart, Woodside, Zhang, Emmons, and Hendrycks]{pan2023machiavelli}
Alexander Pan, Jun~Shern Chan, Andy Zou, Nathaniel Li, Steven Basart, Thomas Woodside, Hanlin Zhang, Scott Emmons, and Dan Hendrycks.
\newblock Do the rewards justify the means? measuring trade-offs between rewards and ethical behavior in the machiavelli benchmark.
\newblock In \emph{Proceedings of the 40th International Conference on Machine Learning}, volume 202 of \emph{Proceedings of Machine Learning Research}, pages 26837--26867. PMLR, 2023.
\newblock URL \url{https://proceedings.mlr.press/v202/pan23a.html}.

\bibitem[Panterino and Fellington(2024)]{panterino_2024_dynamic_moving_target_defens}
Samuel Panterino and Matthew Fellington.
\newblock Dynamic moving target defense for mitigating targeted {LLM} prompt injection, 2024.
\newblock URL \url{http://dx.doi.org/10.36227/techrxiv.171822345.56781952/v1}.

\bibitem[Peng et~al.(2024)Peng, Chen, Chen, Niu, Bi, Liu, Feng, Wang, Yan, Wen, Zhang, Yin, Song, Bao, and Shi]{peng_2024_jailbreaking_and_mitigation}
Benji Peng, Hanxuan Chen, Keyu Chen, Qian Niu, Ziqian Bi, Ming Liu, Pohsun Feng, Tianyang Wang, Lawrence K.~Q. Yan, Yizhu Wen, Yichao Zhang, Caitlyn~Heqi Yin, Xinyuan Song, Riyang Bao, and Jiacheng Shi.
\newblock Jailbreaking and mitigation of vulnerabilities in large language models, 2024.
\newblock URL \url{http://arxiv.org/abs/2410.15236}.

\bibitem[Pink et~al.(2025)Pink, Wu, Vo, Turek, Mu, Huth, and Toneva]{pink_2025_position_episodic_memory_is}
Mathis Pink, Qinyuan Wu, Vy~A. Vo, Javier~S. Turek, Jianing Mu, Alexander~G. Huth, and Mariya Toneva.
\newblock Position: Episodic memory is the missing piece for long-term {LLM} agents, 2025.
\newblock URL \url{http://arxiv.org/abs/2502.06975}.

\bibitem[Podpora et~al.(2025)Podpora, Baranowski, Chopcian, Kwasniewicz, and Radziewicz]{podpora_2025_llm_firewall_using_validator}
Michał Podpora, M~Baranowski, Maciej Chopcian, Lukasz Kwasniewicz, and W.~Radziewicz.
\newblock {LLM} firewall using validator agent for prevention against prompt injection attacks.
\newblock \emph{Applied Sciences}, 2025.
\newblock \doi{10.3390/app16010085}.
\newblock URL \url{https://doi.org/10.3390/app16010085}.

\bibitem[Prakash et~al.(2026)Prakash, Lind, and Cruz]{prakash_2026_hybrid_real_time_framework_f}
Chandra Prakash, Mary Lind, and Elyson De~La Cruz.
\newblock Hybrid real-time framework for detecting adaptive prompt injection attacks in large language models.
\newblock \emph{Journal of Computing Theories and Applications}, 2026.
\newblock \doi{10.62411/jcta.15254}.
\newblock URL \url{https://doi.org/10.62411/jcta.15254}.

\bibitem[Pulipaka et~al.(2026{\natexlab{a}})Pulipaka, Chen, Sharma, Bajwa, Raina, and Sheth]{pulipaka2026persistbench}
Sidharth Pulipaka, Oliver Chen, Manas Sharma, Taaha~S. Bajwa, Vyas Raina, and Ivaxi Sheth.
\newblock Persistbench: When should long-term memories be forgotten by {LLM}s?, 2026{\natexlab{a}}.
\newblock URL \url{https://arxiv.org/abs/2602.01146}.

\bibitem[Pulipaka et~al.(2026{\natexlab{b}})Pulipaka, Hlebik, Raghav, Abdelnabi, Raina, Sheth, and Fritz]{pulipaka2026hidden_memory}
Sidharth Pulipaka, Stanislau Hlebik, Leonidas Raghav, Sahar Abdelnabi, Vyas Raina, Ivaxi Sheth, and Mario Fritz.
\newblock Hidden in memory: Sleeper memory poisoning in llm agents, 2026{\natexlab{b}}.
\newblock URL \url{https://arxiv.org/abs/2605.15338}.

\bibitem[Qi et~al.(2026{\natexlab{a}})Qi, Li, Liu, Shu, Yu, Ma, Cui, Zhao, Chen, Jiang, King, and Xu]{qi2026trustworthy_agentic_ai}
Jinhu Qi, Muzhi Li, Jiahong Liu, Yuqin Shu, Dianzhi Yu, Shicheng Ma, Wenqian Cui, Yiyang Zhao, Yiyi Chen, Ruoxi Jiang, Irwin King, and Zenglin Xu.
\newblock Towards trustworthy agentic {AI}: A comprehensive survey of safety, robustness, privacy, and system security, 2026{\natexlab{a}}.
\newblock URL \url{https://arxiv.org/abs/2605.23989}.

\bibitem[Qi et~al.(2026{\natexlab{b}})Qi, Wu, Guo, Zheng, Lu, He, Qin, and Ren]{qi2026darwin}
Weiwei Qi, Zefeng Wu, Zhilin Guo, Tianhang Zheng, Chaochao Lu, Liang He, Zhan Qin, and Kui Ren.
\newblock {DARWIN}: Evolving jailbreak adversary and guardrail for {LLM} safety evaluation and protection, 2026{\natexlab{b}}.
\newblock URL \url{https://arxiv.org/abs/2607.19829}.

\bibitem[Qi et~al.(2023)Qi, Zeng, Xie, Chen, Jia, Mittal, and Henderson]{qi2023finetuning_safety}
Xiangyu Qi, Yi~Zeng, Tinghao Xie, Pin-Yu Chen, Ruoxi Jia, Prateek Mittal, and Peter Henderson.
\newblock Fine-tuning aligned language models compromises safety, even when users do not intend to!, 2023.
\newblock URL \url{https://arxiv.org/abs/2310.03693}.

\bibitem[Radosevich and Halloran(2025)]{radosevich2025mcp_safety_audit}
Brandon Radosevich and John Halloran.
\newblock {MCP} safety audit: {LLM}s with the model context protocol allow major security exploits, 2025.
\newblock URL \url{https://arxiv.org/abs/2504.03767}.

\bibitem[Rajesh et~al.(2026)Rajesh, Holur, Turali, Duan, and Roychowdhury]{rajesh2026panini}
Shreyas Rajesh, Pavan Holur, Mehmet~Yigit Turali, Chenda Duan, and Vwani Roychowdhury.
\newblock Panini: Continual learning in token space via structured memory, 2026.
\newblock URL \url{https://arxiv.org/abs/2602.15156}.

\bibitem[Rando and Tram{\`e}r(2023)]{rando2024universal_jailbreak_backdoors}
Javier Rando and Florian Tram{\`e}r.
\newblock Universal jailbreak backdoors from poisoned human feedback, 2023.
\newblock URL \url{https://arxiv.org/abs/2311.14455}.

\bibitem[Rao et~al.(2026)Rao, Zhu, Lu, Chen, Niu, Guan, Li, and Xiang]{rao2026fragfusebypassingaccesscontrol}
Zixin Rao, Wentian Zhu, Chan~Aristella Lu, Zhaorun Chen, Wei Niu, Le~Guan, Bo~Li, and Zhen Xiang.
\newblock {FragFuse}: Bypassing access control of large language model agents via memory-based query fragmentation and fusion, 2026.
\newblock URL \url{https://arxiv.org/abs/2606.15609}.

\bibitem[Rathod et~al.(2025)Rathod, Nabavirazavi, Zad, and Iyengar]{rathod_2025_privacy_and_security_challen}
Vishal Rathod, Seyedsina Nabavirazavi, Samira Zad, and Sundararaja~Sitharama Iyengar.
\newblock Privacy and security challenges in large language models, 2025.
\newblock URL \url{https://doi.org/10.1109/ccwc62904.2025.10903912}.

\bibitem[Ravindran(2026)]{ravindran2026portable_agent_memory}
Santhosh~Kumar Ravindran.
\newblock Portable agent memory: A protocol for cryptographically-verified memory transfer across heterogeneous {AI} agents, 2026.
\newblock URL \url{https://arxiv.org/abs/2605.11032}.

\bibitem[Reddy and Gujral(2025)]{reddy_2025_echoleak_the_first_real_worl}
Pavan Reddy and Aditya~Sanjay Gujral.
\newblock Echoleak: The first real-world zero-click prompt injection exploit in a production {LLM} system.
\newblock \emph{Proceedings of the AAAI Symposium Series}, 2025.
\newblock \doi{10.1609/aaaiss.v7i1.36899}.
\newblock URL \url{https://doi.org/10.1609/aaaiss.v7i1.36899}.

\bibitem[Renze and Guven(2024)]{renze_2024_self_reflection_in_llm_agent}
Matthew Renze and Erhan Guven.
\newblock Self-reflection in {LLM} agents: Effects on problem-solving performance, 2024.
\newblock URL \url{http://arxiv.org/abs/2405.06682}.

\bibitem[Romero et~al.(2024)Romero, Zimmerman, Steinfeld, and Tomasic]{romero_2024_synergistic_integration_of_l}
Oscar~J. Romero, John Zimmerman, Aaron Steinfeld, and Anthony Tomasic.
\newblock Synergistic integration of large language models and cognitive architectures for robust {AI}: An exploratory analysis.
\newblock \emph{Proceedings of the AAAI Symposium Series}, 2024.
\newblock \doi{10.1609/aaaiss.v2i1.27706}.
\newblock URL \url{https://doi.org/10.1609/aaaiss.v2i1.27706}.

\bibitem[Sabelfeld and Myers(2003)]{sabelfeld2003language_information_flow}
Andrei Sabelfeld and Andrew~C. Myers.
\newblock Language-based information-flow security.
\newblock \emph{IEEE Journal on Selected Areas in Communications}, 21\penalty0 (1):\penalty0 5--19, 2003.
\newblock \doi{10.1109/JSAC.2002.806121}.
\newblock URL \url{https://doi.org/10.1109/JSAC.2002.806121}.

\bibitem[Salem et~al.(2023)Salem, Paverd, and Köpf]{salem_2023_maatphor_automated_variant_a}
Ahmed Salem, Andrew Paverd, and Boris Köpf.
\newblock Maatphor: Automated variant analysis for prompt injection attacks, 2023.
\newblock URL \url{http://arxiv.org/abs/2312.11513}.

\bibitem[Schulhoff et~al.(2023)Schulhoff, Pinto, Khan, Bouchard, Si, Anati, Tagliabue, Kost, Carnahan, and Boyd‐Graber]{schulhoff_2023_ignore_this_title_and_hackap}
Sander Schulhoff, Jeremy Pinto, Anaum Khan, Louis-François Bouchard, Chenglei Si, Svetlina Anati, Valen Tagliabue, Anson~Liu Kost, Christopher Carnahan, and Jordan Boyd‐Graber.
\newblock Ignore this title and hackaprompt: Exposing systemic vulnerabilities of {LLMs} through a global scale prompt hacking competition.
\newblock In \emph{Proceedings of the 2023 Conference on Empirical Methods in Natural Language Processing}, pages 4945--4977. Association for Computational Linguistics, 2023.
\newblock \doi{10.18653/v1/2023.emnlp-main.302}.
\newblock URL \url{https://aclanthology.org/2023.emnlp-main.302/}.

\bibitem[Shao et~al.(2026)Shao, Ren, Qian, Wei, Guo, Yang, Song, Zhang, Zhang, Liu, and Shao]{shao2026misevolve}
Shuai Shao, Qihan Ren, Chen Qian, Boyi Wei, Dadi Guo, Jingyi Yang, Xinhao Song, Linfeng Zhang, Weinan Zhang, Dongrui Liu, and Jing Shao.
\newblock Your agent may misevolve: Emergent risks in self-evolving {LLM} agents.
\newblock In \emph{International Conference on Learning Representations}, 2026.
\newblock \doi{10.48550/arXiv.2509.26354}.
\newblock URL \url{https://arxiv.org/abs/2509.26354}.

\bibitem[Shapira et~al.(2026)Shapira, Wendler, Yen, Sarti, Pal, Floody, Belfki, Loftus, Jannali, Prakash, Cui, Rogers, Brinkmann, Rager, Zur, Ripa, Sankaranarayanan, Atkinson, Gandikota, Fiotto-Kaufman, Hwang, Orgad, Sahil, Taglicht, Shabtay, Ambus, Alon, Oron, Gordon-Tapiero, Kaplan, Shwartz, Shaham, Riedl, Mirsky, Sap, Manheim, Ullman, and Bau]{shapira2026agents_chaos}
Natalie Shapira, Chris Wendler, Avery Yen, Gabriele Sarti, Koyena Pal, Olivia Floody, Adam Belfki, Alex Loftus, Aditya~Ratan Jannali, Nikhil Prakash, Jasmine Cui, Giordano Rogers, Jannik Brinkmann, Can Rager, Amir Zur, Michael Ripa, Aruna Sankaranarayanan, David Atkinson, Rohit Gandikota, Jaden Fiotto-Kaufman, EunJeong Hwang, Hadas Orgad, P~Sam Sahil, Negev Taglicht, Tomer Shabtay, Atai Ambus, Nitay Alon, Shiri Oron, Ayelet Gordon-Tapiero, Yotam Kaplan, Vered Shwartz, Tamar~Rott Shaham, Christoph Riedl, Reuth Mirsky, Maarten Sap, David Manheim, Tomer Ullman, and David Bau.
\newblock Agents of chaos, 2026.
\newblock URL \url{https://arxiv.org/abs/2602.20021}.

\bibitem[Shen et~al.(2026)Shen, Wei, Niu, Wang, Wu, Chien, Li, Chen, and Li]{shen2026turnlateresponseawaredefense}
Xinjie Shen, Rongzhe Wei, Peizhi Niu, Haoyu Wang, Ruihan Wu, Eli Chien, Bo~Li, Pin-Yu Chen, and Pan Li.
\newblock One turn too late: Response-aware defense against hidden malicious intent in multi-turn dialogue, 2026.
\newblock URL \url{https://arxiv.org/abs/2605.05630}.

\bibitem[Shi et~al.(2026)Shi, Yuan, Tie, Zhou, Gong, and Sun]{shi2026prompt_tool_selection}
Jiawen Shi, Zenghui Yuan, Guiyao Tie, Pan Zhou, Neil~Zhenqiang Gong, and Lichao Sun.
\newblock Prompt injection attack to tool selection in {LLM} agents.
\newblock In \emph{Network and Distributed System Security Symposium}, 2026.
\newblock \doi{10.14722/ndss.2026.230675}.
\newblock URL \url{https://www.ndss-symposium.org/ndss-paper/prompt-injection-attack-to-tool-selection-in-llm-agents/}.

\bibitem[Shi et~al.(2025)Shi, He, Wang, Li, Wu, Guo, and Song]{shi2025progent}
Tianneng Shi, Jingxuan He, Zhun Wang, Hongwei Li, Linyu Wu, Wenbo Guo, and Dawn Song.
\newblock Progent: Securing {AI} agents with privilege control, 2025.
\newblock URL \url{https://arxiv.org/abs/2504.11703}.

\bibitem[Shinn et~al.(2023)Shinn, Cassano, Gopinath, Narasimhan, and Yao]{shinn2023reflexion}
Noah Shinn, Federico Cassano, Ashwin Gopinath, Karthik Narasimhan, and Shunyu Yao.
\newblock Reflexion: Language agents with verbal reinforcement learning.
\newblock In \emph{Advances in Neural Information Processing Systems 36}. Neural Information Processing Systems Foundation, Inc., 2023.
\newblock URL \url{https://proceedings.neurips.cc/paper_files/paper/2023/hash/1b44b878bb782e6954cd888628510e90-Abstract-Conference.html}.

\bibitem[Shu et~al.(2025)Shu, Zhang, Jin, Zhou, and Li]{shu_2025_attackeval_how_to_evaluate_t}
Dong Shu, Chong Zhang, Mingyu Jin, Zihao Zhou, and Lingyao Li.
\newblock Attackeval: How to evaluate the effectiveness of jailbreak attacking on large language models.
\newblock \emph{ACM SIGKDD Explorations Newsletter}, 2025.
\newblock \doi{10.1145/3748239.3748242}.
\newblock URL \url{https://doi.org/10.1145/3748239.3748242}.

\bibitem[Siu et~al.(2026)Siu, He, Montgomery, Wang, Gong, Wang, and Song]{siu2026formalizing_agent_security}
Vincent Siu, Jingxuan He, Kyle Montgomery, Zhun Wang, Neil Gong, Chenguang Wang, and Dawn Song.
\newblock A framework for formalizing {LLM} agent security, 2026.
\newblock URL \url{https://arxiv.org/abs/2603.19469}.

\bibitem[Souza et~al.(2025)Souza, Poteet, Etz, Rosendo, Gueroudji, Shin, Balaprakash, and da~Silva]{souza_2025_llm_agents_for_interactive_w}
Renan Souza, Timothy Poteet, Brian~D. Etz, Daniel Rosendo, Amal Gueroudji, Woong Shin, Prasanna Balaprakash, and Rafael~Ferreira da~Silva.
\newblock {LLM} agents for interactive workflow provenance: Reference architecture and evaluation methodology, 2025.
\newblock URL \url{https://doi.org/10.1145/3731599.3767582}.

\bibitem[Srivastava and He(2025)]{srivastava2025memorygraft}
Saksham~Sahai Srivastava and Haoyu He.
\newblock Memorygraft: Persistent compromise of {LLM} agents via poisoned experience retrieval, 2025.
\newblock URL \url{https://arxiv.org/abs/2512.16962}.

\bibitem[Sun et~al.(2025)Sun, Chen, Luo, Zhang, and Li]{sun2025beyondheuristics}
Changzhi Sun, Xiangyu Chen, Jixiang Luo, Dell Zhang, and Xuelong Li.
\newblock Beyond heuristics: A decision-theoretic framework for agent memory management, 2025.
\newblock URL \url{https://arxiv.org/abs/2512.21567}.

\bibitem[Sun et~al.(2026)Sun, Zhang, Wang, and Wei]{sun2026magnet}
Libo Sun, Jiwen Zhang, Siyuan Wang, and Zhongyu Wei.
\newblock {MAGNET}: Towards adaptive {GUI} agents with memory-driven knowledge evolution.
\newblock In \emph{Proceedings of the 64th Annual Meeting of the Association for Computational Linguistics (Volume 1: Long Papers)}, pages 28181--28206. Association for Computational Linguistics, 2026.
\newblock \doi{10.18653/v1/2026.acl-long.1299}.
\newblock URL \url{https://aclanthology.org/2026.acl-long.1299/}.

\bibitem[Sunil et~al.(2026)Sunil, Sinha, Maheshwari, Todmal, Mallik, and Mishra]{sunil2026memory_poisoning_agents}
Balachandra~Devarangadi Sunil, Isheeta Sinha, Piyush Maheshwari, Shantanu Todmal, Shreyan Mallik, and Shuchi Mishra.
\newblock Memory poisoning attack and defense on memory based llm-agents, 2026.
\newblock URL \url{https://arxiv.org/abs/2601.05504}.

\bibitem[Suo(2024)]{suo_2024_signed_prompt_a_new_approach}
Xuchen Suo.
\newblock Signed-prompt: A new approach to prevent prompt injection attacks against {LLM}-integrated applications, 2024.
\newblock URL \url{https://arxiv.org/abs/2401.07612}.

\bibitem[Syros et~al.(2026)Syros, Suri, Ginesin, Nita-Rotaru, and Oprea]{syros2026saga}
Georgios Syros, Anshuman Suri, Jacob Ginesin, Cristina Nita-Rotaru, and Alina Oprea.
\newblock {SAGA}: A security architecture for governing {AI} agentic systems.
\newblock In \emph{Network and Distributed System Security Symposium}, 2026.
\newblock \doi{10.14722/ndss.2026.230869}.
\newblock URL \url{https://www.ndss-symposium.org/ndss-paper/saga-a-security-architecture-for-governing-ai-agentic-systems/}.

\bibitem[Tabassi(2023)]{tabassi2023ai_rmf}
Elham Tabassi.
\newblock Artificial intelligence risk management framework (ai rmf 1.0).
\newblock \penalty0 (NIST AI 100-1), 2023.
\newblock \doi{10.6028/NIST.AI.100-1}.
\newblock URL \url{https://doi.org/10.6028/NIST.AI.100-1}.

\bibitem[Taheri(2026)]{taheri2026governed_memory}
Hamed Taheri.
\newblock Governed memory: A production architecture for multi-agent workflows, 2026.
\newblock URL \url{https://arxiv.org/abs/2603.17787}.

\bibitem[Tao et~al.(2026)Tao, Zhao, Liu, Xi, Chen, Xu, and Li]{tao2026memconflict}
Zhen Tao, Jinxiang Zhao, Peng Liu, Dinghao Xi, Yanfang Chen, Wei Xu, and Zhiyu Li.
\newblock {MemConflict}: Evaluating long-term memory systems under memory conflicts, 2026.
\newblock URL \url{https://arxiv.org/abs/2605.20926}.

\bibitem[Tao et~al.(2024)Tao, Lin, Chen, Li, Wu, Li, Jin, Huang, Tao, and Zhou]{tao2024self_evolution_llm_survey}
Zhengwei Tao, Ting-En Lin, Xiancai Chen, Hangyu Li, Yuchuan Wu, Yongbin Li, Zhi Jin, Fei Huang, Dacheng Tao, and Jingren Zhou.
\newblock A survey on self-evolution of large language models, 2024.
\newblock URL \url{https://arxiv.org/abs/2404.14387}.

\bibitem[Tong et~al.(2026)Tong, Dai, and Guo]{tong2026aidguardstatefulauthorizationdelegated}
Yingzhe Tong, Leyu Dai, and Songhui Guo.
\newblock {AID-Guard}: Stateful authorization for delegated agent effects, 2026.
\newblock URL \url{https://arxiv.org/abs/2608.21159}.

\bibitem[Torra and Bras-Amor{\'o}s(2026)]{torra2026memorypoisoningmas}
Vicen{\c{c}} Torra and Maria Bras-Amor{\'o}s.
\newblock Memory poisoning and secure multi-agent systems, 2026.
\newblock URL \url{https://arxiv.org/abs/2603.20357}.

\bibitem[Torres et~al.(2026)Torres, Shrestha, and Misra]{torres2026agentsremembermuchmemory}
George Torres, Sharad Shrestha, and Satyajayant Misra.
\newblock When agents remember too much: Memory poisoning attacks on large language model agents, 2026.
\newblock URL \url{https://arxiv.org/abs/2607.06595}.

\bibitem[Triedman et~al.(2025)Triedman, Jha, and Shmatikov]{triedman2025multi}
Harold Triedman, Rishi Jha, and Vitaly Shmatikov.
\newblock Multi-agent systems execute arbitrary malicious code.
\newblock \emph{arXiv preprint arXiv:2503.12188}, 2025.

\bibitem[Triguero et~al.(2023)Triguero, Molina, Poyatos, Ser, and Herrera]{triguero_2023_general_purpose_artificial_i}
Isaac Triguero, Daniel Molina, Javier Poyatos, Javier~Del Ser, and Francisco Herrera.
\newblock General purpose artificial intelligence systems ({GPAIS}): Properties, definition, taxonomy, societal implications and responsible governance.
\newblock \emph{Information Fusion}, 2023.
\newblock \doi{10.1016/j.inffus.2023.102135}.
\newblock URL \url{https://doi.org/10.1016/j.inffus.2023.102135}.

\bibitem[Volta et~al.(2026)Volta, Longo, and Merlo]{volta_2026_experimental_campaign_for_po}
Giorgio Volta, Giacomo Longo, and Alessio Merlo.
\newblock Experimental campaign for political behaviour in {LLM} agents: Experimental insights on governance, global power asymmetries, security risks, and knowledge construction dynamics.
\newblock \emph{Zenodo (CERN European Organization for Nuclear Research)}, 2026.
\newblock \doi{10.5281/zenodo.19052474}.
\newblock URL \url{https://doi.org/10.5281/zenodo.19052474}.

\bibitem[Wang et~al.(2025{\natexlab{a}})Wang, He, Zeng, Xiang, Xing, Tang, and He]{wang2025privacy_agent_memory}
Bo~Wang, Weiyi He, Shenglai Zeng, Zhen Xiang, Yue Xing, Jiliang Tang, and Pengfei He.
\newblock Unveiling privacy risks in llm agent memory.
\newblock In \emph{Proceedings of the 63rd Annual Meeting of the Association for Computational Linguistics (Volume 1: Long Papers)}, pages 25241--25260. Association for Computational Linguistics, 2025{\natexlab{a}}.
\newblock \doi{10.18653/v1/2025.acl-long.1227}.
\newblock URL \url{https://aclanthology.org/2025.acl-long.1227/}.

\bibitem[Wang et~al.(2026{\natexlab{a}})Wang, Li, Liu, Chen, Hou, Qi, Li, Zhang, Ye, Liu, Chen, Zhang, and Yu]{wang2026devil_moltbook}
Chenxu Wang, Chaozhuo Li, Songyang Liu, Zejian Chen, Jinyu Hou, Ji~Qi, Rui Li, Litian Zhang, Qiwei Ye, Zheng Liu, Xu~Chen, Xi~Zhang, and Philip~S. Yu.
\newblock The devil behind moltbook: Anthropic safety is always vanishing in self-evolving {AI} societies, 2026{\natexlab{a}}.
\newblock URL \url{https://arxiv.org/abs/2602.09877}.

\bibitem[Wang et~al.(2023)Wang, Xie, Jiang, Mandlekar, Xiao, Zhu, Fan, and Anandkumar]{wang2023voyager}
Guanzhi Wang, Yuqi Xie, Yunfan Jiang, Ajay Mandlekar, Chaowei Xiao, Yuke Zhu, Linxi Fan, and Anima Anandkumar.
\newblock Voyager: An open-ended embodied agent with large language models, 2023.
\newblock URL \url{https://arxiv.org/abs/2305.16291}.

\bibitem[Wang et~al.(2026{\natexlab{b}})Wang, Zhao, and Zhang]{wang2025madra}
Junjian Wang, Lidan Zhao, and Xi~Sheryl Zhang.
\newblock {MADRA}: Multi-agent debate for risk-aware embodied planning.
\newblock In \emph{Findings of the Association for Computational Linguistics: ACL 2026}, pages 6852--6876. Association for Computational Linguistics, 2026{\natexlab{b}}.
\newblock \doi{10.18653/v1/2026.findings-acl.340}.
\newblock URL \url{https://aclanthology.org/2026.findings-acl.340/}.

\bibitem[Wang et~al.(2026{\natexlab{c}})Wang, Lin, Lou, Zhou, Suvonov, and Li]{wang2026emem}
Kaixiang Wang, Yidan Lin, Jiong Lou, Zhaojiacheng Zhou, Bunyod Suvonov, and Jie Li.
\newblock {E-mem}: Multi-agent based episodic context reconstruction for {LLM} agent memory, 2026{\natexlab{c}}.
\newblock URL \url{https://arxiv.org/abs/2601.21714}.

\bibitem[Wang et~al.(2026{\natexlab{d}})Wang, Lou, Zhou, and Li]{wang2026oep}
Kaixiang Wang, Jiong Lou, Zhaojiacheng Zhou, and Jie Li.
\newblock Oep: Poisoning self-evolving llm agents via locally correct but non-transferable experiences, 2026{\natexlab{d}}.
\newblock URL \url{https://arxiv.org/abs/2605.18930}.

\bibitem[Wang et~al.(2025{\natexlab{b}})Wang, Zhang, Zhou, Wu, Yu, Zhao, Yin, Fu, Yan, Luo, et~al.]{wang2025comprehensive}
Kun Wang, Guibin Zhang, Zhenhong Zhou, Jiahao Wu, Miao Yu, Shiqian Zhao, Chenlong Yin, Jinhu Fu, Yibo Yan, Hanjun Luo, et~al.
\newblock A comprehensive survey in llm (-agent) full stack safety: Data, training and deployment.
\newblock \emph{arXiv preprint arXiv:2504.15585}, 2025{\natexlab{b}}.

\bibitem[Wang et~al.(2025{\natexlab{c}})Wang, Ying, Zhang, Liang, Hu, Zhang, Liu, and Liu]{wang_2025_manipulating_multimodal_agen}
Le~Wang, Zonghao Ying, Tianyuan Zhang, Siyuan Liang, Shengshan Hu, Mingchuan Zhang, Aishan Liu, and Xianglong Liu.
\newblock Manipulating multimodal agents via cross-modal prompt injection, 2025{\natexlab{c}}.
\newblock URL \url{https://doi.org/10.1145/3746027.3755211}.

\bibitem[Wang et~al.(2026{\natexlab{e}})Wang, Cheng, Zhang, Liu, Xu, Lian, Wang, Yu, Yin, Hu, Hu, Zhang, Liu, Chen, and Wang]{wang2026memgovern}
Qihao Wang, Ziming Cheng, Shuo Zhang, Fan Liu, Rui Xu, Heng Lian, Kunyi Wang, Xiaoming Yu, Jianghao Yin, Sen Hu, Yue Hu, Shaolei Zhang, Yanbing Liu, Ronghao Chen, and Huacan Wang.
\newblock Memgovern: Enhancing code agents through learning from governed human experiences, 2026{\natexlab{e}}.
\newblock URL \url{https://arxiv.org/abs/2601.06789}.

\bibitem[Wang et~al.(2024{\natexlab{a}})Wang, Zhu, Liu, Ding, Ye, Zhou, and Yu]{wang2024unique_security_privacy_llm}
Shang Wang, Tianqing Zhu, Bo~Liu, Ming Ding, Dayong Ye, Wanlei Zhou, and Philip~S. Yu.
\newblock Unique security and privacy threats of large language models: A comprehensive survey, 2024{\natexlab{a}}.
\newblock URL \url{https://arxiv.org/abs/2406.07973}.

\bibitem[Wang and Li(2025)]{wang2025openguardrails}
Thomas Wang and Haowen Li.
\newblock Openguardrails: A configurable, unified, and scalable guardrails platform for large language models, 2025.
\newblock URL \url{https://arxiv.org/abs/2510.19169}.

\bibitem[Wang et~al.(2024{\natexlab{b}})Wang, Xue, Zhang, and Qian]{wang_2024_badagent_inserting_and_activ}
Yifei Wang, Dizhan Xue, Shengjie Zhang, and Shengsheng Qian.
\newblock Badagent: Inserting and activating backdoor attacks in {LLM} agents, 2024{\natexlab{b}}.
\newblock URL \url{https://doi.org/10.18653/v1/2024.acl-long.530}.

\bibitem[Wang et~al.(2025{\natexlab{d}})Wang, Chen, Li, Zhang, Zeng, Yang, and Ji]{wang2025aeiouunifieddefenseframework}
Yiming Wang, Jiahao Chen, Qingming Li, Tong Zhang, Rui Zeng, Xing Yang, and Shouling Ji.
\newblock {AEIOU}: A unified defense framework against {NSFW} prompts in text-to-image models, 2025{\natexlab{d}}.
\newblock URL \url{https://arxiv.org/abs/2412.18123}.

\bibitem[Wang et~al.(2026{\natexlab{f}})Wang, Wang, Yao, Li, Yang, Teng, Ma, and Wang]{wang2026agenticeval}
Yixu Wang, Xin Wang, Yang Yao, Xinyuan Li, Xibang Yang, Yan Teng, Xingjun Ma, and Yingchun Wang.
\newblock Agenticeval: Toward agentic and self-evolving safety evaluation of large language models.
\newblock In \emph{Findings of the Association for Computational Linguistics: ACL 2026}, pages 14789--14808. Association for Computational Linguistics, 2026{\natexlab{f}}.
\newblock \doi{10.18653/v1/2026.findings-acl.727}.
\newblock URL \url{https://aclanthology.org/2026.findings-acl.727/}.

\bibitem[Wang et~al.(2026{\natexlab{g}})Wang, Zhai, Jin, Dong, Yang, and Zhang]{wang2026mempot}
Yuhao Wang, Shengfang Zhai, Guanghao Jin, Yinpeng Dong, Linyi Yang, and Jiaheng Zhang.
\newblock Mempot: Defending against memory extraction attack with optimized honeypots, 2026{\natexlab{g}}.
\newblock URL \url{https://arxiv.org/abs/2602.07517}.

\bibitem[Wang et~al.(2026{\natexlab{h}})Wang, Jiang, Liang, Fleming, and Wang]{wang2026magesafeguardingllmagents}
Yuhui Wang, Tanqiu Jiang, Jiacheng Liang, Charles Fleming, and Ting Wang.
\newblock {MAGE}: Safeguarding {LLM} agents against long-horizon threats via shadow memory, 2026{\natexlab{h}}.
\newblock URL \url{https://arxiv.org/abs/2605.03228}.

\bibitem[Wang et~al.(2025{\natexlab{e}})Wang, Siu, Ye, Shi, Nie, Zhao, Wang, Guo, and Song]{wang2025agentvigil}
Zhun Wang, Vincent Siu, Zhe Ye, Tianneng Shi, Yuzhou Nie, Xuandong Zhao, Chenguang Wang, Wenbo Guo, and Dawn Song.
\newblock {AGENTVIGIL}: Automatic black-box red-teaming for indirect prompt injection against {LLM} agents.
\newblock In \emph{Findings of the Association for Computational Linguistics: EMNLP 2025}, pages 23159--23172, Suzhou, China, 2025{\natexlab{e}}. Association for Computational Linguistics.
\newblock \doi{10.18653/v1/2025.findings-emnlp.1258}.
\newblock URL \url{https://aclanthology.org/2025.findings-emnlp.1258/}.

\bibitem[Wang et~al.(2026{\natexlab{i}})Wang, Schiller, Li, Narayana, Nasr, Carlini, Qi, Wallace, Bursztein, Invernizzi, Thomas, Shoshitaishvili, Guo, He, Holz, and Song]{wang2026exploitgym}
Zhun Wang, Nico Schiller, Hongwei Li, Srijiith~Sesha Narayana, Milad Nasr, Nicholas Carlini, Xiangyu Qi, Eric Wallace, Elie Bursztein, Luca Invernizzi, Kurt Thomas, Yan Shoshitaishvili, Wenbo Guo, Jingxuan He, Thorsten Holz, and Dawn Song.
\newblock {ExploitGym}: Can {AI} agents turn security vulnerabilities into real attacks?, May 2026{\natexlab{i}}.
\newblock URL \url{https://arxiv.org/abs/2605.11086}.

\bibitem[Wei et~al.(2025{\natexlab{a}})Wei, Yang, Wang, Li, Li, Yin, Zhan, Holz, Lin, and Wang]{wei2025amemguard}
Qianshan Wei, Tengchao Yang, Yaochen Wang, Xinfeng Li, Lijun Li, Zhenfei Yin, Yi~Zhan, Thorsten Holz, Zhiqiang Lin, and XiaoFeng Wang.
\newblock A-memguard: A proactive defense framework for {LLM}-based agent memory, 2025{\natexlab{a}}.
\newblock URL \url{https://arxiv.org/abs/2510.02373}.

\bibitem[Wei et~al.(2025{\natexlab{b}})Wei, Niu, Shen, Tu, Li, Wu, Chien, Chen, Milenkovic, and Li]{wei2025trojanknowledgebypassingcommercial}
Rongzhe Wei, Peizhi Niu, Xinjie Shen, Tony Tu, Yifan Li, Ruihan Wu, Eli Chien, Pin-Yu Chen, Olgica Milenkovic, and Pan Li.
\newblock The trojan knowledge: Bypassing commercial {LLM} guardrails via harmless prompt weaving and adaptive tree search, 2025{\natexlab{b}}.
\newblock URL \url{https://arxiv.org/abs/2512.01353}.

\bibitem[Wu et~al.(2026{\natexlab{a}})Wu, Li, Zhou, Wang, and Ji]{wu2026comparative}
Baiqi Wu, Qingming Li, Chunyi Zhou, Ting Wang, and Shouling Ji.
\newblock A comparative survey of security risks in ai systems: From llms to ai agents and embodied agents.
\newblock \emph{ACM Computing Surveys}, 58\penalty0 (15):\penalty0 1--38, 2026{\natexlab{a}}.

\bibitem[Wu et~al.(2026{\natexlab{b}})Wu, Shi, Li, Zhao, Li, Adams, Hassan, and Ni]{wu2026evomalselfpoisoningselfevolvingcoding}
Xiaodong Wu, Yu~Shi, Qi~Li, Zhimin Zhao, Xiangman Li, Bram Adams, Ahmed~E. Hassan, and Jianbing Ni.
\newblock {EVOMAL}: Self-poisoning in self-evolving coding agents, 2026{\natexlab{b}}.
\newblock URL \url{https://arxiv.org/abs/2608.25776}.

\bibitem[Xi et~al.(2023)Xi, Chen, Guo, He, Ding, Hong, Zhang, Wang, Jin, Zhou, Zheng, Fan, Wang, Xiong, Zhou, Wang, Jiang, Zou, Liu, Yin, Dou, Weng, Cheng, Zhang, Qin, Zheng, Qiu, Huang, and Gui]{xi_2023_the_rise_and_potential_of_la}
Zhiheng Xi, Wenxiang Chen, Xin Guo, Wei He, Yiwen Ding, Boyang Hong, Ming Zhang, Junzhe Wang, Senjie Jin, Enyu Zhou, Rui Zheng, Xiaoran Fan, Xiao Wang, Limao Xiong, Yuhao Zhou, Weiran Wang, Changhao Jiang, Yicheng Zou, Xiangyang Liu, Zhangyue Yin, Shihan Dou, Rongxiang Weng, Wensen Cheng, Qi~Zhang, Wenjuan Qin, Yongyan Zheng, Xipeng Qiu, Xuanjing Huang, and Tao Gui.
\newblock The rise and potential of large language model based agents: A survey, 2023.
\newblock URL \url{https://arxiv.org/abs/2309.07864}.

\bibitem[Xia et~al.(2024)Xia, Lu, Zhu, Xing, Zhao, and Zhang]{xia_2024_evaluation_driven_developmen}
Boming Xia, Qinghua Lu, Liming Zhu, Zhenchang Xing, Dehai Zhao, and Hao Zhang.
\newblock Evaluation-driven development and operations of {LLM} agents: A process model and reference architecture, 2024.
\newblock URL \url{https://arxiv.org/abs/2411.13768}.

\bibitem[Xiang et~al.(2024)Xiang, Zheng, Li, Hong, Li, Xie, Zhang, Xiong, Xie, Yang, Song, and Li]{xiang2024guardagent}
Zhen Xiang, Linzhi Zheng, Yanjie Li, Junyuan Hong, Qinbin Li, Han Xie, Jiawei Zhang, Zidi Xiong, Chulin Xie, Carl Yang, Dawn Song, and Bo~Li.
\newblock {GuardAgent}: Safeguard llm agents by a guard agent via knowledge-enabled reasoning, 2024.
\newblock URL \url{https://arxiv.org/abs/2406.09187}.

\bibitem[Xie et~al.(2026)Xie, Guo, Zhang, Xia, Yang, Ma, Yan, and Ren]{xie2026memevobench}
Weiwei Xie, Shaoxiong Guo, Fan Zhang, Tian Xia, Xue Yang, Lizhuang Ma, Junchi Yan, and Qibing Ren.
\newblock {MemEvoBench}: Benchmarking safety risks from memory misevolution in {LLM} agents, 2026.
\newblock URL \url{https://arxiv.org/abs/2604.15774}.

\bibitem[Xing et~al.(2024)Xing, Zhang, Xue, Chen, Yang, and Xiao]{xing_2024_understanding_the_weakness_o}
Mingzhe Xing, Rongkai Zhang, Hui Xue, Qi~Chen, Fan Yang, and Zhen Xiao.
\newblock Understanding the weakness of large language model agents within a complex android environment, 2024.
\newblock URL \url{https://doi.org/10.1145/3637528.3671650}.

\bibitem[Xiong et~al.(2026)Xiong, Zhou, Wang, Gu, Tang, Li, Xiong, Wen, and Wen]{xiong2026mapleguardmemoryawarelinkenforcement}
Wenjun Xiong, Yijin Zhou, Jiaqian Wang, Shangding Gu, Bo~Tang, Zhiyu Li, Feiyu Xiong, Ying Wen, and Muning Wen.
\newblock {MAPLE-Guard}: Memory-aware link enforcement against memory-link poisoning in multi-agent systems, 2026.
\newblock URL \url{https://arxiv.org/abs/2608.00426}.

\bibitem[Xu et~al.(2026{\natexlab{a}})Xu, Yang, Hu, Chen, and An]{xu2026memorystaysilentmeasuring}
Lingxiang Xu, Jiaoyun Yang, Min Hu, Hongtu Chen, and Ning An.
\newblock When should memory stay silent: Measuring memory-use boundaries in memory-augmented conversational agents, 2026{\natexlab{a}}.
\newblock URL \url{https://arxiv.org/abs/2606.06055}.

\bibitem[Xu et~al.(2026{\natexlab{b}})Xu, An, Shi, Zhang, Zhou, Li, Du, Fu, Wang, and Ji]{xu2026when}
Naen Xu, Hengyu An, Shuo Shi, Jinghuai Zhang, Chunyi Zhou, Changjiang Li, Tianyu Du, Zhihui Fu, Jun Wang, and Shouling Ji.
\newblock When agents {``Misremember''} collectively: Exploring the mandela effect in {LLM}-based multi-agent systems.
\newblock In \emph{International Conference on Learning Representations}, 2026{\natexlab{b}}.
\newblock \doi{10.48550/arXiv.2602.00428}.
\newblock URL \url{https://openreview.net/forum?id=yIoMqDes7O}.

\bibitem[Xu et~al.(2025)Xu, Liang, Mei, Gao, Tan, and Zhang]{xu_2025_a_mem_agentic_memory_for_llm}
Wujiang Xu, Zujie Liang, Kai Mei, Hang Gao, Juntao Tan, and Yongfeng Zhang.
\newblock A-mem: Agentic memory for {LLM} agents, 2025.
\newblock URL \url{http://arxiv.org/abs/2502.12110}.

\bibitem[Xu et~al.(2024)Xu, Li, Tao, Shen, Cheng, Li, Xu, Tao, and Zhou]{xu_2024_a_survey_on_knowledge_distil}
Xiaohan Xu, Ming Li, Chongyang Tao, Tao Shen, Reynold Cheng, Jinyang Li, Can Xu, Dacheng Tao, and Tianyi Zhou.
\newblock A survey on knowledge distillation of large language models, 2024.
\newblock URL \url{http://arxiv.org/abs/2402.13116}.

\bibitem[Yan et~al.(2026)Yan, Zhang, Li, Zhou, Qi, and Zhang]{yan2026benignaloneharmfultogether}
Bingyu Yan, Xiaoming Zhang, Chaozhuo Li, Ziyi Zhou, Yirui Qi, and Litian Zhang.
\newblock Benign alone, harmful together: Exploiting experience composition in self-evolving {LLM} agents, 2026.
\newblock URL \url{https://arxiv.org/abs/2608.01759}.

\bibitem[Yan et~al.(2025)Yan, Li, Xu, Dong, Zhang, Ren, and Cheng]{yan_2025_on_protecting_the_data_priva}
Biwei Yan, Kun Li, Minghui Xu, Yueyan Dong, Yue Zhang, Zhaochun Ren, and Xiuzhen Cheng.
\newblock On protecting the data privacy of large language models ({LLMs}) and {LLM} agents: A literature review.
\newblock \emph{High-Confidence Computing}, 2025.
\newblock \doi{10.1016/j.hcc.2025.100300}.
\newblock URL \url{https://doi.org/10.1016/j.hcc.2025.100300}.

\bibitem[Yang et~al.(2025{\natexlab{a}})Yang, Liu, Yuan, Wu, Ni, Zhang, and Liu]{yang_2025_synergizing_intelligence_and}
Hongming Yang, Hao Liu, Xin Yuan, Kai Wu, Wei Ni, J.~Andrew Zhang, and Ren~Ping Liu.
\newblock Synergizing intelligence and privacy: A review of integrating internet of things, large language models, and federated learning in advanced networked systems.
\newblock \emph{Applied Sciences}, 15\penalty0 (12):\penalty0 6587, 2025{\natexlab{a}}.
\newblock \doi{10.3390/app15126587}.
\newblock URL \url{https://www.mdpi.com/2076-3417/15/12/6587}.

\bibitem[Yang et~al.(2025{\natexlab{b}})Yang, Feng, Guo, Zhang, Zhang, Ning, Wang, and Mao]{yang_2025_autohma_llm_efficient_task_c}
Tingting Yang, Ping Feng, Qixin Guo, Jindi Zhang, Xiufeng Zhang, Jiahong Ning, Xinghan Wang, and Zhongyang Mao.
\newblock Autohma-{LLM}: Efficient task coordination and execution in heterogeneous multi-agent systems using hybrid large language models.
\newblock \emph{IEEE Transactions on Cognitive Communications and Networking}, 2025{\natexlab{b}}.
\newblock \doi{10.1109/tccn.2025.3528892}.
\newblock URL \url{https://doi.org/10.1109/tccn.2025.3528892}.

\bibitem[Yang et~al.(2026{\natexlab{a}})Yang, Qiu, Zhang, Li, Huang, Feng, Yu, and Dong]{yang2026openclaw_hospital}
Wenxian Yang, Hanzheng Qiu, Bangqun Zhang, Chengquan Li, Zhiyong Huang, Xiaobin Feng, Rongshan Yu, and Jiahong Dong.
\newblock When openclaw meets hospital: Toward an agentic operating system for dynamic clinical workflows, 2026{\natexlab{a}}.
\newblock URL \url{https://arxiv.org/abs/2603.11721}.

\bibitem[Yang et~al.(2026{\natexlab{b}})Yang, He, Ji, Hooi, and Dong]{yang2026zombie_agents}
Xianglin Yang, Yufei He, Shuo Ji, Bryan Hooi, and Jin~Song Dong.
\newblock Zombie agents: Persistent control of self-evolving {LLM} agents via self-reinforcing injections, 2026{\natexlab{b}}.
\newblock URL \url{https://arxiv.org/abs/2602.15654}.

\bibitem[Yao et~al.(2022)Yao, Zhao, Yu, Du, Shafran, Narasimhan, and Cao]{yao2022react}
Shunyu Yao, Jeffrey Zhao, Dian Yu, Nan Du, Izhak Shafran, Karthik Narasimhan, and Yuan Cao.
\newblock React: Synergizing reasoning and acting in language models, 2022.
\newblock URL \url{https://arxiv.org/abs/2210.03629}.

\bibitem[Ye et~al.(2024)Ye, Hu, Deng, Huang, Ma, Zhu, and Wang]{ye_2024_mirai_evaluating_llm_agents}
Chenchen Ye, Ziniu Hu, Yihe Deng, Zijie Huang, Mingyu~Derek Ma, Yanqiao Zhu, and Wei Wang.
\newblock Mirai: Evaluating {LLM} agents for event forecasting, 2024.
\newblock URL \url{http://arxiv.org/abs/2407.01231}.

\bibitem[Ye et~al.(2026)Ye, Zhang, He, Peng, and Tyson]{ye2026behindevomap}
Qiming Ye, Peixain Zhang, Yupeng He, Zifan Peng, and Gareth Tyson.
\newblock Behind {EvoMap}: Characterizing a self-evolving agent-to-agent collaboration network, 2026.
\newblock URL \url{https://arxiv.org/abs/2605.25815}.

\bibitem[Yildirim(2026)]{yildirim2026differentialharm}
Caglar Yildirim.
\newblock Differential harm propensity in personalized {LLM} agents: The curious case of mental health disclosure, 2026.
\newblock URL \url{https://arxiv.org/abs/2603.16734}.

\bibitem[Yin and Du(2026)]{yin2026glove}
Xingkun Yin and Hongyang Du.
\newblock {GLOVE}: Global verifier for {LLM} memory-environment realignment, 2026.
\newblock URL \url{https://arxiv.org/abs/2601.19249}.

\bibitem[Ying et~al.(2026)Ying, Wu, Wu, Zheng, Cheng, Shi, and Guo]{ying2026skilljackpersistentskillbackdoors}
Zonghao Ying, Xiangfan Wu, Huiyu Wu, Xing Zheng, Huangsheng Cheng, Xiaorong Shi, and Jing Guo.
\newblock {SkillJack}: Persistent skill backdoors in self-evolving agents, 2026.
\newblock URL \url{https://arxiv.org/abs/2608.03509}.

\bibitem[Yu et~al.(2025)Yu, Meng, Zhou, Wang, Mao, Pan, Chen, Wang, Li, Zhang, et~al.]{yu2025trustworthy_llm_agents}
Miao Yu, Fanci Meng, Xinyun Zhou, Shilong Wang, Junyuan Mao, Linsey Pan, Tianlong Chen, Kun Wang, Xinfeng Li, Yongfeng Zhang, et~al.
\newblock A survey on trustworthy llm agents: Threats and countermeasures.
\newblock In \emph{Proceedings of the 31st ACM SIGKDD Conference on Knowledge Discovery and Data Mining V. 2}, pages 6216--6226, 2025.

\bibitem[Yu et~al.(2022)Yu, Yin, Li, and Li]{yu_2022_security_preserving_multi_ag}
Xinyi Yu, Xiang Yin, Shaoyuan Li, and Zhaojian Li.
\newblock Security-preserving multi-agent coordination for complex temporal logic tasks.
\newblock \emph{Control Engineering Practice}, 2022.
\newblock \doi{10.1016/j.conengprac.2022.105130}.
\newblock URL \url{https://doi.org/10.1016/j.conengprac.2022.105130}.

\bibitem[Yuan et~al.(2026)Yuan, N{\"o}ther, Jaques, and Radanovi{\'c}]{yuan2026agenticred}
Jiayi Yuan, Jonathan N{\"o}ther, Natasha Jaques, and Goran Radanovi{\'c}.
\newblock Agenticred: Evolving agentic systems for red-teaming, 2026.
\newblock URL \url{https://arxiv.org/abs/2601.13518}.

\bibitem[Yuan et~al.(2023)Yuan, Zhang, Xue, Yin, Chen, Guan, Li, Qian, and Yu]{yuan_2023_robust_multi_agent_coordinat}
Lei Yuan, Ziqian Zhang, Ke~Xue, Hao Yin, Feng Chen, Cong Guan, Lihe Li, Chao Qian, and Yang Yu.
\newblock Robust multi-agent coordination via evolutionary generation of auxiliary adversarial attackers.
\newblock \emph{Proceedings of the AAAI Conference on Artificial Intelligence}, 2023.
\newblock \doi{10.1609/aaai.v37i10.26388}.
\newblock URL \url{https://doi.org/10.1609/aaai.v37i10.26388}.

\bibitem[Yuan et~al.(2024)Yuan, He, Dong, Wang, Zhao, Xia, Xu, Zhou, Li, Zhang, Wang, and Liu]{yuan2024rjudge}
Tongxin Yuan, Zhiwei He, Lingzhong Dong, Yiming Wang, Ruijie Zhao, Tian Xia, Lizhen Xu, Binglin Zhou, Fangqi Li, Zhuosheng Zhang, Rui Wang, and Gongshen Liu.
\newblock R-judge: Benchmarking safety risk awareness for llm agents.
\newblock In \emph{Findings of the Association for Computational Linguistics: EMNLP 2024}, pages 1384--1400, Miami, Florida, USA, 2024. Association for Computational Linguistics.
\newblock \doi{10.18653/v1/2024.findings-emnlp.79}.
\newblock URL \url{https://aclanthology.org/2024.findings-emnlp.79/}.

\bibitem[Zeng et~al.(2026)Zeng, Qin, Li, Jia, Liu, and Jia]{zeng2026colluskilladversarialcrossskillcomposition}
Puyu Zeng, Simeng Qin, Jingzhi Li, Ju~Jia, Zheli Liu, and Xiaojun Jia.
\newblock {ColluSkill}: Adversarial cross-skill composition for evading agent skill scanners, 2026.
\newblock URL \url{https://arxiv.org/abs/2608.09732}.

\bibitem[Zhan et~al.(2024)Zhan, Liang, Ying, and Kang]{zhan_2024_injecagent_benchmarking_indi}
Qiusi Zhan, Zhixiang Liang, Zifan Ying, and Daniel Kang.
\newblock Injecagent: Benchmarking indirect prompt injections in tool-integrated large language model agents.
\newblock In \emph{Findings of the Association for Computational Linguistics: ACL 2024}, pages 10471--10506, Bangkok, Thailand, 2024. Association for Computational Linguistics.
\newblock \doi{10.18653/v1/2024.findings-acl.624}.
\newblock URL \url{https://aclanthology.org/2024.findings-acl.624/}.

\bibitem[Zhan et~al.(2025)Zhan, Fang, Panchal, and Kang]{zhan_2025_adaptive_attacks_break_defen}
Qiusi Zhan, Richard Fang, Henil~Shalin Panchal, and Daniel Kang.
\newblock Adaptive attacks break defenses against indirect prompt injection attacks on llm agents.
\newblock In \emph{Findings of the Association for Computational Linguistics: NAACL 2025}, pages 7116--7132, Albuquerque, New Mexico, 2025. Association for Computational Linguistics.
\newblock \doi{10.18653/v1/2025.findings-naacl.395}.
\newblock URL \url{https://aclanthology.org/2025.findings-naacl.395/}.

\bibitem[Zhan et~al.(2026)Zhan, Zhang, Guo, Zhao, and Liu]{zhan2026memoryauthoritybenchmarkingauthority}
Qiuyang Zhan, Rui Zhang, Sheng Guo, Lepeng Zhao, and Zhuotao Liu.
\newblock When memory becomes authority: Benchmarking authority collapse at the memory consolidation boundary, 2026.
\newblock URL \url{https://arxiv.org/abs/2608.01679}.

\bibitem[Zhang et~al.(2025{\natexlab{a}})Zhang, Tan, Shen, Salem, Backes, Zannettou, and Zhang]{zhang_2024_breaking_agents_compromising}
Boyang Zhang, Yicong Tan, Yun Shen, Ahmed Salem, Michael Backes, Savvas Zannettou, and Yang Zhang.
\newblock Breaking agents: Compromising autonomous {LLM} agents through malfunction amplification.
\newblock In \emph{Proceedings of the 2025 Conference on Empirical Methods in Natural Language Processing}, pages 34952--34964. Association for Computational Linguistics, 2025{\natexlab{a}}.
\newblock \doi{10.18653/v1/2025.emnlp-main.1771}.
\newblock URL \url{https://aclanthology.org/2025.emnlp-main.1771/}.

\bibitem[Zhang et~al.(2026{\natexlab{a}})Zhang, He, He, Sycara, and Xie]{zhang2026echosafe}
Ce~Zhang, Jinxi He, Junyi He, Katia Sycara, and Yaqi Xie.
\newblock Evolving contextual safety in multi-modal large language models via inference-time self-reflective memory.
\newblock In \emph{Proceedings of the IEEE/CVF Conference on Computer Vision and Pattern Recognition}, pages 41182--41192, 2026{\natexlab{a}}.
\newblock URL \url{https://openaccess.thecvf.com/content/CVPR2026/html/Zhang_Evolving_Contextual_Safety_in_Multi-Modal_Large_Language_Models_via_Inference-Time_CVPR_2026_paper.html}.

\bibitem[Zhang et~al.(2024{\natexlab{a}})Zhang, Huang, Mei, Yao, Wang, Zhan, Wang, and Zhang]{zhang_2024_agent_security_bench_asb_for}
Hanrong Zhang, Jingyuan Huang, Kai Mei, Yifei Yao, Zhenting Wang, Chenlu Zhan, Hongwei Wang, and Yongfeng Zhang.
\newblock Agent security bench (asb): Formalizing and benchmarking attacks and defenses in llm-based agents, 2024{\natexlab{a}}.
\newblock URL \url{https://arxiv.org/abs/2410.02644}.

\bibitem[Zhang et~al.(2026{\natexlab{b}})Zhang, Zheng, and Chen]{zhang2026embedding_defenses_fail}
Lingxi Zhang, Guangtao Zheng, and Hanjie Chen.
\newblock When embedding-based defenses fail: Rethinking safety in {LLM}-based multi-agent systems, 2026{\natexlab{b}}.
\newblock URL \url{https://arxiv.org/abs/2605.01133}.

\bibitem[Zhang et~al.(2024{\natexlab{b}})Zhang, Kong, Dewitt, Bräunl, and Hong]{zhang_2024_a_study_on_prompt_injection}
Wenxiao Zhang, Xiangrui Kong, Conan Dewitt, Thomas Bräunl, and Jin~B. Hong.
\newblock A study on prompt injection attack against {LLM}-integrated mobile robotic systems.
\newblock In \emph{2024 IEEE 35th International Symposium on Software Reliability Engineering Workshops (ISSREW)}, pages 361--368. IEEE, 2024{\natexlab{b}}.
\newblock \doi{10.1109/issrew63542.2024.00103}.
\newblock URL \url{https://doi.org/10.1109/issrew63542.2024.00103}.

\bibitem[Zhang et~al.(2026{\natexlab{c}})Zhang, Wang, Fei, Li, Liang, Xiang, Gu, and He]{zhang2026harnesssafeevaluatingsafetypersistent}
Xiao Zhang, Yusheng Wang, Yuhao Fei, Dongyuan Li, Zian Liang, Liuyu Xiang, Hongxun Gu, and Zhaofeng He.
\newblock Harnesssafe: Evaluating safety across persistent carriers in agent harnesses, 2026{\natexlab{c}}.
\newblock URL \url{https://arxiv.org/abs/2608.06984}.

\bibitem[Zhang et~al.(2026{\natexlab{d}})Zhang, Cui, Wang, Li, Qiu, Zhu, and He]{zhang2026library_drift}
Xing Zhang, Yanwei Cui, Guanghui Wang, Ziyuan Li, Wei Qiu, Bing Zhu, and Peiyang He.
\newblock Library drift: Diagnosing and fixing a silent failure mode in self-evolving {LLM} skill libraries, 2026{\natexlab{d}}.
\newblock URL \url{https://arxiv.org/abs/2605.19576}.

\bibitem[Zhang et~al.(2024{\natexlab{c}})Zhang, Xu, Ba, Wang, Hong, Liu, Qin, and Ren]{zhang_2024_privacyasst_safeguarding_use}
Xinyu Zhang, Huiyu Xu, Zhongjie Ba, Zhibo Wang, Yuan Hong, Jian Liu, Zhan Qin, and Kui Ren.
\newblock Privacyasst: Safeguarding user privacy in tool-using large language model agents.
\newblock \emph{IEEE Transactions on Dependable and Secure Computing}, 2024{\natexlab{c}}.
\newblock \doi{10.1109/TDSC.2024.3372777}.
\newblock URL \url{https://doi.org/10.1109/TDSC.2024.3372777}.

\bibitem[Zhang et~al.(2026{\natexlab{e}})Zhang, Zheng, Xu, Zhou, Shen, Ou, Zhang, and Lam]{zhang2026memmorph}
Xuanye Zhang, Yongsen Zheng, Zhuqin Xu, Kaiyu Zhou, Bowen Shen, Haoran Ou, Tianwei Zhang, and Kwok-Yan Lam.
\newblock Memmorph: Tool hijacking in llm agents via memory poisoning, 2026{\natexlab{e}}.
\newblock URL \url{https://arxiv.org/abs/2605.26154}.

\bibitem[Zhang et~al.(2025{\natexlab{b}})Zhang, Ma, Ma, Han, Wu, and Tresp]{zhang_2025_webpilot_a_versatile_and_aut}
Yao Zhang, Zijian Ma, Yunpu Ma, Zhen Han, Yu~Wu, and Volker Tresp.
\newblock Webpilot: A versatile and autonomous multi-agent system for web task execution with strategic exploration.
\newblock \emph{Proceedings of the AAAI Conference on Artificial Intelligence}, 2025{\natexlab{b}}.
\newblock \doi{10.1609/aaai.v39i22.34505}.
\newblock URL \url{https://doi.org/10.1609/aaai.v39i22.34505}.

\bibitem[Zhang et~al.(2026{\natexlab{f}})Zhang, Zhao, Zhang, Zhang, Deng, Liu, Xiao, and Zhang]{zhang2026clawsremembertellstealthy}
Yechao Zhang, Shiqian Zhao, Jiawen Zhang, Jie Zhang, Gelei Deng, Xiaogeng Liu, Chaowei Xiao, and Tianwei Zhang.
\newblock When claws remember but do not tell: Stealthy memory injection in persistent personal agents, 2026{\natexlab{f}}.
\newblock URL \url{https://arxiv.org/abs/2607.05189}.

\bibitem[Zhang et~al.(2025{\natexlab{c}})Zhang, Dai, Bo, Ma, Li, Chen, Zhu, Dong, and Wen]{zhang_2025_a_survey_on_the_memory_mecha}
Zeyu Zhang, Quanyu Dai, Xiaohe Bo, Chen Ma, Rui Li, Xu~Chen, Jieming Zhu, Zhenhua Dong, and Ji-Rong Wen.
\newblock A survey on the memory mechanism of large language model-based agents.
\newblock \emph{ACM Transactions on Information Systems}, 43\penalty0 (6):\penalty0 1--47, 2025{\natexlab{c}}.
\newblock \doi{10.1145/3748302}.
\newblock URL \url{https://doi.org/10.1145/3748302}.

\bibitem[Zhang and Zhang(2026)]{zhang2026agentiauthorizedearned}
Zhaoxi Zhang and Xiaomei Zhang.
\newblock Are you still the agent i authorized? earned authority under a fixed ceiling for evolving agents, 2026.
\newblock URL \url{https://arxiv.org/abs/2607.23586}.

\bibitem[Zhang et~al.(2024{\natexlab{d}})Zhang, Cui, Lu, Zhou, Yang, Wang, and Huang]{zhang_2024_agent_safetybench_evaluating}
Zhexin Zhang, Shiyao Cui, Yida Lu, Jingzhuo Zhou, Junxiao Yang, Hongning Wang, and Minlie Huang.
\newblock Agent-safetybench: Evaluating the safety of llm agents, 2024{\natexlab{d}}.
\newblock URL \url{https://arxiv.org/abs/2412.14470}.

\bibitem[Zhao et~al.(2026{\natexlab{a}})Zhao, Bhaskar, and Dobriban]{zhao2026livepi}
Lei Zhao, Abhay Bhaskar, and Edgar Dobriban.
\newblock {LivePI}: More realistic benchmarking of agents against indirect prompt injection, 2026{\natexlab{a}}.
\newblock URL \url{https://arxiv.org/abs/2605.17986}.

\bibitem[Zhao et~al.(2026{\natexlab{b}})Zhao, Wang, Zhang, Deng, Zhao, Che, Qin, and Liu]{zhao2026faithful_self_evolvers}
Weixiang Zhao, Yingshuo Wang, Yichen Zhang, Yang Deng, Yanyan Zhao, Wanxiang Che, Bing Qin, and Ting Liu.
\newblock Large language model agents are not always faithful self-evolvers, 2026{\natexlab{b}}.
\newblock URL \url{https://arxiv.org/abs/2601.22436}.

\bibitem[Zhao et~al.(2026{\natexlab{c}})Zhao, Zhang, Wang, Deng, Zhao, Zhi, Huang, He, Che, Qin, and Liu]{zhao2026experience_driven_safety}
Weixiang Zhao, Yichen Zhang, Yingshuo Wang, Yang Deng, Yanyan Zhao, Xuda Zhi, Yongbo Huang, Hao He, Wanxiang Che, Bing Qin, and Ting Liu.
\newblock On safety risks in experience-driven self-evolving agents.
\newblock In \emph{Findings of the Association for Computational Linguistics: ACL 2026}, pages 42145--42169. Association for Computational Linguistics, 2026{\natexlab{c}}.
\newblock \doi{10.18653/v1/2026.findings-acl.2091}.
\newblock URL \url{https://aclanthology.org/2026.findings-acl.2091/}.

\bibitem[Zhao et~al.(2026{\natexlab{d}})Zhao, Yuan, Huang, Yuan, Yu, Xu, Hu, Shankarampeta, Huang, Ni, Tian, and Zhao]{zhao2026amabench}
Yujie Zhao, Boqin Yuan, Junbo Huang, Haocheng Yuan, Zhongming Yu, Haozhou Xu, Lanxiang Hu, Abhilash Shankarampeta, Zimeng Huang, Wentao Ni, Yuandong Tian, and Jishen Zhao.
\newblock {AMA}-bench: Evaluating long-horizon memory for agentic applications, 2026{\natexlab{d}}.
\newblock URL \url{https://arxiv.org/abs/2602.22769}.

\bibitem[Zhou et~al.(2025)Zhou, Ni, Lee, and Zhao]{zhou2025backdoor_llm_survey}
Yihe Zhou, Tao Ni, Wei-Bin Lee, and Qingchuan Zhao.
\newblock A survey on backdoor threats in large language models ({LLMs}): Attacks, defenses, and evaluations.
\newblock \emph{Transactions on Artificial Intelligence}, page~3, 2025.
\newblock \doi{10.53941/tai.2025.100003}.
\newblock URL \url{https://www.sciltp.com/journals/tai/articles/2505000595}.

\bibitem[Zhou et~al.(2026)Zhou, Zhang, Cai, Aloqaily, Bouachir, Pang, Mehrotra, Wang, and Wen]{zhou2026mcpshield}
Zhenhong Zhou, Yuanhe Zhang, Hongwei Cai, Moayad Aloqaily, Ouns Bouachir, Linsey Pang, Prakhar Mehrotra, Kun Wang, and Qingsong Wen.
\newblock {MCPShield}: A security cognition layer for adaptive trust calibration in model context protocol agents, 2026.
\newblock URL \url{https://arxiv.org/abs/2602.14281}.

\bibitem[Zhu et~al.(2025)Zhu, Du, Hong, Yang, Guo, Wang, Wang, Qian, Tang, Ji, and You]{zhu_2025_multiagentbench_evaluating_t}
Kunlun Zhu, Hongyi Du, Zhaochen Hong, Xiaocheng Yang, Shu~Qing Guo, Zhe Wang, Zhihua Wang, Cheng Qian, Robert Tang, Heng Ji, and Jiaxuan You.
\newblock Multiagentbench : Evaluating the collaboration and competition of {LLM} agents, 2025.
\newblock URL \url{https://doi.org/10.18653/v1/2025.acl-long.421}.

\bibitem[Zhu et~al.(2026)Zhu, Kellermann, Gupta, Li, Fang, Bindu, and Kang]{yuxuan_2024_teams_of_llm_agents_can_expl}
Yuxuan Zhu, Antony Kellermann, Akul Gupta, Philip Li, Richard Fang, Rohan Bindu, and Daniel Kang.
\newblock Teams of {LLM} agents can exploit zero-day vulnerabilities.
\newblock In \emph{Proceedings of the 19th Conference of the European Chapter of the Association for Computational Linguistics (Volume 1: Long Papers)}, pages 23--35. Association for Computational Linguistics, 2026.
\newblock \doi{10.18653/v1/2026.eacl-long.2}.
\newblock URL \url{https://aclanthology.org/2026.eacl-long.2/}.

\bibitem[Zhuang et~al.(2025)Zhuang, Cao, Yang, Xu, Xu, Wang, and Liu]{zhuang_2025_llm_agents_can_be_choice_sup}
Nan Zhuang, Boyu Cao, Yi~Yang, Jing Xu, Mingda Xu, Yuxiao Wang, and Qi~Liu.
\newblock {LLM} agents can be choice-supportive biased evaluators: an empirical study.
\newblock In \emph{Proceedings of the AAAI Conference on Artificial Intelligence}, volume~39, pages 26436--26444, 2025.
\newblock \doi{10.1609/aaai.v39i25.34843}.
\newblock URL \url{https://doi.org/10.1609/aaai.v39i25.34843}.

\end{thebibliography}

\end{document}